%% file: main.tex
\documentclass[sigconf, nonacm]{acmart}

\newcommand\vldbdoi{XX.XX/XXX.XX}
\newcommand\vldbpages{XXX-XXX}
\newcommand\vldbvolume{17}
\newcommand\vldbissue{11}
\newcommand\vldbyear{2024}
\newcommand\vldbauthors{\authors}
\newcommand\vldbtitle{\shorttitle} 
\newcommand\vldbavailabilityurl{https://github.com/udao-moo/udao-spark-optimizer}
\newcommand\vldbpagestyle{empty} 

\UseRawInputEncoding
\usepackage{graphicx}
\usepackage{subfigure,wrapfig}
\usepackage{caption} 
\usepackage{balance}
\usepackage{color}
\usepackage{tabto}
\usepackage{amsmath}
\usepackage{mathrsfs}
\usepackage{mathtools}
\usepackage{makecell}
\usepackage{multirow}
\usepackage{enumitem}
\usepackage{booktabs}
\newcommand{\ra}[1]{\renewcommand{\arraystretch}{#1}}
\usepackage{algorithmic}
\usepackage[ruled,vlined]{algorithm2e}
\usepackage{bm}
\usepackage{soul}
\usepackage{amsthm} 
\usepackage{tabularx}
\usepackage{pifont}
\usepackage{fancyvrb}

\mathtoolsset{showonlyrefs}
\newcommand{\minip}[1]{\vspace{0.05in} \noindent \textbf{#1}}
\newcommand{\cmark}{\checkmark}%
\newcommand{\xmark}{$\times$}%
\newcommand{\cut}[1]{}
\newcommand{\todo}[1]{{\color{blue}[\textit{#1}]}}

\newcommand{\red}[1]{{\color{red}#1}}
\newcommand{\highlight}[1]{{\color{red}#1}}
\newcommand{\yanlei}[1]{{\color{purple}#1}}

\newcommand{\qi}[1]{{\color{olive}\textit{#1}}}

\newcommand{\sketch}[1]{{\color{cyan}#1}}

\newcommand{\rv}[1]{{\color{black}#1}}
\newcommand{\crv}[1]{{\color{black}#1}}

\newcommand{\po}{Pareto-optimal }
\newcommand{\bs}{\boldsymbol}
\newcommand{\R}{\mathbb{R}}
\newcommand{\LQP}{\mathcal{F}}
\newcommand{\QS}{\mathcal{G}}
\newcommand{\thetabm}{\bm{\theta}}
\newcommand{\Thetabm}{\bm{\Theta}}
\newcommand{\Thetabmp}{\{\bm{\theta_p}\}}
\newcommand{\Thetabms}{\{\bm{\theta_s}\}}
\newcommand{\collap}{$\overline{\mbox{LQP}}$}

\newcommand{\ct}{\texttt{HMOOC3}}
\newcommand{\rt}{\texttt{HMOOC3+}}
\newcommand{\sofw}{\texttt{SO-FW}}
\newcommand{\mows}{\texttt{MO-WS}}

\newtheorem{claim}{Claim}
\newcommand{\restr}[2]{{\ensuremath{\left.#1\right|_{#2}}}}
\DeclarePairedDelimiter\ceil{\lceil}{\rceil}

\newcommand{\code}[1]{\texttt{#1}}

\newboolean{show-tr}
\setboolean{show-tr}{true}
\newcommand{\techreport}[1]{\ifthenelse{\boolean{show-tr}}{{#1}}{{\cite{tech-report}}}}
\newcommand{\techreportrv}[2]{\ifthenelse{\boolean{show-tr}}{{#1}}{{#2} in the techreport~\cite{tech-report}}}
\newcommand{\techreportmain}[2]{\ifthenelse{\boolean{show-tr}}{{#1}}{{#2}}}
\DeclareMathOperator*{\argmin}{arg\,min}

\theoremstyle{definition}
\newtheorem{definition}{Definition}[section] 
\newtheorem{proposition}{Proposition}[section]
\newtheorem{lemma}{Lemma}
\newtheorem{theorem}{Theorem}[section]

\newcommand{\skipzerolist}
    {\begin{list}{\hfil}
        {\topsep 0pt plus 1pt
           \parsep 1pt plus 1pt
           \partopsep 0pt plus 1pt
           \itemsep 0pt plus1pt
           \itemindent -0.25in }
    }

\newcommand{\skipzeroitemize}
    {
      \begin{list}{{$\bullet$}} 
        {
          \topsep 0pt plus 0pt
           \parsep 0pt plus 1pt
           \partopsep 0pt plus 1pt
           \itemsep 0pt plus 1pt}
    }

\begin{document}
\title{A Spark Optimizer for Adaptive, Fine-Grained Parameter Tuning}

\author{Chenghao Lyu}
\affiliation{%
 \institution{University of Massachusetts, Amherst}
}
\email{chenghao@cs.umass.edu}

\author{Qi Fan, Philippe Guyard}
\affiliation{%
 \institution{Ecole Polytechnique}
}
\email{qi.fan@polytechnique.edu}
\email{philippe.guyard@polytechnique.edu}

\author{Yanlei Diao}
\affiliation{%
 \institution{Ecole Polytechnique}
}
\affiliation{%
 \institution{University of Massachusetts, Amherst}
}
\email{yanlei.diao@polytechnique.edu}

\begin{abstract}
 As Spark becomes a common big data analytics platform, its growing complexity makes automatic tuning of numerous parameters critical for performance. Our work on Spark parameter tuning is particularly motivated by two recent trends: Spark's {\em Adaptive Query Execution} (AQE) based on runtime statistics, and the increasingly popular {\em Spark cloud deployments} that make cost-performance reasoning crucial for the end user.
This paper presents our design of {\em a Spark optimizer that controls all tunable parameters 
of each query in the new AQE architecture to explore its performance benefits and, at the same time, casts the tuning problem in the theoretically sound multi-objective optimization (MOO) setting to better adapt to user cost-performance preferences}.   
 To this end, we propose a novel hybrid compile-time/runtime approach to multi-granularity tuning of diverse, correlated Spark parameters, as well as a suite of modeling and optimization techniques to solve the tuning problem in the MOO setting while meeting the stringent time constraint of 1-2 seconds for cloud use. 
Evaluation results using TPC-H and TPC-DS benchmarks demonstrate the superior performance of our approach: ($i$) When prioritizing latency, it achieves \rv{63\% and 65\%} reduction for TPC-H and TPC-DS, respectively, under an average solving time of \rv{0.7-0.8} sec, outperforming the most competitive MOO method that reduces only 18-25\% latency with 2.6-15 sec solving time. ($ii$) When shifting preferences between latency and cost, our approach dominates the solutions of alternative methods, exhibiting superior adaptability to varying preferences.

\end{abstract}







\maketitle

\pagestyle{\vldbpagestyle}
\begingroup\small\noindent\raggedright\textbf{PVLDB Reference Format:}\\
\vldbauthors. 
\vldbtitle. PVLDB, \vldbvolume(\vldbissue): \vldbpages, \vldbyear.\\
\href{https://doi.org/\vldbdoi}{doi:\vldbdoi}
\endgroup
\begingroup
\renewcommand\thefootnote{}\footnote{\noindent
This work is licensed under the Creative Commons BY-NC-ND 4.0 International License. Visit \url{https://creativecommons.org/licenses/by-nc-nd/4.0/} to view a copy of this license. For any use beyond those covered by this license, obtain permission by emailing \href{mailto:info@vldb.org}{info@vldb.org}. Copyright is held by the owner/author(s). Publication rights licensed to the VLDB Endowment. \\
\raggedright Proceedings of the VLDB Endowment, Vol. \vldbvolume, No. \vldbissue\ %
ISSN 2150-8097. \\
\href{https://doi.org/\vldbdoi}{doi:\vldbdoi} \\
}\addtocounter{footnote}{-1}\endgroup

\ifdefempty{\vldbavailabilityurl}{}{
\vspace{.3cm}
\begingroup\small\noindent\raggedright\textbf{PVLDB Artifact Availability:}\\
The source code, data, and/or other artifacts have been made available at \url{\vldbavailabilityurl}.
\endgroup
}

\input{intro}


\input{related_work}

\input{problem_statement}

\input{modeling}

\input{compile-time-opt}
\input{runtime-opt}

\cut{ 
\input{checkpoint-231115/problem_statement}
\input{checkpoint-231115/modeling}

\input{checkpoint-231115/optimization}
\input{checkpoint-231115/Qi_initial_optimizer/initial_optimization.tex}

}

\input{experiments}
\input{conclusion}

\begin{acks}
This work was partially supported by the European Research Council (ERC) Horizon 2020 research and innovation programme (grant
n725561) and China Scholarship Council (CSC). 
We also thank Julien Fontanarava for engineering efforts and Guillaume Lachaud for the discussion and help.
\end{acks}

\clearpage

\balance
\bibliographystyle{ACM-Reference-Format}
\bibliography{refs/bigdata.bib,refs/db.bib,refs/model+opt.bib,refs/optimization.bib}

\techreportmain{
\clearpage
\appendix
\input{appendix/index}
}{}

\end{document}

%% file: intro.tex
\section{Introduction}


Big data query processing has become an integral part of enterprise businesses and many platforms have been developed for this purpose~\cite{BorkarCGOV11,DeanG04,hadoop,fuxi-vldb14,Flink-recovery2015,GatesNCKNORSS09,maxcompute,Murray+13:naiad,scope-vldb12,spark-rdd-nsdi12,ThusooSJSCALWM09,trident-vldb21,Xin+2013:shark}. 
As these systems are becoming increasingly complex, parameter tuning of big data systems has recently attracted a lot of research attention~\cite{KunjirB20,Locat_2022,LiZLG19,vldb23/Li,lite-icde22,spark-moo-icde21}. Take Apache Spark for example. It offers over 180 parameters for governing a {\em mixed set of decisions}, including resource allocation, the degree of parallelism, shuffling behaviors, and SQL-related decisions. 
Our work on parameter tuning of big data query systems is particularly motivated by two recent trends:

{\em Adaptive Query Execution.} Big data query processing systems have undergone architectural changes that distinguish them substantially from traditional DBMSs for the task of parameter tuning. 
A notable feature is that a SQL query is compiled into a physical plan composed of query stages and a query stage is the granularity of scheduling and execution. The stage-based query execution model enables the system to observe the precise statistics of the completed stages at runtime. 
\cut{The recent work~\cite{LyuFSSD22} has explored this opportunity to optimize the resource allocation of each query stage, but is limited to two (CPU and memory) resource parameters of each parallel instance of a stage. }
Recently, Spark has taken a step further to introduce Adaptive Query Execution (AQE), which upon the completion of each query stage, considers runtime statistics and re-optimizes the logical query plan to a new physical plan using parametric rules.  
Spark, however, does not support parameter tuning itself and instead, executes AQE based on the default or pre-specified configuration of the parameters. Hence, it can suffer from suboptimal performance of AQE when the parameters are not set to appropriate values. 
On the other hand, recent work on Spark tuning~\cite{KunjirB20,Locat_2022,LiZLG19,vldb23/Li,lite-icde22,spark-moo-icde21} has limited itself to the traditional setting that the parameters are set at query submission time and then fixed throughout query execution, hence missing the opportunity of exploring AQE to improve the physical query plan. 


{\em Cost-performance reasoning in cloud deployment.}
As big data query processing is increasingly deployed in the cloud,  parameter tuning in the form of cost-performance optimization~\cite{LyuFSSD22,Nathan0024} has become more critical than ever to end users.  
Prior work~\cite{LiZLG19,Zhang:2019:EAC,zhu2017bestconfig} has used fixed weights to combine multiple objectives into a {\em single objective} (SO) and solve the SO problem to return one solution. However, the optimization community has established theory~\cite{marler2004survey} pointing out that solving such a SO problem is unlikely to return a solution that balances the cost-performance as the specified weights intend to express (as we will demonstrate in this work). \rv{The theoretically sound approach to adapting between cost and performance} is to treat it as a {\em multi-objective optimization} (MOO) problem~\cite{marler2004survey,messac2003nc,messac2012from,Emmerich:2018:TMO}, compute the Pareto optimal set, and return one solution from the set that best matches the user preference as reflected by the weights set on the objectives~\cite{spark-moo-icde21,LyuFSSD22}. 


Therefore, our work in this paper aims to {\em design a Spark optimizer that controls all tunable parameters (collectively called a ``configuration'') of each Spark application in the new architecture of adaptive query execution to explore its performance benefits and, at the same time, casts the tuning problem in the multi-objective optimization setting to better adapt to user cost-performance needs}.  
This Optimizer for Parameter Tuning (OPT) complements Spark's cost and rule-based optimization of query plans, where the optimization rules use default or pre-specified values of Spark parameters. 
Our OPT can be implemented as a plugin in the current Spark optimizer framework and runs each time a query is submitted  for execution.  

Designing the optimizer for parameter tuning, as defined above, faces a few salient challenges: 


{\bf Complex control of a mixed parameter space}. One may wonder whether parameter tuning can be conducted solely at runtime, as an augmented AQE process. Unfortunately, Spark parameter tuning is more complex than that due to the need to control a mixed parameter space. More specifically, Spark parameters can be divided into three categories (see Table~\ref{tab:spark-params} for examples): the context parameters, $\thetabm_c$, initialize the Spark context by specifying (shared) resources to be allocated and later controlling various runtime behaviors;  the query plan parameters, $\thetabm_p$, govern the translation from the logical to physical query plan; and the query stage parameters, $\thetabm_s$, govern the optimization of the query stages in the physical plan. The $\thetabm_p$ and $\thetabm_s$ parameters are best tuned at runtime to benefit from precise statistics, but they are strongly correlated with the context parameters, $\thetabm_c$, which control shared resources and must be set at query submission time to initialize the Spark context. 
\cut{For example, Figure~\ref{fig:tpch-9-corr-s5-k1} illustrates that the optimal choice of \code{sql.shuffle.partitions} in $\thetabm_p$ is strongly correlated with the total number of cores configured in $\thetabm_c$. }
How to best tune these mixed parameters, correlated but under different controls in the query lifetime, is a nontrivial issue. 

{\bf Stringent MOO Solving time for cloud use}. The second challenge is solving the MOO problem over a large parameter space of Spark while obeying stringent time constraints for cloud use, e.g., \rv{under 1-2 seconds to avoid delays in starting a Spark application in serverless computing~\cite{LyuFSSD22,SharmaCST16,AgarwalJP15}}. 
Prior work on MOO for Spark tuning~\cite{spark-moo-icde21} has reported the running time of the Evolutional method~\cite{Emmerich:2018:TMO} to be about 5 seconds for query-level control of 12 Spark parameters. In our work, when we allow the $\thetabm_p$ parameters to be tuned separately for different subqueries, the time cost of Evolutional increases beyond 60 seconds for some TPC-H queries, which is unacceptable for cloud use.     

To address the above challenges, we propose a novel approach to multi-granularity tuning of mixed Spark parameters and a suite of modeling and optimization techniques to solve the tuning problem in the MOO setting efficiently and effectively. More specifically, our contributions include the following: 

{\bf 1. A hybrid approach to OPT} (Section~\ref{sec:overview}): Our OPT is designed for multi-granularity tuning of mixed parameters: while the context parameters $\thetabm_c$ configure the Spark context at the {\em query level}, we tune the $\thetabm_p$ and $\thetabm_s$ parameters 
	at the fine-grained {\em subquery level} and {\em query stage level}, respectively, 
	to maximize performance gains. To cope with Spark's different control mechanisms for these parameters, we introduce a new hybrid compile-time/runtime optimization approach to multi-granularity tuning:  the compile-time optimization finds the optimal $\thetabm_c^*$, by leveraging the correlation between $\thetabm_c$ and fine-grained $\{\thetabm_p\}$ and $\{\thetabm_s\}$,  to construct an ideal Spark context for query execution. Then the runtime optimization adjusts fine-grained $\{\thetabm_p\}$ and $\{\thetabm_s\}$ based on the precise statistics of the completed stages. Both compile-time and runtime optimization are cast in the setting of multi-objective optimization. 

{\bf 2. Modeling} (Section~\ref{sec:modeling}): Solving the MOO problem for parameter tuning requires precise models for the objective functions used. Our hybrid approach requires accurate models for both compile-time and runtime optimization, which use different representations of query plans. The Spark execution environment shares resources among parallel stages, which further complicates the modeling problem. To address all of these issues, we introduce a modeling framework that combines a Graph Transformer Network (GTN) embedder of query plans and a regression model that captures the interplay of the tunable parameters (decision variables) and critical contextual factors (non-decision variables) such as query and data characteristics and resource contention. We devise new techniques to construct compile-time and runtime models in this framework.


{\bf 3. MOO Algorithms} (Section~\ref{sec:compile-time}): Solving the MOO problem for multi-granularity tuning needs to conquer the high-dimensionality of the parameter space while obeying the time constraint, which is especially the case at compile-time when we consider the correlation of all the parameters together. We introduce a novel approach for compile-time optimization, named Hierarchical MOO with Constraints (HMOOC): it breaks the optimization problem of a large parameter space into a set of smaller problems, one for each subquery, but subject to the constraint that all subquery-level problems use the same Spark context parameters, $\thetabm_c$.  
Since these subproblems are not independent, we devise a host of techniques to prepare a sufficiently large set of candidate solutions for the subproblems and efficiently aggregate them to build global Pareto optimal solutions. Then our runtime optimization runs as part of AQE to adapt $\thetabm_p$ and $\thetabm_s$ effectively based on precise statistics.   

Evaluation results using TPC-H and TPC-DS benchmarks demonstrate the superior performance of our techniques. 
(1)~Our compile-time MOO algorithm for fine-grained parameter tuning outperforms existing MOO methods with  \rv{4.7\%-54.1\%} improvement in hypervolume (the dominated space by the Pareto front) and \rv{81\%-98.3\%} reduction in solving time. 
(2)~Our compile-time/runtime optimization, when asked to prioritize latency,  achieves \rv{ 63\% and 65\%} reduction for TPC-H and TPC-DS, respectively, under an average solving time of \rv{0.7-0.8} sec, whereas the most competitive MOO method reduces only  18-25\% latency with high solving time of 2.6-15 sec. When shifting preferences between latency and cost, our approach dominates the solutions of alternative methods by a wide margin, exhibiting superior adaptability to varying preferences.

%% file: related_work.tex
\section{Related Work}



\minip{DBMS tuning.} Our problem is related to a body of work on performance tuning for DBMSs. Most DBMS tuning systems employ an \textit{offline}, iterative tuning session for each workload~\cite{VanAken:2017:ADM,Zhang:2019:EAC,ResTune-sigmod21,WangTB21}, which can take long to run (e.g., 15-45 minutes~\cite{VanAken:2017:ADM,Zhang:2019:EAC}).\cut{ Such a long tuning session does not suit the needs of a cloud optimizer that makes an \textit{online} optimization decision (i.e., recommending a configuration) for each arriving dataflow, with a subsecond delay. 
In the setting of offline tuning, }
Ottertune~\cite{VanAken:2017:ADM}  builds a predictive model for each query by leveraging similarities to past queries,  and runs Gaussian Process (GP) exploration to try other configurations to reduce query latency. 
ResTune~\cite{ResTune-sigmod21} accelerates the GP tuning process (with cubic complexity in the number of training samples) by building a meta-learning model to weigh appropriately the base learners trained for  individual tuning tasks. 
CDBTune~\cite{Zhang:2019:EAC} and QTune~\cite{LiZLG19} use Deep Reinforcement Learning (RL) to predict the reward of a configuration, 
as the weighted sum of different objectives,
and explores new configurations to optimize the reward. These methods can take many iterations to achieve good performance~\cite{Locat_2022}. 
UDO~\cite{WangTB21} is an offline RL-based tuner for both database physical design choices and parameter settings.
OnlineTuner~\cite{ZhangW0T0022} tunes workloads in the online setting by exploring a contextual GP to adapt to changing contexts and safe exploration strategies. 
Our work on parameter tuning aims to be part of the Spark optimizer, invoked on-demand for each arriving query, hence different from all the tuning systems that require launching a separate tuning session for each target workload.

\minip{Tuning of big data systems.} 
Among {\em search-based} methods, BestConfig~\cite{zhu2017bestconfig} searches for good configurations by dividing high-dimensional configuration space into subspaces based on samples, but it cold-starts each tuning request. ClassyTune~\cite{zhu2019classytune} solves the optimization problem by classification, 
\rv{and Li et al.~\cite{LiMRSY19} prunes searching space with a running environment independent cost model, both of } 
which cannot be easily extended to the MOO setting.
A new line of work considered Spark parameter tuning for  recurring workloads.  
ReIM~\cite{KunjirB20} tunes memory management decisions online by guiding the GP approach using manually-derived memory models.   
Locat~\cite{Locat_2022} is a data-aware GP approach for tuning Spark queries that repeatedly run with the input data size changing over time. 
While it outperforms prior solutions such as Tuneful~\cite{Tuneful-2020}, ReIM~\cite{KunjirB20}, and QTune~\cite{LiZLG19} in efficiency, it still needs hours to complete. 
Li et al.~\cite{vldb23/Li} further tune periodic Spark jobs using a GP with safe regions and meta-learning from history. 
LITE\cite{lite-icde22} tunes parameters of non-SQL Spark applications  
and relies on stage code analysis to derive predictive models, which is impractical as cloud providers usually have no access to application code under privacy constraints.  
These solutions do not suit our problem as we cannot afford to launch a separate tuning session for each query or target workload, and these methods lack support of adaptive runtime optimization and are limited to single-objective optimization. 



\minip{Resource optimization in big data systems.}
In cluster computing, a resource optimizer (RO) determines the optimal resource configuration {\em on demand} and {\em with low latency} as jobs are submitted. 
\cut{RO for parallel databases~\cite{LiNN14} determines the best data partitioning strategy across different machines to minimize a single objective, latency. Its time complexity of solving the placement problem is quadratic to the number of machines.} 
Morpheus~\cite{JyothiCMNTYMGKK16} 
codifies user expectations as multiple Service-Level Objectives  (SLOs) and enforces them using scheduling methods. 
\cut{Its online packing algorithm minimizes the maximal total allocation with a log-competitive bound.}
However, its optimization focuses on system utilization and predictability, but not cost and latency of Spark queries. 
PerfOrator~\cite{RajanKCK16} 
optimizes latency via an exhaustive search of the solution space while calling its model for predicting the performance of each solution. 
WiseDB~\cite{MarcusP16} manages cloud resources based on a decision tree trained on 
minimum-cost schedules of sample workloads. 
ReLocag\cite{tpds21/ReLocag} presents a predictor to 
find the near-optimal number of CPU cores to minimize job completion time.
Recent work~\cite{LeisK21} proposes a heuristic-based model to recommend a cloud instance that achieves cost optimality for OLAP queries. 
This line of work addresses a smaller set of tunable parameters (only for resource allocation) than the general problem of Spark tuning with large parameter space, and is limited to single-objective optimization.


\minip{Multi-objective optimization} (MOO) computes a set of solutions that are not dominated by any other configuration in all objectives, aka, the Pareto-optimal set (or Pareto front). 
Theoretical MOO solutions  suffer from various performance issues in cloud optimization:
Weighted Sum~\cite{marler2004survey} is known to have \textit{poor coverage} of the Pareto front~\cite{messac2012from}. 
Normalized Constraints~\cite{messac2003nc} lacks in  \textit{efficiency} due to repeated recomputation to return more solutions. 
Evolutionary methods~\cite{Emmerich:2018:TMO} 
approximately compute a Pareto set but suffer from  \textit{inconsistent solutions}.
Multi-objective Bayesian Optimization~\cite{Hernandez-Lobato16,abs-2006-05078} extends the Bayesian approach to modeling an unknown function with an acquisition function for choosing the next point(s) that are likely to be Pareto optimal. But it is shown to take long to run~\cite{spark-moo-icde21} and hence lacks the \textit{efficiency} required by a cloud optimizer.

In the DB literature, MOO for SQL queries~\cite{Hulgeri:2002:PQO,Kllapi:2011:SOD,Trummer:2014,Trummer:2014:ASM,TrummerK15} finds Pareto-optimal query plans by efficiently searching through a large set of plans. The problem, essentially a combinatorial one, differs from MOO for parameter tuning, which is a numerical optimization problem. 
TEMPO~\cite{tan2016tempo} considers multiple SLOs of SQL queries and guarantees max-min fairness when they cannot be all met.
MOO for workflow scheduling~\cite{Kllapi:2011:SOD} assigns operators  to containers to minimize total running time and money cost, but is limited to searching through 20 possible containers and solving a constrained optimization for each option. 

The closest work to ours is UDAO~\cite{spark-moo-icde21,udao-vldb19} that tunes Spark configurations to optimize for multiple objectives. 
It \emph{Progressive Frontier} (PF) method~\cite{spark-moo-icde21} provides the MOO solution for spark parameter tuning with good coverage, efficiency, and consistency. However, the solution is limited to 
coarse-grained query-level control of parameters.  
Lyu et al. extended the MOO solution to serverless computing~\cite{LyuFSSD22} by controlling machine placement and resource allocation to parallel tasks of each query stage. However, its solution only guarantees Pareto optimality for each individual stage, but not the entire query (with potentially many stages).

\cut{ 
None the solutions consider complex structures of big data queries and require modeling end-to-end query latency, which does not permit instance-level optimization.
To do so, our work employs fine-grained modeling and a hierarchical MOO solution. 
However, it works only on the granularity of an entire query and neglects its internal structure. 
Such coarse-grained modeling of latency is unlikely to be accurate for complex big data queries, which leads to poor results of optimization. 
But its MOO works for entire queries, but not fine-grained MOO that suits the complex structure of big data systems. 
}

\cut{
\minip{\todo{Learning and optimization for Spark}} Table~\ref{tab:related-work} lists the tuning properties covered by the state-of-the-arts (SOTAs) for Spark parameter tuning. \todo{More descriptions needed}

\minip{\todo{Learning and optimization in big data analytics}}
(1) granularity of control
Workload-level (Ottertune, CDBTune), query-level (NEO), stage-level (CLEO), instance-level (ali-work), etc.
(2) the resource sharing mechanism
}

%% file: problem_statement.tex
\section{Problem Statement and Overview}
\label{sec:overview}

In this section, \rv{we provide background on Spark including its adaptive query execution extension and present initial results illustrating the benefits and complexity of fine-grained tuning. We then formally define our Spark parameter tuning problem and provide an overview of our compile-time/runtime optimization approach.} 

\cut{
\begin{figure*}[t]
\centering
\captionsetup{justification=centering}
\begin{minipage}{.55\textwidth}
  \raggedright
  \includegraphics[width=.99\linewidth,height=4cm]{figs/sys-overview.pdf}
  \captionof{figure}{\small{System Design}}
  \label{fig:system-overview}
\end{minipage}
\begin{minipage}{.42\textwidth}
  \raggedleft
  \includegraphics[width=.99\linewidth,height=4cm]{figs/sys-aqe.pdf}
  \captionof{figure}{\small{An AQE run with IO}}
  \label{fig:system-aqe}
\end{minipage}
\end{figure*}
}

\begin{figure}[t]
	\setlength{\belowcaptionskip}{0pt}  
	\setlength{\abovecaptionskip}{0pt}
	\subfigure[\small{Mixed control in query lifetime}]
	{\label{fig:param-1}\includegraphics[width=.28\linewidth]{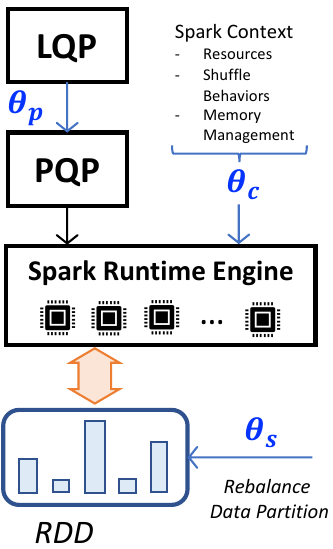}}
	\hfill
	\subfigure[\small{Logical query plan (LQP) and physical query plan (PQP) of TPCH-Q3}]
	{\label{fig:param-2}\includegraphics[width=.66\linewidth]{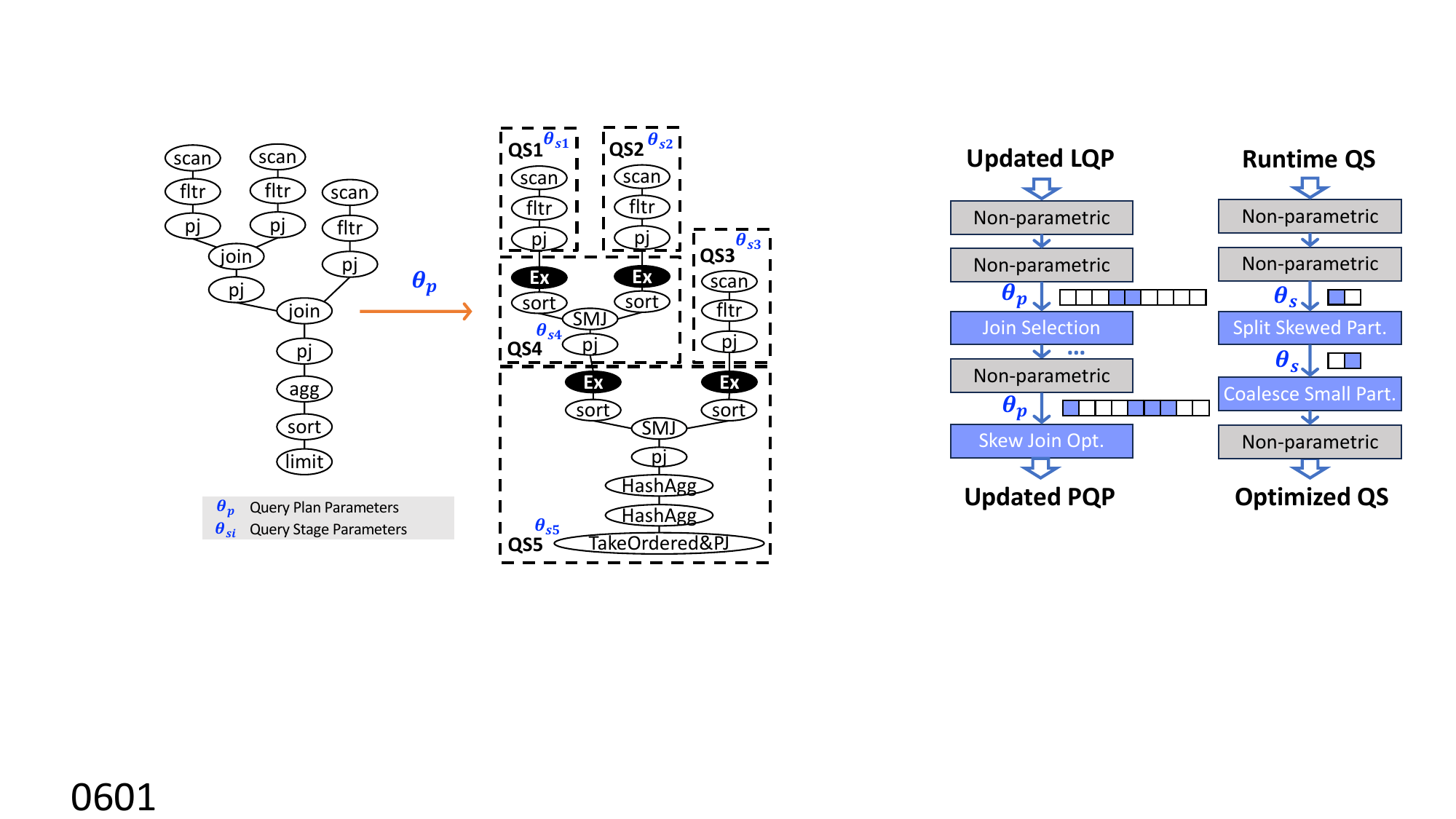}}
\captionof{figure}{\small{Spark parameters provide mixed control through query compilation and execution}}
\label{fig:spark-parameters}
\vspace{-0.15in}
\end{figure}





\subsection{Background on Spark}
\label{subsec:spark}

Apache Spark~\cite{spark-rdd-nsdi12} is an open-source distributed computing system for large-scale data processing and analytics. 
The core concepts of Spark include {\it jobs}, representing computations initiated by actions, and {\it stages}, which are organized based on shuffle dependencies, serving as boundaries that partition the computation graph of a job. 
Stages comprise sets of {\it tasks} executed in parallel, each processing a specific {\it data partition}. {\it Executors}, acting as worker processes, execute these tasks on individual cluster nodes. 

Spark SQL seamlessly integrates relational processing into the Spark framework~\cite{spark-sql}. 
A submitted SQL query undergoes parsing, analysis, and optimization to form a {\it logical query plan} (LQP). In subsequent physical planning, Spark transforms the LQP to one or more {\it physical query plans} (PQP), using physical operators provided by the Spark execution engine. Then it selects one PQP using a cost model, which mostly applies to join algorithms. The physical planner also performs rule-based optimizations, such as pipelining projections or filters into one {\verb|map|} operation. 
The PQP is then divided into a directed acyclic graph (DAG) of {\it query stages} (QSs) based on data exchange dependencies such as shuffling or broadcasting. 
These query stages are then executed in a topological order. 

The execution of a Spark SQL query is configured by three categories of parameters, as shown in Table~\ref{tab:spark-params}, providing different controls in query lifetime. 
As Figure~\ref{fig:param-1} shows, {\bf query plan parameters} $\thetabm_p$ guide the translation from a logical query plan to a physical query plan, influencing the decisions such as the bucket size for file reading  and the join algorithms to use via Spark's parametric optimization rules. 
Figure~\ref{fig:param-2} shows a concrete example of translating a LQP to PQP, where each logical operator is instantiated by specific algorithms (e.g., the first join is  implemented by sorting both input relations and then a merge join of them), additional exchange operators  are injected to realize data exchanges, and query stages are identified at the boundaries of exchange operators. 
Further, {\bf query stage parameters} $\thetabm_s$ control the optimization of a query stage via parametric rules, such as rebalancing data partitions. 
Finally, {\bf context parameters} $\thetabm_c$, specified on the Spark context, control shared resources, shuffle behaviors, and memory management throughout query execution. While they are in effect only at runtime, $\thetabm_c$ must be specified at the query submission time when the Spark context is initialized. 



\begin{table}[t]
	\centering
	\captionsetup{justification=centering}
	\setlength{\belowcaptionskip}{0pt}  
	\setlength{\abovecaptionskip}{0pt}
	\captionof{table}{\crv{Example Spark parameters  in three categories}}
	\label{tab:spark-params}
	\footnotesize
	\ra{0.75}
	\begin{tabular}{@{}ll@{}}
	  \midrule
	  $\thetabm_c$ &{\bf Context Parameters} (\verb|spark.*|)  \\
	  \midrule
	  $k_1$ & \verb|executor.cores| \\
	  $k_2$ & \verb|executor.memory| \\
	  $k_3$ & \verb|executor.instances| \\	  
	  \midrule
	  $\thetabm_p$ &{\bf Logical Query Plan Parameters} (\verb|spark.sql.*|) \\
	  \midrule
	  $s_1$ & \verb|adaptive.advisoryPartitionSizeInBytes| \\
	  $s_3$ & \verb|adaptive.maxShuffledHashJoinLocalMapThreshold| \\
	  $s_4$ & \verb|adaptive.autoBroadcastJoinThreshold| \\	  
	  $s_5$ & \verb|shuffle.partitions| \\
	  \midrule
	  $\thetabm_s$ &{\bf Query Stage Parameters} (\verb|spark.sql.adaptive.*|)  \\
	  \midrule
	  $s_{10}$ & \verb|rebalancePartitionsSmallPartitionFactor| \\
	  \midrule
	\end{tabular}	
	\vspace{-0.22in} 
\end{table}

\begin{figure}[t]
	\centering
	\includegraphics[width=.95\linewidth,height=3.5cm]{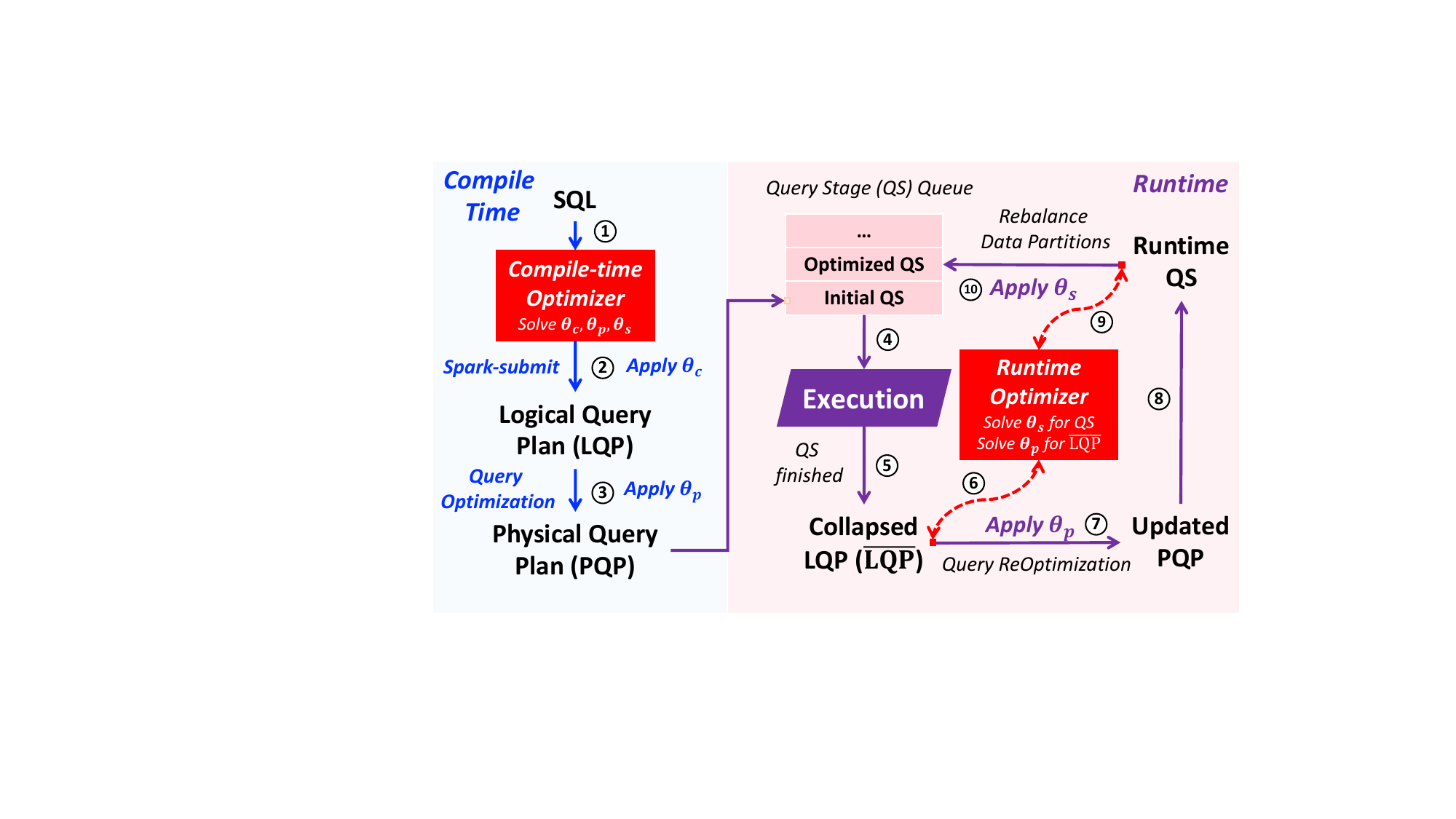}
	\captionof{figure}{\small{Query life cycle with an optimizer for parameter tuning}}
	\label{fig:spark-aqe}
	\vspace{-0.2in}	
\end{figure}

\textbf{Adaptive Query Execution (AQE).}
Cardinality estimation~\cite{NegiMKMTKA21-vldb21,face-vldb21,Sun0021-sigmod21,bayescard,FLAT-vldb21,QiuWYLWZ21-sigmod21,LiuD0Z21-vldb21,LuKKC21-vldb21,WuC21-sigmod21,HasanTAK020-sigmod21,WoltmannOHHL21-vldb21} has been a long-standing issue that impacts the effectiveness of the physical query plan. To address this issue, 
Spark introduced {\it Adaptive Query Execution } (AQE) that enables runtime optimization based on precise statistics collected from completed stages~\cite{spark-aqe}. 
Figure~\ref{fig:spark-aqe} shows the life cycle of a SQL query with the AQE mechanism turned on. 
At compile time, a query is transformed to a LQP and then a PQP through query optimization (step 3). Query stages (QSs) that have their dependencies cleared are then submitted for execution. 
During query runtime, Spark iteratively updates LQP by collapsing completed QSs into dummy operators with observed cardinalities, leading to a so-called collapsed query plan \collap\ (step 5), and re-optimizes the \collap\ (step 7) and the QSs (step 10), until all QSs are completed.
At the core of AQE are runtime optimization rules. Each rule internally traverses the query operators and takes effect on them. 
These rules are categorized as parametric and non-parametric, and 
each parametric rule is configured by a subset of $\thetabm_p$ or $\thetabm_s$ parameters.
The details of those rules are left to \techreport{Appendix~\ref{appendix:optimization-rules}}.

\cut{
\minip{\todo{Default Spark Tuning}} 
(1) dynamic resource allocation (coarse-grained, heuristic-based, still need parameter tuning to decide the range of resources can be used, etc.)
(2) the stage-level tuning (low-level API) does not work with the declared nature of SQL language.
}

%

\begin{figure*}[!htb]
	\setlength{\belowcaptionskip}{0pt}  
	\setlength{\abovecaptionskip}{0pt} 	
	\captionsetup{justification=centering}		
   \begin{minipage}{0.68\textwidth}
     \centering
     \begin{tabular}{lll}
		\subfigure[\small{Latency comparison}]
		{\label{fig:tpch-9-lat}\includegraphics[height=2.5cm,width=.28\textwidth]{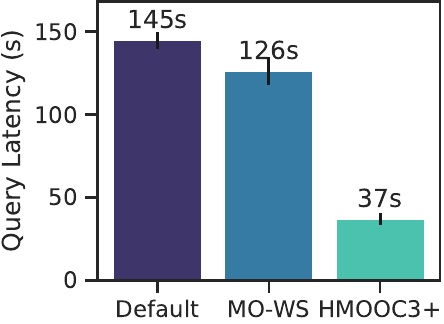}}
		&\hfill
		\subfigure[\small{Physical query plan choices}]
		{\label{fig:tpch-9-pqp}\includegraphics[height=2.5cm,width=.33\textwidth]{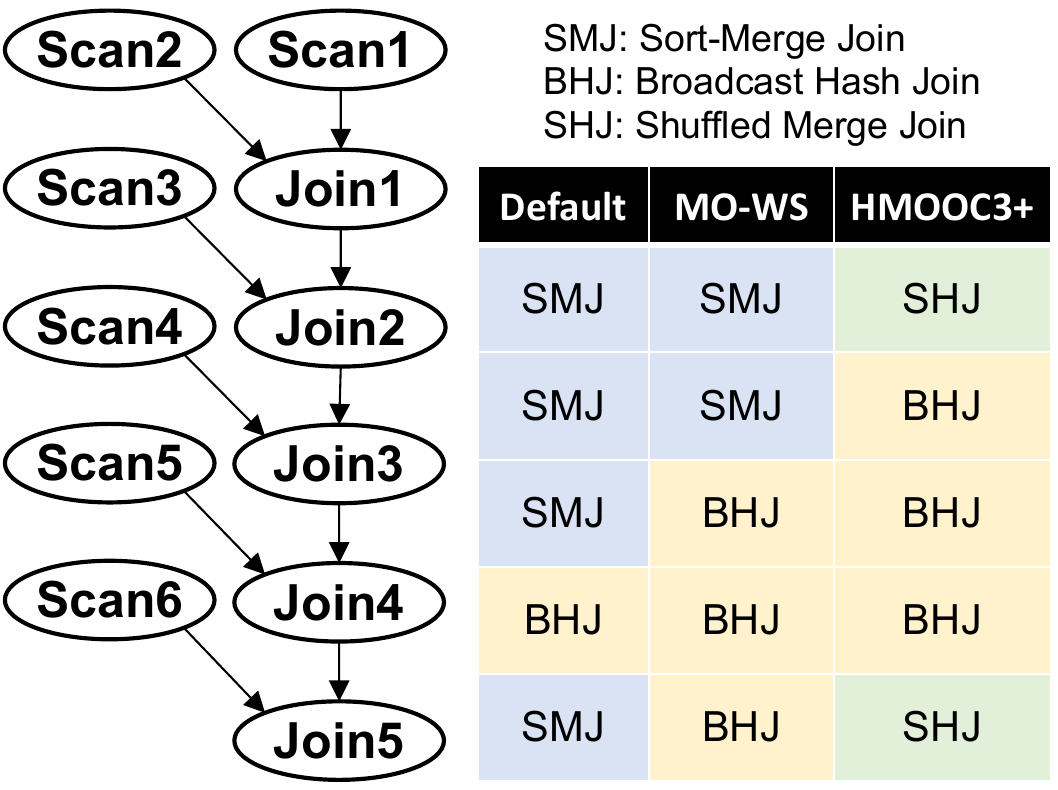}}
		&\hfill
		\subfigure[\small{Correlation in parameters}]
		{\label{fig:tpch-9-corr-s5-k1}\includegraphics[height=2.5cm,width=.28\textwidth]{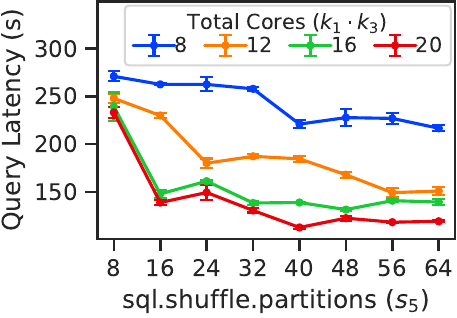}}				
	\end{tabular}
	\caption{\small{Profiling TPCH-Q9 (12 subQs) over different configurations}}
	\label{fig:profiling-tpch-q9}
   \end{minipage}\hfill
   \begin{minipage}{0.3\textwidth}
     \centering
     \includegraphics[height=3cm,width=.9\textwidth]{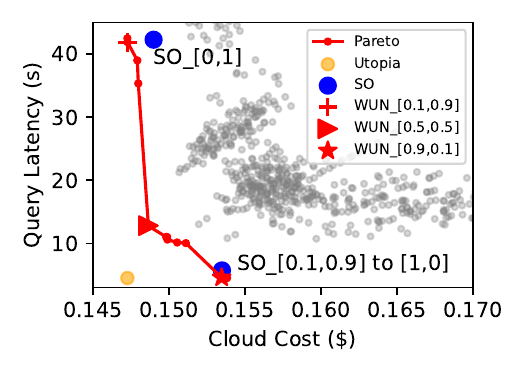} 
      \caption{\small{MOO solutions for TPCH Q2}} 
       \label{fig:eg_moo}
   \end{minipage}
   \vspace{-0.15in}
\end{figure*}

\subsection{Effects of Parameter Tuning}
\label{subsec:tuning-overview}

We next consider the issue of Spark parameter tuning and present initial observations that motivated our approach. 

First, {\em parameter tuning affects performance}. While Spark supports AQE through parametric and non-parametric rules, it does not support parameter tuning itself. The first observation that motivated our work is that tuning over a mixed parameter space is crucial for Spark performance. 
Figure~\ref{fig:tpch-9-lat} shows that for TPCH-Q9, \rv{query-level parameter tuning using a prior MOO method~\cite{spark-moo-icde21} and then running AQE (the middle bar) can already provide a 13\% improvement over AQE with the default configuration (left bar).}

Second, {\em fine-grained tuning has performance benefits over  query-level tuning.} 
\rv{While existing work on Spark parameter tuning~\cite{KunjirB20,Locat_2022,LiZLG19,vldb23/Li,lite-icde22,spark-moo-icde21} focuses on query-level tuning, we show in Figure~\ref{fig:tpch-9-lat}  that adapting $\thetabm_p$ for different collapsed query plans during runtime can  further reduce the latency by 61\% (the right bar).} 
Figure~\ref{fig:tpch-9-pqp} shows the simplified query structure of TPCH-Q9, including 6 scan operators and 5 join operators. \rv{Adapting $\thetabm_p$ for different collapsed query plans with observed statistics allows us to discover a new physical query plan with 3 broadcast hash joins (BHJs) and 2 shuffled hash joins (SHJs), outperforming the query-level tuning result with  2 sort-merge joins (SMJ) + 3 BHJs.} \cut{Specifically, \mows\ broadcasts up to 4.5GB data in Join5 because it finalized its parameter tuning at compile time with underestimated cardinality of Join4, while our fine-grained tuning can derive a better plan by adapting $\thetabm_p$ to runtime statistics.}



Third, {\em the parameters that are best tuned at runtime based on precise statistics are correlated with the parameters that must be set at submission time.} 
While the $\thetabm_p$ and $\thetabm_s$ parameters are best tuned at runtime to benefit from precise statistics, they are strongly correlated with the Spark context parameters, $\thetabm_c$, which control shared resources and must be set at query submission time when the Spark context is initialized.  
Figure~\ref{fig:tpch-9-corr-s5-k1} illustrates that the optimal choice of $s_5$ in $\thetabm_p$ is strongly correlated with the total number of cores $k_1 * k_3$ configured in $\thetabm_c$. Many similar examples exist.

\subsection{Our Parameter Tuning Approach}
\label{subsec:prob-def}

\rv{We next introduce our approach that supports multi-granularity parameter tuning using hybrid compile-time/runtime optimization and formally define the optimization problem in the MOO setting. } 

\subsubsection{Hybrid, Multi-Granularity Tuning}
\rv{The goal of our work is to find, for each Spark query, the optimal configuration of all the $\thetabm_p$,  $\thetabm_s$, and $\thetabm_c$ parameters under {\em multi-granularity tuning}.  
While the context parameters $\thetabm_c$ configure the Spark context at the {\em query level}, we tune other parameters at fine granularity to maximize performance gains, including setting the query plan parameters $\thetabm_p$ distinctly for each {\em collapsed query plan} and the query stage parameters $\thetabm_s$ for each {\em query stage} in the physical plan.}  

To address the correlation between the context parameters $\thetabm_c$ (set at query submission time) and $\thetabm_p$ and $\thetabm_s$ parameters (best tuned at runtime), we introduce a  {\it hybrid compile-time }/ {\it runtime optimization} approach, as depicted by the two red boxes in Figure~\ref{fig:spark-aqe}.

\rv{\underline{Compile-time}: Our goal is to (approximately) 
derive the optimal $\thetabm_c^*$, by leveraging the correlation of all the parameters, to construct an ideal Spark context for query execution. 
Our compile-time optimization uses cardinality estimates by Spark's cost optimizer.}

\rv{\underline{Runtime}: With Spark context fixed, our runtime optimization runs as a plugin of AQE, invoked each time the collapsed query plan ($\overline{\mbox{LQP}}$) is updated from a completed query stage, and adjusts $\thetabm_p$ for the collapsed query plan based on the latest runtime statistics. Then AQE applies $\thetabm_p$ to its parametric rules to generate an updated physical query plan ($\overline{\mbox{PQP}}$). For the query stages in this new physical plan, our runtime optimization kicks in to optimize $\thetabm_s$ parameters based on precise statistics. Then AQE applies parametric rules with the tuned $\thetabm_s$ to optimize data partitions of these stages. }

\subsubsection{Multi-Objective Optimization}
\rv{Targeting cloud use, our optimization problem concerns multiple user objectives such as query latency and cloud cost in terms of CPU hours or a weighted combination of CPU, memory, and IO resources.} 

\rv{Prior work~\cite{LiZLG19,Zhang:2019:EAC,zhu2017bestconfig} used fixed weights to combine multiple objectives into a {\bf single objective} (SO) and solved it to return one solution, denoted as the \sofw\ method. It is a special case of a classical MOO algorithm, \emph{weighted sum} (WS)~\cite{marler2004survey}, that repeatedly applies $n$ weight vectors to create a set of SO problems and returns a solution for each, denoted as the \mows\ method (i.e., $n=1$ in \sofw). It is known from the theory of WS that trying different weights to create SO problems is unlikely to return points that evenly cover the Pareto front unless the objective functions have a very peculiar  shape~\cite{marler2004survey}. Empirically, \mows\ has been reported with sparse coverage of the Pareto front in prior work~\cite{spark-moo-icde21}. 
Figure~\ref{fig:eg_moo} illustrates this for TPCH-Q2 in the 2D space of query latency and cloud cost: 11 SO problems generated from evenly spaced weight vectors return only two distinct solutions (marked by the blue dots), where 10 of them collide to the same bottom point. Increasing to 101 weight vectors still returns only 3 distinct points. 
Such sparse coverage of the Pareto front leads to poor adaptability when the user shifts preference (e.g., from favoring latency to favoring  cost) because there are not enough points on the Pareto front to capture the tradeoffs between the objectives. } 

\rv{For this reason, our work casts the optimization problem in the {\em multi-objective optimization} (MOO) framework~\cite{marler2004survey,messac2003nc,messac2012from,Emmerich:2018:TMO}, which computes the Pareto front properly to capture the tradeoffs and later allows us to recommend one solution that best matches the user preference. }



Formally,  a MOO problem aims to minimize multiple objectives simultaneously, where the objectives are represented as functions  $\bm{f} = (f_1, ..., f_k)$ on all the tunable parameters $\thetabm$.

\begin{definition}\textbf{Multi-Objective Optimization (MOO).}
\begin{eqnarray}
\label{eq:moo_def}
  \arg \min_{\thetabm} & \bm{f}(\thetabm)= & {\left[ f_1(\thetabm),  f_2(\thetabm), \ldots, f_k(\thetabm) \right]} \\
 \nonumber s.t. & &  {\begin{array}{l}
          \thetabm \in \Sigma \subseteq \mathbb{R}^d  \\ 
          \bm{f}(\thetabm) \in \Phi \subseteq \mathbb{R}^k \\ 
          L_i \leq f_i(\thetabm) \leq U_i, \,\,\, i=1,...,k\\
        \end{array}}
\end{eqnarray}
\end{definition}
\noindent{where $\thetabm$ is the configuration with $d$ parameters, 
$\Sigma$ $\subseteq \mathbb{R}^d$ denotes all possible configurations, and 
$\Phi \subseteq \mathbb{R}^k$ denotes the objective space. 
If an objective favors larger values,  we add the minus sign to the objective function to transform it into a minimization problem. 
}
\cut{In general, the MOO problem leads to a set of solutions rather than a single optimal solution. }

\begin{definition}
\label{eq:pareto_opt}
\textbf{Pareto Optimal Set.} In the objective space $\Phi \subseteq \mathbb{R}^k$, a point \bm{$F'$} Pareto-dominates another point \bm{$F''$} iff $\forall i \in [1, k], F_i' \leq F_i''$  and $\exists j \in [1, k], F_j' < F_j''$.
For a given query, solving the MOO problem leads to a Pareto Set (Front) $\mathcal{F}$ that includes all the Pareto optimal solutions $\{(\bm{F}, \thetabm)\}$, where $\bm{F}$ is a Pareto point in the objective space $\Phi$ and $\thetabm$ is its corresponding configuration in $\Sigma$. \cut{We refer to  a Pareto  optimal point  as a {\em reference point}, $F^0_i$, if it achieves the minimum for the objective $f_i$.}
\end{definition}

\rv{Figure~\ref{fig:eg_moo} shows a Pareto front for TPCH-Q2.} 
Most configurations, depicted by the grey dots, are dominated by the Pareto optimal configurations, depicted by the red dots, in both objectives. Hence, the MOO solution allows us to skip the vast set of dominated configurations. 
\rv{Furthermore, the Pareto points themselves represent tradeoffs between the two competing objectives. The optimizer can recommend one of them based on the user preference,} e.g., favoring latency to cost in peak hours with weights 0.9 to 0.1 and vice-versa in off-peak hours. The recommendation can be made based on the Weighted Utopia Nearest (WUN) distance~\cite{spark-moo-icde21} of the Pareto points from the Utopia point $\bm{U}$, which is the hypothetical optimum in all objectives, marked by the orange dot in the figure. \rv{Figure~\ref{fig:eg_moo} shows a few recommendations when we run WUN  on the Pareto front with different weighted preferences on the objectives.} 

We next define the MOO problem for Spark parameter tuning. 

\begin{definition}{\bf Multi-Objective Optimization for Spark SQL}\label{eq:moo-spark}
\begin{align}
	\argmin_{\thetabm_c, \Thetabmp, \Thetabms} & \bm{f}(\thetabm_c, \Thetabmp, \Thetabms) =  {\left[
		\begin{array}{l}
		f_1(\text{LQP}, \thetabm_c, \Thetabmp, \Thetabms, \alpha, \beta, \gamma) \\		... \\
		f_k(\text{LQP}, \thetabm_c, \Thetabmp, \Thetabms, \alpha, \beta, \gamma) \\		\end{array}
		\right]}\\
	\nonumber s.t. & {\begin{array}{l}
	    \thetabm_c \in \Sigma_c, \\
	    \Thetabmp = \{\thetabm_{p1}, \thetabm_{p2}, ..., \thetabm_{pt}, ... \}, \forall \thetabm_{pt} \in \Sigma_p \\
	    \Thetabms = \{\thetabm_{s1}, \thetabm_{s2}, ..., \thetabm_{si}, ...\}, \forall \thetabm_{si} \in \Sigma_s
		\end{array}}
\end{align}
where LQP denotes the logical query plan with operator cardinality estimates, and  
$\thetabm_c, \Thetabmp, \Thetabms$ represent the {\it decision variables} configuring Spark context, LQP transformations, and query stage (QS) optimizations, respectively.
More specifically, $\Thetabmp$ is the collection of all LQP parameters, and $\thetabm_{pt}$ is a copy of $\thetabm_p$ for the $t$-th transformation of the collapsed query plan $\overline{\text{LQP}}$. 
Similarly, $\Thetabms$ is the collection of QS parameters, and $\thetabm_{si}$ is a copy  for optimizing query stage $i$. 
$\Sigma_c, \Sigma_p, \Sigma_s$ are the feasible space for $\thetabm_c$, $\thetabm_p$ and $\thetabm_{s}$, respectively. 
Finally, $\alpha, \beta, \gamma$ are the {\it non-decision variables} (not tunable, but crucial factors that affect model performance), representing the input characteristics, the distribution of partition sizes for data exchange, and resource contention status during runtime.
\end{definition}


\begin{table}[t] 
\ra{0.8}
\footnotesize
\setlength{\belowcaptionskip}{0pt}  
\setlength{\abovecaptionskip}{0pt}
\addtolength{\tabcolsep}{-2pt}
\newrobustcmd{\B}{\bfseries}
\caption{Comparison of Spark parameter tuning methods}
\label{tab:related-work}
\begin{tabular}{lcccc}
\toprule
\multicolumn{1}{c}{} & 
\multicolumn{1}{c}{\begin{tabular}[c]{@{}c@{}}Mixed Param. \\ Space\end{tabular}} & 
\multicolumn{1}{c}{\begin{tabular}[c]{@{}c@{}}Adaptive\\Runtime Opt. \end{tabular}} &
\multicolumn{1}{c}{\begin{tabular}[c]{@{}c@{}}Multi-\\Granularity\end{tabular}} &
\multicolumn{1}{c}{\begin{tabular}[c]{@{}c@{}}Multi-\\Objective\end{tabular}}\\
\midrule  
ReLocag~\cite{tpds21/ReLocag}& \xmark  & \xmark  & \xmark  &  \xmark  \\ \midrule
BestConfig~\cite{zhu2017bestconfig}&   \cmark  & \xmark  & \xmark  &  \xmark  \\
ClassyTune~\cite{zhu2019classytune}&   \cmark  & \xmark  & \xmark  &  \xmark  \\ \midrule
LITE~\cite{lite-icde22}& \cmark  &  \xmark  & \xmark  &  \xmark  \\
LOCAT~\cite{Locat_2022}&   \cmark  & \xmark  & \xmark  &  \xmark  \\
Li et. al~\cite{vldb23/Li}& \cmark  & \xmark  & \xmark  &  \xmark  \\
UDAO~\cite{spark-moo-icde21} & \cmark  & \xmark  &  \xmark &   \cmark \\
\midrule 
\B Ours & \B \cmark & \B \cmark  &\B \cmark  &\B \cmark \\
\bottomrule
\end{tabular}
\vspace{-0.1in}
\end{table}

\subsubsection{Comparison to Existing Approaches} 
We finally summarize our work in relation to existing solutions to Spark parameter tuning in terms of the coverage of the mixed parameter space, adaptive runtime optimization, multi-granularity tuning, and multi-objective optimization (MOO), as shown in Table~\ref{tab:related-work}.
More specifically, ReLocag~\cite{tpds21/ReLocag} 
focuses on individual parameters such as the number of cores but does not cover the broad set of Spark parameters. 
Search-based solutions to parameter tuning~\cite{zhu2017bestconfig,zhu2019classytune} and recent Spark tuning systems~\cite{vldb23/Li,lite-icde22,spark-moo-icde21,Locat_2022} cover mixed parameter space but do not support adaptive runtime optimization or multi-granularity tuning. 
The only system that supports MOO is UDAO~\cite{spark-moo-icde21} but its parameter space is much smaller due to query-level coarse-grained tuning. 
To the best of our knowledge, our work is the first comprehensive solution to Spark parameter tuning, covering the mixed parameter space with multi-granularity tuning by leveraging the Spark AQE mechanism, and best exploring the tradeoffs between objectives in the multi-objective optimization approach.

\cut{
\subsection{Overview of Spark SQL}

Spark SQL seamlessly integrates relational data processing into the Spark framework ~\cite{spark-sql} and recently reinforced itself with the {\it Adaptive Query Execution (AQE)} mechanism that can re-optimize the query plans during runtime based on real-time statistics~\cite{spark-aqe}.

Figure~\ref{fig:spark-aqe} shows the lifecycle of a SQL in Spark. During query compilation, a SQL is extracted as a {\it logical query plan (LQP)}, and subsequently transformed into a {\it physical query plan (PQP)} via the physical query optimization. 
A physical query plan is further divided into a DAG of {\it query stages (QSs)} based on the data exchange dependency (shuffle or broadcast). Each query stage is the unit for scheduling, and the dependency-cleared query stages will first be exported for execution.

During query runtime, Spark SQL iteratively optimizes the remaining query until all stages are done. Specifically, when a query stage finishes, it updates the logical query plan with real-time statistics, constructs a new physical query plan via a set of query optimizations, and exports running query stages from the updated physical query plan. Each runtime query stage is further optimized independently before execution.

\todo{hide AQE}

add parameters to the overview
2. {Motivation of the ...}
3. Runtime optimization.

\cut{
\minip{Spark Context}

\minip{Optimization Rules}

\minip{Cardinality}

\minip{Data Distribution}

\minip{Resource Contention}
}

\subsection{Configuration Parameters}

We divided the Spark configuration parameters into three categories, the context parameters $\thetabm_c$, the logical query plan parameters $\thetabm_p$, and the query stage parameters $\thetabm_s$. Table~\ref{tab:spark-params} shows the selected parameters via the knob selection~\cite{KanellisAV20} and our heuristics.

\subsubsection{Spark Context and $\thetabm_c$} 
Spark context is the entry point of a Spark application configured by the context parameters $\thetabm_c$, which is non-tunable during the SQL execution. 
Specifically, it decides the resource profile (the memory per execution $k_1$, the cores per executor $k_2$, and the number of executors $k_3$), the shuffle behavior (the default degree of the parallelism $k_4$, the memory overhead per reducer task $k_5$, when to avoid merge-sorting data $k_6$, and whether to compress data before shuffling $k_7$), the memory maintenance strategy (the ratio to use memory for storage and execution $k_8$), etc. 

\subsubsection{Optimization Rules and SQL parameters}
At the core of query optimization is a set of optimization rules to optimize a logical query plan or a query stage. Each optimization rule internally traverses the query operators and takes effect on them.
These rules are categorized into parametric and non-parametric based on their configurability.
Each parametric rule is configured by several SQL parameters, which are tunable during runtime. 

Figure~\ref{fig:opt-rules} demonstrates rules applied to a logical query plan and a query stage during runtime.
In the transition to a physical query plan (or an optimized query stage), a logical query plan (or a runtime query stage) passes through a joint list of parametric rules (blue) and non-parametric rules (gray). 
Consequently, we divide the rules into {\it LQP rules} and {\it QS rules}.

\subsubsection{LQP Rules and $\thetabm_p$}
LQP rules include the logical and physical optimization rules, applied to the logical query plan during both the query compilation and runtime. 
Up to Spark 3.5, this comprises 23 non-parametric rules and 4 parametric rules applied in a deterministic order to the logical query plan.

We denote the union of SQL parameters appearing in all parametric LQP rules as $\thetabm_p$. 
Such parametric rules include the {\it dynamic join selection}, configured by an advisory partition size $s_1$ and a broadcast threshold $s_4$ among others. Additionally, {\it skew join optimization} is configured by another set of parameters, including a target partition size $s_1$, the initial partition numbers $s_5$, and skewed partition thresholds $s_6, s_7$. See \techreport{Appendix~\ref{appendix:lpq-rules}} for all LQP rules.

It is worth mentioning that we can use one copy of $\thetabm_p$ across LQP rules for each logical query plan, either at the compile-time or runtime. 
For example, we use $\thetabm_{p0}$ for LQP rules applied to the initial logical query plan, and $\thetabm_{pt}$ for the $t$-th invocation of LQP rules for an updated logical query plan.
Collectively, we denote $\Thetabm_p$ as the set of all $\thetabm_p$ copies during the lifetime of a Spark SQL.


\subsubsection{QS Rules and $\thetabm_s$}
QS rules optimize a query stage during runtime. In Spark 3.5, there are 3 non-parametric rules and 2 parametric rules. 
One of the parametric rules, "rebalancing skewed partitions," splits skewed partitions into a target size $s_1$ and uses a rebalance small partition factor $s_{10}$ to determine whether the remainder after splitting should be merged.
The other parametric rule, {\it coalescing shuffle partitions}, reduces the number of small reduce tasks by coalescing small partitions to reach a target size $s_1$, which should be larger than a minimum partition size $s_{11}$. See \techreport{Appendix~\ref{appendix:qs-rules}} for all QS rules.

We represent the QS parameters as $\thetabm_s = (s_{10}, s_{11})$ and they are independently applied to each query stage as shown in Figure~\ref{fig:spark-parameters}. 
Collectively, we denote $\Thetabm_s$ as the set of $\thetabm_s$ copies across all query stages.
It is important to note that QS rules are designed under the assumption that LQP rules have been applied. Therefore, the target size $s_1$ is configured by LQP parameters and not included in $\thetabm_s$.

\cut{
\subsubsection{Mixed Decision Problem.} \todo{to be re-constructed}
 It is worth noting that the LQP parameters are applied to modify the PQP, and hence are PQP-dependent. In contrast, other parameters, such as the context parameters $\thetabm_c$ and $s_1$-$s_9$, do not impact the PQP and are PQP-independent.
Therefore, configuring Spark SQL involves making decisions in a {\it mixed decision space}, involving both the PQP topology and the PQP-independent parameters. 

Our work bypasses the mixed-decision problem by operating modeling and tuning over the LQP for the following reasons.
Firstly, the PQP-dependent parameters will not affect the topology of a LQP.
Secondly, when the LQP is provided, the PQP is highly correlated to the PQP-dependent parameters. E.g., we can derive the PQP given the LQP and PQP-dependent parameters with a 99.75\% accurate rate in TPC-H over the 5-fold cross-validation.
Lastly, modeling over the LQP plus the PQP-dependent parameters achieves similar performance as modeling over the PQP, indicating that the simplification could be lossless.
See concrete analyses in Appendix~\ref{appendix:mixed-decision-space}.
}

\subsection{General Problem Definition}

Our general problem is a multi-objective optimization (MOO) problem that aims to minimize multiple objectives $\vec{o} = (o_1, ..., o_k)$ by tuning the Spark configuration $\thetabm = (\thetabm_c, \Thetabm_p, \Thetabm_s)$ for running Spark SQL. Each objective $o_i$ is represented by the function $f_i$.

The optimality of the problem relies on the {\bf oracle} information of the three {\it non-decision variables}: 
(1) the cardinality of each query operator $\alpha$,
(2) the distribution of the partition sizes after an operator requires data exchange $\beta$, and 
(3)	the resource contention status during the runtime $\gamma$.

\begin{definition}{\bf Multi-Objective Optimization for Spark SQL}\label{eq:moo-general}
\begin{align}
	\argmin_{\thetabm_c, \Thetabm_p, \Thetabm_s} & \; f(\thetabm_c, \Thetabm_p, \Thetabm_s) =  {\left[
		\begin{array}{l}
		f_1(LQP, \thetabm_c, \Thetabm_p, \Thetabm_s, \alpha, \beta, \gamma) \\
		... \\
		f_k(LQP, \thetabm_c, \Thetabm_p, \Thetabm_s, \alpha, \beta, \gamma) \\
		\end{array}
		\right]}\\
	\nonumber s.t. & {\begin{array}{l}
	    \thetabm_c \in \Sigma_c, \\
	    \Thetabm_p = [\thetabm_{p0}, \thetabm_{p1}, ...], \forall \thetabm_{pt} \in \Sigma_p \\
	    \Thetabm_s = [\thetabm_{s1}, \thetabm_{s2}, ...], \forall \thetabm_{si} \in \Sigma_s
		\end{array}}
\end{align}
\end{definition}
where $LQP$ denotes the logical query plan; 
$\thetabm_c, \Thetabm_p, \Thetabm_s$ represent the {\it decision variables} configuring Spark context, LQP rules, and QS rules across the lifecycle of an SQL;
$\Thetabm_{p}$ is the collection of all LQP parameters $\thetabm_{pt}$, where $\thetabm_{pt}$ is a unique copy of $\thetabm_p$ for the $t$-th invocation of LQP rules;
$\Thetabm_{s}$ is the collection of all QS parameters $\thetabm_{si}$, where $\thetabm_{si}$ is a unique copy of $\thetabm_s$ for optimizing query stage $i$;
$\Sigma_c, \Sigma_p, \Sigma_s$ are the feasible space for $\thetabm_c$, $\thetabm_p$ and $\thetabm_{s}$, respectively; 

\begin{figure}[t]
  \centering
  \includegraphics[width=.99\linewidth]{figs/initial-and-runtime-opt.pdf}
  \captionof{figure}{\small{Compile-time Optimization and Runtime Optimization}}
  \label{fig:opt-framework}
\end{figure}

\subsubsection{Compile Optimization and Runtime Optimization}
Since none of the non-decision factors are precisely provided before query execution, we introduce the concept of {\it initial optimization} during query compilation (compile-time), and then iteratively perform {\it runtime optimization} using real-time statistics after each query stage concludes.

Figure~\ref{fig:opt-framework} shows the inputs and outputs of both optimizations. 
At compile-time, we estimate the cardinality information by Spark's default cost-based optimizer ($\alpha=\alpha_{cbo}$) with the assumption of uniform data distribution ($\beta = \vec{0}$) and zero resource contention ($\gamma = \vec{0}$). 
The initial optimizer then resolves the physical query plan with an optimal choice of LQP parameters $\thetabm_{p0}^*$, as well as the context parameters $\thetabm_c^*$ and the stage parameters $\Thetabm_s^*$.

At runtime when $\thetabm_c$ is fixed, we construct the non-decision variables by the observed cardinality ($\alpha_t$), data distribution ($\beta_t$), and resource usage ($\gamma_t$) at the $t$-th invocation of the LQP rules for an updated logical query plan.
The runtime optimizer returns an updated physical query plan with an optimal $\thetabm_{pt}^*$ and updated optimal $\thetabm_s$ choices for the remaining query stages $\Thetabm_s^*$.

\subsubsection{Challenges}

Our design introduces two main challenges. 

\minip{Multi-granularity control.}
The three decision variables are correlated to provide the {\it multi-granularity control} to the query performance at both compile-time and runtime.
At compile-time, $\thetabm_c$ and $\thetabm_p$ configure the Spark context and the logical query plan respectively for the entire query, providing {\it query-level control}, while $\thetabm_s$ independently configures each query stage, granting {\it stage-level control}, as shown in Figure~\ref{fig:spark-parameters}.
During runtime when the logical query plan is updated and reduced to a subquery, $\thetabm_p$ provides a {\it subquery-level control} with $\thetabm_s$ still offering the stage-level control. 

\minip{Time Constraints.} 
$\thetabm_p$ should always be set ahead of $\Thetabm_s$. Otherwise, the physical query plan remains undetermined, and so do the set of query stages, their corresponding DAG structure, and QS parameters $\Thetabm_s$.

\cut{

The LQP parameters $\thetabm_p$ configure the LQP rules for both the initial and runtime logical query plans.
Accordingly, they provide {\it query-level control} at the compile-time, and {\it subquery-level control} during runtime once the logical query plan is updated and reduced to a subquery.
The QS parameters $\thetabm_s$ independently configure the QS rules, granting the {\it stage-level control}. 
Those parameters are correlated and should be jointly considered during optimization.

Figure~\ref{fig:spark-parameters} illustrates the multi-granularity control through parameters at compile-time, where we can set $\thetabm_c, \thetabm_p, \Thetabm_s$ by solving the initial optimization problem.
During query runtime, $\thetabm_c$ is fixed, but $\thetabm_p$ and $\Thetabm_s$ are still tunable, providing us with {\it time constraints} to tune $\thetabm_c$. 
Specifically, we have to tune $\thetabm_c$ in a joint space with $\thetabm_p$ and $\Thetabm_s$ during initial optimization and tune $\thetabm_p$ and $\Thetabm_s$ during runtime given a determined $\thetabm_c$.

Moreover, it is necessary to set $\thetabm_p$ before tuning $\Thetabm_s$. 
The physical query plan remains undetermined until $\thetabm_p$ is set, which also affects the number of query stages and their corresponding DAG structure.
For instance, increasing the broadcast threshold in $\thetabm_p$ for the query in Figure~\ref{fig:spark-parameters} could transform one of the sort-merge joins (SMJs) into a broadcast hash join (BHJ), leading to a reduction in the number of query stages as well as the size of $\Thetabm_s$.
}

\cut{
\subsection{Overview of Our Approach}

\subsubsection{Choices of Objectives}
In practice, with a special interest in improving performance with a budgetary constraint, our work consistently chooses latency and cloud cost as the two objectives, where the cloud cost is calculated as a weighted sum of CPU-hour and memory-hour~\cite{EMR-serverless}.

\subsubsection{Initial Optimization}

\input{qi-approach-overview}

\subsubsection{Runtime Optimization}
The runtime optimization takes place once the non-decision variables have been calibrated as $\alpha_t, \beta_t, \gamma_t$ using real-time statistics at the entry of $t$-th adaptive query execution (AQE), with the completion of the query stage $t$. 

Since resources have been determined by $\thetabm_c^*$ at the submission time, the cloud cost is proportional to latency, and the MOO problem can be reduced to a single objective optimization problem.

\begin{definition}{\bf Runtime Optimization}
\begin{align}
	\label{eq:moo-runtime}
	\argmin_{\thetabm_p, \Thetabm_s} & \; L = f(LQP^{(t)}, \thetabm_c^*, \thetabm_p, \Thetabm_s, \alpha_t, \beta_t, \gamma_t) \\
	\nonumber s.t. & {\begin{array}{l}
		\Thetabm_s = [\thetabm_{s(t+1)}, \thetabm_{s(t+2)}, ...]
		\end{array}}
\end{align}
where $LQP^{(t)}$ denotes the updated logical query plan at the entry of AQE $t$ and $\Thetabm_s$ is a collection of $\thetabm_s$ for the remaining query stages.
\end{definition}
}

\subsection{Intuitive Results}

We present results based on TPCH-Q5 to show the benefits of multi-granularity tuning over a joint parameter space ($\thetabm_c, \thetabm_p, \thetabm_s$) with compile-time and runtime optimizations. 

First, tuning Spark parameters over a joint parameter space is crucial. 
Figure~\ref{fig:intuition-A} illustrates the latency and cost of 40 randomly sampled configurations (4 knobs from $\thetabm_c$ and 4 knobs from $\thetabm_p$) when the adaptive query execution (AQE) is turned off. 
7 configurations (highlighted in red) dominate most of those configurations, emphasizing the importance of tuning these parameters.

Second, we benefit from multi-granularity control in initial and runtime optimizations over $\thetabm_p$. 
In Figure~\ref{fig:intuition-B}, bars 1,2,4,5 show the latency of tuning $\thetabm_p$ only at compile-time with the query-level control, while bars 3,6 include the runtime optimization for $\thetabm_p$ with additional subquery-level control by increasing the broadcast threshold $s_4$ after the first query stage completes.
Without runtime optimization, the optimal $\thetabm_p$ choice (bar 4) reduces 
11\% 
latency compared to the default (bar 1). 
Without initial optimization, the $\thetabm_p$ runtime tuning (bar 3) outperforms both individual $\thetabm_p$ choices (bar 1,2) by constructing a new physical query plan with 3 broadcast hash joins (BHJs) and 2 sort-merge joins (SMJs), which is not covered by bar 1 (2 BHJs and 3 SMJs with the default $\thetabm_p$) and bar 2 (4 BHJs and 1 SMJ for the updated $\thetabm_p$).
Moreover, jointly tuning $\thetabm_p$ with multi-granularity control at compile-time and runtime (bar 6) yields a total reduction of 22\% latency compared to the default $\thetabm_p$.
\todo{to be replaced by a better example}

Finally, with the initial and runtime optimizations, better objective performances are achieved with multi-granularity control over a joint parameter space.
Configurations from a joint space of $\thetabm_c$ and $\thetabm_p$ can dominate the best result of only tuning $\thetabm_p$ at compile-time and runtime, as shown in Figure~\ref{fig:intuition-C}.

\cut{
\subsubsection{AQE Optimality Analyses} 
\todo{AQE improves the optimality by calibrating cardinality and re-balancing data distribution during the runtime. However, it is sub-optimal with the current static parameter assignments.} 
}
%

\cut{
Spark SQL~\cite{spark-sql} seamlessly integrates SQL-based relational data processing into the Spark framework, inheriting the declarative nature and familiar syntax of SQL, by running over the Spark context.
Its query optimizer efficiently transforms user-defined SQL queries into optimized executable plans and RDD graphs, customized for the specific data and query.
The process involves three main steps:
(1) generating a {\it logical query plan (LQP)} from SQL, 
(2) optimizing the LQP based on a cost-based optimizer, guided by cardinality estimation, 
(3) and converting the optimized LQP to a {\it physical query plan (PQP)} using optimization rules and heuristics. 
The PQP also represents concrete execution steps, including
the selection of physical operators (e.g., \texttt{HashAgg} or \texttt{SortAgg}),
the decision-making on join strategies (e.g., \texttt{SortMergeJoin} or \texttt{BroadcastHashJoin}),
and the injection of the \texttt{Exchange} operators to divide itself into a directed acyclic graph (DAG) of distinct subquery components executed in different stages. 

The {\it Adaptive Query Execution (AQE)}, introduced in Spark 3.0 and set as the default since Spark 3.2, is a new feature in Spark SQL that dynamically optimizes query execution based on real-time runtime statistics~\cite{spark-aqe}. 
When AQE is enabled, Spark partitions the query execution into smaller units called {\it QueryStages} ({\it QS}s), each representing a distinct portion of a PQP bounded by \texttt{Exchange} operators (broadcast or shuffle).
The workflow begins by executing the leaf QSs, and iteratively triggers AQE runs upon QS materialization until all QSs are processed.
In each AQE run, runtime statistics from the preceding QSs are leveraged to re-optimize the remaining query plans, create QSs with cleared dependency, and apply further optimization rules for each created QS.
With its adaptability and utilization of runtime statistics, AQE significantly enhances query execution performance, making it a popular feature for efficient data processing in Spark SQL applications.

\subsection{Spark Configuration with Mixed Decisions}
Spark provides a variety of tunable configuration parameters, called knobs, enabling users to customize Spark applications. 

\subsubsection{Spark Configuration}
Based on the range of the parameter effects, we divide the Spark configuration $\thetabm$ into two distinct categories: the context configuration $\thetabm_c$ and the runtime SQL configuration $\thetabm_r$.
Our work focuses on 17 selected parameters~\cite{KanellisAV20}, covering both categories, as shown in Table~\ref{tab:spark-params}.

\minip{The context configuration $\thetabm_c$} characterizes the Spark context and is unchangeable throughout the execution of a SQL query in Spark.
It includes (1) the resource parameters $k_1$-$k_3$ to determine the number of executors, the number of cores per executor, and the memory size per executor, (2) the shuffle parameters $k_4$-$k_7$ to affect the shuffle behavior such as the maximum output size of mappers and whether to compress data for shuffling, and (3) the memory management parameter $k_8$ to control the split of memory for execution and storage.

\minip{The runtime SQL configuration $\thetabm_r$} governs the optimization rules applied to different procedures in AQE runs. It is designed to be adjustable within the Spark context, allowing modification during query execution. 
Specifically, AQE's logical optimizer applies parameters $s_1$-$s_4$ in a dynamic join selection rule to compose internal hints for each logical join operator;
after getting an initial PQP, parameters $s_1, s_5$-$s_7$ tweak the threshold for identifying a skewed join in a skewed join optimization rule; 
once QSs are created, parameters $s_1$, $s_8$-$s_9$ are applied on each individual QS to control their partition distributions.
In our work, we use $\thetabm_r^{(0)}$ to denote the SQL configuration at the submission time and use $\thetabm_r^{(t)}$ to denote the SQL configuration at the $t^{th}$ AQE run.

\subsubsection{Mixed Decision Problem}
 It is important to note that some parameters, such as $s_1$-$s_7$, can be applied in the rules to modify the PQP, and hence are PQP-dependent. In contrast, other parameters, such as the context parameters and $s_8$-$s_9$, do not impact the PQP and are PQP-independent.
Therefore, configuring Spark SQL involves making decisions in a {\it mixed decision space}, involving both the PQP topology and the PQP-independent parameters. See details in \ref{appendix:mixed-decision-space}.

Our work bypasses the mixed-decision problem by operating modeling and tuning over the LQP for the following reasons.
Firstly, the PQP-dependent parameters will not affect the topology of a LQP.
Secondly, when the LQP is provided, the PQP is highly correlated to the PQP-dependent parameters. E.g., we can derive the PQP given the LQP and PQP-dependent parameters with a 99.75\% accurate rate in TPCH over the 5-fold cross-validation.
Lastly, modeling over the LQP plus the PQP-dependent parameters achieves similar performance as modeling over the PQP, indicating that the simplification could be lossless.
See concrete analyses in Appendix~\ref{appendix:mixed-decision-space-simplification}.
}

\cut{
\subsection{Extended Spark Overview}


We present our configuration tuning design, which extends Spark with three key processes as shown in Figure~\ref{fig:system-overview}:

{\bf The bootstrap tuning process} involves a LQP extractor and the {\it Initial Optimizer (IO)}. The LQP extractor converts a submitted SQL to the initial LQP and feeds it to the IO. 
The IO then recommends the optimal context configuration $\thetabm_c$ and the initial SQL configuration $\thetabm_r^{(0)}$ based on the LQP by solving a MOO problem.

{\bf The runtime optimization process} is triggered by the event of query stage completions. The {\it Runtime Optimizer (RO)} fine-tunes the runtime SQL configuration $\thetabm_r$ within each AQE run by leveraging the runtime statistics and the updated LQP.

{\bf The model update process} involves a {\it Trace Collector} and a {\it Model Server}. The trace collector asynchronously gathers Spark traces (Step a), including the LQP and QS topologies and their corresponding performance data from AQE runs. The model server builds and updates performance models for both LQPs and QSs (Steps b,c), supporting bootstrap tuning and runtime optimization.

\subsubsection{Life Cycle of a SQL in Our Tuning Design.} 
A SQL initiates the bootstrap tuning process to get the recommended configuration ($\thetabm_c$ and $\thetabm_r^{(0)}$) for Spark submission (Steps 1-3 in Figure~\ref{fig:system-overview}). 
After the Spark submission and the creation of the Spark context, the default query optimizer translates the SQL to the initial executable plan, which is then directed to the AQE framework (Steps 4-5).
The AQE progressively executes the dependency-cleared QSs within the execution plan (Steps 6-7) until all QSs are completed. Each AQE run is triggered by QS completion and, with real-time statistics, involves SQL parameter fine-tuning by the RO and subsequent query re-optimization by Spark (Steps 8-9). 

\subsubsection{An AQE run with RO} \todo{todo}

%

%
%
%
}

\cut{
\subsection{MOO Problem Definition} 

Our problem is designed to optimize multiple query-level objectives by searching a joined space in parameters and physical plans.

\cut{
\begin{definition}{\bf Multi-Objective Optimization for Spark SQL}
\begin{align}
	\label{eq:moo-general}
	\argmin_{\thetabm_c, \thetabm_p, \Thetabm_s} & \; f(\thetabm_c, \thetabm_p, \Thetabm_s) =  {\left[
		\begin{array}{l}
		f_1(LQP, \thetabm_c, \thetabm_p, \Thetabm_s, \mathcal{P}, \delta) \\ 
		... \\
		f_k(LQP, \thetabm_c, \thetabm_p, \Thetabm_s, \mathcal{P}, \delta)
		\end{array}
		\right]}\\
	\nonumber s.t. & {\begin{array}{l}
	    \thetabm_c \in \Sigma_c, \thetabm_p \in \Sigma_p \\
	    \Thetabm_s = [\thetabm_{s1}, \thetabm_{s2}, ...], \thetabm_{si} \in \Sigma_s \\
	    LQP \in \mathcal{LQP}, \; \delta \in \Delta
		\end{array}}
\end{align}
\end{definition}
}

\begin{definition}{\bf Multi-Objective Optimization for Spark SQL}
\begin{align}
	\label{eq:moo-general}
	\argmin_{\thetabm_c, \thetabm_r} & \; f(\thetabm_c, \thetabm_r) =  {\left[
		\begin{array}{l}
		f_1(LQP, \thetabm_c, \thetabm_r, \mathcal{P}, \delta) \\ 
		... \\
		f_k(LQP, \thetabm_c, \thetabm_r, \mathcal{P}, \delta)
		\end{array}
		\right]}\\
	\nonumber s.t. & {\begin{array}{l}
	    \thetabm_c \in \Sigma_c, \thetabm_r \in \Sigma_r \\
	    LQP \in \mathcal{LQP}, \; \delta \in \Delta
		\end{array}}
\end{align}
\end{definition}
where 
\begin{enumerate}
	\item $\thetabm_c$ and $\thetabm_r$ denote the parameters to configure Spark context and the rule-based plans in Spark SQL, respectively.
	\item $F_1,...,F_k$ denote $k$ query-level target objectives.
	\item $LQP$ denotes the optimized logical query plan in the form of a graph given by Spark, where each node in the graph has the information of the operator type and cardinality.
	\item $\mathcal{P}$ denotes the rules in Spark's physical query planner and derives the physical query plan $PQP = \mathcal{P}_{\thetabm_r}(LQP)$. 
	\item $\delta$ denotes other factors that affect the performance, such as the machine system states and the input data information.
	\item $\Sigma_c$, $\Sigma_r$ are the feasible spaces for $\thetabm_c$ and $\thetabm_r$. $\mathcal{LQP}, \Delta$ are feasible spaces for $LQP$ and $\delta$.
\end{enumerate}

\minip{A na\"ive approach given the oracle cardinality.} \todo{(1) give the enum approaches. (2) challenges - card and complexity; (3) the chance to improve the issue; (4) clarify that improving the card est is an orthogonal issue to our work}

\newpage 
\subsubsection{Static Meta Optimization}
Suppose $f$ is the query latency predictive model for the initial LQP before running, and $g$ is the closed-form function to summarize the unit price of the resources. 

\begin{definition}{\bf Static Meta Optimization}
\begin{align}
	\label{eq:moo-static}
	\argmin_{\thetabm_c, \thetabm_q, \Thetabm_s} &  {\left[
		\begin{array}{l}
		L(\thetabm_c, \thetabm_q, \Thetabm_s) = f(LQP^{(0)}, \thetabm_c, \thetabm_q, \Thetabm_s, \mathcal{P}, \alpha_0, \beta_0, \gamma_0) \\ 
		C(\thetabm_c, \thetabm_q, \Thetabm_s) = L(\thetabm_c, \thetabm_q, \Thetabm_s) \cdot g(\thetabm_c)\\
		\end{array}
		\right]}\\
	\nonumber s.t. & {\begin{array}{l}
		\thetabm_c \in \Sigma_c, \thetabm_q \in \Sigma_q \\
		\Thetabm_s = [\thetabm_{s1}, \thetabm_{s2}, ...], \thetabm_{si} \in \Sigma_s \\
	    LQP^{(0)} \in \mathcal{LQP}
		\end{array}}
\end{align}
\end{definition}

\begin{definition}{\bf MOO over LQP (query level)}
\begin{align}
	\label{eq:moo-static}
	\argmin_{\thetabm_c, \thetabm_q, \thetabm_s} &  {\left[
		\begin{array}{l}
		L(\thetabm_c, \thetabm_q, \thetabm_s) = \color{blue}{f_{LQP}(LQP^{(0)}, \thetabm_c, \thetabm_q, \thetabm_s, \delta)} \\ 
		C(\thetabm_c, \thetabm_q, \thetabm_s) = L(\thetabm_c, \thetabm_q, \thetabm_s) \cdot g(\thetabm_c)\\
		\end{array}
		\right]}\\
	\nonumber s.t. & {\begin{array}{l}
		\thetabm_c \in \Sigma_c, \thetabm_q \in \Sigma_q, \thetabm_s \in \Sigma_s \\
	    LQP^{(0)} \in \mathcal{LQP}
		\end{array}}
\end{align}
\end{definition}

\begin{definition}{\bf MOO over PQP (stage level)}
\begin{align}
	\label{eq:moo-static}
	\argmin_{\Thetabm_s} &  {\left[
		\begin{array}{l}
		L(\thetabm_c^*, \Thetabm_s) = \color{red}{f_{PQP}(PQP^{(0)}, \thetabm_c^*, \Thetabm_s, \delta)} \\ 
		C(\thetabm_c^*, \Thetabm_s) = L(\thetabm_c^*,  \Thetabm_s) \cdot g(\thetabm_c^*)\\
		\end{array}
		\right]}\\
	\nonumber s.t. & {\begin{array}{l}
		\Thetabm_s = [\thetabm_s^1, \thetabm_s^2, ...], \thetabm_s \in \Sigma_s \\
	    |\Thetabm_s| = \textbf{num\_of\_qs}(PQP^{(0)})
		\end{array}}
\end{align}
\end{definition}

\begin{definition}{\bf Runtime Optimization}
\begin{align}
	\label{eq:moo-static}
	\argmin_{\thetabm_p, \Thetabm_s} &  {\left[
		\begin{array}{l}
		L = f(LQP^{(t)}, \thetabm_c^*, \thetabm_p, \Thetabm_s, \mathcal{P}, \delta_t) \\ 
		C = L \cdot g(\thetabm_c^*)\\
		\end{array}
		\right]}\\
	\nonumber s.t. & {\begin{array}{l}
		\thetabm_p \in \Sigma_p \\
		\Thetabm_s = [\thetabm_{s1}, \thetabm_{s2}, ...], \thetabm_{si} \in \Sigma_s
		\end{array}}
\end{align}
\end{definition}

where 
\begin{enumerate}
	\item $L$ and $C$ are two examples of objectives at the query level, latency and cloud cost.
	\item $LQP^{(0)}$ is the LQP with the initial cardinality estimations.
\end{enumerate}

\subsubsection{Adaptive Meta Optimization.} 
%
%
%

\begin{definition}{\bf Runtime Optimization}
\begin{align}
	\label{eq:moo-adaptive}
	\argmin_{\thetabm_p, \Thetabm_s} & \; L = f(LQP^{(t)}, \thetabm_c^*, \thetabm_p, \Thetabm_s, \mathcal{P}, \alpha_t, \beta_t, \gamma_t)\\
	\nonumber s.t. & {\begin{array}{l}
		\thetabm_p \in \Sigma_p \\
		\Thetabm_s = [\thetabm_{s1}, \thetabm_{s2}, ...], \thetabm_{si} \in \Sigma_s
		\end{array}}
\end{align}
\end{definition}

\begin{definition}{\bf Runtime Optimization}
\begin{align}
	\label{eq:moo-adaptive}
	\argmin_{\thetabm_p, \thetabm_s} & \; L = f(LQP^{(t)}, \thetabm_c^*, \thetabm_p, \thetabm_s, \mathcal{P}, \alpha_t, \beta_t, \gamma_t)\\
	\nonumber s.t. & {\begin{array}{l}
		\thetabm_p \in \Sigma_p \\
		\thetabm_s = \thetabm_{s1} = \thetabm_{s2} = ... \in \Sigma_s
		\end{array}}
\end{align}
\end{definition}

\begin{definition}{\bf Qurey Stage Optimization}
\begin{align}
	\label{eq:moo-adaptive}
	\argmin_{\thetabm_{si}} & \; L_{QS} = g(QS_{i}, \thetabm_c^*, \thetabm_{si}, \alpha_i, \beta_i, \gamma_i)\\
	\nonumber s.t. & {\begin{array}{l}
		\thetabm_{si} \in \Sigma_s
		\end{array}}
\end{align}
\end{definition}

\begin{definition}{\bf Adaptive Meta Optimization}
\begin{align}
	\label{eq:moo-adaptive}
	\argmin_{\thetabm_p, \thetabm_s} & \; L(\thetabm_p, \thetabm_s) = \color{blue}{f_{LQP}(LQP^{(t)}, \thetabm_c^*, \thetabm_p, \thetabm_s, \delta_t)}\\
	\nonumber s.t. & {\begin{array}{l}
	    LQP^{(t)} \in \mathcal{LQP}
		\end{array}}
\end{align}
\end{definition}

\begin{definition}{\bf Adaptive Meta Optimization}
\begin{align}
	\label{eq:moo-adaptive}
	\argmin_{\Thetabm_s} & \; L(\Thetabm_s) = \color{red}{f_{PQP}(PQP^{(t)}, \thetabm_c^*, \Thetabm_s, \delta_t)}\\
	\nonumber s.t. & {\begin{array}{l}
	    LQP^{(t)} \in \mathcal{LQP}
		\end{array}}
\end{align}
\end{definition}

where 
\begin{enumerate}
	\item $L$ denotes the latency of $LQP^{(t)}$, ranging from the entry of the AQE to the completion of the entire query. 
	\item $\thetabm_c^*$ is from the solutions of static meta optimization.
	\item $S_t$ and $\delta_t$ are the shared-resource execution status for this job and other systematic factors at the AQE entry $t$.
\end{enumerate}


\begin{definition}{\bf Multi-Objective Optimization for Spark SQL}
\begin{align}
	\label{eq:general-moo}
	\argmin_{\thetabm_c, \thetabm_r^{(0)}, \Thetabm_r} & \; F(\thetabm_c, \thetabm_r^{(0)}, \Thetabm_r) =  {\left[
		\begin{array}{l}
		F_1(LQP^{(0)}, \thetabm_c, \thetabm_r^{(0)}, \Thetabm_r, \mathcal{P}, \delta) \\ 
		... \\
		F_k(LQP^{(0)}, \thetabm_c, \thetabm_r^{(0)}, \Thetabm_r, \mathcal{P}, \delta)
		\end{array}
		\right]}\\
	\nonumber s.t. & {\begin{array}{l}
	    \thetabm_c \in \Sigma_c, \thetabm_r^{(0)} \in \Sigma_r \\
	    \Thetabm_r = [ \thetabm_r^{(1)}, \thetabm_r^{(2)}, ...], \; \thetabm_r^{(t)} \in \Sigma_r,\; \\
	    LQP^{(0)} \in \mathcal{LQP}, \; \delta \in \Delta
		\end{array}}
\end{align}
\end{definition}
where 
\begin{enumerate}
	\item $F_1,...,F_k$ denote the $k$ target objectives; 
		  $LQP^{(0)}$ is the initial logical query 
		  $\thetabm_c$ and $\thetabm_r^{(0)}$ denote the context configuration and the initial runtime SQL configuration;
		  $\Thetabm_r$ denotes the set of runtime SQL configurations at each AQE entry.
		  $\mathcal{P}$ denotes the planning rules in Spark's query optimizer;
		  $\delta$ denotes the dynamic systematic factors, including the most recent machine system states and the input data information.
	\item $\Sigma_c$, $\Sigma_r$ are the feasible spaces for $\thetabm_c$ and $\thetabm_r^{(t)} $. $\mathcal{LQP}, \Delta$ are feasible spaces for $LQP^{(0)}$ and $\delta$. 
	\item $t$ indicates the AQE runs during execution.
	\end{enumerate}

\begin{definition}{\bf Static Meta Optimization}
\begin{align}
	\label{eq:static-moo}
	\argmin_{\thetabm_c, \thetabm_r^{(0)}, \Thetabm_r} & \; F(\thetabm_c, \thetabm_r^{(0)}, \Thetabm_r) =  {\left[
		\begin{array}{l}
		\hat{F}_1(LQP^{(0)}, \thetabm_c, \thetabm_r^{(0)}, \Thetabm_r, \mathcal{P}, \delta) \\ 
		... \\
		\hat{F}_k(LQP^{(0)}, \thetabm_c, \thetabm_r^{(0)}, \Thetabm_r, \mathcal{P}, \delta)
		\end{array}
		\right]}\\
	\nonumber s.t. & {\begin{array}{l}
	    \thetabm_c \in \Sigma_c, \thetabm_r^{(0)} \in \Sigma_r \\
	    \Thetabm_r = [ \thetabm_r^{(1)}, \thetabm_r^{(2)}, ...], \; \thetabm_r^{(t)} \in \Sigma_r,\; \\
	    LQP^{(0)} \in \mathcal{LQP}, \; \delta \in \Delta
		\end{array}}
\end{align}
\end{definition}

\begin{definition}{\bf Multi-Objective Optimization for Spark SQL}
\begin{align}
	\label{eq:general-moo}
	\argmin_{\thetabm_c, \Thetabm_r, \Thetabm_s} & \; F(\thetabm_c, \Thetabm_r, \Thetabm_s) =  {\left[
		\begin{array}{l}
		F_1(LQP^{(0)}, \thetabm_c, \Thetabm_r, \Thetabm_s, \delta) \\ 
		... \\
		F_k(LQP^{(0)}, \thetabm_c, \Thetabm_r, \Thetabm_s, \delta)
		\end{array}
		\right]}\\
	\nonumber s.t. & {\begin{array}{l}
	    \thetabm_c \in \Sigma_c \\
	    \Thetabm_r = [ \thetabm_r^{(0)}, \thetabm_r^{(1)}, \thetabm_r^{(2)}, ...], \; \thetabm_r^{(t)} \in \Sigma_r,\; \\
	    \Thetabm_s = [ \thetabm_s^{(0)}, \thetabm_s^{(1)}, \thetabm_s^{(2)}, ...], \; \thetabm_s^{(i)} \in \Sigma_s,\; \\
	    LQP \in \mathcal{LQP}, \; \delta \in \Delta
		\end{array}}
\end{align}
\end{definition}
where 
\begin{enumerate}
	\item $F_1,...,F_k$ denote the $k$ target objectives; $LQP^{(0)}$ is the initial logical query plan; $\thetabm_c$ and $\thetabm_r^{(0)}$ denote the context configuration and the initial runtime SQL configuration; $\delta$ denotes the dynamic systematic factors including the most recent machine system states and the input data information.
	\item $\Sigma_c$, $\Sigma_r$, $\Sigma_s$ are the feasible spaces for $\thetabm_c$, $\thetabm_r^{(t)} $, and $\thetabm_s^{(i)}$. $\mathcal{LQP}, \Delta$ are feasible spaces for $LQP$, $\delta$. 
	\item $t$ and $i$ indicate the number of AQE runs and the id of QSs. 
	\end{enumerate}

\begin{definition}{\bf Multi-Objective Optimization for Spark SQL}
\begin{align}
	\label{eq:general-moo}
	\argmin_{\thetabm_c, \thetabm_r^{(0)}, \pi} & \; F(\thetabm_c, \thetabm_r^{(0)}, \pi) =  {\left[
		\begin{array}{l}
		F_1(LQP^{(0)}, \thetabm_c, \thetabm_r^{(0)}, \pi, \delta) \\ 
		... \\
		F_k(LQP^{(0)}, \thetabm_c, \thetabm_r^{(0)}, \pi, \delta)
		\end{array}
		\right]}\\
	\nonumber s.t. & {\begin{array}{l}
	    \thetabm_c \in \Sigma_c \\
	    \thetabm_r^{(0)} \in \Sigma_r \\
	    \pi:(\mathcal{LQP}, \Sigma_c, \Delta) \rightarrow \Sigma_r
		\end{array}}
	\end{align}
\end{definition}
where 
\begin{enumerate}
	\item $F_1,...,F_k$ denote the $k$ target objectives; $LQP^{(0)}$ is the initial logical query plan; $\thetabm_c$ and $\thetabm_r^{(0)}$ denote the context configuration and the initial runtime SQL configuration; $\delta$ denotes the dynamic systematic factors including the most recent machine system states and the input data information.
	\item $\Sigma_c$ and $\Sigma_r$ are the feasible spaces for $\thetabm_c$ and $\thetabm_r^{t} (t\geq 0)$. $\mathcal{LQP}, \Delta$ are feasible spaces for $LQP$, $\delta$. 
	\item $\pi$ is the strategy to optimize the runtime SQL parameters at each AQE run. Specifically, in the $t^{th}$ AQE run, $\theta_r^{(t)} = \pi(LQP^{(t-1)}, \theta_c, \delta)$, where $\delta = \delta^{(t)}$ for this particular run. 
	\end{enumerate}

In practice, our initial optimizer resolves the initial configuration $\thetabm_0$ and our RO applies the optimization strategy $\pi$ at runtime to configuration $\thetabm_r$ at each AQE run.

\subsubsection{Initial Optimizer} {\ }\\
\input{qi-initial-optimizer}

\subsubsection{Runtime Optimizer}
Note that we have particular interests in optimizing latency and cloud cost in our MOO problem. In the production environment, cloud cost is calculated as the weighted sum of CPU-hours and memory-hours. 
When $\thetabm_c$ is fixed with CPU and memory determined, cloud cost becomes directly proportional to query latency. 
As a result, the MOO problem concerning latency and cloud cost can be reduced to a single objective optimization problem during the runtime.

\begin{definition}{\bf Runtime Optimization over $LQP^{(t)} $(\texttt{RO-AQE})}
\begin{align}
	\label{eq:rt-pi1}
	\thetabm_{AB}^*, \Thetabm_{s}^* &= \argmin_{\thetabm_{AB}, \Thetabm_{s}} L_{LQP}(\thetabm_{AB}, \Thetabm_{s}, LQP^{(t)}, \thetabm_c, \delta) \\
	\nonumber s.t. & {\begin{array}{l}
	    \Thetabm_{s} = ||_{\forall QS^{(i)} \in LQP^{(t)}} \thetabm_s^{(i)}, \; \thetabm_s^{(i)} \in \Sigma_s \\
	    \thetabm_{s.comm}^{(i)} \in h_i(\thetabm_{AB})
		\end{array}}
	\end{align}
\end{definition}

\begin{definition}{\bf Runtime Optimization over LQP (\texttt{RO-1})}
\begin{align}
	\label{eq:rt-pi1}
	\thetabm_{r}^*	&= \argmin_{\thetabm_{r}} L_{LQP}(\thetabm_{r}, LQP^{(t)}, \thetabm_c, \mathcal{P}, \delta) \\
	\nonumber s.t. & {\begin{array}{l}
	    \thetabm_{r} \in \Sigma_{r}
		\end{array}}
	\end{align}
\end{definition}

where $\thetabm_{r}^*$ is the optimal configuration for the runtime LQP in this AQE run, $L_{LQP}$ is the latency of a runtime LQP ranging from the entry time of this AQE to the completion of the entire query, and $\thetabm_{r}^*$ is assumed to be fixed for the remaining AQE runs.

\begin{definition}{\bf Runtime Optimization over QS (\texttt{RO-2})}
\begin{align}
	\label{eq:rt-pi2}
	\thetabm_{s}^* &= \argmin_{\thetabm_s = \thetabm_{s.comm} \cup \thetabm_{s.indp}} L_{QS}(\thetabm_s, QS^{(i)}, \thetabm_c, \delta) \\
	\nonumber s.t. & {\begin{array}{l}
	    \thetabm_{s.comm} \in h(\thetabm_r^*) \\
	    \thetabm_{s} = \thetabm_{s.comm} \cup \thetabm_{s.indp} \in \Sigma_{s} \\
	    QS \in \mathcal{QS}
		\end{array}}
	\end{align}
\end{definition}

where $\thetabm_{C}^*$ is the optimal configuration for a QS, including the QS-independent parameters $\thetabm_{qsIndp}$ that are only applied for the QS (in Step $\mathcal{C}$) and the common parameters $\thetabm_{comm}$ shared with other steps in $\mathcal{A}$ and $\mathcal{B}$, $L_{QS}$ is the latency of the QS, and $\mathcal{QS}$ and $\Sigma_C$ are the feasible spaces for QS and $\thetabm_C$ respectively.
Notice that in some scenarios of $\tilde{\thetabm_r}$, $\thetabm_{comm}$ can pick among a range of values without changing the outcomes of the applied logical and physical rules in the procedure $\mathcal{A}$ and $\mathcal{B}$, and we define such value range by the function $h(\tilde{\thetabm_r})$. 
In our case, we have $\thetabm_{comm} = \{s_1\}$, $\thetabm_{qsIndp} = \{s_8, s_9\}$ and $h(\tilde{\thetabm})$ as a decision tree shown in Figure~\todo{add}.
}


}

%% file: checkpoint-231115/Qi_initial_optimizer/qi-approach-overview.tex
\todo{be brief in this overview}

The initial optimization is addressed at compile time, which optimizes $\bm{\theta_c}$, $\bm{\theta_p}$ and $\bm{\Theta_s}$ jointly as they are correlated. We assume that $\alpha$ and $\beta$ could be well estimated at compile time.

The initial optimization is non-trivial due to not only the multi-granularity control of parameters but also the timing constraints. The $\bm{\theta_p}$ decides the PQP and QSs, which should proceed the tuning of $\bm{\theta_c}$ and $\bm{\theta_s}$ as each QS takes both $\bm{\theta_c}$ and $\bm{\theta_{si}}, i \in [1,m]$, where $m$ is the number of QSs. Moreover, Spark forces fixing $\bm{\theta_c}$ during query submission. 

After a PQP is given (i.e. under one $\bm{\theta_p}$ sample), the total parameter space of optimization is $d_c + m \times d_s$, which could be high when $m$ is large, where $d_c$ and $d_s$ are the number of parameters of $\bm{\theta_c}$ and $\bm{\theta_{si}}, i \in [1,m]$ respectively. We propose an algorithm design with the divide-and-conquer framework, which tunes each QS independently under the constraint that $\bm{\theta_{c}}$ is identical among all QSs to get the local QS-level Pareto optimal, and then compose those local Pareto solutions into optimal query-level solutions intelligently over a DAG of QSs. In summary, it includes two sequential steps named \textbf{QS-tuning} and \textbf{DAG optimization} respectively. More details can be found in later sections.

%% file: modeling.tex
\section{Modeling}
\label{sec:modeling}

In this section, we introduce our modeling methods that support both compile-time optimization and runtime optimization with fine-grained parameter tuning.

\rv{
\subsection{Compile-time and Runtime Models}\label{subsec:ct-rt-models}
We first devise models for fine-grained parameter tuning at both compile-time and runtime. 

\textbf{Runtime models.}
The Spark optimizer offers the collapsed logical query plan (\collap) and query stages (QS) in the physical plan at runtime (see Section~\ref{subsec:spark}). Hence, we collect these data structures and build runtime \collap\ and QS models to enable fine-grained tuning of query plan ($\thetabm_p$) and query stage ($\thetabm_s$) parameters, respectively.

\textbf{Compile-time model.} At compile time, Spark provides a logical query plan (LQP) and then a physical plan (PQP), but no other data structures that would suit our goal of fine-grained tuning.  
Therefore, we introduce the notion of {\em compile-time subquery} ({subQ}) to denote a group of logical operators that will correspond to a query stage (QS) when the logical plan is translated to a physical plan. In other words, it is a sub-structure of the logical plan that is reversely mapped from a query stage in a physical plan. Figure~\ref{fig:param-2} illustrates the LQP of TPCH-Q3, which can be divided into five subQs, each corresponding to one QS.
As an enhancement, our work can sample multiple physical plans for each query at compile time, which will lead to different subQ structures. We collect all of these subQ structures, and develop a predictive  model for each subQ to enable fine-grained tuning of $\thetabm_p$ at compile-time. 
}

\subsection{Modeling Objectives}


\begin{figure}[t]
  \setlength{\belowcaptionskip}{0pt}  
	\setlength{\abovecaptionskip}{0pt} 	
  \centering
  \captionsetup{justification=centering}
  \begin{minipage}{.48\linewidth}
    \includegraphics[width=.99\linewidth,height=2.8cm]{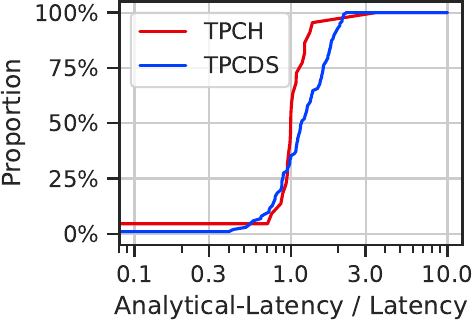}
    \captionof{figure}{\small{CDF of analytical latency over actual latency}}
    \label{fig:analytical-lat-cdf}
  \end{minipage}
  \hfill
  \begin{minipage}{.48\linewidth}
    \centering
      \includegraphics[width=.99\linewidth,height=2.8cm]{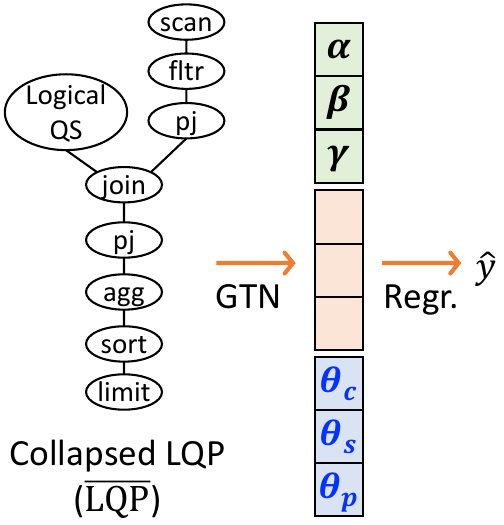}
      \captionof{figure}{\small{Model structure (GTN embedder + regressor) for \collap}}
      \label{fig:model-struct-collap}
  \end{minipage}
  \vspace{-0.2in}
\end{figure}


With the objective of optimizing latency and cost, our modeling work seeks to make these metrics more robust and predictable.

Query latency in Spark, defined as the end-to-end duration to execute a query, benefits from Spark's cluster manager that ensures a dedicated allocation of cores and memory to the entire query. Such resource isolation enhances the predictability of latency, making it a suitable target for optimizing a query or a collapsed query plan.

However, within a query, Spark shares resources among parallel query stages, raising two issues in modeling latency at the stage level.
First, the end-to-end latency of a set of parallel stages 
often leads to a longer latency than their maximum due to resource contention. Prior work~\cite{LyuFSSD22} assumed ample resources in industry-scale clusters and simplified this issue by taking the \texttt{max}  latency among parallel tasks, which is not applicable in the Spark environment of shared resources.
Second, predicting the latency of each stage directly is very hard due to its variability in a shared-resource setting, where performance fluctuates based on resource contention.

To address these issues, we propose the concept of {\it analytical latency}, calculated as the sum of the task latencies across all data partitions divided by the total number of cores. \crv{
  \cut{This approach yields two significant advantages.}
  This yields two advantages.
}\crv{
  \cut{Firstly, it establishes a direct link between the latency of a query and its constituent stages, enabling the computation of query-level latency at compile time through a \texttt{sum} aggregator over the task latencies of all subQs.}
  First, it directly links the query latency to its constituent stages, enabling the computation of query-level latency at compile time through a \texttt{sum} aggregator over task latencies of all subQs.
}
Second, it enhances the predictability of query stage latency by excluding the variability introduced by resource wait times, thus offering a more consistent basis for latency prediction.
To validate the efficacy of the analytical latency, we
\crv{
  compare it with actual query latencies using
}TPC-H and TPC-DS benchmarks under the default Spark configuration.
The results demonstrate a robust correlation between analytical and actual latencies, with Pearson correlation coefficients of 97.2\% for TPC-H and 87.6\% for TPC-DS. \crv{
  \cut{Furthermore, Figure~\ref{fig:analytical-lat-cdf} shows that the analytical latency closely mirrors the actual execution time, with the ratio close to 1 for most of the tested queries. }
  As shown in Figure~\ref{fig:analytical-lat-cdf}, analytical latency closely mirrors actual execution time, with the ratios close to 1 for most of the tested queries. 
}

Cloud costs are mainly based on the consumption of resources, such as CPU-hour, memory-hour, IO and \rv{shuffle sizes~\cite{dataflow-pricing}}. 
We model all of these costs to support multi-objective optimization. 

To summarize,  \rv{our models capture 1) end-to-end latency and cost for collapsed query plans (\collap) at runtime, 2) analytical latency and cost for query stages (QS) at runtime, in the face of resource sharing, and 3) analytical latency and cost for subQs at compile time---the latter two ensure both to be robust targets for modeling.}

\subsection{Model Formulation for Optimization}

We now introduce the methodology for building models for subQ, \collap, and \textit{QS}, which will enable their respective fine-graining later. 

{\bf Feature Extraction.} We extract features to capture \crv{query characteristics and the dynamics of their execution environment, configured by both decision and non-decision variables.} 
First, we extract the query plan as a DAG of vectors, where each query operator is \crv{encoded by concatenating}
i) operator type via one-hot encoding, 
ii) its cardinality, represented by row count and size in bytes, and 
iii) the average of the word embeddings~\cite{word2vec} from its predicates, 
providing a rich, multidimensional representation of the operator's functional and data characteristics.
Second, we capture critical contextual factors as non-decision variables, including 
i) input characteristics \(\alpha\), aggregated from the statistics of leaf operators, 
ii) data distribution \(\beta\), quantifying the size distribution of input partitions with metrics like standard deviation-to-average ratio (\(\frac{\sigma}{\mu}\)), skewness ratio (\(\frac{\max-\mu}{\mu}\)), and range-to-average ratio (\(\frac{\max-\min}{\mu}\)), and
iii) runtime contention \(\gamma\), capturing the statistics of parallel stages in a numeric vector, tracking  their tasks in running and waiting states, and aggregating statistics of completed tasks to characterize their behaviors.
Third, we convert the tunable parameters as decision variables into a numeric vector to represent the Spark behavior.

{\bf Model Structures.}
The hybrid data structure of the query plan, with a DAG of operator encodings, and other tabular features, poses a challenge in model formulation.
To tackle this, we adopt a multi-channel input framework~\cite{LyuFSSD22} that incorporates a Graph Transformer Network (GTN)~\cite{gtn-aaai21} and a regressor to predict our objectives, as shown in Figure~\ref{fig:model-struct-collap}.
We first derive the query embedding using \crv{the GTN~\cite{gtn-aaai21}, which handles non-linear and non-sequential relationships by using Laplacian positional encoding to encode positional information and attention mechanisms to capture operator correlations}.
These embeddings are then concatenated with other tabular data and processed through a regressor, \crv{capturing} the interplay among the query characteristics, critical contextual factors, and tunable parameters.
Figure~\ref{fig:model-struct-collap} illustrates the architecture of the \collap\ model, which has the largest number of feature factors.

%% file: compile-time-opt.tex
\section{Compile-time/Runtime Optimization}
\label{sec:compile-time}

In this section, we present our hybrid compile-time/runtime optimization approach to multi-granularity parameter tuning in the multi-objective optimization setting.

\input{Qi_compile_time_opt/div_and_conq_moo.tex}

\input{Qi_compile_time_opt/LQP_tuning.tex}
\input{Qi_compile_time_opt/dag_opt.tex}

\input{compile-time-opt-improvement}

%% file: Qi_compile_time_opt/div_and_conq_moo.tex
\subsection{Hierarchical MOO with Constraints}
\label{subsec:hmooc}

\begin{figure*}
	\setlength{\belowcaptionskip}{0pt}  
	\setlength{\abovecaptionskip}{0pt} 	
	\begin{minipage}{0.75\textwidth}
		\subfigure[\small{Example subQ}]{\label{fig:eg_w_3_subq}\includegraphics[height=3.5cm,width=0.22\textwidth]{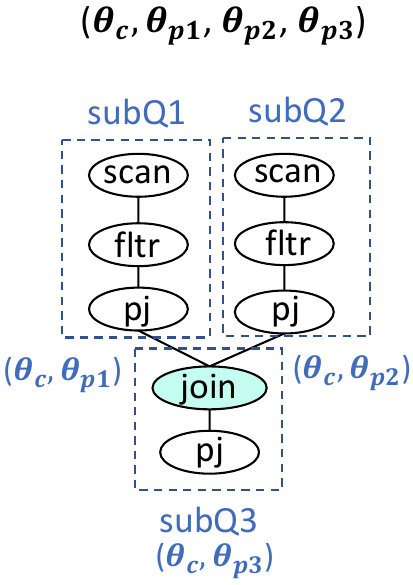}}
		\hfill
		\subfigure[\small{Approach Overview}]{\label{fig:qs_tuning_s1}\includegraphics[height=3.7cm,width=0.76\textwidth]{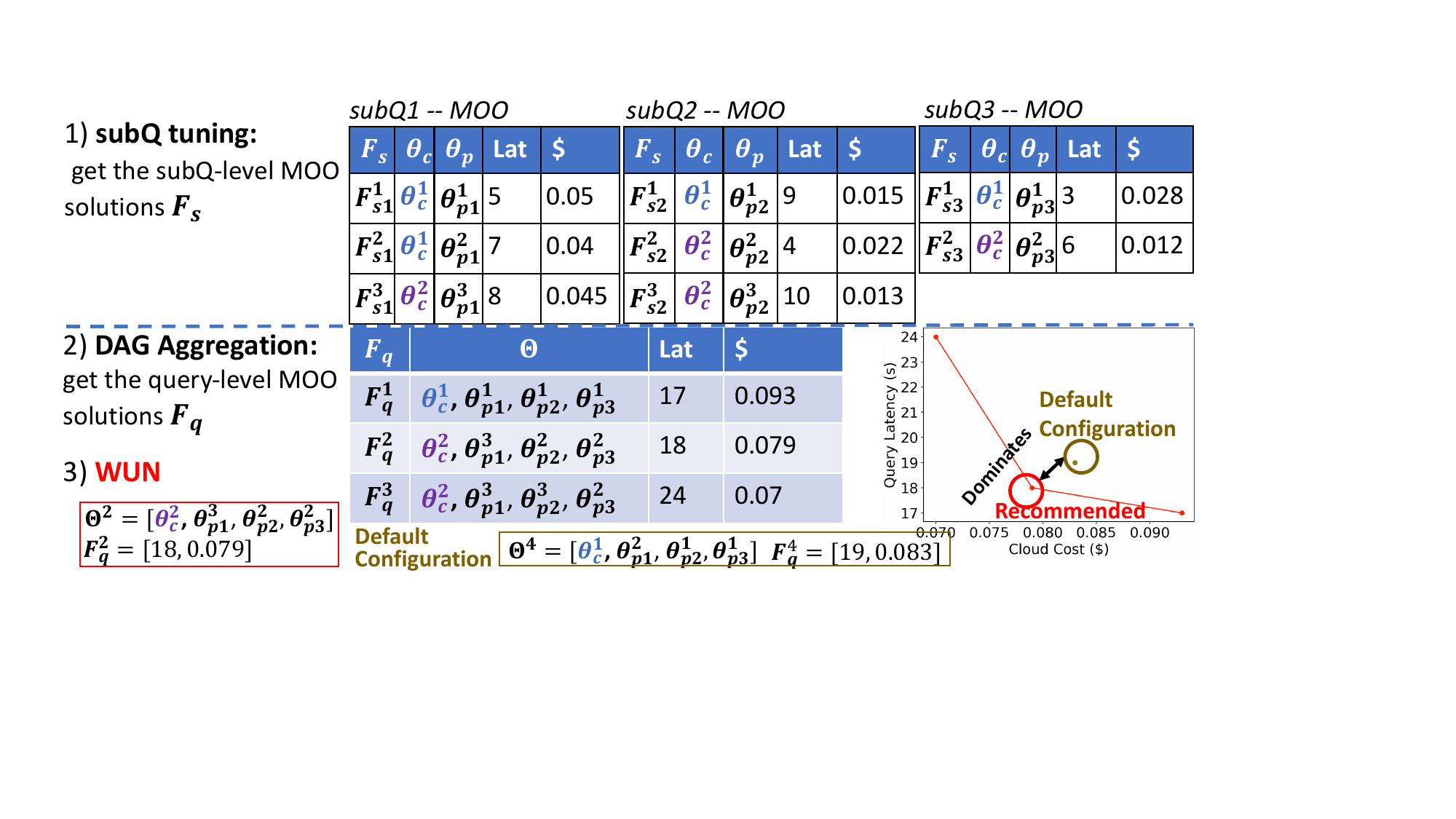}}
		\caption{\small Example of the Compile-time optimization of TPCH Q3}
		\label{fig:eg_compile_opt}
	\end{minipage} 
	\hfill
	\begin{minipage}{0.22\textwidth}
	\includegraphics[height=3.5cm,width=.99\textwidth]{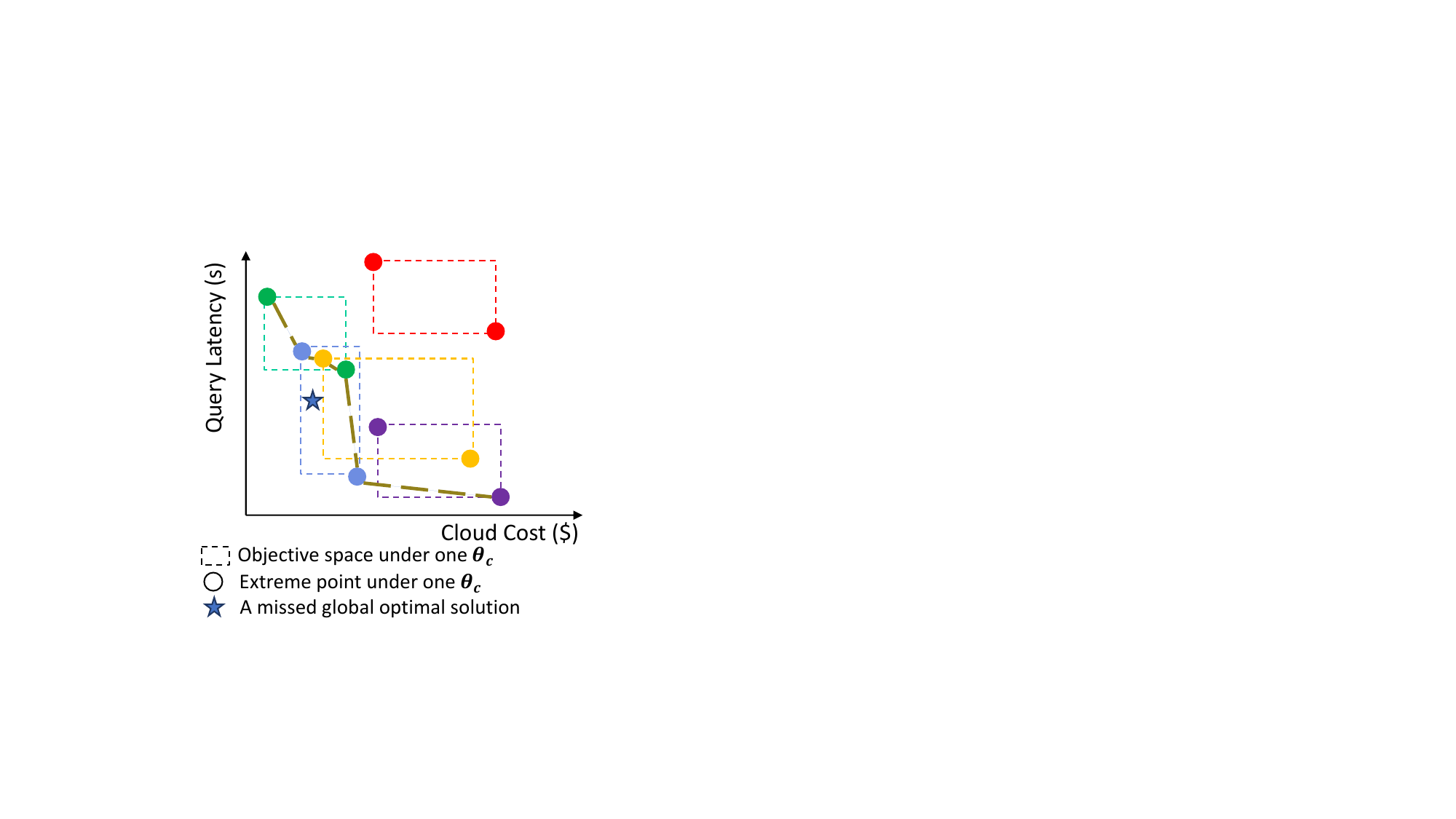} 
		\caption{\small{Boundary approximation of DAG optimization}} 
		\label{fig:approx_dag_opt}
	\end{minipage}
\vspace{-0.1in}
\end{figure*}

Our compile-time optimization finds the optimal configuration $\thetabm_c^*$ of the context parameters to construct an ideal Spark context for query execution. 
We do so by exploring the correlation of $\thetabm_c$ with fine-grained $\bm{\theta_p}$ and $\bm{\theta_s}$ parameters for different subqueries (subQs), under the modeling constraint that the cardinality estimates are based on Spark's cost-based optimizer. 
Despite the modeling constraint, capturing the correlation between the mixed parameter space allows us to find a better Spark context for query execution. 


The multi-objective optimization problem in Def.~\ref{eq:moo-spark} provides fine-grained control of $\bm{\theta_{p}}$ and $\bm{\theta_{s}}$ at the subQ/query stage level, besides the query level control of $\bm{\theta_{c}}$. 
This leads to a large parameter space, linear in the number of query stages in the plan, which defeats most existing MOO methods when 
	the solving time must be kept under the constraint of 1-2 seconds for cloud use.  

To combat the high-dimensional parameter space, we propose a new approach named \textit{Hierarchical MOO with Constraints} (HMOOC). It follows a divide-and-conquer framework to break a large optimization problem on ($\bm{\theta_{c}}$, $\{\bm{\theta_{p}}\}$, $\{\bm{\theta_{s}}$\})  to a set of smaller problems on ($\bm{\theta_{c}}$, $\bm{\theta_{p}}$, $\bm{\theta_{s}}$), one for subQ of the logical query plan. However, these smaller problems are not independent as they must obey the constraint that all the subproblems must choose the same $\bm{\theta_{c}}$ value. 
More specifically,  the problem for HMOOC is defined as follows:

{\definition{\textbf{Hierarchical MOO with Constraints (HMOOC)}
		\begin{eqnarray}
		\label{eq:div_query_moo_def}
		\arg \min_{\bm{\theta}} & \bm{f}(\bm{\theta})={\left[
			\begin{array}{l}
			f_1(\bm{\theta}) = \Lambda(  \phi_1(\text{LQP}_1, \bm{\theta_c},\bm{\theta_{p1}}, \bm{\theta_{s1}}),\dots,\\ 
				 \,\,\,\,\,\,\,\,\,\,\,\,\,\,\,\,\,\,\,\,\,\,\,\,\,\, \phi_1(\text{LQP}_m, \bm{\theta_c},\bm{\theta_{pm}}, \bm{\theta_{sm}})) \\ 
			{\centering \hspace{2.2cm} \vdots}\\
			f_k(\bm{\theta}) = \Lambda(  \phi_k(LQP_1, \bm{\theta_c},\bm{\theta_{p1}}, \bm{\theta_{s1}}),\dots,\\
			 \,\,\,\,\,\,\,\,\,\,\,\,\,\,\,\,\,\,\,\,\,\,\,\,\,\, 
			 	\phi_k(LQP_m, \bm{\theta_c},\bm{\theta_{pm}}, \bm{\theta_{sm}}))
			\end{array}
			\right]} \\
		\nonumber s.t. & \bm{\theta_c} \in \Sigma_c \subseteq \mathbb{R}^{d_c}, \quad \bm{\theta_{pi}} \in  \Sigma_p \subseteq \mathbb{R}^{d_p}, \\ 
		& \bm{\theta_{si}} \in \Sigma_s \subseteq \mathbb{R}^{d_{s}}, \quad i = 1,\dots, m
		\end{eqnarray}
}}\noindent{where LQP$_i$ denotes the $i$-th subQ of the logical plan query,  $\bm{\theta_i} = (\bm{\theta_c},\bm{\theta_{pi}}, \bm{\theta_{si}})$ denotes its configuration, with $i=1,\ldots,m$, and $m$ is the number of subQs. Most notably, all the subQs share the same $\bm{\theta_c}$, but can use different values of $\bm{\theta_{pi}}$ and $\bm{\theta_{si}}$. Additionally, $\phi_j$ is the subQ predictive model of the $j$-th objective, where $j=1,\ldots,k$. The function $\Lambda$ is the mapping from subQ-level objective values to query-level objective values, which can be aggregated using \code{sum} based on our choice of analytical latency and cost metrics.
}

\cut{Regarding the non-decision variables, the compile-time optimization sets the non-decision variable $\alpha$ for cardinality estimates based on Spark's cost-based optimizer, the data distribution variable $\beta$ to be uniform across partitions, and the resource contention variable $\gamma$ to be zero.  
For the sake of simplicity, we omit the non-decision variables $\alpha$, $\beta$, and $\gamma$ in the following discussion of compile-time optimization.}

\crv{Our main idea} is to tune each subQ independently under the constraint that $\bm{\theta_{c}}$ is identical among all subQ's. By doing so, we aim to get the local subQ-level solutions, and then recover the query-level Pareto optimal solutions by composing these local solutions efficiently. In brief, it includes three sequential steps: (1)~\textbf{subQ tuning}, (2)~\textbf{DAG aggregation}, and (3)~\textbf{WUN recommendation}.



Figure \ref{fig:eg_compile_opt} illustrates an example of compile-time optimization for TPCH-Q3 under the latency and cost objectives. For simplicity, we show only the first three subQ's in this query and omit $\bm{\theta_{s}}$ in this example. In subQ-tuning, we obtain subQ-level solutions with configurations of $\bm{\theta_{c}}$ and $\bm{\theta_{p}}$, where $\bm{\theta_{c}}$ has the same set of two values ($\bm{\theta_{c}^1}$, $\bm{\theta_{c}^2}$) among all subQ's, but $\bm{\theta_{p}}$ values vary. Subsequently in the DAG aggregation step, the query-level latency and cost are computed as the sum of the three subQ-level latency and cost values, and only the Pareto optimal values of latency and cost are retained. Finally, in the third step, we use the WUN (weighted Utopia nearest) policy to recommend a configuration from the Pareto front.




%

%% file: Qi_compile_time_opt/LQP_tuning.tex
\subsubsection{Subquery (subQ) Tuning}
\label{subsec:subq-tuning}


\begin{figure}
	\setlength{\belowcaptionskip}{0pt}  
	\setlength{\abovecaptionskip}{0pt}	
	\centering
	\includegraphics[width=.95\linewidth]{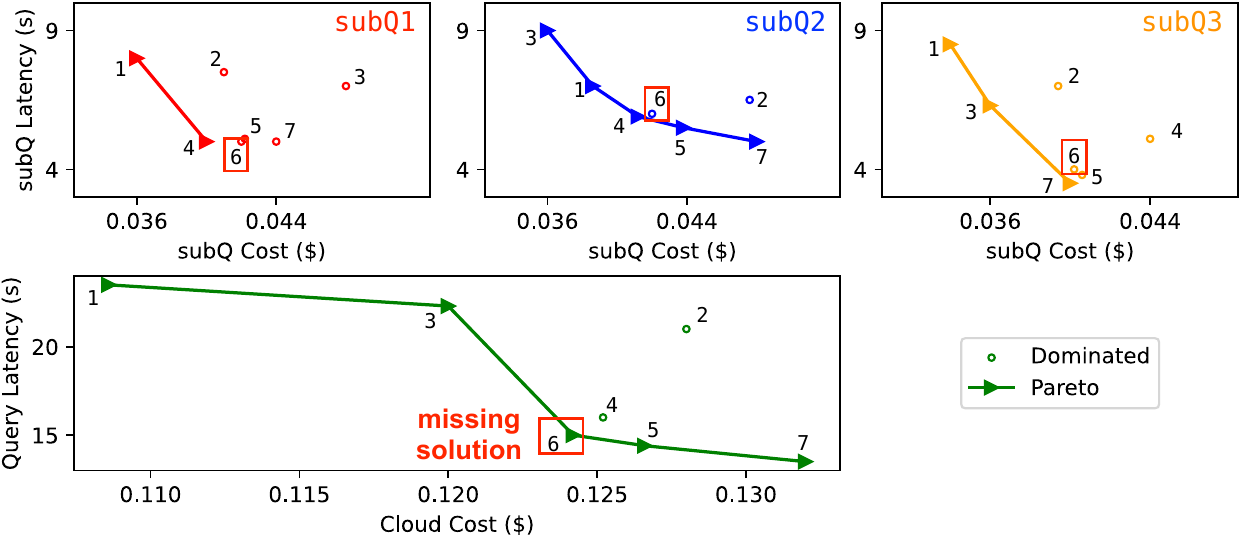}
	\captionof{figure}{\small{Example of missed global optimal solutions in TPCH Q3}}
	\label{fig:eg_qs_tuning-new}
	\vspace{-0.15in}
\end{figure}


Subquery (subQ) tuning aims to generate an effective set of local solutions of ($\bm{\theta_{c}}$, $\bm{\theta_{p}}$, $\bm{\theta_{s}}$) for each subQ while obeying the constraint that all the subQs share the same $\bm{\theta_{c}}$. 
For simplicity, we focus on ($\bm{\theta_{c}}$, $\bm{\theta_{p}}$) in the following discussion as $\bm{\theta_{s}}$ is treated the same way as $\bm{\theta_{p}}$.

One may wonder whether it is sufficient to generate only the local Pareto solutions of ($\bm{\theta_{c}}$, $\bm{\theta_{p}}$) of each subQ.  Unfortunately, this will lead to missed global Pareto optimal solutions due to the constraint on $\bm{\theta_{c}}$. 
Figure~\ref{fig:eg_qs_tuning-new} illustrates an example with 3 subQs, where solutions sharing the same index fall under the same $\bm{\theta_{c}}$ configuration and have achieved optimal $\bm{\theta_{p}}$ under that $\bm{\theta_{c}}$ value. 
The first row in Figure~\ref{fig:eg_qs_tuning-new} showcases
subQ-level solutions, where triangle points represent subQ-level optima and circle points denote dominated solutions. 
The second row in Figure~\ref{fig:eg_qs_tuning-new}
displays the global query-level values, where both latency and cost are the sums of subQ-level latency and cost. Notably, solution 6 is absent from the local subQ-level Pareto optimal solutions across all subQs. Due to the identical $\bm{\theta_{c}}$ constraint and the sum from subQ-level values to query-level values, the sum of solution 6's subQ-level latency and cost becomes better than solution 4 (a subQ-level Pareto optimal solution) and is a query-level Pareto point. 

\minip{1. Effective $\bm{\theta_c}$ Candidates.}
To minimize the chance of missing global solutions, we seek to construct a diverse, effective set of  $\bm{\theta_{c}}$ configurations to be considered across all subQs.
$\bm{\theta_c}$ can be initialized by random sampling or grid-search over its domain of values.
Then, we  enrich  the $\bm{\theta_c}$ set using a few methods. Drawing inspiration from the evolutionary algorithms~\cite{Emmerich:2018:TMO}, we introduce a {\em crossover} operation over the existing $\bm{\theta_c}$ population to generate new candidates. 
If crossover cannot generate more candidates, e.g., for some grid search methods used for initial sampling of $\bm{\theta_c}$ candidates, then we add random sampling to discover new candidates. 


\minip{2. Optimal $\bm{\theta_{p}}$ Approximation.}
Next,  under each $\bm{\theta_{c}}$ candidate, we show that it is crucial to keep track of the local Pareto optimal $\bm{\theta_{p}}$ within each subQ. The following proposition explains why. 


\begin{proposition} \label{prop:need_opt_p}
	Under any specific value $\bm{\theta_{c}}^j$, 
	only subQ-level Pareto optimal solutions $(\bm{\theta_{c}}^j,{\bm{\theta_{p}}^*})$ contribute to the query-level Pareto optimal solutions.
\end{proposition}

In the interest of the space, all the proofs in this paper are deferred to \techreport{Appendix~\ref{appendix:proof}}.

The above result allows us to restrict our search of $\bm{\theta_p}$ to only the local Pareto optimal ones. However, given the large, diverse set of $\bm{\theta_c}$ candidates, it is computationally expensive to solve the MOO problem for $\bm{\theta_p}$ repeatedly, once for each $\bm{\theta_c}$  candidate. 
We next introduce a clustering-based approximation to reduce the computation complexity. It is based on the hypothesis that, within the same subQ, similar $\bm{\theta_c}$ candidates entail  similar optimal $\bm{\theta_p}$ values in the tuning process.
By clustering similar $\bm{\theta_c}$ values into a small number of groups (based on their Euclidean distance), we then solve the MOO problem of $\bm{\theta_p}$ for a single $\bm{\theta_c}$ representative of each group. 
\cut{To expedite the repeated solving of $\bm{\theta_p}$ for different $\bm{\theta_c}$ representatives, we maintain a pool of samples of $\bm{\theta_p}$ and among them find the Pareto optimal values for each  $\bm{\theta_c}$ representative. }
We then use the optimized $\bm{\theta_p}$  as the estimated optimal solution for other $\bm{\theta_c}$ candidates within each group.




\minip{Algorithm.} Algorithm \ref{alg:effective_set_generation} describes the steps for obtaining an effective solution set of $(\bm{\theta_c}, \bm{\theta_p})$ for each subQ.
\rv{Line 1 initiates  by generating $C \times P$ samples, where $C$ and $P$ are the numbers of  distinct values of $\bm{\theta_c}$ and $\bm{\theta_p}$, respectively.}
The $\bm{\theta_c}$ candidates are then grouped using a clustering approach (Line 2), where $rep\_c\_list$ constitutes the list of $\bm{\theta_c}$ representatives for the $n$ groups, $C\_list$ includes the members within all $n$ groups, and $\kappa$ represents the clustering model.
In Line 3, $\bm{\theta_p}$ optimization is performed for each representative $\bm{\theta_c}$ candidate (using the samples from Line 1). 
Subsequently, the optimal $\bm{\theta_p}$ of the  representative $\bm{\theta_c}$ is assigned to all members within the same group and is fed to the predictive models to get objective values (Line 4). After that, the initial effective set is obtained, where $\Omega^{(0)}$ represents the subQ-level objective values under different $\bm{\theta_c}$, 
and $\Theta^{(0)}$ represents the corresponding configurations. 
Line 5 further enriches $\bm{\theta_c}$ using the crossover method or random sampling, which expands the initial effective set to generate new $\bm{\theta_c}$ candidates.
Afterwards,  the cluster model $\kappa$ assigns the new $\bm{\theta_c}$ candidates with their group labels (Line 6). The previous optimal $\bm{\theta_p}$ values are then assigned to the new members within the same group, resulting in their corresponding subQ-level values as the enriched set (Line 7).
Finally, the initial set and the enriched set are combined as the final effective set of subQ tuning (Line 8). 




\rv{
\minip{Sampling methods.} We next detail the sampling methods used in Line 1 of the algorithm. 
(1) We include \underline{basic sampling methods}, including {\it random sampling} and {\it  Latin-hypercube sampling} (LHS)~\cite{LHS/McKayBC00} as a grid-search method. 
(2) To reduce dimensionality, we introduce  \underline{feature importance score} (FIS) based parameter filtering: we sort the parameters by the FIS value from the trained model and leverage the long-tail distribution to drop the parameters at the tail. Precisely, many parameters at the tail have a low cumulative FIS and we apply a threshold (e.g., 5\% model loss) to remove them from sampling. 
(3) We further propose an \underline{adaptive grid search} method with FIS-based parameter filtering. Given the sample budget ($C$ or $P$), it goes down the FIS ranking list and progressively covers one more parameter, including its min, median and max values. If it reaches the budget before adding all parameters, it ignores those uncovered ones. Otherwise, it loops over the list again to add more sampled values of each parameter. 
\cut{When all important parameters are covered, it further increases the search space by adding the $1^{st}$ and $3^{rd}$ quartiles to each important parameter with the same order. When the parameter space scales up but the sampling rates are limited due to the time constraints, the adaptive grid search can adapt to a lower sampling rate by reducing or dropping the choices of the parameters in the ascending order of FIS.}
(4) We conduct hyperparameter tuning to derive low, medium, and high values for $C$ and $P$. We also employ a simple runtime adaptive scheme to adjust the sampling budget based on the predicted latency under the default configuration. 
See \techreport{Appendix~\ref{appendix:adaptive-grid-search} and~\ref{appendix:fis-filtering}} for more details.
}

\begin{algorithm}[t]
	\caption{Effective Set Generation}
	\label{alg:effective_set_generation}
	\footnotesize
	\begin{algorithmic}[1]  
		\REQUIRE {$Q$, $\phi_i, n$, $\forall i\in [1,k], \alpha, \beta, \gamma$, $C$, $P$}.
		\STATE $\bm{\Theta_c^{(0)}}, \bm{\Theta_p^{(0)}}$ = sampling($C$, $P$)
        \STATE $rep\_c\_list$, $C\_list$, 	$\kappa$ = cluster($\bm{\Theta_c^{(0)}}$, $n$)
		\STATE $\bm{\Theta_p^{*}}$ = optimize\_p\_moo($\bm{\Theta_p^{(0)}}$, $rep\_c\_list$, $\phi$, $\alpha, \beta, \gamma$, $Q$)
		\STATE $\bm{\Omega^{(0)}}$, $\bm{\Theta^{(0)}}$ = assign\_opt\_p($C\_list$, $rep\_c\_list$, $\bm{\Theta_p^{*}}$, $\phi$, $\alpha, \beta, \gamma$, $Q$)
		\STATE \crv{$\bm{\Theta_c^{'}}$} = enrich\_c($\bm{\Omega^{(0)}}$, $\bm{\Theta^{(0)}}$)
        \STATE \crv{$C\_list^{'}$} = assign\_cluster(\crv{$\bm{\Theta_c^{'}}$}, $rep\_c\_list$, $\kappa$)
        \STATE \crv{$\bm{\Omega^{'}}$, $\bm{\Theta^{'}}$} = assign\_opt\_p(\crv{$C\_list^{'}$}, $rep\_c\_list$, $\bm{\Theta_p^{*}}$, $\phi$, $\alpha, \beta, \gamma$, $Q$)
        \STATE $\bm{\Omega}$, $\bm{\Theta}$ = union($\bm{\Omega^{(0)}}$, $\bm{\Theta^{(0)}}$, \crv{$\bm{\Omega^{'}}$, $\bm{\Theta^{'}}$})
        \RETURN $\bm{\Omega}$, $\bm{\Theta}$
	\end{algorithmic}
\end{algorithm}

%% file: Qi_compile_time_opt/dag_opt.tex
\subsubsection{DAG Aggregation}
\label{subsec:dag-opt}
DAG aggregation aims to recover query-level Pareto optimal solutions from subQ-level solutions. It is a combinatorial MOO problem, as each subQ must select a solution from its non-dominated solution set while satisfying the  constraint of sharing the $\bm{\theta_c}$ configuration among all subQs.
The complexity of this process can be exponential in the number of subQs.  
Our approach below addresses this challenge by providing optimality guarantees and reducing the computation complexity.

\minip{Simplified DAG.} A crucial observation that has enabled our efficient methods is that our optimization problem over a DAG structure can be simplified to an optimization problem over a list structure. This is due to our choice of analytical latency and cost metrics, where the query-level objective can be computed as the sum of subQ-level objectives\cut{, which applies to the analytical latency, IO cost, CPU cost, etc., as explained in the previous section}. 
The MOO problem over a DAG can be simulated with a list structure for computing query-level objectives.




\minip{HMOOC1: Divide-and-Conquer}.
Under a fixed $\bm{\theta_c}$, i.e., satisfying the constraint inherently, we propose a divide-and-conquer method to compute the Pareto set of the simplified DAG, which is reduced to a list of subQs. The idea is to (repeatedly) partition the list into two halves,  solve their respective subproblems, and merge their solutions to global optimal ones. The merge operation enumerates all the combinations of solutions of the two subproblems, sums up their objective values, and retains only the Pareto optimal ones.
Our proof (available in \techreport{Appendix~\ref{appendix:proof}}) shows that this method returns a full set of query-level Pareto optimal solutions\cut{ as it enumerates those combinations of subQ-level solutions that have a chance to be global Pareto optimal}.

\minip{HMOOC2: WS-based Approximation}. Our second technique approximates the MOO solution over a list structure. 
For each fixed $\bm{\theta_c}$, we apply the weighted sum (WS) method to generate evenly spaced weight vectors. Then for each weight vector, we obtain the (single) optimal solution for each subQ and sum the solutions of subQ's to get the query-level optimal solution. It can be proved that this WS method over a list of subQs guarantees to return a subset of query-level Pareto solutions (see \techreport{Appendix~\ref{appendix:proof}}).

%

\input{Qi_compile_time_opt/approx_dag_opt.tex}

\cut{Overall, theoretically, among \textbf{General Divide-and-conquer (GD)}, \textbf{WS-based Approximation (WSA)} and \textbf{Boundary-based Approximation (BA)}, from the view of Pareto optimalily, \textbf{GD} performs better than \textbf{WSA} and \textbf{BA} performs the worst. While from the view of efficiency, \textbf{BA} achieves the lowest time comlexity and the \textbf{GD} and \textbf{WSA} achieve the higher.}

%% file: Qi_compile_time_opt/approx_dag_opt.tex
\minip{HMOOC3: Boundary-based Approximation}.
\cut{Given that DAG aggregation under each $\bm{\theta_c}$ candidate operates independently, it is inefficient to do so repeatedly when we have a large number of $\bm{\theta_c}$ candidates. }
Our next approximate technique stems from the idea that the objective space of DAG aggregation under each $\bm{\theta_c}$ can be approximated by $k$ \textit{extreme points}, where $k$ is the number of objectives.  In our context, the \textit{extreme point} under a fixed $\bm{\theta_c}$ is the Pareto optimal point with the best query-level value for any objective.
\cut{Then, the approximate query-level Pareto set is determined by the non-dominated extreme points among all $\bm{\theta_c}$ points.}
The rationale behind this approximation lies in the observation that solutions from different $\bm{\theta_c}$ candidates correspond to distinct regions on the query-level Pareto front. This arises from the fact that each $\bm{\theta_c}$ candidate determines the total resources allocated to the query, and a diverse set of $\bm{\theta_c}$ candidates ensures good coverage across these resources. Varying total resources, in turn, lead to different objectives of query performance, hence resulting in good coverage of the Pareto front of cost-performance tradeoffs. 


Therefore, we consider the degenerated \textit{extreme points} to symbolize the boundaries of different (resource) regions within the query-level Pareto front.
Figure \ref{fig:approx_dag_opt} illustrates an example. Here, the dashed rectangles with their extreme points under different colors represent the objective space of query-level solutions under various $\bm{\theta_c}$ candidates. The brown dashed line represents the approximate query-level Pareto front derived by filtering the dominated solutions from the collection of extreme points. The star solution indicates a missed query-level Pareto solution, as it cannot be captured from the extreme points.
 
The algorithm works as follows. For each $\bm{\theta_c}$ candidate, for each objective, we select the subQ-level solution with the best value for that objective for each subQ, and then sum up the objective values of such solutions from all subQs to form one query-level \textit{extreme point}. 
Repeating this procedure will lead to a maximum of $kn$ query-level solutions, where  $k$ is the number of objectives and $n$ is the number of $\bm{\theta_c}$  candidates. 
An additional filtering step will retain the non-dominated solutions from the $kn$ candidates, using an existing method of complexity $O(k n \log(k n))$~\cite{kung1975finding}. 


\cut{Our formal results include the following:}  
 
\begin{proposition} \label{prop:bounded_obj_space}
	Under a fixed $\bm{\theta_c}$ candidate, the query-level objective space of Pareto optimal solutions is bounded by its extreme points in a 2D objective space.
\end{proposition}

\begin{proposition} \label{prop:k_pareto}
	Given subQ-level solutions, our boundary approximation method guarantees to include at least $k$ query-level Pareto optimal solutions for a MOO problem with $k$ objectives.
\end{proposition}


%% file: compile-time-opt-improvement.tex
\rv{
\subsubsection{Multiple Query Plan Search}\label{subsec:multi-pqp-search}
Our compile-time MOO algorithm so far has considered only one physical query plan (with the corresponding subQs) based on the default configuration. Since at runtime, AQE may generate a very different physical plan, we further enhance our compile-time optimization by considering multiple physical plans.   
Initially, we sample plan-related parameters ($s_3$ and $s_4$) in $\thetabm_p$ to collect different physical query plans. We then rank these plans based on their predicted query latency under the default configuration. Subsequently, we trigger compile-time optimization (the HMOOC algorithm) for each of the top-k fastest query plans, run them  in parallel, and merge their Pareto solutions into the same objective space to obtain the final Pareto set.
}

%% file: runtime-opt.tex
\subsection{Runtime Optimization}
\label{sec:runtime-opt}

\rv{The compile-time optimization relied on the estimated cardinality and assumption of uniform data distributions. 
However, its true value  is to recommend the optimal context parameters $\thetabm_c^*$ by considering the correlations with  $\thetabm_p$ and $\thetabm_s$. 
Then, our runtime optimization addresses the remaining problems, adapting $\thetabm_p$ and $\thetabm_s$ based on actual runtime statistics and plan structures. } 



To start the runtime process, we need to first suit the constraint that 
Spark accepts only one copy of $\thetabm_p$ and $\thetabm_s$ at the query submission time.  
To do so, we intelligently aggregate the fine-grained $\thetabm_p$ and $\thetabm_s$ from compile-time optimization to initialize the runtime process. In particular, Spark AQE can convert a sort-merge join (SMJ) to a shuffled hash join (SHJ) or a broadcast hash join (BHJ), but not vice versa. Thus, imposing high thresholds ($s_3, s_4$ in Table~\ref{tab:spark-params}) to force SHJ or BHJ based on inaccurate compile-time cardinality can result in suboptimal plans.
\rv{In the example of Figure~\ref{fig:tpch-9-pqp}, when the cardinality of \texttt{Join4} is underestimated at the compile time, MO-WS returns a query plan broadcasting the output of \texttt{Join4}. At runtime, when the actual output size of \texttt{Join4} is observed to be 4.5GB, Spark cannot switch the BHJ back to other joins but broadcasting the 4.5GB data, leading to suboptimal performance.}
\rv{On the other hand, \crv{setting $s_3, s_4$ to zeros initially might overlook opportunities to apply} BHJs, especially for joins on base tables with small input sizes.}
To mitigate this, we initialize $\thetabm_p$ with the smallest threshold among all join-based subQs, enabling more effective runtime decisions. 
Other details of aggregating $\thetabm_p$ and $\thetabm_s$ are in \techreport{Appendix~\ref{appendix:runtime-optimization}}.


Runtime optimization then operates within a client-server model. The client, integrated with the Spark driver, dispatches optimization requests---including runtime statistics and plan structures---when a collapsed logical query plan (\collap) or a runtime query stage (QS) necessitates optimization (Steps 6, 9 in Figure~\ref{fig:spark-aqe}).
The server, hosted on a GPU-enabled node and supported by the learned models and a library of MOO algorithms~\cite{spark-moo-icde21}, processes these requests over a high-speed network connection.

Complex queries can trigger numerous optimization requests every time when a collapsed logical plan or a runtime QS is produced, significantly impacting overall latency. For instance, TPC-DS queries, with up to 47 subQs, may generate up to nearly a hundred requests throughout a query's lifecycle.
To address this, we established rules to prune unnecessary requests based on the runtime semantics of parametric rules
as detailed in \techreport{Appendix~\ref{appendix:runtime-optimization}}.
By applying these rules, we substantially reduce the total number of optimization calls by 86\% and 92\% for TPC-H and TPC-DS respectively.


%% file: checkpoint-231115/problem_statement.tex
\section{Problem Statement}

\begin{figure*}[t]
\centering
\captionsetup{justification=centering}
\begin{minipage}{.50\textwidth}
  \centering
  \includegraphics[width=.99\linewidth,height=3cm]{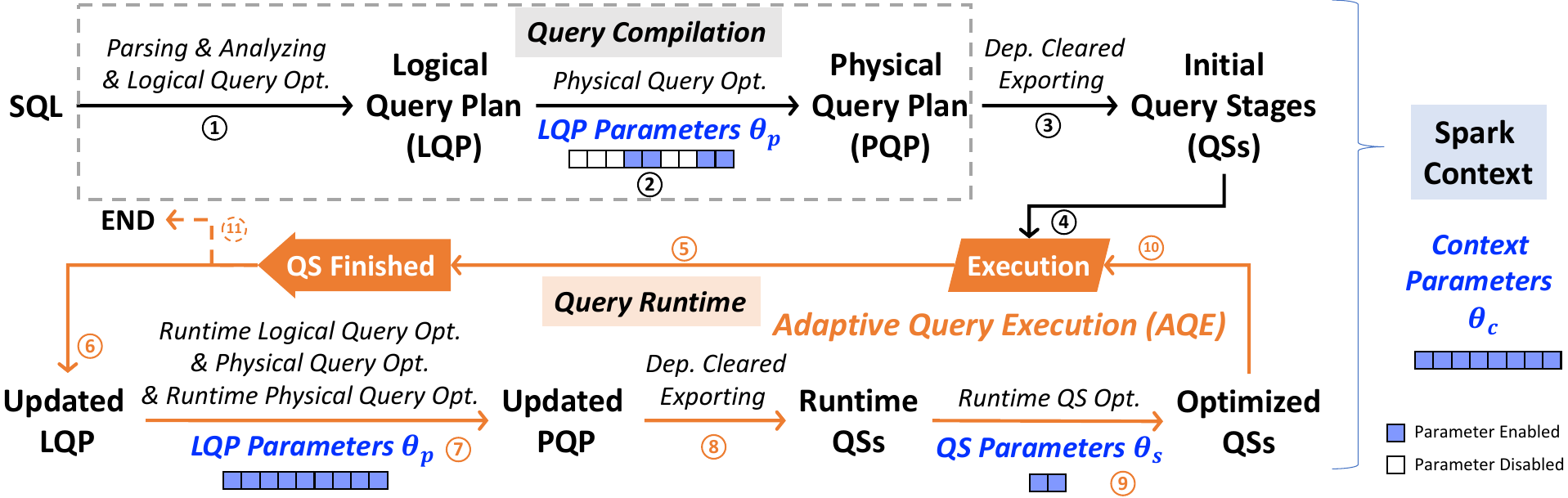}
  \captionof{figure}{\small{Spark SQL with Adaptive Query Execution (AQE)}}
  \label{fig:spark-aqe}
\end{minipage}
\hfill
\begin{minipage}{.21\textwidth}
  \centering
  \includegraphics[width=.99\linewidth,height=3cm]{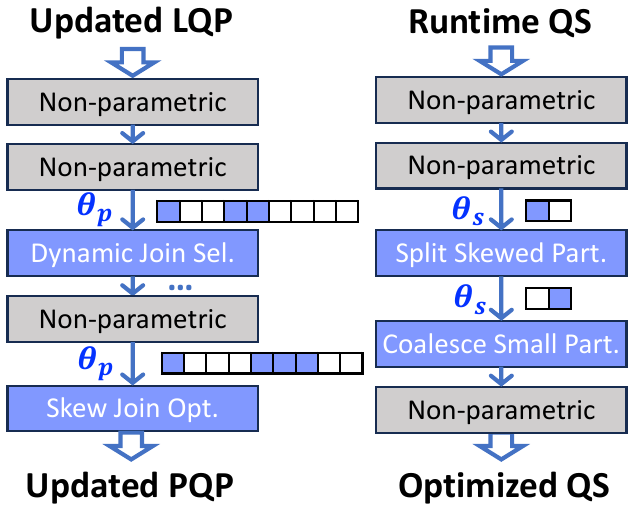}
  \captionof{figure}{\small{Runtime Opt. Rules}}
  \label{fig:opt-rules}
\end{minipage}
\hfill
\begin{minipage}{.25\textwidth}
  \centering
  \includegraphics[width=.99\linewidth,height=3cm]{figs/parameters.pdf}
  \captionof{figure}{\small{Spark Parameters}}
  \label{fig:spark-parameters}
\end{minipage}
\end{figure*}

\cut{
\begin{figure*}[t]
\centering
\captionsetup{justification=centering}
\begin{minipage}{.55\textwidth}
  \raggedright
  \includegraphics[width=.99\linewidth,height=4cm]{figs/sys-overview.pdf}
  \captionof{figure}{\small{System Design}}
  \label{fig:system-overview}
\end{minipage}
\begin{minipage}{.42\textwidth}
  \raggedleft
  \includegraphics[width=.99\linewidth,height=4cm]{figs/sys-aqe.pdf}
  \captionof{figure}{\small{An AQE run with IO}}
  \label{fig:system-aqe}
\end{minipage}
\end{figure*}
}

\begin{table}[t]
\ra{1.1}
\footnotesize
\centering
\caption{Parameter Table}
\label{tab:spark-params}
\begin{tabular}{ll}
\midrule
$\thetabm_c$ &{\bf Context Parameters}  \\
\midrule
$k_1$ & \verb|spark.executor.memory| \\
$k_2$ & \verb|spark.executor.cores| \\
$k_3$ & \verb|spark.executor.instances| \\
$k_4$ & \verb|spark.default.parallelism| \\
$k_5$ & \verb|spark.reducer.maxSizeInFlight| \\
$k_6$ & \verb|spark.shuffle.sort.bypassMergeThreshold| \\
$k_7$ & \verb|spark.shuffle.compress| \\
$k_8$ & \verb|spark.memory.fraction| \\
\midrule
\cut{
$\thetabm_r$ &{\bf Runtime SQL Parameters} \\
\midrule
$s_1$ & \verb|spark.sql.adaptive.advisoryPartitionSizeInBytes| \\
$s_2$ & \verb|spark.sql.adaptive.nonEmptyPartitionRatioForBroadcastJoin| \\
$s_3$ & \verb|spark.sql.adaptive.maxShuffledHashJoinLocalMapThreshold| \\
$s_4$ & \verb|spark.sql.adaptive.autoBroadcastJoinThreshold| \\
$s_5$ & \verb|spark.sql.shuffle.partitions| \\
$s_6$ & \verb|spark.sql.adaptive.skewJoin.skewedPartitionThresholdInBytes| \\
$s_7$ & \verb|spark.sql.adaptive.skewJoin.skewedPartitionFactor| \\
$s_8$ & \verb|spark.sql.adaptive.rebalancePartitionsSmallPartitionFactor| \\
$s_9$ & \verb|spark.sql.adaptive.coalescePartitions.minPartitionSize| \\
\midrule
}
$\thetabm_p$ &{\bf Logical Query Plan Parameters} \\
\midrule
$s_1$ & \verb|spark.sql.adaptive.advisoryPartitionSizeInBytes| \\
$s_2$ & \verb|spark.sql.adaptive.nonEmptyPartitionRatioForBroadcastJoin| \\
$s_3$ & \verb|spark.sql.adaptive.maxShuffledHashJoinLocalMapThreshold| \\
$s_4$ & \verb|spark.sql.adaptive.autoBroadcastJoinThreshold| \\
$s_5$ & \verb|spark.sql.shuffle.partitions| \\
$s_6$ & \verb|spark.sql.adaptive.skewJoin.skewedPartitionThresholdInBytes| \\
$s_7$ & \verb|spark.sql.adaptive.skewJoin.skewedPartitionFactor| \\
$s_8$ & \verb|spark.sql.files.maxPartitionBytes| \\
$s_9$ & \verb|| \\
\midrule
$\thetabm_s$ &{\bf Query Stage Parameters} \\
\midrule
$s_{10}$ & \verb|spark.sql.adaptive.rebalancePartitionsSmallPartitionFactor| \\
$s_{11}$ & \verb|spark.sql.adaptive.coalescePartitions.minPartitionSize| \\
\midrule
\end{tabular}
\end{table}



\cut{
\subsection{Spark Background}

Apache Spark~\cite{spark-rdd-nsdi12} is an open-source distributed computing system for large-scale data processing and analytics. 
Spark's underlying Massively Parallel Processing (MPP) architecture enables efficient parallel computation across cluster nodes. 
The core concepts of Spark include {\it jobs}, representing computations initiated by actions, and {\it stages}, which are organized based on shuffle dependencies, serving as boundaries that partition the computation graph of a job. 
Stages comprise sets of {\it tasks} executed in parallel, each processing a specific {\it data partition}. {\it Executors}, acting as worker processes, execute these tasks on individual cluster nodes. 

}

\subsection{Overview of Spark SQL}

Spark SQL seamlessly integrates relational data processing into the Spark framework ~\cite{spark-sql} and recently reinforced itself with the {\it Adaptive Query Execution (AQE)} mechanism that can re-optimize the query plans during runtime based on real-time statistics~\cite{spark-aqe}.

Figure~\ref{fig:spark-aqe} shows the lifecycle of a SQL in Spark. During query compilation, a SQL is extracted as a {\it logical query plan (LQP)}, and subsequently transformed into a {\it physical query plan (PQP)} via the physical query optimization. 
A physical query plan is further divided into a DAG of {\it query stages (QSs)} based on the data exchange dependency (shuffle or broadcast). Each query stage is the unit for scheduling, and the dependency-cleared query stages will first be exported for execution.

During query runtime, Spark SQL iteratively optimizes the remaining query until all stages are done. Specifically, when a query stage finishes, it updates the logical query plan with real-time statistics, constructs a new physical query plan via a set of query optimizations, and exports running query stages from the updated physical query plan. Each runtime query stage is further optimized independently before execution.

\todo{hide AQE}

add parameters to the overview
2. {Motivation of the ...}
3. Runtime optimization.

\cut{
\minip{Spark Context}

\minip{Optimization Rules}

\minip{Cardinality}

\minip{Data Distribution}

\minip{Resource Contention}
}

\subsection{Configuration Parameters}

We divided the Spark configuration parameters into three categories, the context parameters $\thetabm_c$, the logical query plan parameters $\thetabm_p$, and the query stage parameters $\thetabm_s$. Table~\ref{tab:spark-params} shows the selected parameters via the knob selection~\cite{KanellisAV20} and our heuristics.

\subsubsection{Spark Context and $\thetabm_c$} 
Spark context is the entry point of a Spark application configured by the context parameters $\thetabm_c$, which is non-tunable during the SQL execution. 
Specifically, it decides the resource profile (the memory per execution $k_1$, the cores per executor $k_2$, and the number of executors $k_3$), the shuffle behavior (the default degree of the parallelism $k_4$, the memory overhead per reducer task $k_5$, when to avoid merge-sorting data $k_6$, and whether to compress data before shuffling $k_7$), the memory maintenance strategy (the ratio to use memory for storage and execution $k_8$), etc. 

\subsubsection{Optimization Rules and SQL parameters}
At the core of query optimization is a set of optimization rules to optimize a logical query plan or a query stage. Each optimization rule internally traverses the query operators and takes effect on them.
These rules are categorized into parametric and non-parametric based on their configurability.
Each parametric rule is configured by several SQL parameters, which are tunable during runtime. 

Figure~\ref{fig:opt-rules} demonstrates rules applied to a logical query plan and a query stage during runtime.
In the transition to a physical query plan (or an optimized query stage), a logical query plan (or a runtime query stage) passes through a joint list of parametric rules (blue) and non-parametric rules (gray). 
Consequently, we divide the rules into {\it LQP rules} and {\it QS rules}.

\subsubsection{LQP Rules and $\thetabm_p$}
LQP rules include the logical and physical optimization rules, applied to the logical query plan during both the query compilation and runtime. 
Up to Spark 3.5, this comprises 23 non-parametric rules and 4 parametric rules applied in a deterministic order to the logical query plan.

We denote the union of SQL parameters appearing in all parametric LQP rules as $\thetabm_p$. 
Such parametric rules include the {\it dynamic join selection}, configured by an advisory partition size $s_1$ and a broadcast threshold $s_4$ among others. Additionally, {\it skew join optimization} is configured by another set of parameters, including a target partition size $s_1$, the initial partition numbers $s_5$, and skewed partition thresholds $s_6, s_7$. See Appendix~\ref{appendix:lpq-rules} for all LQP rules.

It is worth mentioning that we can use one copy of $\thetabm_p$ across LQP rules for each logical query plan, either at the compile time or runtime. 
For example, we use $\thetabm_{p0}$ for LQP rules applied to the initial logical query plan, and $\thetabm_{pt}$ for the $t$-th invocation of LQP rules for an updated logical query plan.
Collectively, we denote $\Thetabm_p$ as the set of all $\thetabm_p$ copies during the lifetime of a Spark SQL.


\subsubsection{QS Rules and $\thetabm_s$}
QS rules optimize a query stage during runtime. In Spark 3.5, there are 3 non-parametric rules and 2 parametric rules. 
One of the parametric rules, "rebalancing skewed partitions," splits skewed partitions into a target size $s_1$ and uses a rebalance small partition factor $s_{10}$ to determine whether the remainder after splitting should be merged.
The other parametric rule, {\it coalescing shuffle partitions}, reduces the number of small reduce tasks by coalescing small partitions to reach a target size $s_1$, which should be larger than a minimum partition size $s_{11}$. See Appendix~\ref{appendix:qs-rules} for all QS rules.

We represent the QS parameters as $\thetabm_s = (s_{10}, s_{11})$ and they are independently applied to each query stage as shown in Figure~\ref{fig:spark-parameters}. 
Collectively, we denote $\Thetabm_s$ as the set of $\thetabm_s$ copies across all query stages.
It is important to note that QS rules are designed under the assumption that LQP rules have been applied. Therefore, the target size $s_1$ is configured by LQP parameters and not included in $\thetabm_s$.

\cut{
\subsubsection{Mixed Decision Problem.} \todo{to be re-constructed}
 It is worth noting that the LQP parameters are applied to modify the PQP, and hence are PQP-dependent. In contrast, other parameters, such as the context parameters $\thetabm_c$ and $s_1$-$s_9$, do not impact the PQP and are PQP-independent.
Therefore, configuring Spark SQL involves making decisions in a {\it mixed decision space}, involving both the PQP topology and the PQP-independent parameters. 

Our work bypasses the mixed-decision problem by operating modeling and tuning over the LQP for the following reasons.
Firstly, the PQP-dependent parameters will not affect the topology of a LQP.
Secondly, when the LQP is provided, the PQP is highly correlated to the PQP-dependent parameters. E.g., we can derive the PQP given the LQP and PQP-dependent parameters with a 99.75\% accurate rate in TPCH over the 5-fold cross-validation.
Lastly, modeling over the LQP plus the PQP-dependent parameters achieves similar performance as modeling over the PQP, indicating that the simplification could be lossless.
See concrete analyses in Appendix~\ref{appendix:mixed-decision-space}.
}

\subsection{General Problem Definition}

Our general problem is a multi-objective optimization (MOO) problem that aims to minimize multiple objectives $\vec{o} = (o_1, ..., o_k)$ by tuning the Spark configuration $\thetabm = (\thetabm_c, \Thetabm_p, \Thetabm_s)$ for running Spark SQL. Each objective $o_i$ is represented by the function $f_i$.

The optimality of the problem relies on the {\bf oracle} information of the three {\it non-decision variables}: 
(1) the cardinality of each query operator $\alpha$,
(2) the distribution of the partition sizes after an operator requires data exchange $\beta$, and 
(3)	the resource contention status during the runtime $\gamma$.

\begin{definition}{\bf Multi-Objective Optimization for Spark SQL}\label{eq:moo-general}
\begin{align}
	\argmin_{\thetabm_c, \Thetabm_p, \Thetabm_s} & \; f(\thetabm_c, \Thetabm_p, \Thetabm_s) =  {\left[
		\begin{array}{l}
		f_1(LQP, \thetabm_c, \Thetabm_p, \Thetabm_s, \alpha, \beta, \gamma) \\
		... \\
		f_k(LQP, \thetabm_c, \Thetabm_p, \Thetabm_s, \alpha, \beta, \gamma) \\
		\end{array}
		\right]}\\
	\nonumber s.t. & {\begin{array}{l}
	    \thetabm_c \in \Sigma_c, \\
	    \Thetabm_p = [\thetabm_{p0}, \thetabm_{p1}, ...], \forall \thetabm_{pt} \in \Sigma_p \\
	    \Thetabm_s = [\thetabm_{s1}, \thetabm_{s2}, ...], \forall \thetabm_{si} \in \Sigma_s
		\end{array}}
\end{align}
\end{definition}
where $LQP$ denotes the logical query plan; 
$\thetabm_c, \Thetabm_p, \Thetabm_s$ represent the {\it decision variables} configuring Spark context, LQP rules, and QS rules across the lifecycle of an SQL;
$\Thetabm_{p}$ is the collection of all LQP parameters $\thetabm_{pt}$, where $\thetabm_{pt}$ is a unique copy of $\thetabm_p$ for the $t$-th invocation of LQP rules;
$\Thetabm_{s}$ is the collection of all QS parameters $\thetabm_{si}$, where $\thetabm_{si}$ is a unique copy of $\thetabm_s$ for optimizing query stage $i$;
$\Sigma_c, \Sigma_p, \Sigma_s$ are the feasible space for $\thetabm_c$, $\thetabm_p$ and $\thetabm_{s}$, respectively; 

\begin{figure}[t]
  \centering
  \includegraphics[width=.99\linewidth]{figs/initial-and-runtime-opt.pdf}
  \captionof{figure}{\small{Compile-time Optimization and Runtime Optimization}}
  \label{fig:opt-framework}
\end{figure}

\subsubsection{Compile Optimization and Runtime Optimization}
Since none of the non-decision factors are precisely provided before query execution, we introduce the concept of {\it initial optimization} during query compilation (compile time), and then iteratively perform {\it runtime optimization} using real-time statistics after each query stage concludes.

Figure~\ref{fig:opt-framework} shows the inputs and outputs of both optimizations. 
At compile time, we estimate the cardinality information by Spark's default cost-based optimizer ($\alpha=\alpha_{cbo}$) with the assumption of uniform data distribution ($\beta = \vec{0}$) and zero resource contention ($\gamma = \vec{0}$). 
The initial optimizer then resolves the physical query plan with an optimal choice of LQP parameters $\thetabm_{p0}^*$, as well as the context parameters $\thetabm_c^*$ and the stage parameters $\Thetabm_s^*$.

At runtime when $\thetabm_c$ is fixed, we construct the non-decision variables by the observed cardinality ($\alpha_t$), data distribution ($\beta_t$), and resource usage ($\gamma_t$) at the $t$-th invocation of the LQP rules for an updated logical query plan.
The runtime optimizer returns an updated physical query plan with an optimal $\thetabm_{pt}^*$ and updated optimal $\thetabm_s$ choices for the remaining query stages $\Thetabm_s^*$.

\subsubsection{Challenges}

Our design introduces two main challenges. 

\minip{Multi-granularity control.}
The three decision variables are correlated to provide the {\it multi-granularity control} to the query performance at both compile time and runtime.
At compile time, $\thetabm_c$ and $\thetabm_p$ configure the Spark context and the logical query plan respectively for the entire query, providing {\it query-level control}, while $\thetabm_s$ independently configures each query stage, granting {\it stage-level control}, as shown in Figure~\ref{fig:spark-parameters}.
During runtime when the logical query plan is updated and reduced to a subquery, $\thetabm_p$ provides a {\it subquery-level control} with $\thetabm_s$ still offering the stage-level control. 

\minip{Time Constraints.} 
$\thetabm_p$ should always be set ahead of $\Thetabm_s$. Otherwise, the physical query plan remains undetermined, and so do the set of query stages, their corresponding DAG structure, and QS parameters $\Thetabm_s$.

\cut{

The LQP parameters $\thetabm_p$ configure the LQP rules for both the initial and runtime logical query plans.
Accordingly, they provide {\it query-level control} at the compile time, and {\it subquery-level control} during runtime once the logical query plan is updated and reduced to a subquery.
The QS parameters $\thetabm_s$ independently configure the QS rules, granting the {\it stage-level control}. 
Those parameters are correlated and should be jointly considered during optimization.

Figure~\ref{fig:spark-parameters} illustrates the multi-granularity control through parameters at compile time, where we can set $\thetabm_c, \thetabm_p, \Thetabm_s$ by solving the initial optimization problem.
During query runtime, $\thetabm_c$ is fixed, but $\thetabm_p$ and $\Thetabm_s$ are still tunable, providing us with {\it time constraints} to tune $\thetabm_c$. 
Specifically, we have to tune $\thetabm_c$ in a joint space with $\thetabm_p$ and $\Thetabm_s$ during initial optimization and tune $\thetabm_p$ and $\Thetabm_s$ during runtime given a determined $\thetabm_c$.

Moreover, it is necessary to set $\thetabm_p$ before tuning $\Thetabm_s$. 
The physical query plan remains undetermined until $\thetabm_p$ is set, which also affects the number of query stages and their corresponding DAG structure.
For instance, increasing the broadcast threshold in $\thetabm_p$ for the query in Figure~\ref{fig:spark-parameters} could transform one of the sort-merge joins (SMJs) into a broadcast hash join (BHJ), leading to a reduction in the number of query stages as well as the size of $\Thetabm_s$.
}

\cut{
\subsection{Overview of Our Approach}

\subsubsection{Choices of Objectives}
In practice, with a special interest in improving performance with a budgetary constraint, our work consistently chooses latency and cloud cost as the two objectives, where the cloud cost is calculated as a weighted sum of CPU-hour and memory-hour~\cite{EMR-serverless}.

\subsubsection{Initial Optimization}

\input{qi-approach-overview}
\subsubsection{Runtime Optimization}
The runtime optimization takes place once the non-decision variables have been calibrated as $\alpha_t, \beta_t, \gamma_t$ using real-time statistics at the entry of $t$-th adaptive query execution (AQE), with the completion of the query stage $t$. 

Since resources have been determined by $\thetabm_c^*$ at the submission time, the cloud cost is proportional to latency, and the MOO problem can be reduced to a single objective optimization problem.

\begin{definition}{\bf Runtime Optimization}
\begin{align}
	\label{eq:moo-runtime}
	\argmin_{\thetabm_p, \Thetabm_s} & \; L = f(LQP^{(t)}, \thetabm_c^*, \thetabm_p, \Thetabm_s, \alpha_t, \beta_t, \gamma_t) \\
	\nonumber s.t. & {\begin{array}{l}
		\Thetabm_s = [\thetabm_{s(t+1)}, \thetabm_{s(t+2)}, ...]
		\end{array}}
\end{align}
where $LQP^{(t)}$ denotes the updated logical query plan at the entry of AQE $t$ and $\Thetabm_s$ is a collection of $\thetabm_s$ for the remaining query stages.
\end{definition}
}

\subsection{Intuitive Results}

We present results based on TPCH-Q5 to show the benefits of multi-granularity tuning over a joint parameter space ($\thetabm_c, \thetabm_p, \thetabm_s$) with compile time and runtime optimizations. 

\begin{figure*}
\centering
\captionsetup{justification=centering}
\begin{minipage}{.25\linewidth}
  \centering
  \includegraphics[width=.99\linewidth,height=3cm]{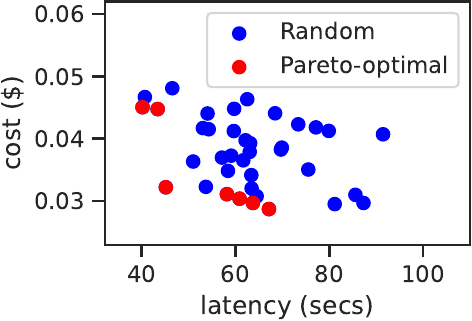}
  \captionof{figure}{\small{40 random sampled configurations (from $\theta_c$ and $\theta_p$) at compile time, in 2D objective space, when AQE is off}}
  \label{fig:intuition-A}
\end{minipage}
\hfill
\begin{minipage}{.42\linewidth}
  \centering
  \includegraphics[width=.99\linewidth,height=3cm]{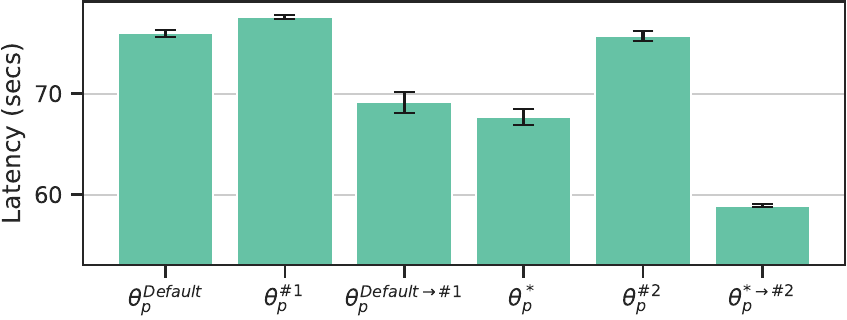}
  \captionof{figure}{\small{Latency against different $\theta_p$ choices at compile time and runtime, with default $\theta_c, \theta_s$ and AQE on. \\ Bars 1,2,4,5 do not change $\theta_p$ during runtime; \\ Bars 3 and 6 tune $\theta_p$ during runtime by increasing $s_4$.}}
  \label{fig:intuition-B}
\end{minipage}
\hfill
\begin{minipage}{.27\linewidth}
  \centering
  \includegraphics[width=.99\linewidth,height=3cm]{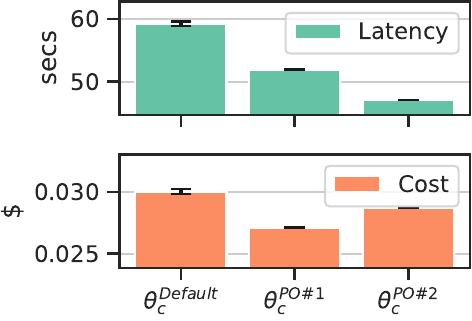}
  \captionof{figure}{\small{Better objectives achieved by tuning $\theta_c$ and $\theta_p$ jointly, at different granularities, with default $\theta_s$ and AQE on}}
  \label{fig:intuition-C}
\end{minipage}
\end{figure*}

First, tuning Spark parameters over a joint parameter space is crucial. 
Figure~\ref{fig:intuition-A} illustrates the latency and cost of 40 randomly sampled configurations (4 knobs from $\thetabm_c$ and 4 knobs from $\thetabm_p$) when the adaptive query execution (AQE) is turned off. 
7 configurations (highlighted in red) dominate most of those configurations, emphasizing the importance of tuning these parameters.

Second, we benefit from multi-granularity control in initial and runtime optimizations over $\thetabm_p$. 
In Figure~\ref{fig:intuition-B}, bars 1,2,4,5 show the latency of tuning $\thetabm_p$ only at compile time with the query-level control, while bars 3,6 include the runtime optimization for $\thetabm_p$ with additional subquery-level control by increasing the broadcast threshold $s_4$ after the first query stage completes.
Without runtime optimization, the optimal $\thetabm_p$ choice (bar 4) reduces 
11\% 
latency compared to the default (bar 1). 
Without initial optimization, the $\thetabm_p$ runtime tuning (bar 3) outperforms both individual $\thetabm_p$ choices (bar 1,2) by constructing a new physical query plan with 3 broadcast hash joins (BHJs) and 2 sort-merge joins (SMJs), which is not covered by bar 1 (2 BHJs and 3 SMJs with the default $\thetabm_p$) and bar 2 (4 BHJs and 1 SMJ for the updated $\thetabm_p$).
Moreover, jointly tuning $\thetabm_p$ with multi-granularity control at compile time and runtime (bar 6) yields a total reduction of 22\% latency compared to the default $\thetabm_p$.
\todo{to be replaced by a better example}

Finally, with the initial and runtime optimizations, better objective performances are achieved with multi-granularity control over a joint parameter space.
Configurations from a joint space of $\thetabm_c$ and $\thetabm_p$ can dominate the best result of only tuning $\thetabm_p$ at compile time and runtime, as shown in Figure~\ref{fig:intuition-C}.

\cut{
\subsubsection{AQE Optimality Analyses} 
\todo{AQE improves the optimality by calibrating cardinality and re-balancing data distribution during the runtime. However, it is sub-optimal with the current static parameter assignments.} 
}
%

\cut{
Spark SQL~\cite{spark-sql} seamlessly integrates SQL-based relational data processing into the Spark framework, inheriting the declarative nature and familiar syntax of SQL, by running over the Spark context.
Its query optimizer efficiently transforms user-defined SQL queries into optimized executable plans and RDD graphs, customized for the specific data and query.
The process involves three main steps:
(1) generating a {\it logical query plan (LQP)} from SQL, 
(2) optimizing the LQP based on a cost-based optimizer, guided by cardinality estimation, 
(3) and converting the optimized LQP to a {\it physical query plan (PQP)} using optimization rules and heuristics. 
The PQP also represents concrete execution steps, including
the selection of physical operators (e.g., \texttt{HashAgg} or \texttt{SortAgg}),
the decision-making on join strategies (e.g., \texttt{SortMergeJoin} or \texttt{BroadcastHashJoin}),
and the injection of the \texttt{Exchange} operators to divide itself into a directed acyclic graph (DAG) of distinct subquery components executed in different stages. 

The {\it Adaptive Query Execution (AQE)}, introduced in Spark 3.0 and set as the default since Spark 3.2, is a new feature in Spark SQL that dynamically optimizes query execution based on real-time runtime statistics~\cite{spark-aqe}. 
When AQE is enabled, Spark partitions the query execution into smaller units called {\it QueryStages} ({\it QS}s), each representing a distinct portion of a PQP bounded by \texttt{Exchange} operators (broadcast or shuffle).
The workflow begins by executing the leaf QSs, and iteratively triggers AQE runs upon QS materialization until all QSs are processed.
In each AQE run, runtime statistics from the preceding QSs are leveraged to re-optimize the remaining query plans, create QSs with cleared dependency, and apply further optimization rules for each created QS.
With its adaptability and utilization of runtime statistics, AQE significantly enhances query execution performance, making it a popular feature for efficient data processing in Spark SQL applications.

\subsection{Spark Configuration with Mixed Decisions}
Spark provides a variety of tunable configuration parameters, called knobs, enabling users to customize Spark applications. 

\subsubsection{Spark Configuration}
Based on the range of the parameter effects, we divide the Spark configuration $\thetabm$ into two distinct categories: the context configuration $\thetabm_c$ and the runtime SQL configuration $\thetabm_r$.
Our work focuses on 17 selected parameters~\cite{KanellisAV20}, covering both categories, as shown in Table~\ref{tab:spark-params}.

\minip{The context configuration $\thetabm_c$} characterizes the Spark context and is unchangeable throughout the execution of a SQL query in Spark.
It includes (1) the resource parameters $k_1$-$k_3$ to determine the number of executors, the number of cores per executor, and the memory size per executor, (2) the shuffle parameters $k_4$-$k_7$ to affect the shuffle behavior such as the maximum output size of mappers and whether to compress data for shuffling, and (3) the memory management parameter $k_8$ to control the split of memory for execution and storage.

\minip{The runtime SQL configuration $\thetabm_r$} governs the optimization rules applied to different procedures in AQE runs. It is designed to be adjustable within the Spark context, allowing modification during query execution. 
Specifically, AQE's logical optimizer applies parameters $s_1$-$s_4$ in a dynamic join selection rule to compose internal hints for each logical join operator;
after getting an initial PQP, parameters $s_1, s_5$-$s_7$ tweak the threshold for identifying a skewed join in a skewed join optimization rule; 
once QSs are created, parameters $s_1$, $s_8$-$s_9$ are applied on each individual QS to control their partition distributions.
In our work, we use $\thetabm_r^{(0)}$ to denote the SQL configuration at the submission time and use $\thetabm_r^{(t)}$ to denote the SQL configuration at the $t^{th}$ AQE run.

\subsubsection{Mixed Decision Problem}
 It is important to note that some parameters, such as $s_1$-$s_7$, can be applied in the rules to modify the PQP, and hence are PQP-dependent. In contrast, other parameters, such as the context parameters and $s_8$-$s_9$, do not impact the PQP and are PQP-independent.
Therefore, configuring Spark SQL involves making decisions in a {\it mixed decision space}, involving both the PQP topology and the PQP-independent parameters. See details in \ref{appendix:mixed-decision-space}.

Our work bypasses the mixed-decision problem by operating modeling and tuning over the LQP for the following reasons.
Firstly, the PQP-dependent parameters will not affect the topology of a LQP.
Secondly, when the LQP is provided, the PQP is highly correlated to the PQP-dependent parameters. E.g., we can derive the PQP given the LQP and PQP-dependent parameters with a 99.75\% accurate rate in TPCH over the 5-fold cross-validation.
Lastly, modeling over the LQP plus the PQP-dependent parameters achieves similar performance as modeling over the PQP, indicating that the simplification could be lossless.
See concrete analyses in Appendix~\ref{appendix:mixed-decision-space-simplification}.
}

\cut{
\subsection{Extended Spark Overview}


We present our configuration tuning design, which extends Spark with three key processes as shown in Figure~\ref{fig:system-overview}:

{\bf The bootstrap tuning process} involves a LQP extractor and the {\it Initial Optimizer (IO)}. The LQP extractor converts a submitted SQL to the initial LQP and feeds it to the IO. 
The IO then recommends the optimal context configuration $\thetabm_c$ and the initial SQL configuration $\thetabm_r^{(0)}$ based on the LQP by solving a MOO problem.

{\bf The runtime optimization process} is triggered by the event of query stage completions. The {\it Runtime Optimizer (RO)} fine-tunes the runtime SQL configuration $\thetabm_r$ within each AQE run by leveraging the runtime statistics and the updated LQP.

{\bf The model update process} involves a {\it Trace Collector} and a {\it Model Server}. The trace collector asynchronously gathers Spark traces (Step a), including the LQP and QS topologies and their corresponding performance data from AQE runs. The model server builds and updates performance models for both LQPs and QSs (Steps b,c), supporting bootstrap tuning and runtime optimization.

\subsubsection{Life Cycle of a SQL in Our Tuning Design.} 
A SQL initiates the bootstrap tuning process to get the recommended configuration ($\thetabm_c$ and $\thetabm_r^{(0)}$) for Spark submission (Steps 1-3 in Figure~\ref{fig:system-overview}). 
After the Spark submission and the creation of the Spark context, the default query optimizer translates the SQL to the initial executable plan, which is then directed to the AQE framework (Steps 4-5).
The AQE progressively executes the dependency-cleared QSs within the execution plan (Steps 6-7) until all QSs are completed. Each AQE run is triggered by QS completion and, with real-time statistics, involves SQL parameter fine-tuning by the RO and subsequent query re-optimization by Spark (Steps 8-9). 

\subsubsection{An AQE run with RO} \todo{todo}

%

%
%
%
}

\cut{
\subsection{MOO Problem Definition} 

Our problem is designed to optimize multiple query-level objectives by searching a joined space in parameters and physical plans.

\cut{
\begin{definition}{\bf Multi-Objective Optimization for Spark SQL}
\begin{align}
	\label{eq:moo-general}
	\argmin_{\thetabm_c, \thetabm_p, \Thetabm_s} & \; f(\thetabm_c, \thetabm_p, \Thetabm_s) =  {\left[
		\begin{array}{l}
		f_1(LQP, \thetabm_c, \thetabm_p, \Thetabm_s, \mathcal{P}, \delta) \\ 
		... \\
		f_k(LQP, \thetabm_c, \thetabm_p, \Thetabm_s, \mathcal{P}, \delta)
		\end{array}
		\right]}\\
	\nonumber s.t. & {\begin{array}{l}
	    \thetabm_c \in \Sigma_c, \thetabm_p \in \Sigma_p \\
	    \Thetabm_s = [\thetabm_{s1}, \thetabm_{s2}, ...], \thetabm_{si} \in \Sigma_s \\
	    LQP \in \mathcal{LQP}, \; \delta \in \Delta
		\end{array}}
\end{align}
\end{definition}
}

\begin{definition}{\bf Multi-Objective Optimization for Spark SQL}
\begin{align}
	\label{eq:moo-general}
	\argmin_{\thetabm_c, \thetabm_r} & \; f(\thetabm_c, \thetabm_r) =  {\left[
		\begin{array}{l}
		f_1(LQP, \thetabm_c, \thetabm_r, \mathcal{P}, \delta) \\ 
		... \\
		f_k(LQP, \thetabm_c, \thetabm_r, \mathcal{P}, \delta)
		\end{array}
		\right]}\\
	\nonumber s.t. & {\begin{array}{l}
	    \thetabm_c \in \Sigma_c, \thetabm_r \in \Sigma_r \\
	    LQP \in \mathcal{LQP}, \; \delta \in \Delta
		\end{array}}
\end{align}
\end{definition}
where 
\begin{enumerate}
	\item $\thetabm_c$ and $\thetabm_r$ denote the parameters to configure Spark context and the rule-based plans in Spark SQL, respectively.
	\item $F_1,...,F_k$ denote $k$ query-level target objectives.
	\item $LQP$ denotes the optimized logical query plan in the form of a graph given by Spark, where each node in the graph has the information of the operator type and cardinality.
	\item $\mathcal{P}$ denotes the rules in Spark's physical query planner and derives the physical query plan $PQP = \mathcal{P}_{\thetabm_r}(LQP)$. 
	\item $\delta$ denotes other factors that affect the performance, such as the machine system states and the input data information.
	\item $\Sigma_c$, $\Sigma_r$ are the feasible spaces for $\thetabm_c$ and $\thetabm_r$. $\mathcal{LQP}, \Delta$ are feasible spaces for $LQP$ and $\delta$.
\end{enumerate}

\minip{A na\"ive approach given the oracle cardinality.} \todo{(1) give the enum approaches. (2) challenges - card and complexity; (3) the chance to improve the issue; (4) clarify that improving the card est is an orthogonal issue to our work}

\newpage 
\subsubsection{Static Meta Optimization}
Suppose $f$ is the query latency predictive model for the initial LQP before running, and $g$ is the closed-form function to summarize the unit price of the resources. 

\begin{definition}{\bf Static Meta Optimization}
\begin{align}
	\label{eq:moo-static}
	\argmin_{\thetabm_c, \thetabm_q, \Thetabm_s} &  {\left[
		\begin{array}{l}
		L(\thetabm_c, \thetabm_q, \Thetabm_s) = f(LQP^{(0)}, \thetabm_c, \thetabm_q, \Thetabm_s, \mathcal{P}, \alpha_0, \beta_0, \gamma_0) \\ 
		C(\thetabm_c, \thetabm_q, \Thetabm_s) = L(\thetabm_c, \thetabm_q, \Thetabm_s) \cdot g(\thetabm_c)\\
		\end{array}
		\right]}\\
	\nonumber s.t. & {\begin{array}{l}
		\thetabm_c \in \Sigma_c, \thetabm_q \in \Sigma_q \\
		\Thetabm_s = [\thetabm_{s1}, \thetabm_{s2}, ...], \thetabm_{si} \in \Sigma_s \\
	    LQP^{(0)} \in \mathcal{LQP}
		\end{array}}
\end{align}
\end{definition}

\begin{definition}{\bf MOO over LQP (query level)}
\begin{align}
	\label{eq:moo-static}
	\argmin_{\thetabm_c, \thetabm_q, \thetabm_s} &  {\left[
		\begin{array}{l}
		L(\thetabm_c, \thetabm_q, \thetabm_s) = \color{blue}{f_{LQP}(LQP^{(0)}, \thetabm_c, \thetabm_q, \thetabm_s, \delta)} \\ 
		C(\thetabm_c, \thetabm_q, \thetabm_s) = L(\thetabm_c, \thetabm_q, \thetabm_s) \cdot g(\thetabm_c)\\
		\end{array}
		\right]}\\
	\nonumber s.t. & {\begin{array}{l}
		\thetabm_c \in \Sigma_c, \thetabm_q \in \Sigma_q, \thetabm_s \in \Sigma_s \\
	    LQP^{(0)} \in \mathcal{LQP}
		\end{array}}
\end{align}
\end{definition}

\begin{definition}{\bf MOO over PQP (stage level)}
\begin{align}
	\label{eq:moo-static}
	\argmin_{\Thetabm_s} &  {\left[
		\begin{array}{l}
		L(\thetabm_c^*, \Thetabm_s) = \color{red}{f_{PQP}(PQP^{(0)}, \thetabm_c^*, \Thetabm_s, \delta)} \\ 
		C(\thetabm_c^*, \Thetabm_s) = L(\thetabm_c^*,  \Thetabm_s) \cdot g(\thetabm_c^*)\\
		\end{array}
		\right]}\\
	\nonumber s.t. & {\begin{array}{l}
		\Thetabm_s = [\thetabm_s^1, \thetabm_s^2, ...], \thetabm_s \in \Sigma_s \\
	    |\Thetabm_s| = \textbf{num\_of\_qs}(PQP^{(0)})
		\end{array}}
\end{align}
\end{definition}

\begin{definition}{\bf Runtime Optimization}
\begin{align}
	\label{eq:moo-static}
	\argmin_{\thetabm_p, \Thetabm_s} &  {\left[
		\begin{array}{l}
		L = f(LQP^{(t)}, \thetabm_c^*, \thetabm_p, \Thetabm_s, \mathcal{P}, \delta_t) \\ 
		C = L \cdot g(\thetabm_c^*)\\
		\end{array}
		\right]}\\
	\nonumber s.t. & {\begin{array}{l}
		\thetabm_p \in \Sigma_p \\
		\Thetabm_s = [\thetabm_{s1}, \thetabm_{s2}, ...], \thetabm_{si} \in \Sigma_s
		\end{array}}
\end{align}
\end{definition}

where 
\begin{enumerate}
	\item $L$ and $C$ are two examples of objectives at the query level, latency and cloud cost.
	\item $LQP^{(0)}$ is the LQP with the initial cardinality estimations.
\end{enumerate}

\subsubsection{Adaptive Meta Optimization.} 
%
%
%

\begin{definition}{\bf Runtime Optimization}
\begin{align}
	\label{eq:moo-adaptive}
	\argmin_{\thetabm_p, \Thetabm_s} & \; L = f(LQP^{(t)}, \thetabm_c^*, \thetabm_p, \Thetabm_s, \mathcal{P}, \alpha_t, \beta_t, \gamma_t)\\
	\nonumber s.t. & {\begin{array}{l}
		\thetabm_p \in \Sigma_p \\
		\Thetabm_s = [\thetabm_{s1}, \thetabm_{s2}, ...], \thetabm_{si} \in \Sigma_s
		\end{array}}
\end{align}
\end{definition}

\begin{definition}{\bf Runtime Optimization}
\begin{align}
	\label{eq:moo-adaptive}
	\argmin_{\thetabm_p, \thetabm_s} & \; L = f(LQP^{(t)}, \thetabm_c^*, \thetabm_p, \thetabm_s, \mathcal{P}, \alpha_t, \beta_t, \gamma_t)\\
	\nonumber s.t. & {\begin{array}{l}
		\thetabm_p \in \Sigma_p \\
		\thetabm_s = \thetabm_{s1} = \thetabm_{s2} = ... \in \Sigma_s
		\end{array}}
\end{align}
\end{definition}

\begin{definition}{\bf Qurey Stage Optimization}
\begin{align}
	\label{eq:moo-adaptive}
	\argmin_{\thetabm_{si}} & \; L_{QS} = g(QS_{i}, \thetabm_c^*, \thetabm_{si}, \alpha_i, \beta_i, \gamma_i)\\
	\nonumber s.t. & {\begin{array}{l}
		\thetabm_{si} \in \Sigma_s
		\end{array}}
\end{align}
\end{definition}

\begin{definition}{\bf Adaptive Meta Optimization}
\begin{align}
	\label{eq:moo-adaptive}
	\argmin_{\thetabm_p, \thetabm_s} & \; L(\thetabm_p, \thetabm_s) = \color{blue}{f_{LQP}(LQP^{(t)}, \thetabm_c^*, \thetabm_p, \thetabm_s, \delta_t)}\\
	\nonumber s.t. & {\begin{array}{l}
	    LQP^{(t)} \in \mathcal{LQP}
		\end{array}}
\end{align}
\end{definition}

\begin{definition}{\bf Adaptive Meta Optimization}
\begin{align}
	\label{eq:moo-adaptive}
	\argmin_{\Thetabm_s} & \; L(\Thetabm_s) = \color{red}{f_{PQP}(PQP^{(t)}, \thetabm_c^*, \Thetabm_s, \delta_t)}\\
	\nonumber s.t. & {\begin{array}{l}
	    LQP^{(t)} \in \mathcal{LQP}
		\end{array}}
\end{align}
\end{definition}

where 
\begin{enumerate}
	\item $L$ denotes the latency of $LQP^{(t)}$, ranging from the entry of the AQE to the completion of the entire query. 
	\item $\thetabm_c^*$ is from the solutions of static meta optimization.
	\item $S_t$ and $\delta_t$ are the shared-resource execution status for this job and other systematic factors at the AQE entry $t$.
\end{enumerate}


\begin{definition}{\bf Multi-Objective Optimization for Spark SQL}
\begin{align}
	\label{eq:general-moo}
	\argmin_{\thetabm_c, \thetabm_r^{(0)}, \Thetabm_r} & \; F(\thetabm_c, \thetabm_r^{(0)}, \Thetabm_r) =  {\left[
		\begin{array}{l}
		F_1(LQP^{(0)}, \thetabm_c, \thetabm_r^{(0)}, \Thetabm_r, \mathcal{P}, \delta) \\ 
		... \\
		F_k(LQP^{(0)}, \thetabm_c, \thetabm_r^{(0)}, \Thetabm_r, \mathcal{P}, \delta)
		\end{array}
		\right]}\\
	\nonumber s.t. & {\begin{array}{l}
	    \thetabm_c \in \Sigma_c, \thetabm_r^{(0)} \in \Sigma_r \\
	    \Thetabm_r = [ \thetabm_r^{(1)}, \thetabm_r^{(2)}, ...], \; \thetabm_r^{(t)} \in \Sigma_r,\; \\
	    LQP^{(0)} \in \mathcal{LQP}, \; \delta \in \Delta
		\end{array}}
\end{align}
\end{definition}
where 
\begin{enumerate}
	\item $F_1,...,F_k$ denote the $k$ target objectives; 
		  $LQP^{(0)}$ is the initial logical query 
		  $\thetabm_c$ and $\thetabm_r^{(0)}$ denote the context configuration and the initial runtime SQL configuration;
		  $\Thetabm_r$ denotes the set of runtime SQL configurations at each AQE entry.
		  $\mathcal{P}$ denotes the planning rules in Spark's query optimizer;
		  $\delta$ denotes the dynamic systematic factors, including the most recent machine system states and the input data information.
	\item $\Sigma_c$, $\Sigma_r$ are the feasible spaces for $\thetabm_c$ and $\thetabm_r^{(t)} $. $\mathcal{LQP}, \Delta$ are feasible spaces for $LQP^{(0)}$ and $\delta$. 
	\item $t$ indicates the AQE runs during execution.
	\end{enumerate}

\begin{definition}{\bf Static Meta Optimization}
\begin{align}
	\label{eq:static-moo}
	\argmin_{\thetabm_c, \thetabm_r^{(0)}, \Thetabm_r} & \; F(\thetabm_c, \thetabm_r^{(0)}, \Thetabm_r) =  {\left[
		\begin{array}{l}
		\hat{F}_1(LQP^{(0)}, \thetabm_c, \thetabm_r^{(0)}, \Thetabm_r, \mathcal{P}, \delta) \\ 
		... \\
		\hat{F}_k(LQP^{(0)}, \thetabm_c, \thetabm_r^{(0)}, \Thetabm_r, \mathcal{P}, \delta)
		\end{array}
		\right]}\\
	\nonumber s.t. & {\begin{array}{l}
	    \thetabm_c \in \Sigma_c, \thetabm_r^{(0)} \in \Sigma_r \\
	    \Thetabm_r = [ \thetabm_r^{(1)}, \thetabm_r^{(2)}, ...], \; \thetabm_r^{(t)} \in \Sigma_r,\; \\
	    LQP^{(0)} \in \mathcal{LQP}, \; \delta \in \Delta
		\end{array}}
\end{align}
\end{definition}

\begin{definition}{\bf Multi-Objective Optimization for Spark SQL}
\begin{align}
	\label{eq:general-moo}
	\argmin_{\thetabm_c, \Thetabm_r, \Thetabm_s} & \; F(\thetabm_c, \Thetabm_r, \Thetabm_s) =  {\left[
		\begin{array}{l}
		F_1(LQP^{(0)}, \thetabm_c, \Thetabm_r, \Thetabm_s, \delta) \\ 
		... \\
		F_k(LQP^{(0)}, \thetabm_c, \Thetabm_r, \Thetabm_s, \delta)
		\end{array}
		\right]}\\
	\nonumber s.t. & {\begin{array}{l}
	    \thetabm_c \in \Sigma_c \\
	    \Thetabm_r = [ \thetabm_r^{(0)}, \thetabm_r^{(1)}, \thetabm_r^{(2)}, ...], \; \thetabm_r^{(t)} \in \Sigma_r,\; \\
	    \Thetabm_s = [ \thetabm_s^{(0)}, \thetabm_s^{(1)}, \thetabm_s^{(2)}, ...], \; \thetabm_s^{(i)} \in \Sigma_s,\; \\
	    LQP \in \mathcal{LQP}, \; \delta \in \Delta
		\end{array}}
\end{align}
\end{definition}
where 
\begin{enumerate}
	\item $F_1,...,F_k$ denote the $k$ target objectives; $LQP^{(0)}$ is the initial logical query plan; $\thetabm_c$ and $\thetabm_r^{(0)}$ denote the context configuration and the initial runtime SQL configuration; $\delta$ denotes the dynamic systematic factors including the most recent machine system states and the input data information.
	\item $\Sigma_c$, $\Sigma_r$, $\Sigma_s$ are the feasible spaces for $\thetabm_c$, $\thetabm_r^{(t)} $, and $\thetabm_s^{(i)}$. $\mathcal{LQP}, \Delta$ are feasible spaces for $LQP$, $\delta$. 
	\item $t$ and $i$ indicate the number of AQE runs and the id of QSs. 
	\end{enumerate}

\begin{definition}{\bf Multi-Objective Optimization for Spark SQL}
\begin{align}
	\label{eq:general-moo}
	\argmin_{\thetabm_c, \thetabm_r^{(0)}, \pi} & \; F(\thetabm_c, \thetabm_r^{(0)}, \pi) =  {\left[
		\begin{array}{l}
		F_1(LQP^{(0)}, \thetabm_c, \thetabm_r^{(0)}, \pi, \delta) \\ 
		... \\
		F_k(LQP^{(0)}, \thetabm_c, \thetabm_r^{(0)}, \pi, \delta)
		\end{array}
		\right]}\\
	\nonumber s.t. & {\begin{array}{l}
	    \thetabm_c \in \Sigma_c \\
	    \thetabm_r^{(0)} \in \Sigma_r \\
	    \pi:(\mathcal{LQP}, \Sigma_c, \Delta) \rightarrow \Sigma_r
		\end{array}}
	\end{align}
\end{definition}
where 
\begin{enumerate}
	\item $F_1,...,F_k$ denote the $k$ target objectives; $LQP^{(0)}$ is the initial logical query plan; $\thetabm_c$ and $\thetabm_r^{(0)}$ denote the context configuration and the initial runtime SQL configuration; $\delta$ denotes the dynamic systematic factors including the most recent machine system states and the input data information.
	\item $\Sigma_c$ and $\Sigma_r$ are the feasible spaces for $\thetabm_c$ and $\thetabm_r^{t} (t\geq 0)$. $\mathcal{LQP}, \Delta$ are feasible spaces for $LQP$, $\delta$. 
	\item $\pi$ is the strategy to optimize the runtime SQL parameters at each AQE run. Specifically, in the $t^{th}$ AQE run, $\theta_r^{(t)} = \pi(LQP^{(t-1)}, \theta_c, \delta)$, where $\delta = \delta^{(t)}$ for this particular run. 
	\end{enumerate}

In practice, our initial optimizer resolves the initial configuration $\thetabm_0$ and our RO applies the optimization strategy $\pi$ at runtime to configuration $\thetabm_r$ at each AQE run.

\subsubsection{Initial Optimizer} {\ }\\
\input{qi-initial-optimizer}

\subsubsection{Runtime Optimizer}
Note that we have particular interests in optimizing latency and cloud cost in our MOO problem. In the production environment, cloud cost is calculated as the weighted sum of CPU-hours and memory-hours. 
When $\thetabm_c$ is fixed with CPU and memory determined, cloud cost becomes directly proportional to query latency. 
As a result, the MOO problem concerning latency and cloud cost can be reduced to a single objective optimization problem during the runtime.

\begin{definition}{\bf Runtime Optimization over $LQP^{(t)} $(\texttt{RO-AQE})}
\begin{align}
	\label{eq:rt-pi1}
	\thetabm_{AB}^*, \Thetabm_{s}^* &= \argmin_{\thetabm_{AB}, \Thetabm_{s}} L_{LQP}(\thetabm_{AB}, \Thetabm_{s}, LQP^{(t)}, \thetabm_c, \delta) \\
	\nonumber s.t. & {\begin{array}{l}
	    \Thetabm_{s} = ||_{\forall QS^{(i)} \in LQP^{(t)}} \thetabm_s^{(i)}, \; \thetabm_s^{(i)} \in \Sigma_s \\
	    \thetabm_{s.comm}^{(i)} \in h_i(\thetabm_{AB})
		\end{array}}
	\end{align}
\end{definition}

\begin{definition}{\bf Runtime Optimization over LQP (\texttt{RO-1})}
\begin{align}
	\label{eq:rt-pi1}
	\thetabm_{r}^*	&= \argmin_{\thetabm_{r}} L_{LQP}(\thetabm_{r}, LQP^{(t)}, \thetabm_c, \mathcal{P}, \delta) \\
	\nonumber s.t. & {\begin{array}{l}
	    \thetabm_{r} \in \Sigma_{r}
		\end{array}}
	\end{align}
\end{definition}

where $\thetabm_{r}^*$ is the optimal configuration for the runtime LQP in this AQE run, $L_{LQP}$ is the latency of a runtime LQP ranging from the entry time of this AQE to the completion of the entire query, and $\thetabm_{r}^*$ is assumed to be fixed for the remaining AQE runs.

\begin{definition}{\bf Runtime Optimization over QS (\texttt{RO-2})}
\begin{align}
	\label{eq:rt-pi2}
	\thetabm_{s}^* &= \argmin_{\thetabm_s = \thetabm_{s.comm} \cup \thetabm_{s.indp}} L_{QS}(\thetabm_s, QS^{(i)}, \thetabm_c, \delta) \\
	\nonumber s.t. & {\begin{array}{l}
	    \thetabm_{s.comm} \in h(\thetabm_r^*) \\
	    \thetabm_{s} = \thetabm_{s.comm} \cup \thetabm_{s.indp} \in \Sigma_{s} \\
	    QS \in \mathcal{QS}
		\end{array}}
	\end{align}
\end{definition}

where $\thetabm_{C}^*$ is the optimal configuration for a QS, including the QS-independent parameters $\thetabm_{qsIndp}$ that are only applied for the QS (in Step $\mathcal{C}$) and the common parameters $\thetabm_{comm}$ shared with other steps in $\mathcal{A}$ and $\mathcal{B}$, $L_{QS}$ is the latency of the QS, and $\mathcal{QS}$ and $\Sigma_C$ are the feasible spaces for QS and $\thetabm_C$ respectively.
Notice that in some scenarios of $\tilde{\thetabm_r}$, $\thetabm_{comm}$ can pick among a range of values without changing the outcomes of the applied logical and physical rules in the procedure $\mathcal{A}$ and $\mathcal{B}$, and we define such value range by the function $h(\tilde{\thetabm_r})$. 
In our case, we have $\thetabm_{comm} = \{s_1\}$, $\thetabm_{qsIndp} = \{s_8, s_9\}$ and $h(\tilde{\thetabm})$ as a decision tree shown in Figure~\todo{add}.
}


%% file: checkpoint-231115/modeling.tex
\section{Modeling}

We provide the logical query plan (LQP) and query stage (QS) models to serve the multi-granularity control in optimizations.

\subsection{Feature Extraction}

\subsubsection{Operator Encoding} 
The features of each operator are concatenated by 
(1) the operator type (one-hot encoding),
(2) the cardinality estimation (row count, size in bytes); 
(3) the predicate embedding constructed by a pre-trained word embedding model that transforms the operator predicates (e.g., the column predicates) into a numeric vector.
\todo{to be polished}



\subsubsection{Query Embedding} 
We provide a DAG-based embedder to transform a query plan (LQP/QS) as a DAG of operators into a vector. 
\todo{More descriptions needed}

\subsubsection{Other Non-decision Variables Encoding} 
We introduce encoding methods for input size ($\alpha$),  data distribution ($\beta$), and runtime contention ($\gamma$).

$\alpha$ is encoded in the query level as the input characteristics, by \red{aggregating the row count and sizes from the plan's leaf operators}.

$\beta$ is the distribution of the partition sizes after an operator requires data exchange, such as shuffling. It is encoded as a numeric vector that captures 
\begin{itemize}
	\item the std-avg ratio ($\frac{\sigma}{\mu}$)
	\item the skewness ratio ($\frac{\max-\mu}{\mu}$)
	\item the range-avg ratio ($\frac{\max-\min}{\mu}$)
\end{itemize}
In the case of a uniform distribution, we set $\beta = \vec{0}$.


Finally, we represent the runtime contention $\gamma$ by analyzing the real-time resource usage. This includes 
\begin{itemize}
	\item the number of running tasks
	\item the number of finished tasks
	\item the total running time of finished tasks
	\item various statistics of finished task running times, such as the minimum, $1^{st}$ quartile, median, $3^{rd}$ quartile, and the maximum.
\end{itemize}

\subsection{Model Structure}

\todo{apply the multi-channel input framework}

\subsubsection{LQP Model}

\todo{to prepare LQP model $\LQP$ at the query level}

\subsubsection{QS Model}

\todo{to prepare QS model $\QS$ at the stage level}

\begin{figure*}
\centering
\captionsetup{justification=centering}
\begin{minipage}{.24\textwidth}
  \centering
  \includegraphics[height=3.5cm,width=3.3cm]{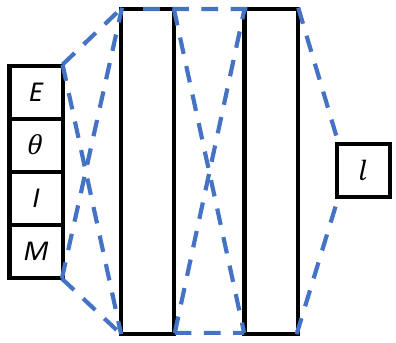}
  \captionof{figure}{\small{An MLP Module as \\the Neural Network Regressor}}
  \label{fig:model-nnr}
\end{minipage}
\begin{minipage}{.48\textwidth}
  \centering
  \includegraphics[height=3.5cm]{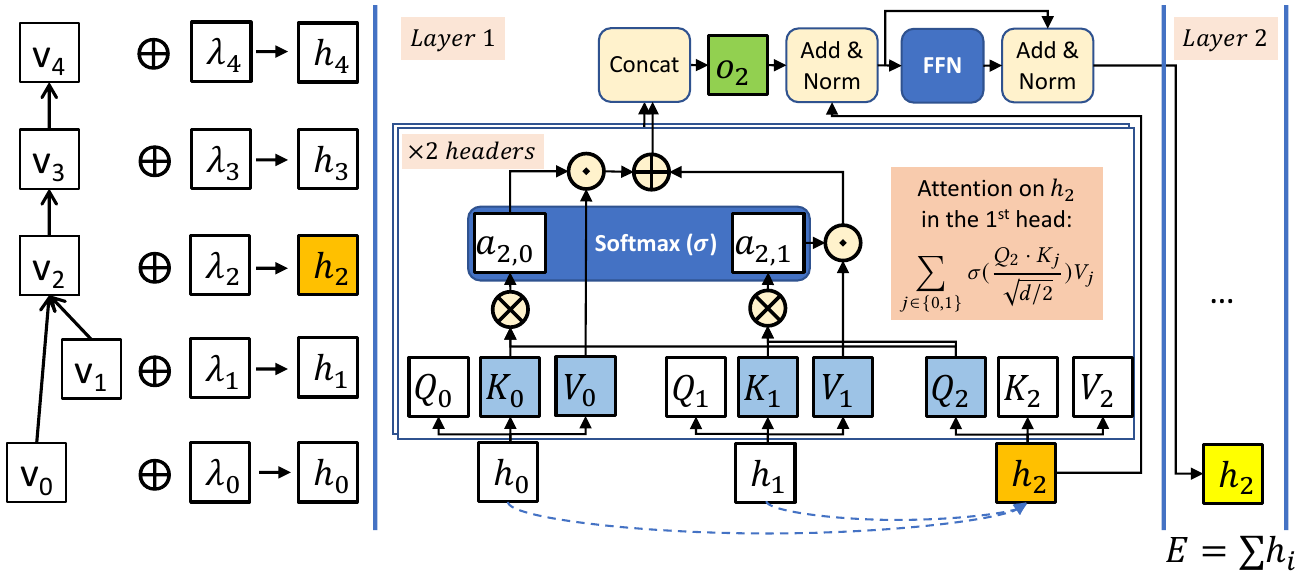}
  \captionof{figure}{\small{A GTN Example with 2 Heads and 2 Layers, \\with the focus on the forward path for $v_2$}}
  \label{fig:model-gtn-example}
\end{minipage}
\begin{minipage}{.24\textwidth}
  \centering
  \includegraphics[height=3.5cm,width=3.3cm]{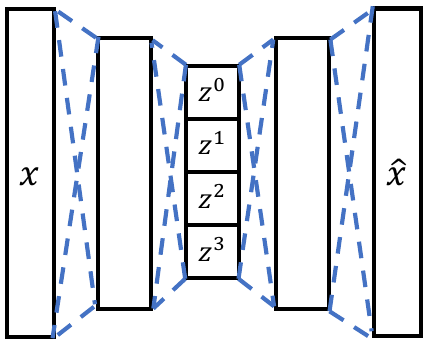}
  \captionof{figure}{\small{An Auto-encoder \\ as the Query Embedder}}
  \label{fig:model-ae}
\end{minipage}
\end{figure*}

\cut{
\subsection{Model Structure Overview}

Modeling the latency over big data analytics systems needs to consider factors like the query characteristics, the input data, machine systems states, and runtime parameters (a.k.a. configurations). 
Therefore, we apply the multi-channel input framework~\cite{RO-alibaba} to build a {\bf neural network regressor} as follows\footnote{we simplified our model by not considering the factor of hardware properties because our Spark cluster is deployed over nodes with identical hardware.}
\begin{align}
	l &= f(E, I, M, \bs{\theta})
\end{align}
where we denote $E$ as the query embedding that captures the characteristics of the query $Q$, $I$ as the input meta information from the distributed storage layer, $M$ as the machine system state before executing $Q$, and $\bs{\theta}$ as the configuration.
As shown in Figure~\ref{fig:model-nnr}, the multi-channel input framework takes input as the concatenation of features from multiple sources. It applies a multilayer perceptron (MLP) for inferencing.

While factors like M, S, and $\bs{\theta}$ are accessible directly from the system, capturing the query embedding $E$ is non-trivial. One needs to take not only the query plan structure but also the characteristics of each physical query operator into consideration.

Therefore, we design tailored {\bf query embedders} for pure SQLs, SQLs with UDFs, and SQLs with ML tasks.
Specifically, we build the graph transformer network (GTN)~\cite{gtn-aaai21} as the embedder for pure SQLs and an auto-encoder (AE) as the embedder for SQLs with UDFs \todo{and SQLs with ML tasks}.

\begin{table}[t]
\ra{1.2}
\small
\caption{Notations}
\label{tab:spark-notations}
\centering
	\begin{tabular}{cl}\toprule
	Symbol & Description \\\midrule
$Q$ & a query (Spark SQL) \\
$E$ & the query plan embedding for $Q$ \\
$I$ & input meta information for $Q$\\
$M$ & system states before running $Q$\\
$\bs{\theta}$ & the configuration \\
\midrule\
$S=\{s_i\}$ & a set of stages (subqueries) in query $Q$ \\
$|S|$ & the number of stages in $Q$ \\
$G$ & the stage topology in $Q$, a DAG \\
$E_i$ & embedding for subquery plan for stage $s_i$ \\
$I_i$ & input meta information for $s_i$\\
$M_i$ & system states before running $s_i$ \\
\midrule
$l$, $c$ & the e2e latency and cost of running query $Q$ \\
$l_{s_i}$, $c_{s_i}$ & the e2e latency and cost of running stage $s_i$ \\
$t_{s_i}$ & the latency summation of all the tasks in stage $s_i$ \\
$f$ & the e2e latency model \\
$g$ & the cost model \\
$h$ & \begin{tabular}{@{}l@{}}the latency model for the summation of all task \\ latencies in a stage  \end{tabular} \\
$\phi$ & \begin{tabular}{@{}l@{}}map the latency from a DAG of stages to the query \\ when the internal Spark scheduling mode is FIFO/FAIR \end{tabular}\\
\bottomrule
	\end{tabular}
\end{table}
}

%% file: checkpoint-231115/optimization.tex
\section{Optimization}

\todo{reorg for multi-granularity control}

\subsection{Compile-time Optimization}

\begin{definition}{\bf Compile-time Coarse-grained Optimization}\label{eq:moo-compile-step1}
\begin{align}
	\argmin_{\thetabm_c, \thetabm_p, \thetabm_s} & \; 
	{\left[
		\begin{array}{l}
		\phi_{lat}(LQP, \thetabm_c, \thetabm_p, \thetabm_s, \alpha, \beta, \gamma) \\
		\phi_{cost}(LQP, \thetabm_c, \thetabm_p, \thetabm_s, \alpha, \beta, \gamma) \\
		\end{array}
		\right]}\\
	\nonumber s.t. & {\begin{array}{l}
	    \thetabm_c \in \Sigma_c, \thetabm_{p} \in \Sigma_p, \thetabm_{s} \in \Sigma_s
		\end{array}}
\end{align}
\end{definition}

\begin{definition}{\bf Compile-time Fine-grained Optimization}\label{eq:moo-compile-step2-op1}
\begin{align}
	\argmin_{\thetabm_{si}} & \; \psi_{lat}(QS_i, \thetabm^*_c, \thetabm_{si}, \alpha, \beta, \gamma) \\
	\nonumber s.t. & {\begin{array}{l}
	    \thetabm_{si} \in \Sigma_s
		\end{array}}
\end{align}
\end{definition}

\begin{definition}{\bf Compile-time Fine-grained Optimization}\label{eq:moo-compile-step2-opt2}
\begin{align}
	\argmin_{\thetabm_c, \Thetabm_s} & \; {\left[
		\begin{array}{l}
		\sum_i \psi_{lat}(QS_i, \thetabm_c, \thetabm_{si}, \alpha, \beta, \gamma) \\
		\sum_i \psi_{cost}(QS_i, \thetabm_c, \thetabm_{si}, \alpha, \beta, \gamma) \\
		\end{array}
		\right]}\\
	\nonumber s.t. & {\begin{array}{l}
	    \thetabm_c \in \Sigma_c \\
	    \Thetabm_s = [\thetabm_{s1}, \thetabm_{s2}, ...], \forall \thetabm_{si} \in \Sigma_s
		\end{array}}
\end{align}
\end{definition}

\subsection{Runtime Optimization}

\begin{definition}{\bf Runtime Coarse-grained Optimization}\label{eq:moo-runtime-step1}
\begin{align}
	\argmin_{\thetabm_q, \thetabm_{s}} & \; \phi_{lat}(LQP^{(t)}, \thetabm^*_c, \thetabm_p, \thetabm_{s}, \alpha, \beta, \gamma) \\
	\nonumber s.t. & {\begin{array}{l}
		\thetabm_{p} \in \Sigma_p, 
	    \thetabm_{s} \in \Sigma_s
		\end{array}}
\end{align}
\end{definition}

\cut{
\begin{definition}{\bf The latency-cost bi-objective optimization}
	\begin{align}
		\label{eq:1-boo}
		\argmin_{\thetabm = [\thetabm_q, \thetabm_{s_1}, ..., \thetabm_{s_n}]} & {\left[
			\begin{array}{l}
			L(\thetabm) = f(\dot{E}, \dot{I}, \dot{M}, \thetabm) \\ 
			C(\thetabm) = L(\thetabm) \cdot c(\thetabm_q)
			\end{array}
			\right]}\\
		\nonumber s.t. & {\begin{array}{l}
		    \thetabm_q \in \Sigma_q \subset \R^{d_1} \\
		    \thetabm_{s_i} \in \Sigma_s \subset \R^{d_2}, \;\; \forall i \in [n] \\
		    \thetabm \in \Sigma \subset \R^{d_1 + n\cdot d_2}
			\end{array}}
	\end{align}
	
where $c: \R^{d_1} \rightarrow \R$ is the mapping from the resource parameters in $\thetabm$ to the resource price per time unit;
$t$ denotes the number of join operators in a query $Q$;
$n$ denotes the number of stages in a query $Q$, which depend on the choices of some plan-dependent parameters in each $\thetabm_s$;
 $\dot{E}$, $\dot{I}$ and $\dot{M}$ denote the encoding of $Q$, the input data information and most recent machine system states; $\Sigma_q$ and $\Sigma_s$ are the feasible parameter values for the selected general Spark parameters $\thetabm_q$ ($d_1$-dim) and the selected runtime query parameters $\thetabm_{s_i}$ ($d_2$-dim) respectively.
\end{definition}

\begin{definition}{\bf The latency-cost optimization with a proxy configuration $\tilde{\thetabm} = [\thetabm_q, \thetabm_s]$}
	\begin{align}
		\label{eq:1-boo}
		\argmin_{\tilde{\thetabm} = [\thetabm_q, \thetabm_{s}]} & {\left[
			\begin{array}{l}
			L(\tilde{\thetabm}) = f(\dot{E}, \dot{I}, \dot{M}, [\thetabm_q, \underbrace{\thetabm_{s}, ..., \thetabm_{s}}_n]) \\ 
			C(\tilde{\thetabm}) = L(\tilde{\thetabm}) \cdot c(\thetabm_q)
			\end{array}
			\right]}\\
		\nonumber s.t. & {\begin{array}{l}
		    \thetabm_q \in \Sigma_q \subset \R^{d_1} \\
		    \thetabm_{s} \in \Sigma_s \subset \R^{d_2}
			\end{array}}
	\end{align}
	
where $\tilde{\thetabm} = (\thetabm_q, \thetabm_{s})$ is the proxy configuration of $\thetabm$ such that $\thetabm_{s} = \thetabm_{s_1} = ... = \thetabm_{s_n}$.
Notice that $L(\tilde{\thetabm}) \leq L(\thetabm)$\todo{add a proof in appendix}; Therefore, this step minimizes the upper bound of the query latency and cost.
\end{definition}
}

\begin{definition}{\bf The latency-cost bi-objective optimization}
	\begin{align}
		\label{eq:1-boo}
		\argmin_{\thetabm = [\thetabm_q, \thetabm_{s}^0, ..., \thetabm_{s}^{T}]} & {\left[
			\begin{array}{l}
			L(\thetabm) = f(\dot{E}, \dot{I}, \dot{M}, \thetabm) \\ 
			C(\thetabm) = L(\thetabm) \cdot c(\thetabm_q)
			\end{array}
			\right]}\\
		\nonumber s.t. & {\begin{array}{l}
		    \thetabm_q \in \Sigma_q \subset \R^{d_1} \\
		    \thetabm_{s}^t \in \Sigma_s \subset \R^{d_2}, \;\; \forall t \in [T] 
			\end{array}}
	\end{align}
	
where $c: \R^{d_1} \rightarrow \R$ is the mapping from the resource parameters in $\thetabm$ to the resource price per time unit;
$T$ denotes the number of times the adaptive query execution (AQE) is performed for runtime optimization, and $t$ is the index of AQE iteration;
 $\dot{E}$, $\dot{I}$ and $\dot{M}$ denote the encoding of $Q$, the input data information and most recent machine system states; $\Sigma_q$ and $\Sigma_s$ are the feasible parameter values for the selected general Spark parameters $\thetabm_q$ ($d_1$-dim) and the selected runtime query parameters $\thetabm_{s}^{t}$ ($d_2$-dim) respectively. The $\thetabm_{s}^t$ includes three sets of runtime parameters $(\thetabm_{s_A}^t, \thetabm_{s_B}^t, \thetabm_{s_C}^t)$ aiming at tuning (A) the remaining logical query plan, (B) the remaining physical query plan, and (C) the next query stage respectively.
\end{definition}

\begin{definition}{\bf The latency-cost optimization with a proxy configuration $\tilde{\thetabm} = [\thetabm_q, \thetabm_s]$}
	\begin{align}
		\label{eq:2-boo-proxy}
		\argmin_{\tilde{\thetabm} = [\thetabm_q, \thetabm_{s}]} & {\left[
			\begin{array}{l}
			L(\tilde{\thetabm}) = f(\dot{E}, \dot{I}, \dot{M}, [\thetabm_q, \underbrace{\thetabm_{s}, ..., \thetabm_{s}}_{T+1}]) \\ 
			C(\tilde{\thetabm}) = L(\tilde{\thetabm}) \cdot c(\thetabm_q)
			\end{array}
			\right]}\\
		\nonumber s.t. & {\begin{array}{l}
		    \thetabm_q \in \Sigma_q \subset \R^{d_1} \\
		    \thetabm_{s} \in \Sigma_s \subset \R^{d_2}
			\end{array}}
	\end{align}
	
where $\tilde{\thetabm} = (\thetabm_q, \thetabm_{s})$ is the proxy configuration of $\thetabm$ such that $\thetabm_{s} = \thetabm_{s}^0 = ... = \thetabm_{s}^T$.
Notice that $L(\tilde{\thetabm}) \leq L(\thetabm)$\todo{add a proof in appendix}; Therefore, this step minimizes the upper bound of the query latency and cost.
\end{definition}

\newpage
\begin{definition}{\bf runtime optimization over the updated logical query plan $LQP_t$.}
	\begin{align}
		\label{eq:3-boo-rt-A}
		\argmin_{\thetabm_{s_A}} \;\;\;\;& L(\thetabm_{s_A}) = f(\dot{E}_{LQP_t}, \dot{I}_{LQP_t}, \dot{M}_{LQP_t}, \thetabm_{s_A})
			\\
		\nonumber s.t. \;\;\;\;& \thetabm_{s_A} \in \Sigma_{s_A} \subset \R^{d_{21}}
	\end{align}
\end{definition}

\begin{definition}{\bf runtime optimization over the updated physical query plan $PQP_t$.}
	\begin{align}
		\label{eq:3-boo-rt-B}
		\argmin_{\thetabm_{s_B}} \;\;\;\;& L(\thetabm_{s_B}) = f(\dot{E}_{PQP_t}, \dot{I}_{PQP_t}, \dot{M}_{PQP_t}, \thetabm_{s_B})
			\\
		\nonumber s.t. \;\;\;\;& \thetabm_{s_B} \in \Sigma_{s_B} \subset \R^{d_{22}}
	\end{align}
\end{definition}

\begin{definition}{\bf runtime optimization over the created query stage $QS$.}
	\begin{align}
		\label{eq:3-boo-rt-C}
		\argmin_{\thetabm_{s_C}} \;\;\;\;& L(\thetabm_{s_C}) = f(\dot{E}_{QS}, \dot{I}_{QS}, \dot{M}_{QS}, \thetabm_{s_C})
			\\
		\nonumber s.t. \;\;\;\;& \thetabm_{s_C} \in \Sigma_{s_C} \subset \R^{d_{23}}
	\end{align}
	
where $\thetabm_s = [\thetabm_s^A, \thetabm_s^B, \thetabm_s^C] \in \Sigma_s$ and $d_2 = d_{21} + d_{22} + d_{23}$.
\end{definition}

We aim to tune the configurations for running a specific query to optimize multiple objectives, e.g., minimizing job-level latency and cloud cost. Based on our tuning experiences, we picked the 12 most important knobs in Table~\ref{tab:spark-knobs} and further grouped the 12 knobs into three categories in Table~\ref{tab:spark-knob-category}.

\begin{table}[t]
\ra{1.2}
\small
\caption{Selected Spark Knobs}
\label{tab:spark-knobs}
\centering
	\begin{tabular}{cl}\toprule
\verb|k1| & spark.executor.memory \\
\verb|k2| & spark.executor.cores \\
\verb|k3| & spark.executor.instances \\
\verb|k4| & spark.defalut.parallelism \\
\verb|k5| & spark.reducer.maxSizeInFlight \\
\verb|k6| & spark.shuffle.sort.bypassMergeThreshold \\
\verb|k7| & spark.shuffle.compress \\
\verb|k8| & spark.memory.fraction \\
\verb|s1| & spark.sql.inMemoryColumnarStorage.batchSize \\
\verb|s2| & spark.sql.files.maxPartitionBytes \\
\verb|s3| & spark.sql.autoBroadcastJoinThreshold \\
\verb|s4| & spark.sql.shuffle.partitions
\\\bottomrule
	\end{tabular}
\end{table}

\begin{table}[t]
\ra{1.2}
\small
\caption{Spark Knob Category}
\label{tab:spark-knob-category}
\centering
	\begin{tabular}{cl}\toprule
	Symbol & Description \\\midrule
$\bs{x}$ & the list of the resource knobs (\verb|k1, k2, k3|) \\
$\bs{y}$ & the list of the other non SQL knobs (\verb|k4-k8|) \\
$\bs{z}$ & the list of the SQL knobs (\verb|s1, s2, s3, s4|) \\
$\bs{\theta} = [\bs{x}, \bs{y}, \bs{z}]$ & the list of the all 12 important knobs \\
\bottomrule
	\end{tabular}
\end{table}

\subsection{Treat Query as a Whole}

\sketch{\minip{\texttt{opt1}}}: apply the model aiming at the job-level objectives and recommend its configuration $\bs{\theta}^*$ by PF-AP + MOGD + WUN.

\subsection{Treat Query as a DAG of Stages}

Notice that modeling at the query level involves more uncertainties than at the stage level. For example, the query-level latency $l_Q$ could involve the pending time for triggering each stage. The pending time is hard to be predicted and hence increases more variance for modeling the query-level objective.

In this subsection, we aim to recommend more robust configurations based on the objective models at the stage level. 

\todo{$\Lambda$. how to get the job-level latency from the stages? 
The resources applied to a stage can change during its life cycle when multiple stages run in parallel. Consider two ways, (1) from $l_{s_i}$, (2) from $t_{s_i}$.
}

\subsubsection{solve $\bs{\theta}$ based on the recommended choice for each stage}
{\ }
\sketch{\minip{\texttt{opt2}}}: pick the optimal Spark configuration $\bs{\theta}^*$ from the recommended configurations for each stage. Notice that Spark deploys the same configuration for all stages by default, although the recommended configuration differs among different stages. 

\begin{enumerate}
	\item Apply \texttt{opt1} to get the job-level recommendation $\bs{\theta}_0 = [\bs{x}_0, \bs{y}_0, \bs{z}_0]$.
	\item Fix $\bs{y}_0, \bs{z}_0$, and get the recommended resource knobs $\bs{x}_i^*$ for each stage $s_i$ by \red{PF-AP + MOGD + WUN}.
	\item Get $\bs{x}^*$ for all stages by minimizing both the latency and cost for the query $Q$
		\begin{align}
			\bs{x}^* = \mathop{\arg\min}_{\red{\bs{x} \in \{\bs{x}_i^* |_{i=1,...,|S|}\}}} \begin{cases} 
				\Lambda(G, f_1(\bs{x}), f_2(\bs{x}), ..., f_{|S|}(\bs{x}))\\ 
				\sum_{i=1}^{|S|} g(E_i, I_i, M_i, [\bs{x}, \bs{y}_0, \bs{z}_0])
			\end{cases}
		\end{align}
		where we have $f_i(\bs{x}) = f(E_i, I_i, M_i, [\bs{x}, \bs{y}_0, \bs{z}_0])$.
	\item $\bs{\theta}^* = [\bs{x}^*, \bs{y}_0, \bs{z}_0]$.
\end{enumerate}

\subsubsection{solve $\bs{\theta}$ based on the Pareto-optimal choices for each stage}
{\ }
\sketch{\minip{\texttt{opt3}}}: pick the optimal Spark configuration $\bs{\theta}^*$ from the \po configurations for each stage. 
\begin{enumerate}
	\item Apply \texttt{opt1} to get the job-level recommendation $\bs{\theta}_0 = [\bs{x}_0, \bs{y}_0, \bs{z}_0]$.
	\item Fix $\bs{y}_0, \bs{z}_0$, and get a set of \po resource knobs $\red{\bs{X}_i^* = \{ \bs{x}_{i,1}^*, ..., \bs{x}_{i,p_i}^*\}}$ for each stage $s_i$ by \red{PF-AP + MOGD}, where $p_i$ is the number of \po solutions for $s_i$.
	\item Get $\bs{x}^*$ for all stages by minimizing both the latency and cost for the query $Q$
		\begin{align}
			\bs{x}^* = \mathop{\arg\min}_{\red{\bs{x} \in \bigcup_{i=1}^{|S|}\bs{X}_i^* }} \begin{cases} 
				\Lambda(G, f_1(\bs{x}), f_2(\bs{x}), ..., f_{|S|}(\bs{x}))\\ 
				\sum_{i=1}^{|S|} g(E_i, I_i, M_i, [\bs{x}, \bs{y}_0, \bs{z}_0])
			\end{cases}
		\end{align}
		where we have $f_i(\bs{x}) = f(E_i, I_i, M_i, [\bs{x}, \bs{y}_0, \bs{z}_0])$.
\end{enumerate}

\subsubsection{Spark extension and the further optimization}
{\ }\\

Spark default does not distinguish the configuration (12 knobs) among different stages in a submitted application (a Spark SQL).
To further optimize the query performance, our work adds customized rules via the Spark extension API such that the SQL knobs ($\bs{z}$) can be tuned at the stage level when AQE is on.

With our extension, \texttt{opt2} and \texttt{opt3} can be extended accordingly.

\sketch{\minip{\texttt{opt2*} and \texttt{opt3*}}}: pick the optimal Spark configuration with Spark extension, where SQL knobs $\bs{z}$ can be set differently among stages.
\begin{enumerate}
	\item Get $\bs{x}^*, \bs{y}_0, \bs{z}_0$ from \texttt{opt2} or \texttt{opt3}.
	\item Fix $\bs{x}^*, \bs{y}_0$, and get the optimal SQL knobs $\bs{z}_i$ for each stage $s_i$ by minimizing $l_{s_i}$ (and hence $c_{s_i}$)
		\begin{align}
			\bs{z}_i^* = \arg\min_{\bs{z}}f(E_i, I_i, M_i, [\bs{x}^*, \bs{y}_0, \bs{z}]), \forall i = 1,...,|S|
		\end{align}
	\item the optimal configuration is $[\bs{x}^*, \bs{y}_0, \{ \bs{z}_i^* |_{i = 1,...,|S|}\}]$
\end{enumerate}
It is worth mentioning that when the resource knobs are fixed as $\bs{x}^*$, the cost is linear to the latency. Therefore, minimizing latency will also minimize the cost for $Q$.

\cut{
\subsection{Technical Issues}

\minip{Issue 1}.
We have to trigger two Spark applications for running one Spark SQL unless we assume the query plans of all queries are calculated in advance (which is hard to guarantee in the production workload).
\begin{itemize}
	\item To get the physical query plan, one needs to trigger the Spark application to create a Spark context and go through the parser, analyzer, optimizer and planner in the Spark engine. \color{red}{solved}.
	\item Once the Spark context (where we specified all the general knobs) is created for a Spark application, its resource-related knobs are no longer changeable. \color{red}{solved}.
	\item It is discouraged that running multiple Spark Context in one Spark submission\footnote{https://issues.apache.org/jira/browse/SPARK-2243}.
\end{itemize}
}

\cut{
\subsection{Per-workload Tuning}

\todo{a potential extension to another multi-granularity tuning problem (workload-query)}

\minip{Workload Discussion.} In a production environment, a significant number of queries are injected daily, including both ad-hoc queries and periodic queries. It is important to consider a workload as a collection of periodic queries injected within the same period or session. This approach ensures that the insights derived from analyzing the historical workload remain relevant and applicable in the near future.

\minip{Comparison with Related Work.} Previous projects such as Ottertune~\cite{VanAken:2017:ADM} and CDBTune~\cite{Zhang:2019:EAC} have primarily focused on tuning database systems at the workload level. However, these approaches have overlooked several opportunities to optimize queries at the query-level and stage-level in the context of big data systems. \todo{In our work, we ...}
}

%% file: checkpoint-231115/Qi_initial_optimizer/initial_optimization.tex
\section{Initial Optimization}

In this section, we present the definitions, optimization problems and the corresponding algorithm design for \textit{Initial Optimizer} in Figure \ref{fig:model-nnr}. 

\input{Qi_initial_optimizer/definitions.tex}

\input{Qi_initial_optimizer/overview_ideas.tex}

\input{Qi_initial_optimizer/overall_algo.tex}

%% file: checkpoint-231115/Qi_initial_optimizer/definitions.tex
\subsection{Definitions}

We introduce our associated terminology here.

{\definition{\textbf{Tuning Parameter}}: A tuning parameter is a configuration variable that decides the performance of a big data system for an application.} 

A Spark physical plan of a query could be represented as a Directed Acyclic Graph (DAG) of stages. Any arbitrary stage $s$ has two types of tuning parameters, i.e. \textbf{Shared Parameters} for the Spark context $\bm{\theta_c}$ and \textbf{Stage-specific Parameters} $\bm{\theta_s}$, where $\bm{\theta_c} \in \mathbb{R}^{d_c}$ should be the same among all stages, and $\bm{\theta_s} \in \mathbb{R}^{d_2}$ could be different for different stages. So, the parameter space of a stage includes $d_c + d_2$ parameters, and the parameter space of a query includes $d_c + |S| * d_2$ parameters when there are $|S|$ stages. 

{\definition{\textbf{Configuration}}: A configuration is a constant vector, which sets a specific value for each tuning parameter under a set of fixed and ordered parameters. The \textbf{Configuration Space} is the set including all possible configurations.} 

For example, a configuration for the query with $|S|$ stages is a ($d_c + |S| * d_2$)-dim constant vector and a configuration for an arbitrary stage is a ($d_c + d_2$)-dim constant vector. 

Our goal is to find the Pareto optimal configurations for a query by considering multiple user-specified objectives. This query-level MOO problem could be addressed in either \textbf{Query-level tuning} or \textbf{Stage-level tuning}.

{\definition{\textbf{Query-level tuning}}: Query-level tuning solves the query-level MOO problem directly, which treats ($\bm{\theta_c}, \bm{\theta_1}, ..., \bm{\theta_{|S|}}$) for all $|S|$ stages in a query as the input, and its output is the Pareto optimal configurations for this query. When $|S|$ is large, its parameter space $d_c + |S| * d_2$ could be large (e.g. hundreds or thousands of parameters).}

By query-level tuning, we can define a general query-level MOO problem as follows:

{\definition{\textbf{General Query-level Multi-Objective Optimization (MOO).}
		\begin{eqnarray}
		\label{eq:1_general_query_mult_obj_opt_def}
		\arg \min_{\bm{\theta}} &F(\bm{\theta})=& {\left[
			\begin{array}{l}
			F_1(\bm{\theta}) = \Psi_1(\bm{\alpha}, \bm{\theta_c},\bm{\theta_1},...,\bm{\theta_{|S|}}) \\ 
			...\\
			F_k(\bm{\theta}) = \Psi_k(\bm{\alpha}, \bm{\theta_c},\bm{\theta_1},...,\bm{\theta_{|S|}})
			\end{array}
			\right]} \\
		\nonumber s.t. & 
		& \bm{\theta_c} \in  \mathbb{R}^{d_c} \quad \bm{\theta_s} \in  \mathbb{R}^{d_2}, \quad s = 1,..., |S|\\
		\end{eqnarray}
}}

\noindent{ 
where we denote $\bm{\alpha}$ as the non-decision variables required in the query-level predictive model, and $\bm{\theta} = [\bm{\theta_c},\bm{\theta_1},...,\bm{\theta_{|S|}}]$ is configurations for the whole query, $\Psi_j$ is the query-level predictive model for $j$-th objective, where $j=1,...,k$.}

{\definition{\textbf{Stage-level tuning}}: Stage-level tuning addresses the MOO problem within a stage, and its output is the Pareto optimal configurations for each stage, where its parameters space is fixed with $d_c + d_2$ parameters. To achieve the query-level Pareto optimal configurations, it is necessary to design an algorithm to optimally compose stage-level configurations into the query level.}

By stage-level tuning, we can define the above query-level MOO in \ref{eq:1_general_query_mult_obj_opt_def} as the following:

{\definition{\textbf{Query-level Multi-Objective Optimization (MOO).}
		\begin{eqnarray}
		\label{eq:sep_query_mult_obj_opt_def}
		\arg \min_{\bm{\theta}} &F(\bm{\theta})={\left[
			\begin{array}{l}
			F_1(\bm{\theta}) = \Lambda_1(\Phi_1(\bm{\alpha_1}, \bm{\theta_c}, \bm{\theta_1}),...,\Phi_1(\bm{\alpha_{|S|}}, \bm{\theta_c},\bm{\theta_{|S|}})) \\ 
			...\\
			F_k(\bm{\theta}) = \Lambda_k(\Phi_k(\bm{\alpha_1}, \bm{\theta_c},\bm{\theta_1}),...,\Phi_k(\bm{\alpha_{|S|}},\bm{\theta_c},\bm{\theta_{|S|}}))
			\end{array}
			\right]} \\
		\nonumber s.t. & \bm{\theta_c} \in  \mathbb{R}^{d_c} \quad\bm{\theta_s} \in  \mathbb{R}^{d_2}, s = 1, ..., |S|\\
		\end{eqnarray}
}}

\noindent{
where we denote $\bm{\alpha_s}$ as the non-decision variables required in the predictive model of the stage $s$, $\bm{\theta} = [\bm{\theta_c},\bm{\theta_1},...,\bm{\theta_{|S|}}]$ is configurations for all stages within a query, $\Phi_j$ is the stage-level predictive model of  $j$-th objective, where $j=1,..,k$. $\Lambda_j$ is the general function mapping from the stage-level values to the $j$-th query-level objective values, which follows the calculation of the topological order of DAG.}

For an arbitrary stage $s$, after stage-level tuning, its Pareto optimal configurations for ($\bm{\theta_c}, \bm{\theta_s}$) have the following properties.

\minip{Property 1:} For optimal $\bm{\theta_c}$, under different stages, optimal $\bm{\theta_c}$ after \textbf{stage-level tuning} could be different as the non-decision variables $\bm{\alpha_s}$ of different stages are different.

\minip{Property 2:} For optimal $\bm{\theta_s}$ under a fixed $\bm{\theta_c}$ configuration,

\begin{itemize}
	\item Optimal $\bm{\theta_s}$ could have multiple configurations as tuning $\bm{\theta_s}$ under a fixed $\bm{\theta_c}$ configuration is a MOO problem.
	\item For the same stage, due to different $\bm{\theta_c}$ configurations, optimal $\bm{\theta_s}$ configurations could be different.
	\item Under different stages with a fixed $\bm{\theta_c}$, optimal $\bm{\theta_s}$ configurations could be different as the non-decision variables $\bm{\alpha_s}$ of different stages are different.
\end{itemize}

The above properties and the requirement that $\bm{\theta_c}$ configurations should be the same among all the stages make
\textbf{stage-level tuning} more complex with different control on $\bm{\theta_c}$ and $\bm{\theta_s}$.
While this could be addressed by an efficient algorithm design, named \textbf{2-step divided and conquer MOO approach}, which will be explained in later sections. The following section will illustrate the intuitions related to the \textbf{stage-level tuning} with theoretical analysis.

%% file: checkpoint-231115/Qi_initial_optimizer/overview_ideas.tex
\subsection{Overview}

This subsection introduces an overview of the challenges and main ideas for the algorithm design.

\subsubsection{Challenges} ~

\textbf{Query-level tuning} solves the parameter tuning of the whole query directly, while it suffers from the high-dimensional parameter space which increases with the increasing of the number of stages. \textbf{Stage-level tuning} could address the high-dimensional problem by solving the subproblems for each stage in the query, where the parameter space is fixed by $d_c + d_2$. 

Due to the properties of \textbf{stage-level tuning}, the algorithm design faces the following three challenges regarding the optimality and efficiency of being Pareto optimal in the query level.

\minip{Challenge 1: $\bm{\theta_c}$ candidates generation} It is important to generate the $\bm{\theta_c}$ candidates likely to be the Pareto optimal in the query level as all stages in the query should follow the same $\bm{\theta_c}$ configuration. Since optimal $\bm{\theta_c}$ returned by  \textbf{stage-level tuning} could be different for different stages, a strategy is needed to ensure that all stages should have the same $\bm{\theta_c}$ configuration.   Moreover, solutions may be missed if only considering optimal $\bm{\theta_c}$ configurations from \textbf{stage-level tuning}. For example, in Figure \ref{fig:eg_missing_solutions}, solution 6 does not include the same $\bm{\theta_c}$ configuration of any stage after \textbf{stage-level tuning}, while it could dominate the solution 4 obtained based on the \textbf{stage-level tuning}. Overall, it is challenging to find the $\bm{\theta_c}$ configurations contributed to the query-level Pareto optimal. 

\minip{Challenge 2: $\bm{\theta_s}$ tuning} For each $\bm{\theta_c}$ configuration, tuning the corresponding $\bm{\theta_s}$ for a stage $s$ is a MOO problem. For each $\bm{\theta_c}$ configuration, only solutions with optimal $\bm{\theta_s}$ are able to contribute to the query-level Pareto optimal. This property is proved in \ref{s_tuning}, which indicates that the query-level Pareto optimality actually relies on the optimality of $\bm{\theta_c}$ candidates generation when their corresponding $\bm{\theta_s}$ configuration is optimal. This also explains why $\bm{\theta_c}$ candidates generation is cruitial for the algorithm design. While $\bm{\theta_c}$ selection could possibly keep hundreds or thousands of $\bm{\theta_c}$ candidates, to achieve a better efficiency, it is challenging to just keep configurations likely to be the query-level Pareto optimal.

\minip{Challenge 3: Optimization over DAG} A query-plan is represented as a DAG graph of stages. For all stages, given solutions with optimal $\bm{\theta_s}$ of all $\bm{\theta_c}$ candidates, the complexity of getting the optimal query-level Pareto optimal could be exponential to the number of stages. For example, we denote the maximnum number of solutions with optimal $\bm{\theta_s}$ for one arbitrary $\bm{\theta_c}$ among all $|S|$ stages in a query as $p_{max}$. The complexity could be  $p_{max}^{|S|}$ with enumeration, which is not acceptable in practice.  Thus, it is non-trivial to solve the optimization over the DAG of stages even given the optimal configurations of each stage due to the efficiency.

\subsubsection{Main ideas} ~

Our proposed algorithm, named \textbf{2-step divided and conquer MOO approach}, is designed to solve the above challenges. As we discussed previous in Challenge 3, even given the optimal configurations $\bm{\theta_c}$ with corresponding $\bm{\theta_s}$, the optimization over DAG of stages is  non-trivial. So in our algorithm design, instead of going over the DAG iteratively to find optimal $\bm{\theta_c}$ and $\bm{\theta_s}$ configurations, we proposed to generate all possibly optimal $\bm{\theta_c}$ and $\bm{\theta_s}$ first, and then visit the graph just once to get the query-level Pareto optimal configuration.

The algorithm design mainly includes three components.

\minip{Component 1:} This part deals with the Challenge 1, which aims to generate $\bm{\theta_c}$ candidates which are likely to be Pareto optimal in the query-level. First, we union the $\bm{\theta_c}$ from stage-level Pareto optimal solutions of all stages together to form the initial $\bm{\theta_c}$ candidates. In section \ref{q_selection}, we proved that $\bm{\theta_c}$ configurations unioned from all stages could provide a warm-start of optimal $\bm{\theta_c}$ candidates. Seconly, inspired by the crossover operation of evolutionary algorithms, we treat the initial $\bm{\theta_c}$ candidates as the population with good genes to do crossover further. The crossover operation finally is able to generate new samples with new $\bm{\theta_c}$ configurations.
Details will be discussed in the subsection \ref{q_selection}. 

\minip{Component 2:} This part focuses on the Challenge 2, which aims to get optimal $\bm{\theta_s}$ of all $\bm{\theta_c}$ candidates for all stages efficiently. 

Related work \cite{SinhaMD15} indicates a possible direction to deal with this efficiency issue, which aims to reduce the number of $\bm{\theta_s}$ tuning that results in computational savings. It proposed to estimate the optimal $\bm{\theta_s}$ by using quadratic approximation for each variable of $\bm{\theta_s}$ under an unknown $\bm{\theta_c}$, rather than to tune it. That is, for each stage, $\bm{\theta_s}$ tuning is made only for a subset of all $\bm{\theta_c}$ candidates, and then based on the optimal $\bm{\theta_s}$ and its corresponding $\bm{\theta_c}$ configurations, the relationship between each variable of $\bm{\theta_s}$ and $\bm{\theta_c}$ are captured by a quadratic function $f_c$, i.e. $\bm{\theta^z_s}=f_c(\bm{\theta_c})$, where $z \in [1, d_c]$. The estimation of all optimal $\bm{\theta_s}$ variables under an unknown $\bm{\theta_c}$ configuration is obtained by the quadratic functions. However, it is not available in our problem as fitting quadratic functions already introduces extra complexity. Apart from that, the accuracy of the quadratic functions also infludence the performance a lot, which is not controllable. 

To tackle the efficiency issue, we proposed a heuristic method called \textit{tolerance-bound}. The aim is to filter the $\bm{\theta_c}$ candidates not likely to be the query-level Pareto optimal.
Details will be discussed in the subsection \ref{s_tuning}. 

\minip{Component 3:} Challenge 3 is addressed in this part. Given the optimal configurations (with all optimal $\bm{\theta_c}$ and the corresponding $\bm{\theta_s}$ configurations) of all stages, we proposed an efficient algorithm which guarantees to be query-level Pareto optimal with the given input. It is inspired by the observation that parallel and list stages can be compressed to one pseudo stage efficiently while keeping optimality.
Details will be discussed in section \ref{dag_opt}. 

\cut{
\todo{todo: whether properties exist to find optimal $\bm{\theta_c}$ (component 1) and eliminate similar $\bm{\theta_c}$ (component 2)}
}

%% file: checkpoint-231115/Qi_initial_optimizer/overall_algo.tex
\subsection{Algorithm design}


Our proposed algorithm, named \textbf{2-step divided and conquer MOO approach}, is designed to solve the above challenges. As we discussed in the previous section, even given the optimal configurations $\bm{\theta_c}$ with corresponding $\bm{\theta_s}$, the optimization over DAG of stages is  non-trivial. So in our algorithm design, instead of going over the DAG iteratively to find optimal $\bm{\theta_c}$ and $\bm{\theta_s}$ configurations, we proposed to generate all possibly optimal $\bm{\theta_c}$ and $\bm{\theta_s}$ first, and then visit the graph just once to get the query-level Pareto optimal configuration.

\input{Qi_initial_optimizer/q_selection.tex}

\input{Qi_initial_optimizer/s_tuning.tex}

\input{Qi_initial_optimizer/dag_opt.tex}

%% file: checkpoint-231115/Qi_initial_optimizer/q_selection.tex
\subsubsection{${\theta_c}$ Candidates Generation} \label{q_selection} ~

Tuning each stage is a general MOO problem, which can be regarded as one subproblem to be solved by existing MOO algorithms. The aim of tuning each stage is to generate $\bm{\theta_c}$ candidates which possibly to be the Pareto optimal in the query level. Since tuning each stage is independent, it can be executed parallelly.

The following shows the corresponding intuitions with theoretical analysis.

\minip{Intuition 1:} \textbf{Stage-level tuning} could provide a warm-start to achieve the query-level Pareto optimal.

\begin{figure}
  \centering
  \begin{tabular}{ll}
		\subfigure[\small{Stage 1}]
		{\label{fig:local_pareto_stage_1}\includegraphics[width=.45\linewidth]{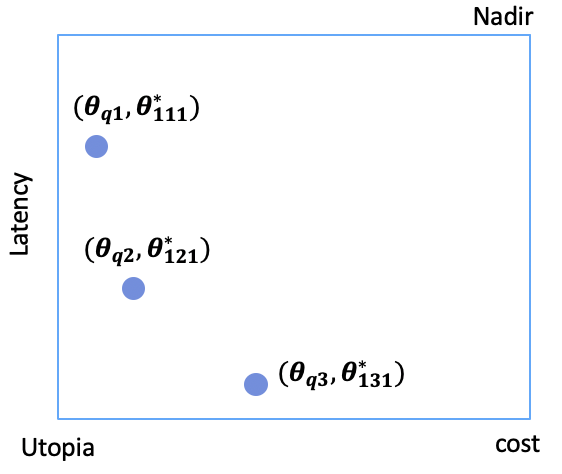}}
		\subfigure[\small{Stage 2}]
		{\label{fig:local_pareto_stage_2}\includegraphics[width=.45\linewidth]{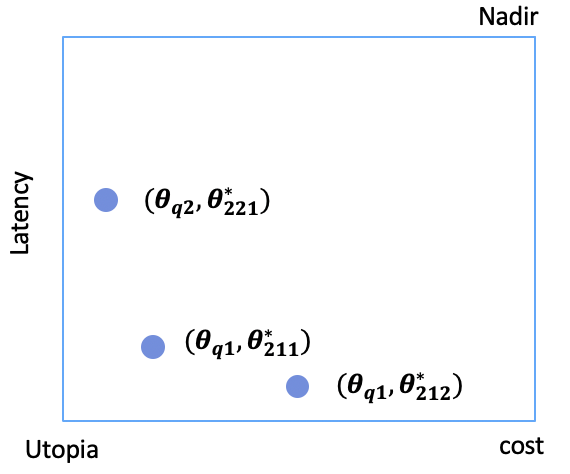}} 					  
  \end{tabular}
  \vspace{-0.2in}
  \captionof{figure}{\small{Example of stage-level Pareto solutions after Stage-level tuning}}
  \label{fig:eg_stage_tuning}
\end{figure}

\minip{Example} ~

Figure \ref{fig:eg_stage_tuning} shows an example of a query with 2 stages. Each subplot shows the objective space of a stage, which is bounded by Utopia and Nadir points, where Utopia achieves the minimum values in both latency and cost objectives, and Nadir achieves the maximum. Each blue point is a Pareto optimal point returned after stage-level tuning. 

We denote the $\bm{\theta_c}$ configurations from the Pareto optimal solutions of stage $s$ as the $\bm{cq_s}$, where $s=1,2$ in Figure \ref{fig:eg_stage_tuning}. For example, in stage 1, $\bm{cq_1}=\{\bm{\theta_{c1}}, \bm{\theta_{c2}}\}$; in stage 2, $\bm{cq_2}=\{\bm{\theta_{c1}}, \bm{\theta_{c2}}, \bm{\theta_{c3}}\}$. 
As the $\bm{\theta_c}$ configurations should be the same among all stages, we denote the union of $\bm{cq_s}$ as a $\bm{\theta_c}$ candidate set $\bm{\Theta^{cand}_c}=\{ \bm{\theta^*_{ci}}, i=1,...,|\bm{\Theta^{cand}_c}| \}$, which provides a global set for all stages. In Figure \ref{fig:eg_stage_tuning}, $\bm{\Theta^{cand}_c}=\{ \bm{\theta_{c1}}, \bm{\theta_{c2}}, \bm{\theta_{c3}} \}$, where $\bm{\theta^*_{c1}}=\bm{\theta_{c1}}, \bm{\theta^*_{c2}}=\bm{\theta_{c2}}, \bm{\theta^*_{c3}}=\bm{\theta_{c3}}$. 
We denote the $j$-th optimal stage-specific configuraion under $\bm{\theta^*_{ci}}$ for stage $s$ as $\bm{\theta^*_{sij}}$, where $j=1,...,p_i$, and $p_i$ is the number of optimal stage-level specific configurations under $\bm{\theta^*_{ci}}$. For example, in stage 1, under $\bm{\theta_{c1}}$, $p_1=2$, the second optimal $\bm{\theta_s}$ configuration is $\bm{\theta^*_{112}}$.

\begin{figure}
  \centering
  \begin{tabular}{ll}
		\subfigure[\small{Stage 1}]
		{\label{fig:global_pareto_stage_1}\includegraphics[width=.45\linewidth]{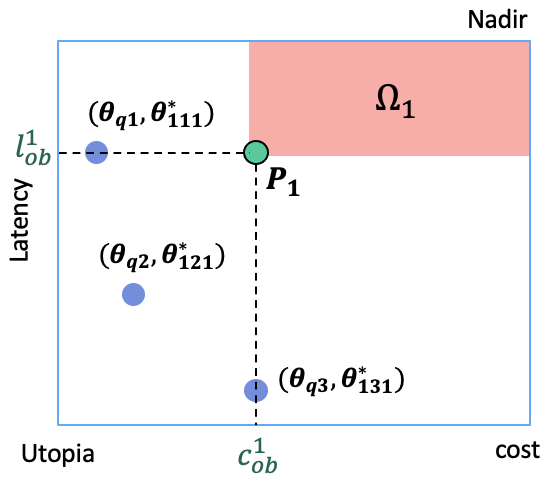}}
		\subfigure[\small{Stage 2}]
		{\label{fig:global_pareto_stage_2}\includegraphics[width=.45\linewidth]{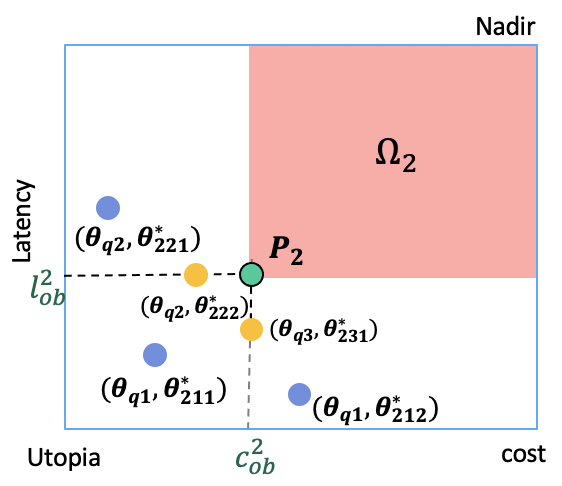}} 					  
  \end{tabular}
  \vspace{-0.2in}
  \captionof{figure}{\small{Example of stage-level Pareto solutions under $\bm{\Theta^{cand}_c}$}}
  \label{fig:eg_stage_space_division}
\end{figure}

Since all the stages should share the same $\bm{\theta_c}$ configurations, the above example is updated in Figure \ref{fig:eg_stage_space_division} under $\bm{\Theta^{cand}_c}$. Based on the $\bm{\Theta^{cand}_c}$ and their corresponding optimal $\bm{\theta_s}$ configurations, the objective space could be divided into different regions (e.g. the red region in Figure \ref{fig:global_pareto_stage_1}) by considering query-level optimality. 

The query-level optimality is decided by an \textbf{Optimality-boundary point} in each stage. The intuition of \textbf{Optimality-boundary point} is to identify and prune sub-regions which is impossible to contribute to the query-level Pareto optimal configurations. For example, the green point $P_1$ in Figure \ref{fig:global_pareto_stage_1} is the \textbf{Optimality-boundary point} of stage 1. The optimality proof related to the divided regions is presented in the later sub-section.

\textbf{Optimality-boundary point} of a stage is determined by \textbf{Optimality-boundary latency of stage $s$} (i.e. $lat_{ob}^s$) and \textbf{Optimality-boundary cost of stage $s$} (i.e. $cost_{ob}^s$),
where $lat_{ob}^s=\max(f_{lat}(\min(\bm{\theta^*_{ci}}, \bm{\theta^*_{sij}}|j=1,...,p_i)|i=1,...,|\bm{\Theta^{cand}_c}|)$, $cost_{ob}^s=\max(f_{cost}(\min(\bm{\theta^*_{ci}}, \bm{\theta^*_{sij}}|j=1,...,p_i)|i=1,...,|\bm{\Theta^{cand}_c}|)$ with stage-level latency and cost predictive models $f_{lat}$ and $f_{cost}$ respectively.

For example, in stage 1, under a specific $\bm{\theta^*_{ci}}$ configuration, we denote the latency and cost of one of the points as 

\begin{equation}
	\begin{split}
	lat_{sij} & = f_{lat}(\bm{\theta^*_{ci}}, \bm{\theta^*_{sij}}), j=1,...,p_i\\
	cost_{sij}& = f_{cost}(\bm{\theta^*_{ci}}, \bm{\theta^*_{sij}}), j=1,...,p_i
	\end{split}
\end{equation}
	
%

Then the $lat_{ob}^1$ and $cost_{ob}^1$ are as the following:

\begin{equation}
	\begin{split}
	lat_{ob}^1 & = \max(\min(lat_{111}),\min(lat_{121}),\min(lat_{131}))\\
	& = lat_{111}
	\end{split}
\end{equation}

\begin{equation}
	\begin{split}
	cost_{ob}^1 & = \max(\min(cost_{111}),\min(cost_{121}),\min(cost_{131}))\\
	& = cost_{131}
	\end{split}
\end{equation}
%

Where the point $P_1$ is determined as the \textbf{Optimality-boundary point} since it achieves both $lat_{ob}^1$ and $cost_{ob}^1$.

Similarly in stage 2, the $lat_{ob}^2$ and $cost_{ob}^2$ are as the following:

\begin{equation}
	\begin{split}
	lat_{ob}^2 & = \max(\min(lat_{211},lat_{212}),\min(lat_{221},lat_{222}), \min(lat_{231}))\\
	& = \max(lat_{211}, lat_{222}, lat_{231})\\
	& = lat_{222}
	\end{split}
\end{equation}

\begin{equation}
	\begin{split}
	cost_{ob}^2 & = \max(\min(cost_{211},cost_{212}),\min(cost_{221},cost_{222}), \min(cost_{231}))\\
	& = \max(cost_{211}, cost_{221}, cost_{231})\\
	& = cost_{231}
	\end{split}
\end{equation}

%

Where the point $P_2$ is determined as the \textbf{Optimality-boundary point} since it achieves both $lat_{ob}^2$ and $cost_{ob}^2$.

Based on the \textbf{Optimality-boundary point}, the objective space of each stage could be split into different sub-regions, where solutions located in the red regions (e.g. $\Omega_1$ and $\Omega_2$ in Figure \ref{fig:eg_stage_space_division}) can be proved to be dominated in the query-level. Details can be found in the theorectical analysis later this section.

\todo{intuitions related to "missing solutions", the current proposal is heuristic, which could be improved further with solid properties}

\minip{Intuition 2:} Solutions constructed from the \textbf{stage-level tuning} may be dominated by other solutions $\bm{\mathbb{F}}$.

Here we denote the solutions $\bm{\mathbb{F}}$ as \textit{missing solutions}, which have two features. 

\begin{enumerate}
	\item Dominated in each stage with different $\bm{\theta_c}$ configurations from all stage-leve Pareto optimal $\bm{\theta_c}$ candidates.
	\item Non-dominated in the query-level, which at least domimates one solution constructed from the stage-leve Pareto optimal $\bm{\theta_c}$ candidates.
\end{enumerate}

\begin{figure*}
  \centering
  \begin{minipage}{.99\textwidth}
  	\begin{tabular}{ll}
  		\subfigure[\small{Query}]
		{\label{fig:eg_query}\includegraphics[width=1.1cm,height=3cm]{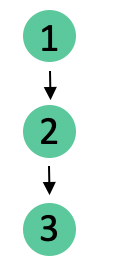}} 
		\subfigure[\small{Solutions in Stage 1}]
		{\label{fig:eg_solutions_stage_1}\includegraphics[width=4.1cm,height=3cm]{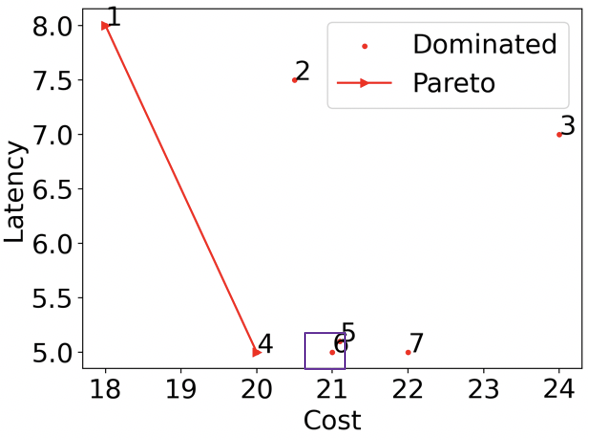}} 
		\subfigure[\small{Solutions in Stage 2}]
		{\label{fig:eg_solutions_stage_2}\includegraphics[width=4.1cm,height=3cm]{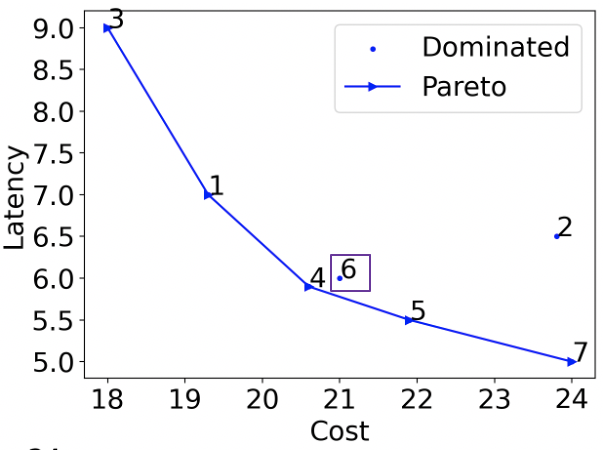}} 	
		\subfigure[\small{Solutions in Stage 3}]
		{\label{fig:eg_solutions_stage_3}\includegraphics[width=4.1cm,height=3cm]{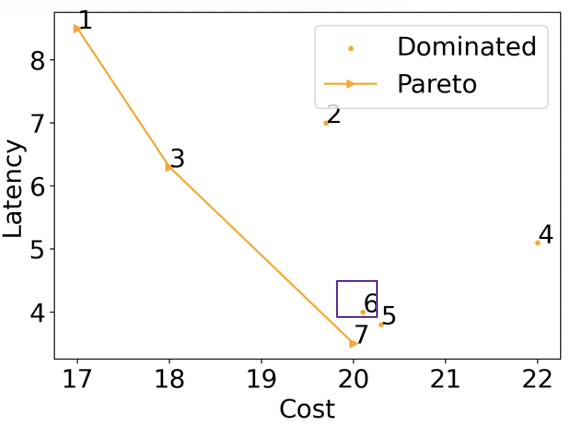}} 
		\subfigure[\small{Query-level solutions}]
		{\label{fig:eg_solutions_query}\includegraphics[width=4.1cm,height=3cm]{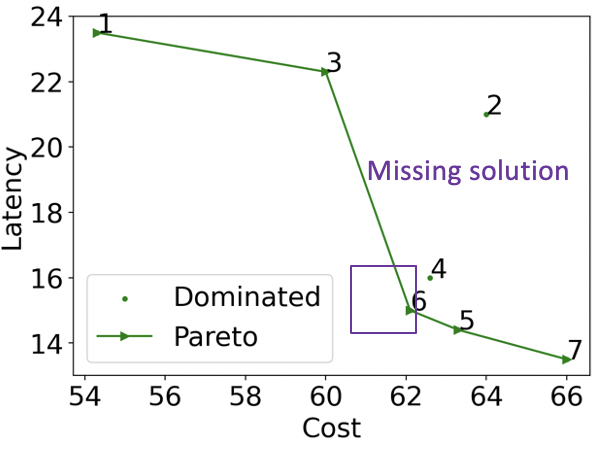}} 				  
  \end{tabular}
  \end{minipage}
  
  \vspace{-0.2in}
  \captionof{figure}{\small{Example of a simple query with 3 stages and the missing solutions}}
  \label{fig:eg_missing_solutions}
\end{figure*}

\minip{Example}

Figure \ref{fig:eg_missing_solutions} shows a simple example with 3 list stages (e.g. Figure \ref{fig:eg_solutions_stage_1}). Figures \ref{fig:eg_solutions_stage_1}, \ref{fig:eg_solutions_stage_2},\ref{fig:eg_solutions_stage_3} and \ref{fig:eg_solutions_query} show solutions of stages and the query respectively, where the solutions with the same index means they are under the same $\bm{\theta_c}$ configuration. In this example, $\bm{\theta_c}$ configurations of solutions 1, 3, 4, 5 and 7 are from \textbf{stage-level tuning}, which is obtained by unioning $\bm{\theta_c}$ configurations of Pareto solutions of all stages (where we denote their corresponding $\bm{\theta_c}$ configuration as stage-level Pareto $\bm{\theta_c}$), and the others with indices of 2 and 6 are dominated in all stages (where we denote their corresponding $\bm{\theta_c}$ configuration as stage-level dominated $\bm{\theta_c}$). In the query level, it is observed that solution 6 (from the stage-level dominated $\bm{\theta_c}$ configuration) dominates solution 4 ((from the stage-level Pareto $\bm{\theta_c}$ configuration). So, solution 6 is the \textit{missing solution}, as it comes from the stage-level dominated $\bm{\theta_c}$ configuration, and it is non-dominated and dominates solution 4 constructed from the stage-level Pareto $\bm{\theta_c}$ configuration.

Typically, MOO algorithms search for a set of the non-dominated discrete solutions.  The \textit{missing solutions} come from two sources. One is from the samples using in the \textbf{stage-level tuning}, and the other is from the unknown samples which are never explored during \textbf{stage-level tuning}.

To tackle this problem, \todo{TODO: for the first case, find properties to show solid theoretical foundation}

\begin{figure}
  \centering
  \includegraphics[width=.9\linewidth]{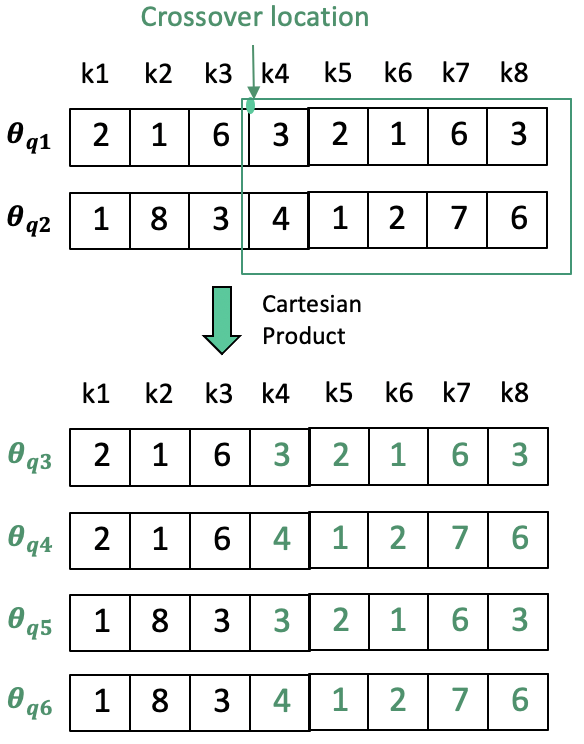}
  \captionof{figure}{\small{Example of $\bm{\theta_c}$ crossover}}
  \label{fig:crossover_q}
\end{figure}

To find other unexplored $\bm{\theta_c}$ configurations, inspired by the crossover operation in evolutionary algorithms, we propose a heuristic method called \textit{$\bm{\theta_c}$ crossover}. The main idea is to firstrandomly generate a crossover location, where the initial $\bm{\theta_c}$ configurations could be splitted into two parts. Then a Cartisian Product is implemented for these two parts to generate new $\bm{\theta_c}$ candidates.

\minip{Example}

In Figure \ref{fig:crossover_q}, $\bm{\theta_{c1}}$ and $\bm{\theta_{c2}}$ are initial $\bm{\theta_{c}}$ candidates obtained from the stage-level tuning, where there are 8 variables in each $\bm{\theta_{c}}$ (e.g. $k1, ..., k8$). Each value within the box represents the configurations of $\bm{\theta_{c}}$. The green point is a crossover location generated randomly. Based on this location, the $\bm{\theta_{c1}}$ and $\bm{\theta_{c2}}$ are divided into two parts, i.e. $k1, k2, k3$  and $k4, k5, k6, k7, k8$, marked by a green rectangle. Then, a Cartisian Product is applied for the two parts. In this example, there are two distinct configurations for $k1, k2, k3$ and 2 distinct configurations for $k4, k5, k6, k7, k8$. So after Cartisian Product, there are 4 configurations generated as $\bm{\theta_{c3}}$ to $\bm{\theta_{c6}}$. It is noted that \textit{$\bm{\theta_c}$ crossover} generates new  $\bm{\theta_c}$ configurations (e.g. $\bm{\theta_{c4}}$, $\bm{\theta_{c6}}$) without losing the initial $\bm{\theta_c}$ candidates (e.g. $\bm{\theta_{c3}},\bm{\theta_{c5}}$).

\minip{Theoretical Analysis} ~

\begin{proposition}
	For any arbitrary stage $s$, the optimality-boundary point $P_s$ is dominated by stage-level optimal solutions constructed by all $\bm{\theta_c} \in \bm{\Theta^{cand}_c}$.
\end{proposition}

\begin{proof}
In any stage $s$ of an arbitrary DAG, the optimality-boundary point $P_s$ is dominated by stage-level optimal solutions $f^s$ constructed by any $\bm{\theta_c} \in \bm{\Theta^{cand}_c}$. 

Suppose $P_s$ is dominant or non-dominated with $f^s$. There must be $P_s$ has lower latency and cost than $f^s$, which is impossible as $P_s$ achieves maximum latency and cost among stage-level optimal solutions constructed by all $\bm{\theta_c} \in \bm{\Theta^{cand}_c}$.
Then the proof is concluded.  

\end{proof}

\begin{proposition}
	For all stages, solutions located in the red region cannot contribute to query-level Pareto optimal solutions.
\end{proposition}

\begin{proof}
In any stage $s$ of an arbitrary DAG, the dominated solution $f^s$ (red area) is dominated by ${f'}^s$ with any arbitrary $\bm{\theta_c}$; query-level solution $F$ built from $f^s$, $F$ built from ${f'}^s$. 

Suppose $F$ is non-dominated with $F'$ or dominates $F'$. There must be at least one $f^s$ with lower latency or cost than ${f'}^s$, which is impossible and concludes the proof.  

\end{proof}

%% file: checkpoint-231115/Qi_initial_optimizer/s_tuning.tex
\subsubsection{${\theta_s}$ Tuning} \label{s_tuning}~

For an arbitrary ${\theta_c}$, tuning ${\theta_s}$ for the stage $s$ is a general MOO problem. While all stages should be tuned under the same ${\theta_c}$ candidates, assuming there are $\bm{|\theta_c|}$ candidates and $|S|$, $\bm{\theta_s}$ should be tuned $\bm{|\theta_c|}*|S|$ time. Due to the efficiency requirement, it is not affordable to tune ${\theta_s}$ for all $\bm{|\theta_c|}$ candidates of each stage, especially when $\bm{|\theta_c|}$ and $|S|$ are very large.

Furthermore, for an arbitrary $\bm{\theta_c}$, tuning $\bm{\theta_s}$ for the stage $s$ is necessary as only the solutions with optimal $\bm{\theta_s}$ configurations can contribute to the query-level Pareto optimal.

\begin{proposition}
	Under any $\bm{\theta_{ci}}$, $i \in [1, |s|]$, only stage-level Pareto optimal solutions $(\bm{\theta_{ci}},\bm{\theta_{s}}^∗)$ contribute to the query-level Pareto optimal.
\end{proposition}

\begin{proof}
Under an arbitrary $\bm{\theta_c}$, given $F$ and $F'$ are non-dominated, where $F'$ includes stage-level dominated solutions in an arbitrary stage $s$ ($f'_s$), and $F$ only includes stage-level non-dominated solutions in all stages ($f_s$). 

For the arbitrary stage $s$ including dominated solutions in $F'$, there is at least another non-dominated stage-level solution (e.g. $f_s$) to reduce all the objective values (e.g. both latency and cost) further. In other words, there must be another query-level solution $F''$ dominating $F'$ with lower latency and cost, which takes $f_s$ for $s$ in $F'$. 

This contradicts the statement that $F'$ is non-dominated, and concludes the proof.

\end{proof}

\cut{
Related work \cite{SinhaMD15} indicates a possible direction to deal with this efficiency issue, which aims to reduce the number of $\bm{\theta_s}$ tuning that results in computational savings. It proposed to estimate the optimal $\bm{\theta_s}$ by using quadratic approximation for each variable of $\bm{\theta_s}$ under an unknown $\bm{\theta_c}$, rather than to tune it. That is, for each stage, $\bm{\theta_s}$ tuning is made only for a subset of all $\bm{\theta_c}$ candidates, and then based on the optimal $\bm{\theta_s}$ and its corresponding $\bm{\theta_c}$ configurations, the relationship between each variable of $\bm{\theta_s}$ and $\bm{\theta_c}$ are captured by a quadratic function $f_c$, i.e. $\bm{\theta^z_s}=f_c(\bm{\theta_c})$, where $z \in [1, d_c]$. The estimation of all optimal $\bm{\theta_s}$ variables under an unknown $\bm{\theta_c}$ configuration is obtained by the quadratic functions. However, it is not available in our problem as fitting quadratic functions already introduces extra complexity. Apart from that, the accuracy of the quadratic functions also infludence the performance a lot, which is not controllable. 
}

To tackle the efficiency issue, we proposed a heuristic method called \textit{tolerance-bound}. The aim is to filter the $\bm{\theta_c}$ candidates not likely to be the query-level Pareto optimal. Intuitively, as all stages share the same $\bm{\theta_c}$ candidates, if similar $\bm{\theta_c}$ configurations appear under different query-level Pareto optimal solutions (with different $\bm{\theta_s}$), it means the those $\bm{\theta_c}$ are likely to be optimal in the query-level. It includes two hyperparamters, the number of $\bm{\theta_s}$ samples $N$ and the tolerance $R$, where $N$ indicates the number of $\bm{\theta_s}$ used under all $\bm{\theta_c}$ candidates, and $R$ indicates the priority of tolerance on different $\bm{\theta_c}$ varibales. The detials and examples are shown in the following example. 

The Pseudo code is represented in Algorithm \ref{alg:tolerance_bound}.

\begin{algorithm}[t]
	\caption{Tolerance-bound}
	\label{alg:tolerance_bound}
	\small
	\begin{algorithmic}[1]  
		\REQUIRE {$\Phi_s, s \in [1, |S|]$, $\bm{\theta^*_c}$, $N, R$}.
		\STATE tol = $zeros$(size($\bm{\theta_c}$))
		\STATE $\bm{\Theta_s}$ = $sample(\bm{\theta_s}, N)$
		\STATE [$\bm{\theta^{*1}_c}, ..., \bm{\theta^{*N}_c}$] = $query\_opt$($\bm{\theta^*_c}, \bm{\Theta_s}$)
		\STATE $\bm{\Theta_c}$ = []
		\WHILE {$len(\bm{\Theta_c}) == 0$}
			\STATE $\bm{c}$ = $tol\_intersect$(tol, [$\bm{\theta^{*1}_q}, ..., \bm{\theta^{*N}_c}$])
			\IF {$len(\bm{c}) == 0$}
				\STATE tol = $update\_tol$(tol, $R$)
			\ELSE
				\STATE $\bm{\Theta_c}.append$($\bm{c}$)
				\STATE break
			\ENDIF
		\ENDWHILE
        \RETURN $\bm{\Theta_c}$
	\end{algorithmic}
\end{algorithm}

This algorithm is implemented for each stage, where $\Phi_s$ is the predcitive model of stage $s$. The input $\bm{\theta^*_c}$ is the $\bm{\theta_c}$ candidates from $\bm{\theta_c}$ generation; $N$ and $R$ are two hyperparameters as the number of samples and the priority vector. For example, assume we have $N=2$, and priority $R=[0,0,0,1,1,1,1,1]$ for 8 variables in $\bm{\theta_c}$, where 1 represents high priority and 0 represents low priority to update tolerance. 

Line 1 initializes tolerance of all variables in $\bm{\theta_c}$ as 0, which is the most strict tolerance to let all optimal $\bm{\theta_c}$ configurations the same under different $\bm{\theta_s}$ samples.  Line 2 samples the $\bm{\theta_s}$ condigurations randomly. For example, assume there are 4 variables in $\bm{\theta_s}$, and $\bm{\theta_{s1}}=[4,6,0,1], \bm{\theta_{s2}}=[2,3,1,1]$ are two samples. Assume there are 100 $\bm{\theta_q}$ candidates, then the input of line 3 is $100 \times 2$ configuration vectors to get the query-level solutions $F$, where all stages use the same $\bm{\theta_s}$ samples, e.g. $F(\bm{\theta_{ci}}, \bm{\theta_{s1}}, ..., \bm{\theta_{s1}}), i \in [1, 100]$. Line 3 generates the optimal query-level solutions. For example, there are 2 optimal query-level configurations generated, where their $\bm{\theta_c}$ configurations are $\bm{\theta^{*1}_c}=[[2,4,4,3,5,0,0,75],[2,5,4,4,6,0,0,74],[3,5,4,3,7,1,0,60]]$ under $\bm{\theta_{s1}}$, and $\bm{\theta^{*2}_c}=[[2,5,4,4,5,1,0,75],[3,5,4,3,6,1,1,75],[4,5,$  $5,3,8,1,0,60],[6,5,10,3,8,1,0,60]]$ under $\bm{\theta_{s2}}$. We focus on their $\bm{\theta_c}$ configurations as they should be similar among different $\bm{\theta_s}$ configurations.  Then line 4 initializes the similar $\bm{\theta_c}$ configuration set as empty $\bm{\Theta_c}$. If it is empty, line 6 keeps the similar $\bm{\theta_c}$ configurations under the current tolerance. For example, currently, the tolerance is $[0,0,0,0,0,0,0,0]$ for all 8 variables in $\bm{\theta_c}$. It means that the same $\bm{\theta_c}$ configuration should appear in all $\bm{\theta^{*1}_c}$ and $\bm{\theta^{*2}_c}$. While there are no overlap of the $\bm{\theta_c}$ configuration between $\bm{\theta^{*1}_c}$ and $\bm{\theta^{*2}_c}$. So Line 7 is true and the tolerance is updated in line 8. The pripority $R=[0,0,0,1,1,1,1,1]$, which means when update the tolerance, the values of $k4, k5, k6, k7, k8$ are gradually increased by 1. After the tolerance of $\bm{\theta_c}$ variables with high priories are updated to the upper bound of the variable range, the tolerance of $\bm{\theta_c}$ variables with low priories starts to update. So in this example, the tolerance is updated as $[0,0,0,1,1,1,1,1]$ as $k4, k5, k6, k7, k8$ are with high pripority. Then it loops again as there are no $\bm{\theta_c}$ configuration added in $\bm{\Theta_c}$. Based on the updated tolerance, now there are similar configuraions between $\bm{\theta^{*1}_c}$ and $\bm{\theta^{*2}_c}$, which are $[2,5,4,4,6,0,0,74]$ and $[2,5,4,4,5,1,0,75]$ repsectively. As their $k1, k2, k3$ values are the same, and $k4, k5, k6, k7, k8$ values vary with the tolerance 1. Then, these two configurations are added into $\bm{\Theta_c}$ (line 10). It breaks the loop (line 11) and returns the $\bm{\Theta_c}$.

\todo{TODO: the current algorithm is heuristic, to have a more solide theoretical fundation, to think whether there are theoretical properties of filtering $\theta_c$ candidates}

%% file: checkpoint-231115/Qi_initial_optimizer/dag_opt.tex
\subsubsection{DAG optimization} \label{dag_opt} ~

For the query with DAG of stages, it is non-trivial to get its Pareto optimal solutions while optimizing multiple objectives, even it is given the optimal solutions of each stage. Intuitively, under an arbitrary $\bm{\theta_c}$, the complexity of finding the query-level Pareto optimal solutions could be $p_{max}^{|S|}$, which exponential to the number of stages $|S|$, where $p_{max}$ is the maximum number of optimal solutions among all stages. To address this problem with the requirements of efficiency and optimality, we proposed an algorithm named \textit{DAG compressing}, which compresses serial and parallel stages, and simplfies the DAG by the compressed stages.

Compressing DAG with a tree structure and a graph structure is different. We first explain the main idea and propose the algorithm for DAG with a tree structure.

\begin{figure*}
  \centering
  \begin{minipage}{.99\linewidth}
  	\begin{tabular}{ll}
  		\subfigure[\small{TPCH Q10}]
		{\label{fig:eg_compressing_tpchq10}\includegraphics[width=3.1cm,height=4cm]{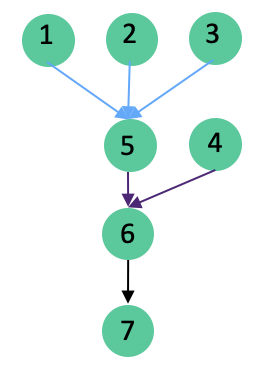}} 
		\hspace{0.1in}
		\subfigure[\small{Step 1}]
		{\label{fig:eg_compressing_step1}\includegraphics[width=2.1cm,height=3.3cm]{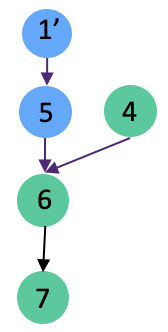}} 
		\hspace{0.1in}
		\subfigure[\small{Step 2}]
		{\label{fig:eg_compressing_step2}\includegraphics[width=2.1cm,height=2.8cm]{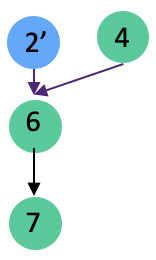}}
		\hspace{0.1in}
		\subfigure[\small{Step 3}]
		{\label{fig:eg_compressing_step3}\includegraphics[width=1.3cm,height=2.8cm]{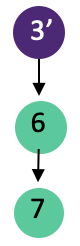}}
		\hspace{0.1in}
		\subfigure[\small{Step 4}]
		{\label{fig:eg_compressing_step4}\includegraphics[width=1.5cm,height=1.3cm]{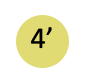}}				  
  \end{tabular}
  \end{minipage}
  
  \vspace{-0.2in}
  \captionof{figure}{\small{Example of compressing a DAG with a tree structure}}
  \label{fig:eg_compressing_tree}
\end{figure*}

\minip{Intuition of DAG compressing} ~

Figure \ref{fig:eg_compressing_tree} shows an example of the main idea of \textit{DAG compressing} with a tree structure. Figure \ref{fig:eg_compressing_tpchq10} is the original query structure of TPCH query with 7 stages, where each node represents one stage with stage-level Pareto optimal solutions under all $\bm{\theta_c}$ candidates. The main idea of the compressing is to deal with the parallel stages (e.g. stages 1, 2, 3) and the serial stages (e.g. stages 6, 7). After compressing, the compressed node is treated as a Pseudo-stage (e.g. Pseudo-stage 1', 2', 3' and 4') to be computed further. Finally, the DAG with a tree structure is compressed as one node, where the optimal solutions of this node is the final optimal solutions of the query. Thus, intuitively, if the compressing is guaranteed to be efficient and maintains the optimality, it guarantees that the optimization over the DAG meets the requirement of the optimality and efficiency.

\cut{
To illustrate with the simplest case, we assume there are only one $\bm{\theta_c}$ candidate, and their corresponding optimal stage-level solutions are as following. 
}

The Algorithm \ref{alg:dag_compressing_tree} is the Pseudo-code of the overall idea.

\begin{algorithm}[t]
    \caption{DAG\_compressing\_tree}
    \label{alg:dag_compressing_tree}
    \small
    \begin{algorithmic}[1]  
        \REQUIRE {$Q$, $\bm{\Theta_c}$, $f\_{dict}$, $conf\_{dict}$}.
        \STATE topo\_order = $topo\_sort(Q)$
        \WHILE {len(topo\_order) > 1}
        	\STATE $n$ = topo\_order[0]
        	\STATE $n_{succ}=find\_succ(Q, n)$
        	\STATE join\_flag, para\_nodes = $is\_terminal\_join(Q, n_{succ})$
        	\IF {join\_flag}
        		\STATE $n_{compress}$ = $compress\_para$(para\_nodes, $\bm{\Theta_c}$, $f\_{dict}$, $conf\_{dict}$)
        		\STATE $Q'$ = update\_dag($Q$, $n$, para\_nodes, $n_{compress}$)
        	\ELSE
        		\STATE serial\_nodes =  get\_serial\_nodes($Q$, $n$)
        		\IF {len(serial\_nodes) > 1}
        			\STATE $n_{compress}$ = $compress\_serial$(serial\_nodes, $\bm{\Theta_c}$, $f\_{dict}$, $conf\_{dict}$)
        		\ELSE 
        			\STATE $n_{compress}$ = $n$
        		\ENDIF
        		\STATE $Q'$ = update\_dag($Q$, $n$, serial\_nodes, $n_{compress}$)
        	\ENDIF
        	\STATE topo\_order = $topo\_sort(Q')$
        \ENDWHILE
        \RETURN topo\_order[0]
    \end{algorithmic}
\end{algorithm}

Line 1 of Algorithm \ref{alg:dag_compressing_tree} gets the topological order of stages in the query. For example, its topological order probably be [1, 2, 3, 4, 5, 6, 7], refered to Figure \ref{fig:eg_compressing_tree}. Lines 2-20 compressing the stages and updating the DAG with its topological order within a while-loop. Firstly, it will check whether the current DAG is compressed to one node or not. If so, the while loop ends and return this compressed node with its configuration. Otherwise, line 3 pops out the first node $n$ in the stages after topological sorting, e.g. stage 1. And the successor node of $n$ is found in line 4. For example, the successor stage of stage 1 is stage 5. Line 5 checks whether this successor node is a join node or not. Here join node indicates whether it has multiple predecessor nodes or not. For example, stage 5 and 6 are join nodes while the others are not. If true, the join flag is returned as true, and the parallel nodes (i.e. predecessor nodes) are returned as well. For example, the current returned parallel nodes are stage 1, 2, and 3, and the join flag of the successor node (i.e. stage 5) is true. Then it executes lines 7 and 8 to get compressed node (e.g. pseudo-stage 1') of stages 1, 2, 3 and update the DAG as shown in Figure \ref{fig:eg_compressing_step1}. After that, a topological order of the updated DAG is computed in Line 17 and enters the while-loop again. For example, the updated topological  order is [1', 5, 4, 6, 7].

Then, with this new topological order, line 3 returns pseudo-stage 1', where its successor node is stage 5. Now stage 5 only has one predecessor (i.e. pseudo-stage 1'), so line 5 returns a false flag, and it goes to line 9 next. Lines 10 gets the current serial nodes. In this example, as stage 6 is a join node with multiple predecessor nodes, the serial nodes should be stages 1' and 5. If there are more than 1 serial nodes, those serial nodes should be compressed by line 12. Otherwise, it keeps the node itself as there is no need to compress just one 1 node (line 14). Then, the DAG is updated with the compressed node (i.e. pseudo-stage 2') as shown in Figure \ref{fig:eg_compressing_step2}. After that, the topological order is updated again to enter while-loop. In the same way, finally, the DAG is compressed into one node (e.g. pseudo-stage 4'), and its optimal solution is the final query-elevel Pareto optimal solutions.

It is noted that the compressing is non-trivial which could be exponential to the number of parallel or serial nodes in order to achieve the Pareto optimal solutions. Compressing serial and parallel nodes are different due to the different objectives. For example, let's denote the objective value of parallel stages as the \textit{compressed objective value}. Based on parallel stages, For the objective like latency, its compressed latency takes the \textit{max} among all stage-level latencies. While for the objective like cost, its compressed cost takes the \textit{sum} among all stage-level cost. Based on serial stages, no matter for objectives like latency or cost, its compressed objective value is the sum over all stages. Thus, the difference of compressing between parallel and serial stages are the way to obtain the compressed objective values. For simplicity of representation, the rest denotes both the "stage" and "pseudo-stage" as the \textit{node}.

Assume in a general case, there are $v1$ objectives with \textit{max} operator and $v2$ objectives with \textit{sum} operator.

\begin{algorithm}[t]
    \caption{Compressing\_serial\_nodes}
    \label{alg:compress_serial}
    \small
    \begin{algorithmic}[1]  
        \REQUIRE {nodes\_list, ws\_pairs}.
        \STATE PO = [ ], conf = \{\}
        \FOR{$[w_l, w_c]$ \textbf{in} ws\_pairs}
            \STATE po\_n = [ ], conf\_n = \{\}
            \FOR{$\bm{n}$ \textbf{in} nodes\_list}
                \STATE ${\bm{n}.po}'$ = normalize($\bm{n}.po$)
                \STATE opt\_ind, opt\_ws = minimize\_ws($[w_l, w_c]$, ${\bm{n}.po}'$)
                \STATE results$.append$($\bm{n}.po$[opt\_ind])
                \STATE conf\_n[$\bm{n}.id$] = $\bm{n}.conf$
            \ENDFOR
            \STATE PO$.append$(sum(results)), conf$.append$(conf\_n)
        \ENDFOR
        \RETURN PO, conf
    \end{algorithmic}
\end{algorithm}

\minip{Compressing serial nodes}

By applying weighted sum in compressing serial nodes, like shown in Algorithm \ref{alg:compress_serial}, the compressing guarantees to return a subset of Pareto optimal solutions for the compressed node.

\todo{example, optimality and complexity analysis could refer to the proof in VLDB2022 paper}

\todo{
\minip{Compressing parallel nodes} with lower complexity
}


\todo{optimize over a graph}

%% file: experiments.tex
\section{Experimental Evaluation}\label{sec:expt}

In this section, we evaluate our modeling and fine-grained compile-time/runtime optimization techniques. We further present an end-to-end evaluation against the SOTA tuning methods.

{\bf Spark setup.} 
We perform SQL queries at two 6-node Spark 3.5.0 clusters with runtime optimization plugins. 
Our optimization focuses on 19 parameters, including 8 for $\thetabm_c$, 9 for $\thetabm_p$, and 2 for $\thetabm_s$, selected based on feature selection profiling~\cite{KanellisAV20} and best practices from the Spark documentation. 
More details are in \techreport{Appendix~\ref{appendix:expt-more-setup}}.

{\bf Workloads.} We generate datasets from the TPC-H and TPC-DS benchmarks with a scale factor of 100. We use the default 22 TPC-H and 102 TPC-DS queries for the optimization analyses and end-to-end evaluation. To collect training data, we further treat these queries as templates to generate 50k distinct parameterized queries for TPC-H and TPC-DS, respectively. We run each query under one configuration sampled via Latin Hypercube Sampling~\cite{LHS/McKayBC00}.

\rv{
{\bf Implementation.} We implement our optimizer using plugins into Spark.
(1) The \underline{trace collector} is implemented as customized Spark listeners to track runtime plan structures and statistics, costing an average of 0.02s per query.
(2) The \underline{compile-time optimizer} operates as a standalone module, costing an average of 0.4s. 
(3) The \underline{runtime optimizer} operates in the server-client model, where the client is wrapped in customized Spark query plan rules, and the server is on a standalone machine, costing an average of 0.3-0.4s per query.
(4) The \underline{model server} maintains up-to-date models on the same server as the optimizer. 
The TPC-H and TPC-DS traces were split into 8:1:1 for training, validation, and testing. It took 6-12 hours to train one model and 2 weeks for hyperparameter tuning  on a GPU node with 4 NVIDIA A100 cards.
}




\input{experiments/expt-model}

\input{experiments/expt-fine-grained-control}
\input{experiments/expt-integration}

%% file: experiments/expt-model.tex
\subsection{Model Evaluation}
\label{subsec:model-eval}
We trained separate models for subQ, QS, and \collap\ to support compile-time/runtime optimization. 
We evaluate each model using the weighted mean absolute percentage error (WMAPE), median and 90th percentile errors (P50 and P90), Pearson correlation (Corr), and inference throughput (Xput).

\begin{table}[t] 
    \ra{0.8}
    \setlength{\belowcaptionskip}{0pt}  
	\setlength{\abovecaptionskip}{0pt} 	   
    \newrobustcmd{\BL}{\color{blue}}
    \newrobustcmd{\B}{\bfseries}
    \centering
    \addtolength{\tabcolsep}{-2.5pt}
    \footnotesize
    \caption{\small{Model performance with Graph Transofmer Network (GTN) + Regressor}}
    \begin{tabular}{c|c|cccc|cccc|c}
    \toprule
    & \multirow{2}{*}{Target}       & \multicolumn{4}{c|}{Ana-Latency/{\BL Latency} (s)}    & \multicolumn{4}{c|}{\rv{Shuffle Size (MB)}}          & Xput \\
    & & \verb|WMAPE| & \verb|P50| & \verb|P90| & \verb|Corr| &\verb|WMAPE| & \verb|P50| & \verb|P90| & \verb|Corr| & K/s        \\ 
    \midrule
    \multicolumn{1}{c|}{\multirow{3}{*}{TPC-H}} 
    & subQ & 0.131& 0.029& 0.292& 0.99& 0.025& 0.006& 0.045& 1.00& 70 \\ 
    & QS & 0.149& 0.027& 0.353& 0.98& 0.002& 3e-05& 0.004& 1.00& 86 \\
    & \collap & \BL 0.164& \BL 0.060& \BL 0.337& \BL 0.95& 0.010& 8e-05& 0.002& 1.00& 146 \\ 
    \midrule
    \multicolumn{1}{c|}{\multirow{3}{*}{TPC-DS}} 
    & subQ & 0.249& 0.030& 0.616& 0.95& 0.098& 0.016& 0.134& 0.99& 60 \\
    & QS &  0.279& 0.060& 0.651& 0.95& 0.028& 4e-04& 0.023& 1.00& 79 \\
    & \collap & \BL  0.223& \BL 0.095& \BL 0.459& \BL 0.93& 0.107& 0.028& 0.199& 0.99& 462 \\
    \midrule\midrule
    \multicolumn{1}{c|}{\multirow{3}{*}{\rv{\makecell{TPC-H* \\ (TPC-DS)}}}}
    & subQ & 0.139 & 0.018 & 0.346 & 0.984 & 0.019 & 0.002 & 0.034 & 0.996 & 45 \\
    & QS & 0.143 & 0.017 & 0.346 & 0.982 & 0.003 & 0.001 & 0.006 & 1.000 & 72 \\
    & \collap & \BL 0.148 & \BL 0.046 & \BL 0.292 & \BL 0.924 & 0.019 & 0.005 & 0.028 & 0.998 & 242 \\    
    \bottomrule
    \end{tabular}
    \label{tab:model-perf-tpcx}
    \vspace{-0.2in}
\end{table}

\underline{Expt 1:} {\it Model Performance.}
We present the performance of our best-tuned models for TPC-H and TPC-DS in the first 6 rows of Table~\ref{tab:model-perf-tpcx}.
First, our models can provide highly accurate prediction in latency and analytical latency for Spark queries for different compile-time and runtime targets, achieving WMAPEs of 13-28\%, P50 of 3-10\%, and P90 of 29-65\%, alongside a correlation range of 93-99\% with the ground truth.
Second, the \rv{shuffle size} is more predictable than latency, evidenced by a WMAPE of 0.2-11\% and almost perfect 99-100\% correlation with the actual \rv{shuffle size}, attributed to its consistent performance across configurations.
Third, the models show high inference throughput, ranging from 60-462K queries per second, which enables efficient solving time of our compile-time and runtime optimizations.

\begin{figure}[t]
    \centering
    \captionsetup{justification=centering}
    \setlength{\belowcaptionskip}{0pt}  
    \setlength{\abovecaptionskip}{0pt} 	
    \begin{minipage}{.58\linewidth}
        \centering
        \includegraphics[width=.92\linewidth]{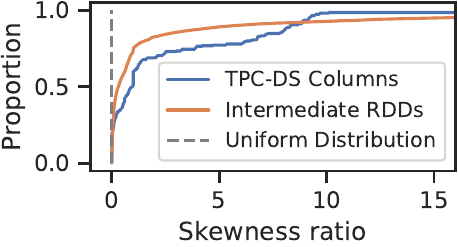}
        \captionof{figure}{\small{\rv{CDF of TPC-DS skewness}}}
        \label{fig:skew-cdf-tpcds}
    \end{minipage}       
    \hfill
    \begin{minipage}{.38\linewidth}
        \ra{0.8}
        \centering
        \addtolength{\tabcolsep}{-3pt}        
        \captionof{table}{\small{\rv{Correlation between model errors and skewness ratios in intermediate RDDs.}}}
        \label{tab:skew-corr}
        \footnotesize
        \begin{tabular}{@{}lccc@{}}
            \toprule
            Lat. range & >1s & >$L_{P90}$ & > $L_{P99}$ \\ \midrule
            TPC-H  & 0.001 & 0.019 & -0.026 \\
            TPC-DS & -0.017 & -0.015 & 0.048 \\
            \bottomrule
        \end{tabular}        
    \end{minipage}    
    \vspace{-0.1in}
\end{figure}

\rv{
\underline{Expt 2:} {\it Impact of Skewness.} We now investigate the impact of data skewness on our latency models. Following recent work~\cite{KoutrisS16}, we define the skewness ratio as the gap between the maximum and average partition size divided by the average. 
Figure~\ref{fig:skew-cdf-tpcds} shows the skewness ratio in base tables and intermediate RDDs in the TPC-DS benchmark, which exceeds 100\% for over 40\% of columns in base tables and for over 26\% of the intermediate RDDs.
Table~\ref{tab:skew-corr} further shows that the correlations between model errors and skewness are close to 0 during runtime optimization, across queries in different latency ranges. 
This indicates that once the skewness is captured at runtime, the model can work quite  well regardless of the skewness level. Additional results are reported in \techreport{Appendix~\ref{appendix:more-skewness}}.
}

\underline{Expt 3:} {\it Model Generalizability.} 
We further explore the generalizability of our trained model to unseen workloads. The first approach applies the latency model trained on TPC-DS directly to TPC-H, or vice versa. However, it leads to high errors due to significant differences in the execution environments. The second approach transfers the graph embedding through the GTN model and retrains only the regressor for latency prediction for a different workload. 
Our results indicate that graph embeddings trained on a workload with a broader range of query operators (e.g., TPC-DS) can be transferred effectively. \crv{As shown in the last three rows of Table~\ref{tab:model-perf-tpcx}, applying the GTN learned from TPC-DS to TPC-H queries and retraining the regressor for TPC-H leads to a modest 0.008 WMAPE increase for the subQ model and a 0.006 and 0.016 WMAPE decrease for QS and \collap, respectively}. More details can be found in \techreport{Appendix~\ref{appendix:more-generalization}}.


\cut{
\underline{Expt 1:}  {\it Impact of operator encoding.} We train separate models with different operator encoding methods, including (1) averaging the word embeddings over each word in the query operator statement~\cite{RAAL};
(2) using the CBO's features~\cite{RO-alibaba}, and (3) leveraging both the cardinality estimation from CBO and the operator statements based on a \texttt{doc2vec} model.

\underline{Expt 2:}  {\it Availability of our unified attention model.} Our super model unifies different ways to encode the query plan structures and use attention mechanisms.

\underline{Expt 3:} {\it Comparison over different model structure.} We train the query latency model based on (1) a holistic model, (2) a close-form composed model based on a set of stage latency predictions, and (3) a learned composed model based on a set of stage latency predictions.

\underline{Expt 4:} {\it Query with UDF/ML.}
}

%% file: experiments/expt-fine-grained-control.tex
\subsection{Compile-time MOO Methods}

We next evaluate our compile-time MOO methods against existing MOO methods. The objectives include query latency and cloud cost as a weighted sum of cpu hours, memory hours, shuffle sizes. 
 
\begin{figure*}
	\centering
	\setlength{\belowcaptionskip}{0pt}  
	\setlength{\abovecaptionskip}{0pt}  
	\begin{tabular}{ccc}
		\subfigure[\small{Solving time of DAG aggregation methods}]
		{\label{fig:solving_time_new_grids_dag_opt}\includegraphics[height=2.5cm,width=0.3\textwidth]{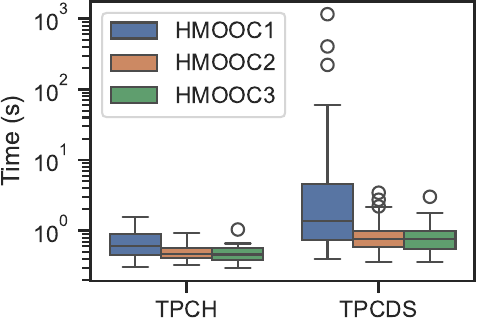}}
		&
		\subfigure[\rv{\small{Mean Hypervolume of \texttt{HMOOC3} and baselines}}]
		{\label{fig:hv_tpch_tpcds}\includegraphics[height=2.5cm,width=0.3\textwidth]{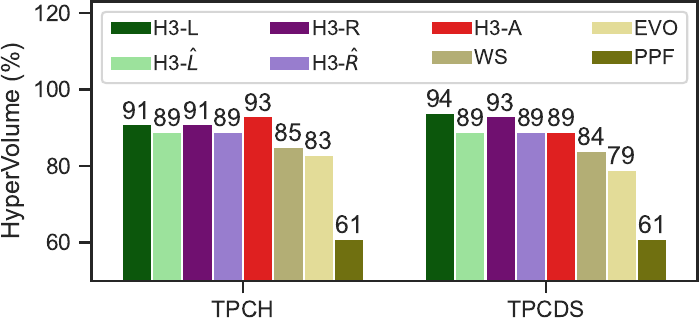}}
		&	
		\subfigure[\rv{\small{Solving time of \texttt{HMOOC3} and baselines}}]
		{\label{fig:time_tpch_tpcds}\includegraphics[height=2.5cm,width=0.3\textwidth]{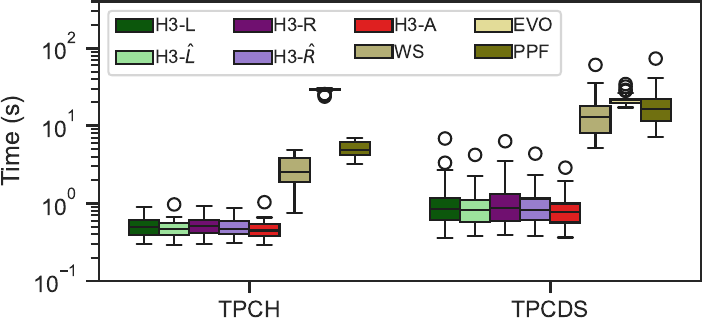}}
	\end{tabular}
	\caption{\small{Accuracy and efficiency of our compile-time MOO algorithms, compared to existing MOO methods}}
	\label{fig:expr-in-analyses-1}
	\vspace{-0.1in}
\end{figure*}

\underline{Expt 4:} {\it DAG Aggregation methods.}
We first compare the three DAG aggregation methods (\S\ref{subsec:dag-opt}) in the HMOOC framework in terms of accuracy and efficiency. Hypervolume (HV) is a standard measure of the dominated space of a Pareto set in the objective space. \rv{All three methods provide similar HV and hence we omit the plot in the interest of space. For efficiency, Figure~\ref{fig:solving_time_new_grids_dag_opt} shows that Boundary-based Approximation (\texttt{HMOOC3}) is the most efficient for both benchmarks, achieving the mean solving time of 0.5-0.8s. Therefore, we use \texttt{HMOOC3} in the remaining experiments.}

\begin{figure*}[t]
	\centering
	\setlength{\belowcaptionskip}{0pt}  
	\setlength{\abovecaptionskip}{0pt}  
	\begin{tabular}{ccc}
		\subfigure[\small{\rv{Latency of queries in TPC-H}}]
		{\label{fig:all_tpch}\includegraphics[height=2.5cm,width=0.25\textwidth]{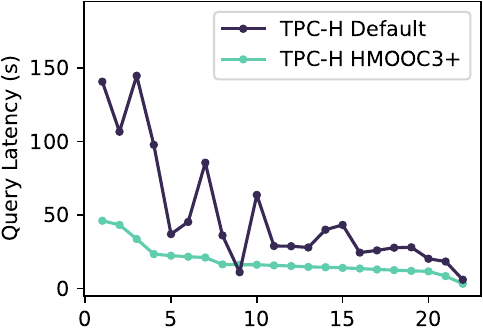}}
		&
		\subfigure[\small{\rv{Latency of queries in TPC-DS}}]
		{\label{fig:all_tpcds}\includegraphics[height=2.5cm,width=0.25\textwidth]{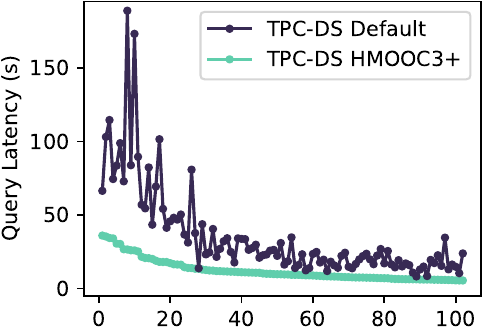}}
		&
		\subfigure[\small{\rv{Long-running queries in TPC-H and DS}}]
		{\label{fig:expt9-e2e-long}\includegraphics[height=2.5cm,width=0.4\textwidth]{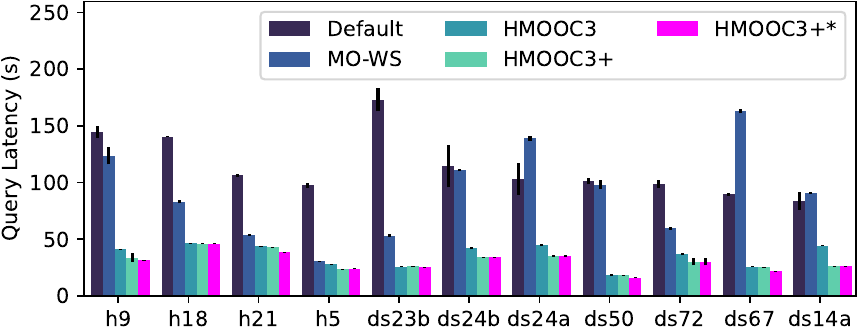}}		
	\end{tabular}
	\caption{\small{End-to-end performance of our algorithm, compared to the state-of-the-art (SOTA) methods}}
	\label{fig:expr-in-analyses-2}
	\vspace{-0.1in}
\end{figure*}

\rv{\underline{Expt 5:} {\it Sampling Methods}. We next compare different sampling methods used with \texttt{HMOOC3}, where \texttt{H3-R} and \texttt{H3-L}  denote \underline{R}andom sampling and \underline{L}atin Hypercube Sampling (LHS), and \texttt{H3-$\hat{R}$} and \texttt{H3-$\hat{L}$} denote their variants with feature importance score (FIS) based parameter filtering. Further,  \texttt{H3-A} denotes our \underline{A}daptive grid-search with FIS-based parameter filtering. The sampling rate is set uniformly $(C=54) \times (P=243)$ for all methods. 
Figures~\ref{fig:hv_tpch_tpcds}-\ref{fig:time_tpch_tpcds} report on HV and solving time, in the first 5 bars in each group. 
In terms of HV, \texttt{H3-A} is the best for TPC-H, and \texttt{H3-L} and \texttt{H3-R} slightly outperform the other three with parameter filtering for TPC-DS. Considering solving time, \texttt{H3-A} achieves the lowest solving time, finishing all TPC-H queries  in 1s and 100/102 TPC-DS queries in  2s. The methods without parameter filtering can have the solving time exceeding 6 seconds for some TPC-DS queries. Overall,  \texttt{H3-A} achieves a good balance between HV and solving time. }



\underline{Expt 6:} {\it Comparison with SOTA MOO methods.}
We next compare with SOTA MOO methods, 
WS~\cite{marler2004survey} (with tuned hyperparameters of 10k samples and 11 pairs of weights),
Evo ~\cite{Emmerich:2018:TMO} (with a population size of 100 and 500 function evaluations),
and PF~\cite{spark-moo-icde21}, for fine-grained tuning of parameters based on Def.~\ref{eq:moo-spark}.
\rv{Figures~\ref{fig:hv_tpch_tpcds}-\ref{fig:time_tpch_tpcds}
report their HV and solving time, in the last three bars of each group. 
\texttt{HMOOC3} outperforms other MOO methods with 4.7\%-54.1\% improvement in HV and 81\%-98.3\% reduction in solving time.}
These results stem from HMOOC's hierarchical framework, which addresses a smaller search space with only one set of $\bm{\theta_c}$ and $\bm{\theta_p}$ at a time, and uses efficient DAG aggregation to recover query-level values from subQ-level ones. In contrast, other methods solve the optimization problem using the global parameter space, including $m$ sets of $\bm{\theta_p}$, where $m$ is the number of subQs in a query. 

%% file: experiments/expt-integration.tex
\subsection{End-to-End Evaluation}

We now extend our best compile-time optimization method \ct\ with runtime optimization,  denoted as \rt, \rv{which includes the adaptive sampling rate scheme introduced in Section~\ref{subsec:hmooc} to achieve a better balance between accuracy and efficiency.} 
We compare it with existing methods in actual execution time when Spark AQE is enabled.
To account for model errors, we refine the search range for each Spark parameter by avoiding the extreme values of the parameter space that could make the predictions less reliable.

\begin{table}[t]
    \setlength{\belowcaptionskip}{0pt}  
	\setlength{\abovecaptionskip}{0pt} 	    
    \ra{0.8}
    \newrobustcmd{\B}{\bfseries}
    \newrobustcmd{\RD}{\color{black}}    
    \footnotesize
    \centering
    \addtolength{\tabcolsep}{-2.5pt}
    \caption{\small{Latency reduction with a strong speed preference}} 
    \label{tab:expt9-e2e}
        \begin{tabular}{l|rcl|rcl}
        \toprule
        & \multicolumn{3}{c|}{TPC-H} & \multicolumn{3}{c}{TPC-DS} \\ 
        & \texttt{MO-WS}    & \texttt{HMOOC3}  & \texttt{HMOOC3+}  & \texttt{MO-WS} & \texttt{HMOOC3}  &\texttt{HMOOC3+} \\
        \midrule
        
        
        
        Total Lat Reduction 
        & 18\% & \RD 61\% & \RD 63\% & 25\% & \RD 61\% & \RD 65\%\\
        
        Avg Lat Reduction 
        & -1\% & \RD 53\% & 52\% & 34\% & \RD 56\% & \RD 59\%  \\
        \midrule
        
         Avg Solving Time (s) 
        &  2.6 &  \RD 0.41 &  \RD 0.70 & 15 &  \RD 0.41 &  \RD 0.80 \\
        
        \RD P50 Solving Time (s)
        & \RD 2.3 & \RD 0.38 & \RD 0.69 & \RD 14 & \RD 0.35 & \RD 0.65 \\

        \RD P90 Solving Time (s)
        & \RD 4.0 & \RD 0.51 & \RD 0.90 & \RD 26 & \RD 0.63 & \RD 1.4 \\

        Max Solving Time (s) 
        & 4.5 & \RD 0.81 & \RD 1.4 & 68 & \RD 1.4 & \RD 2.9 \\
        
        \bottomrule
    \end{tabular} 
    \vspace{-0.1in}
\end{table}

\rv{
\underline{Expt 7:} {\it Benefits over Default Configuration.}
Figure~\ref{fig:all_tpch} and ~\ref{fig:all_tpcds} show the per-query latency using  \rt\ or the default configuration. 
Among the 22 TPC-H and 102  TPC-DS queries, \rt\ significantly outperforms the default configuration for most of the queries,  while being similar (within 2s latency improvement) for three short-running queries (\texttt{h6}, \texttt{ds8}, and \texttt{ds20}). \rt\ loses to the default configuration only for one short-running query   (\texttt{h14}), mainly due to the model error in predicting its latency. As we use WMAPE as the training loss to give more weight to long-running queries, short queries can have inaccurate predictions and hence experience a suboptimal query plan or insufficient resource allocation.
}

\underline{Expt 8:} {\it Benefits over Query-level MOO.}
We next show the advantages of our methods (\ct\ and \rt) over the best-performing MOO method identified in the previous study, i.e., WS for query-level MOO, denoted as \mows. 
Here, we prioritize latency over cost with a preference vector of $(0.9, 0.1)$ on latency and cost.
The results in Table~\ref{tab:expt9-e2e} show the improvement over the default Spark configuration.
First, {\em fine-grained tuning significantly enhances performance} (major result \textbf{R1}), cutting latency by \rv{61}\% for both benchmarks with compile-time optimization (\ct), and by \rv{63-65}\% with runtime optimization (\rt). They both outperform \mows\ with only 18-25\% reductions and in some cases, worse than the default configuration.
Second, {\em \mows, even limited to query-level tuning, suffers in efficiency}, \rv{with 14s and 26s as the P50 and P90 solving time}, respectively, for TPC-DS. In contrast, our approach solves MOO with \rv{a P50 solving time  of 0.65-0.69s and P90 time of 0.9-1.4s for the two benchmarks}. 
\cut{
Third, when we consider a new efficiency measure, defined as latency reduction per unit of solving time, 
our method vastly outperforms \mows, {\em achieving a 1-2 order-of-magnitude improvement in efficiency for latency reduction} (\textbf{R2})}

\underline{Expt 9:} {\it Benefits for Long-running Queries.} 
We further consider long-running queries that are hard to optimize and often suffer from suboptimal plans with $\thetabm_p$ tuned based on compile-time cardinality estimates. 
Figure~\ref{fig:expt9-e2e-long} shows the top long-running queries from TPC-H and TPC-DS. While \ct\ already offers significant latency reduction, \rt\ achieves up to a 22\% additional latency reduction over the default configuration. \rv{We finally turned on multi-query plan search, running up to 3 query plan structures in parallel at compile-time optimization. This method, \rt*, further improves \texttt{h9}, \texttt{h21},  \texttt{ds50}, etc. Overall, {\em \rt* reduces latency by 64-85\% for long-running queries} (\textbf{R2}).}   


\begin{table}[t]
    \setlength{\belowcaptionskip}{0pt}  
	\setlength{\abovecaptionskip}{0pt} 	
    \ra{0.8}
    \newrobustcmd{\RD}{\color{black}}       
    \footnotesize
    \centering
    \addtolength{\tabcolsep}{-0pt}
    \caption{\small{Latency and cost adapting to preferences}}
    \label{tab:expt10-adapt}
    \begin{tabular}{l|cc|cc}
    \toprule   
    Prefs.            & \multicolumn{2}{c|}{TPC-H} & \multicolumn{2}{c}{TPC-DS} \\ 
    Lat/Cost    & \texttt{SO-FW}    & \rt*    & \texttt{SO-FW}         & \rt*       \\\midrule

    (0.0, 1.0) & 20\% / -11\% & \RD -15\% /  \RD -10\%  & -6\% / 64\% & \RD  -45\% /  \RD -22\% \\
    (0.1, 0.9) & 1\% / 1\% & \RD -31\% /  \RD -5\% & -28\% / 105\% & \RD  -57\% /  \RD -7\% \\
    (0.5, 0.5) & -1\% / 25\% &\RD -46\% /  \RD -3\%  & -28\% / 128\% & \RD -59\% / \RD  29\% \\
    (0.9, 0.1) & -13\% / 27\% &\RD -51\% / \RD  0\% & -34\% / 139\% & \RD -59\% / \RD  55\% \\
    (1.0, 0.0) & -14\% / 44\% &\RD -55\% / \RD  3\% & -26\% / 144\% & \RD -59\% / \RD  64\% \\

    \bottomrule
    \end{tabular}
    \vspace{-0.2in}
\end{table}

\underline{Expt 10:} {\it Adaptability Comparison to SO with fixed weights.} 
As \mows\ is too slow for cloud use due to its inefficiency, we now compare the adaptability of our approach against the common, practical approach that combines multiple objectives into a single objective using fixed weights~\cite{LiZLG19,Zhang:2019:EAC,zhu2017bestconfig}, denoted as \sofw.
Table~\ref{tab:expt10-adapt} shows the average reduction rates in latency and cost relative to the default configurations across a range of preference vectors.
First, {\em \rv{\rt*} dominates \sofw\ with more latency and cost reductions in most cases} (\textbf{R3}), achieving up to \cut{52-64\%}\rv{55-59\%} latency reduction and \cut{9-22\%} \rv{10-22\%} cost reduction in both benchmarks, while \sofw\ gets at most 1-34\% average latency reduction and in most cases, increases the cost compared to default.
Second, {\em our approach demonstrates superior adaptability to varying preferences} (\textbf{R4}), enhancing latency reductions progressively as preferences shift towards speed. In contrast, \sofw\ does not make meaningful recommendations: Under a cost-saving preference of $(0.0, 1.0)$, \sofw\ struggles to lower costs in TPC-DS, instead increasing the average cost by 64\% across all queries. Despite this cost increase, it achieves a merely 6\% reduction in latency. In contrast, \rv{\rt*} achieves a \rv{45\%} reduction in latency alongside a 22\% cost saving, underscoring its effectiveness and adaptability to the specified cost performance preference.

%% file: conclusion.tex
\section{Conclusions}


This paper presented a Spark optimizer for fine-grained parameter tuning in the new AQE architecture based on a hybrid compile-time/runtime optimization approach. 
Our approach employed sophisticated modeling techniques to capture different compile-time and runtime modeling targets, and a suite of techniques tailored for multi-objective optimization (MOO) while meeting the stringent solving time constraint of 1-2 seconds. 
Evaluation results using TPC-H and TPC-DS benchmarks show that ($i$) when prioritizing latency, our approach achieves \rv{63}\% and \rv{65}\% latency reduction on average for TPC-H and TPC-DS, respectively, under the solving time of \rv{0.7-0.8} sec, outperforming the most competitive MOO method with 18-25\% latency reduction and high solving time of 2.6-15 sec; ($ii$) when shifting preferences between latency and cost, our approach dominates the solutions from alternative methods by a wide margin.
In the future, we plan to extend our tuning approach to support diverse (e.g., machine learning) workloads.  
\rv{In addition, we plan to extend our approach to other big data/DMBS systems like   Presto~\cite{Presto}, Greenplum~\cite{Greenplum}, and MaxCompute~\cite{maxcompute}, which can observe runtime statistics and support runtime adaptability.}

%% file: appendix/index.tex
\section{Additional Materials for MOO}
\input{appendix/app-compile-time-opt}

\input{appendix/app-runtime-opt}

\input{appendix/app-experiments}

\input{appendix/app-related-work}

%% file: appendix/app-compile-time-opt.tex

\input{appendix/proof}

\input{appendix/appendix_compile_time_opt}

%% file: appendix/proof.tex
\subsection{Algorithms and Proofs}
\label{appendix:proof}

In this section, we include the algorithms, proofs and complexity analysis of our optimization techniques. 

\subsubsection{Subquery (subQ) Tuning}

Below, we provide the proof of Proposition~\ref{prop:need_opt_p}. 

\noindent {\bf{Proposition \ref{prop:need_opt_p}}} Under any specific value $\bm{\theta_{c}}^j$, 
	only subQ-level Pareto optimal solutions $(\bm{\theta_{c}}^j,{\bm{\theta_{pi}}^*})$ for the $i$-th subQ contribute to the query-level Pareto optimal solutions $(\bm{\theta_{c}}^j,{\bm{\{\theta_{p}}^*\}})$.

\begin{proof}

Let $F_q^j$ be a query-level Pareto optimal solution for $\theta_c^j$.
It can be expressed as $F_q^j = \sum_{i=1}^m F^j_{si}$. Assume that there exists at least one $i$, e.g., $i_1$, such that $F^j_{si_1}$ is not optimal for the $i_1$-th subQ.
Let $F^{j\prime}_{q} = F^{j\prime}_{si_1} +  \sum_{i=1, i \neq i_1}^m F^j_{si}$ where $F^{j\prime}_{si_1}$ is Pareto optimal for the $i_1$-th subQ.

We have 
\begin{equation}
	F_q^j - F^{j\prime}_{q} = F^j_{si_1} - F^{j\prime}_{si_1}
\end{equation} 

Since $F^{j\prime}_{si_1}$ is optimal for the $i_1$-th subQ, we have $ F^j_{si_1} > F^{j\prime}_{si_1}$. This means that $F^j_{si_1}$ is dominated by $F^{j\prime}_{si_1}$, which contradicts our hypothesis. Therefore, a Pareto optimal solution for the query-level can only contain subQ-level Pareto optimal solutions under a fixed $\theta^j_c$. 

\end{proof}

{\minip{Complexity analysis of model inference related to the sampling rate}}

{The sampling rate includes the number of $\bm{\theta_c}$ and $\bm{\theta_p}$ samples, denoted as $N_c$ and $N_p$ respectively. The model is called three times in line 3, 4 and 7 respectively.}

{
Line 3 calls the model to provide the predictions of $C$ cluster-representative $\bm{\theta_c}$ and $N_p$ $\bm{\theta_p}$ samples. For $m$ subQs, there is $m \times C \times N_p$ predictions. And it filters the dominated solutions of each $\bm{\theta_c}$ under each subQ to ensure the $\bm{\theta_c}$ are identical among all subQs while keeping the optimality of $\bm{\theta_p}$ as illustrated in Proposition \ref{prop:need_opt_p}.}

{Line 4 assigns the optimal $\bm{\theta_p}$ solutions to $N_c$ initialized $\bm{\theta_c}$ samples for all subQs and provides the corresponding predictions. 
Since the number of optimal solutions of a $\bm{\theta_c}$ sample for different subQs could be different, for simplicity, we use $p_{avg}$ to denote the average number of optimal solutions among all $\bm{\theta_c}$ samples and all subQs. 
The total number of predictions is $N_c \times p_{avg} \times m$.
}

{Line 7 functions the same as line 4, where the only difference is to assign the optimal $\bm{\theta_p}$ solutions (returned from line 3) to the newly generated $\bm{\theta_c}$ samples for all subQs. 
Assume $N^{new}_c$ new $\bm{\theta_c}$ samples are generated, the total number of predictions is $N^{new}_c \times p_{avg} \times m$.}

\subsubsection{DAG Aggregation}

We provide further details of three methods for DAG aggregation. 

\minip{HMOOC1: Divide-and-Conquer}. 
This DAG aggregation method is described in Algorithm \ref{alg:general_div_and_conq}. The $\Omega, \Theta$ are the effective set of all subQs, where $\Omega$ represents subQ-level objective values, and $\Theta$ represents the corresponding configurations, including $\{\thetabm_c, \Thetabmp, \Thetabms\}$. If there is only one subQ, it returns the $\Omega, \Theta$ (lines 1-2). Otherwise, it follows a divide-and-conquer framework (lines 4-8).

 The main idea is a merging operation, which is described in Algorithm \ref{alg:merge}. The input includes the subQ-level objective values (e.g. $\mathcal{F}^h$ is a Pareto frontier) and its configurations ($\mathcal{C}^r$) for the two nodes to be merged, where $h$ and $r$ denote they are two different nodes. It merges two nodes into a pseudo node by enumerating all the combinations of solutions in the two nodes (lines 2-3), summing up their objective values (lines 4-5) and taking its Pareto frontier as the solutions of this pseudo node (line 8). 


\begin{algorithm}[t]
	\caption{General\_Divide\_and\_conquer}
	\label{alg:general_div_and_conq}
	\small
	\begin{algorithmic}[1]  
		\REQUIRE {subQ-level values $\Omega$, subQ-level configurations $\Theta$}.
		\IF {$|\Omega|$ == 1} 
			\RETURN $\Omega$, $\Theta$
		\ELSE
			\STATE $\Omega^h$, $\Theta^h$ = first\_half($\Omega$, $\Theta$)
			\STATE $\Omega^r$, $\Theta^r$ = second\_half($\Omega$, $\Theta$)
			\STATE $\mathcal{F}^h, \mathcal{C}^h$ = General\_Divide\_and\_conquer($\Omega^h$, $\Theta^h$)
			\STATE $\mathcal{F}^r, \mathcal{C}^r$ = General\_Divide\_and\_conquer($\Omega^r$, $\Theta^r$)
			\RETURN merge($\mathcal{F}^h, \mathcal{C}^h, \mathcal{F}^r, \mathcal{C}^r$)
		\ENDIF
	\end{algorithmic}
\end{algorithm}

\begin{algorithm}[t]
	\caption{merge}
	\label{alg:merge}
	\small
	\begin{algorithmic}[1]  
		\REQUIRE {$\mathcal{F}^h, \mathcal{C}^h, \mathcal{F}^r, \mathcal{C}^r$}.
		\STATE $\mathcal{F}, \mathcal{C}$ = $\emptyset, \emptyset$
		\FOR {$(F_1, F_2), (c_1, c_2) \in (\mathcal{F}^h, \mathcal{C}^h)$}
            \FOR{$(F_1', F_2'), (c_1', c_2') \in (\mathcal{F}^r, \mathcal{C}^r)$}
                \STATE $\mathcal{F} = \mathcal{F} \cup \{ (F_1 + F_1', F_2 + F_2') \}$
                \STATE $\mathcal{C} = \mathcal{C} \cup \{ (c_1, c_1'), (c_2, c_2') \}$       
            \ENDFOR 
        \ENDFOR
        \RETURN $\mathcal{F}^*, \mathcal{C}^*$ = filter\_dominated($\mathcal{F}, \mathcal{C}$)		
	\end{algorithmic}
\end{algorithm}



From the view of optimality, Algorithm \ref{alg:general_div_and_conq} is proved to return a full set of the query-level Pareto optimal solutions as it enumerates over all subQ-level solutions. 
The complexity of \textit{merge} function is $O(M*N) + O((M*N)log(M*N))$ if there are $M$ and $N$ solutions in two nodes, where the enumeration takes $O(M*N)$ and filtering dominated solutions takes $O((M*N)log(M*N))$. While after merging several times, $M$ and $N$ could be high. Thus the total complexity could be high.


\input{appendix/appendix-theoretical_ana_hmooc1.tex}

\minip{HMOOC2: WS-based Approximation}. We propose a second technique to approximate the MOO solution over a list structure. 

For each fixed $\bm{\theta_c}$, we apply the weighted sum (WS) method to generate evenly spaced weight vectors. Then for each weight vector, we obtain the (single) optimal solution for each subQ and sum the solutions of subQ's to get the query-level optimal solution. It can be proved that this WS method over a list of subQs guarantees to return a subset of query-level Pareto solutions.

Algorithm \ref{alg:compress_list} describes the full procedures. The input includes a subQ\_list, which includes both subQ-level objective values and the corresponding configurations of all subQs. ws\_paris are the weight pairs, e.g. $[[0.1, 0.9], [0.2, 0.8], \dots]$ for latency and cost. Line 1 initializes the query-level objective values and configurations. Lines 3-9 address the Weighted Sum (WS) method to generate the query-level optimal solution for each weight pair. Specifically, Lines 5-8 apply the WS method to obtain the optimal solution choice for each subQ, and sum all subQ-level values to get the query-level values. Upon iterating through all weights, a Pareto solution set is derived after the necessary filtering (Line 12).

\begin{algorithm}[t]
    \caption{Compressing\_list\_nodes}
    \label{alg:compress_list}
    \small
    \begin{algorithmic}[1]  
        \REQUIRE {subQ\_list, ws\_pairs}.
        \STATE PO = [ ], conf = [ ]
        \FOR{$[w_l, w_c]$ \textbf{in} ws\_pairs}
            \STATE po\_n = [ ], conf\_n = \{\}
            \FOR{subQ\_po, subQ\_conf \textbf{in} subQ\_list}
                \STATE subQ\_po\_norm = normalize\_per\_subQ(subQ\_po)
                \STATE opt\_ind = minimize\_ws($[w_l, w_c]$, subQ\_po\_norm)
                \STATE po\_n$.append$(subQ\_po[opt\_ind])
                \STATE conf\_n$.append$(subQ\_conf[opt\_ind])
            \ENDFOR
            \STATE PO$.append$(sum(po\_n)), conf$.append$(conf\_n)
        \ENDFOR
        \RETURN filter\_dominated(PO, conf)
    \end{algorithmic}
\end{algorithm}

\input{appendix/appendix-theoretical_ana_hmooc2.tex}

\minip{HMOOC3: Boundary-based Approximation}.


We now show the proof of Proposition~\ref{prop:bounded_obj_space}.

\noindent {\bf{Proposition \ref{prop:bounded_obj_space}}} Under a fixed $\bm{\theta_c}$ candidate, the query-level objective space of Pareto optimal solutions is bounded by its extreme points in a 2D objective space.
	
\begin{proof}
	~\\
	\indent Assume that $F_q^{1}=[F_q^{1*}, F_q^{2-}]$ and $F_q^{2}=[F_q^{1-}, F_q^{2*}]$ are two \textit{extreme points}  under a fixed $\bm{\theta_c}$, recalling that the \textit{extreme point} under a fixed $\bm{\theta_c}$ is the Pareto optimal point with the best query-level value for any objective. Here the superscript \{$1*$\} means it achieves the best in objective 1 and \{$2*$\} means it achieves the best in objective 2.
	The two extreme points form an objective space as a rectangle. 
	
	Suppose that an existing query-level Pareto optimal solution $F_q^{'} = [F_q^{1'}, F_q^{2'}]$ is outside this rectangle, which includes 2 scenarios. In scenario 1, it has $F_q^{1'} < F_q^{1*}$ or $F_q^{2'}< F_q^{2*}$, which is impossible as extreme points already achieves the minimum values of two objectives. In scenario 2, it has $F_q^{1'} > F_q^{1-}$ or $F_q^{2'} > F_q^{2-}$, which is impossible as $F_q^{'}$ is dominated by any points inside the rectangle.
	
 So there is no Pareto optimal solution $F_q^{'}$ existing outside the rectangle and it concludes the proof.
\end{proof}


We next show the proof of Proposition~\ref{prop:k_pareto}.

\noindent {\bf{Proposition \ref{prop:k_pareto}}} Given subQ-level solutions, our boundary approximation method guarantees to include at least $k$ query-level Pareto optimal solutions for a MOO problem with $k$ objectives.

\begin{proof}
	~\\
	\indent Assume that $F_q^{1},... F_q^{k}$ are $k$ extreme points, which are Pareto optimal and achieve the best (e.g. the lowest in the minimization problem) query-level values of objectives $1,...,k$ among all $\bm{\theta_c}$ configurations. 
	Suppose that an existing Pareto optimal solution $F_q^{'}$, distinct from the extreme points, dominates any point in $F_q^{1},... F_q^{k}$. $F_q^{'}$ must achieve a better value than $F_q^{1*},... F_q^{k*}$ in any objectives $1,...,k$, where the superscript \{$1*$\} means it achieves the best in objective 1 and \{$k*$\} means it achieves the best in objective $k$. 
	which is impossible as the extreme points already achieves the best.
	
 So these $k$ extreme points cannot be dominated by any other solutions. Thus, they are Pareto optimal and this concludes the proof.
\end{proof}

\cut{
\noindent \underline{Complexity}
Among all $N$ $\bm{\theta_c}$ candidates, we assume that the maximum number of corresponding optimal solutions among all the subQs is $p_{max}$. Finding the extreme points of $N$ $\bm{\theta_c}$ candidates takes $O(p_{max} \cdot m \cdot N)$, where $m$ is the number of subQs. This process returns a maximum of $k \cdot N$ query-level solutions for the MOO with $k$ objectives. The complexity of filtering dominated solutions is $O(k \cdot N \cdot \log(k \cdot N))$ \cite{kung1975finding}.
}

%% file: appendix/appendix-theoretical_ana_hmooc1.tex
The core operation in HMOOC1 is the \textit{merge}, which enumerates over all subQ-level solutions. The following are the theoretical proof. For the sake of simplicity, we consider the case with two nodes.

\begin{proposition}
	Algorithm \ref{alg:general_div_and_conq} always output the full Pareto front of the simplified DAG.
\end{proposition}

\begin{proof}
Let $D$ and $G$ be two nodes (e.g., subQs or aggregated subQs). Let
$\oplus$ be the Minkowski sum,
i.e., $D \oplus G = \{ F_D + F_G, F_D\in D, F_G \in G\}$. Let $Pf$ denote the
Pareto Front of a node.

\begin{equation}
     Pf(Pf(D) \oplus Pf(G)) = Pf(D \times G)
\end{equation}

Let $E_D: \mathcal{C}_D \to \mathbb{R}$ be the evaluation function of node
$D$, where
$\mathcal{C}_D$ is the set of configurations for node $D$. We define $E_G$ in a
similar manner. We also define $E : (c_D, c_G) \mapsto
E_D(c_D) +
E_G(c_G)$, where $c_D$ and $c_G$ are one configuration in $\mathcal{C}_D$ and $\mathcal{C}_G$ respectively.

Let $p \in Pf(Pf(D) \oplus Pf(G))$. Then $p$ can be addressed as a sum
of two
terms: one from $Pf(D)$ and the other from $Pf(G)$, i.e.,

\begin{equation}
     p = p_D + p_G,\qquad p_D \in Pf(D), p_G \in Pf(G)
\end{equation}

Let $c_D \in Pf^{-1}(p_D)$ and $c_G \in Pf^{-1}(p_G)$, i.e.,
$c_D$ is chosen such as $E_D(c_D) = p_D$, and likewise for
$c_G$.
Then we have $p = E(c_D, c_G)$, so $p$ belongs to the
objective space
of $D \times G$. Suppose $p$ doesn't belong to $Pf(D \times G)$. This
means that
there exists $p^{\prime}$ in $Pf(D \times G)$ that dominates $p$.
$p^{\prime}$
can be expressed as $p^{\prime} = p^{\prime}_D + p^{\prime}_G$ with
$p^{\prime}_D \in Pf(D)$ and $p^{\prime}_G \in Pf(G)$. $p^{\prime}$
dominates
$p$ can be expressed as $p^{\prime}_D < p_D$ and $p^{\prime}_G \leq p_G$ or $p^{\prime}_D \leq p_D$ and $p^{\prime}_G < p_G$ in our optimization problem with minimization, which
contradicts the definition of $p$ as belonging to the Pareto Front.
Therefore
$p$ must belong to $Pf(D \times G)$ and thus $Pf(Pf(D) \oplus Pf(G))
\subseteq
Pf(D \times G)$.

Let us now suppose that $p \in Pf(D\times G)$. Then there exists
$c$ in $D
\times G$, i.e., $c = (c_D, c_G)$ with $c_D \in D,
c_G
\in G$, such that $p = E(c)$. By setting $p_D = E_D(c_D)$ and
$p_G =
E_G(c_G)$, we have $p = p_D + p_G$. By definition, $p_D$ belongs to
$Pf(D)$
and $p_G$ belongs to $Pf(G)$. Hence, $p \in Pf(D) \oplus Pf(G)$.

Suppose that $p$ doesn't belong to $Pf(Pf(D) \oplus Pf(G))$ and that there
exists $p^{\prime}$ in $Pf(Pf(D) \oplus Pf(G))$ that dominates $p$. We
showed
above that $p^{\prime}$ must belong to $Pf(D \times G)$. Thus both $p$ and
$p^{\prime}$ belong to $Pf(D \times G)$ and $p^{\prime}$ dominates $p$,
which is
impossible. Therefore $p^{\prime}$ must belong to $Pf(Pf(D) \oplus
Pf(G))$ and
$Pf(Pf(D) \oplus Pf(G)) \subseteq Pf(D \times G)$.

By combining the two inclusions, we obtain that $Pf(Pf(D) \oplus Pf(G))
= Pf(D \times G)$, where the left side is the Pareto optimal solution from the Algorithm \ref{alg:general_div_and_conq} and the right side is the Pareto optimal solution over the whole configuration space of two nodes. Thus, Algorithm \ref{alg:general_div_and_conq} returns a full set of Pareto solutions.

\end{proof}

%% file: appendix/appendix-theoretical_ana_hmooc2.tex
\minip{Theoretical Analysis}

\begin{lemma} \label{lemma: lemma_proof} 
	For a DAG aggregation problem with $k$ objectives that use the \texttt{sum} operator only, Algorithm~\ref{alg:compress_list} guarantees to find a non-empty subset of query-level Pareto optimal points under a specified $\bm{\theta_c}$ candidate.
\end{lemma}

In proving Lemma \ref{lemma: lemma_proof}, we observe that Algorithm~\ref{alg:compress_list} is essentially a Weighted Sum procedure over Functions (WSF). Indeed we will prove the following two Propositions: 1) each solution returned by WSF is Pareto optimal; 2) the solution returned by the Algorithm~\ref{alg:compress_list} is equivalent to the solution returned by WSF. Then it follows that the solution returned by Algorithm~\ref{alg:compress_list} is Pareto optimal.

To introduce WSF, we first introduce the indicator variable $x_{ij}$, $i \in [1, ..., m], j \in [1, ..., p_i]$, to indicate that the $j$-th solution in $i$-th subQ is selected to contribute to the query-level solution. $\sum_{j=1}^{p_i} x_{ij} = 1$ means that only one solution is selected for each subQ. Then $x = [x_{1j_1}, ... x_{mj_m}]$ represents the 0/1 selection for all $m$ subQs to construct a query-level solution. Similarly, $f = [f^1_{1j_1}, 
\dots, f^k_{mj_m}]$ represents the value of the objectives associated with $x$.

So for the $v$-th objective, its query-level value could be represented as the function $H$ applied to $x$: 
\begin{equation}
\begin{aligned}
& F_v = H_v (x; f) = \sum_{i=1}^{m} \sum_{j=1}^{p_i} x_{ij} \times f_{ij}^v, \\
& \quad where \sum_{j=1}^{p_i} x_{ij} = 1, i \in [1, ..., m], 
 v \in [1, ... k]
\end{aligned}
\end{equation}

For simplicity, we refer to $H_v(x;f)$ as $H_v(x)$ when there is no confusion.
Now we introduce the Weighted Sum over Functions (WSF) as: 

\begin{align}
& \arg\!\min_{x} (\sum_{v=1}^{k} w_v * H_v(x)) \\
\mbox{s.t.} & \;\; \sum_{v=1}^{k} w_v = 1, 
\quad w_v \geq 0 \text{ for } v=1,\dots, k  
\end{align}

Where $w_v$ is the weight value for objective $v$.
Next, we prove for Lemma \ref{lemma: lemma_proof}. As stated before, It is done in two steps.

\begin{proposition}
	The solution constructed using $x$ returned by WSF is Pareto optimal.
\end{proposition}

\begin{proof}
	~\\
	\indent Assume that $x^*$ (corresponding to $[F_1^*, ..., F_{k}^*]$ ) is the solution of WSF.
	Suppose that another solution $[F_1^{'}, ..., F_{k}^{'}]$ (corresponding to $x^{\prime}$)  dominates $[F_1^*, ..., F_{k}^*]$. This means that $\sum_{v=1}^{k} w_v * H_v({x^{\prime}})$ is smaller than that of $x^*$. \par
	This contradict that  $x^*$ is the solution of WSF. So there is no $[F_1^{'}, ..., F_{k}^{'}]$ dominating $[F_1^*, ..., F_{k}^*]$. Thus, $[F_1^*, ..., F_{k}^*]$ is Pareto optimal.
\end{proof}

\begin{proposition}
	The optimal solution returned by the Algorithm~\ref{alg:compress_list} is equivalent to the solution constructed using $x$ returned by WSF.
\end{proposition}

\begin{proof}
	~\\
	\indent Suppose $x^{\prime}$ is returned by WSF. The corresponding query-level solution is $[F_1^{'}, ..., F_{k}^{'}]$
	\begin{equation}
	\begin{split}
	x^{\prime} & = \arg\!\min_x(\sum_{v=1}^{k} w_v \times H_v(x))\\
	& = \arg\!\min(\sum_{v=1}^{k} w_v \times (\sum_{i=1}^{m} \sum_{j=1}^{p_i} x_{ij} \times f^v_{ij}))\\
	& = \arg\!\min(\sum_{i=1}^{m} (\sum_{v=1}^{k} \sum_{j=1}^{p_i} (w_v \times f^v_{ij}) \times x_{ij}))
	\end{split}
	\end{equation}
	
	For the solution $[F_1^{''}, ..., F_{k}^{''}]$ returned by Algorithm~\ref{alg:compress_list}, $x^{\prime\prime}$ represents the corresponding selection. It is achieved by minimizing the following formula:
	
	\begin{equation}
	\begin{split}
	& \sum_{i=1}^{m} \min_{j\in [1, p_i]} (WS_{ij})\\
	& = \sum_{i=1}^{m} \min_{j\in [1, p_i]} (\sum_{v=1}^{k} w_v \times f^v_{ij})\\
	& = \sum_{i=1}^{m} \min_{j\in [1, p_i]} (\sum_{v=1}^{k} \sum_{j=1}^{p_i}
	(w_v \times f^v_{ij}) \times x_{ij})\\
	\end{split}
	\end{equation} 
	where $WS_{ij} = \sum_{v=1}^{k} w_v \times f^v_{ij}$. Given a fixed $i$, $x_{ij}$ can only be positive (with value $1$) for one value of $j$.
	
	So, $x^{\prime\prime}$ must solve:
	\begin{equation}
	\begin{split}
	x^{\prime\prime} & = (\sum_{i=1}^{m} \arg\!\min_{x_{ij}}(\sum_{v=1}^{k} \sum_{j=1}^{p_i} (w_v \times f^v_{ij}) \times x_{ij}))\\
	 & = \arg\!\min_{x}(\sum_{i=1}^{m} (\sum_{v=1}^{k} \sum_{j=1}^{p_i} (w_v \times f^v_{ij}) \times x_{ij}))
	\end{split}
	\end{equation}
	\noindent 
	Here, optimizing for each subQ is independent of the optimization of the other subQs, so we can invert the sum over $i$ and the $\arg\min$.
	 Thus, $x^{\prime} = x^{\prime \prime}$. Therefore, WSF and Algorithm \ref{alg:compress_list} are equivalent.
	
\end{proof}
 
With these two propositions, we finish the proof of Lemma \ref{lemma: lemma_proof}.

Algorithm \ref{alg:compress_list} varies $w$ weight vectors to generate multiple query-level solutions. And under each weight vector, it takes $O(m \cdot p_{max})$ to select the optimal solution for each LQP-subtree based on WS, where $p_{max}$ is the maximum number of solutions among $m$ subQs. Thus, the overall time complexity of one $\bm{\theta_c}$ candidate is $O(w \cdot (m \cdot p_{max}))$.   

%% file: appendix/appendix_compile_time_opt.tex
\subsection{Compile-time optimization} \label{appd: compile-time_opt}

\subsubsection{$\theta_c$ enrichment}

\begin{figure}[t] 
\centering
\includegraphics[height=3.5cm,width=9cm]{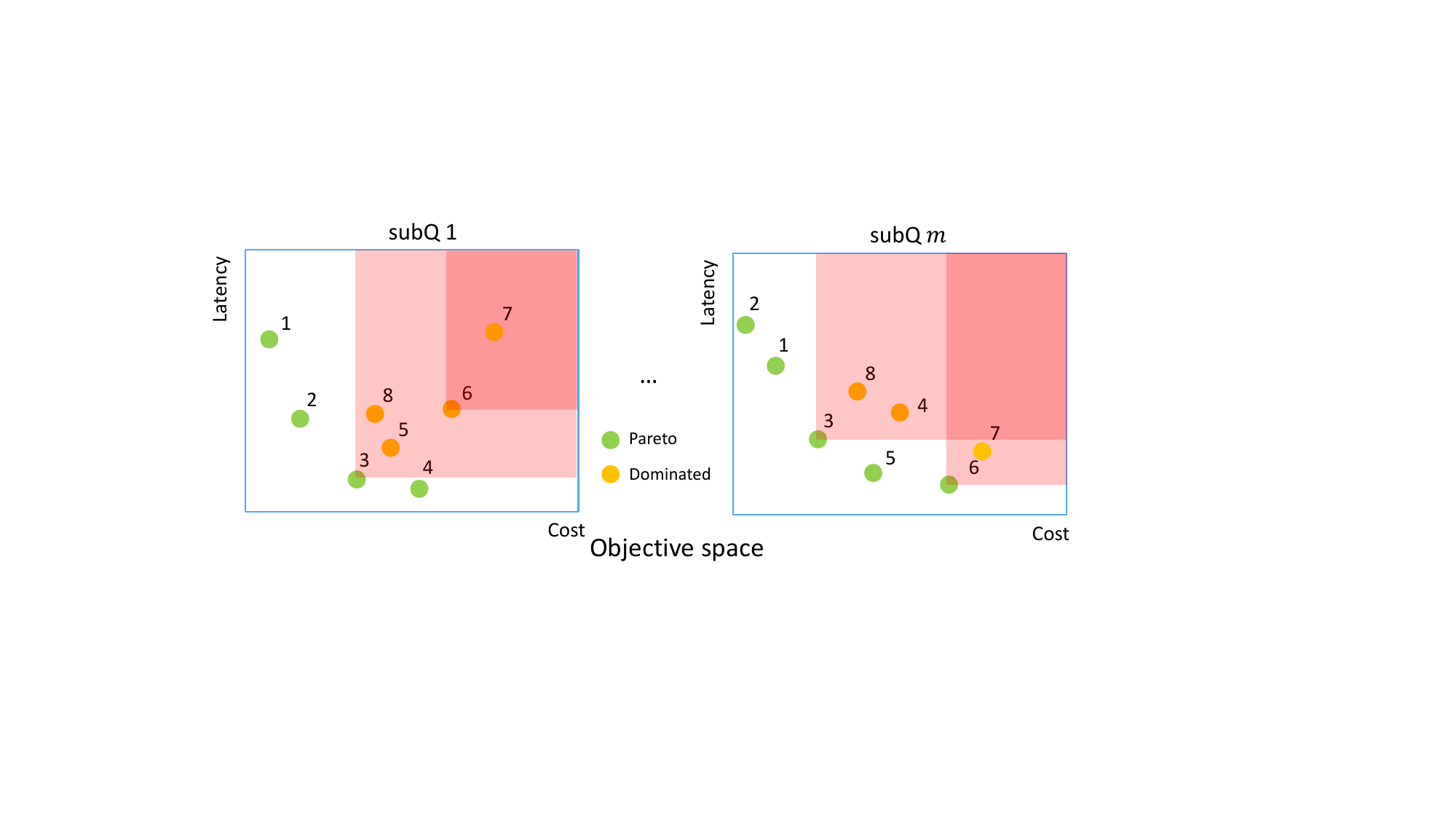} 
\caption{Objective space} 
\label{fig:red_region}
\end{figure}

\begin{figure}[t] 
\centering
\includegraphics[height=6cm,width=6cm]{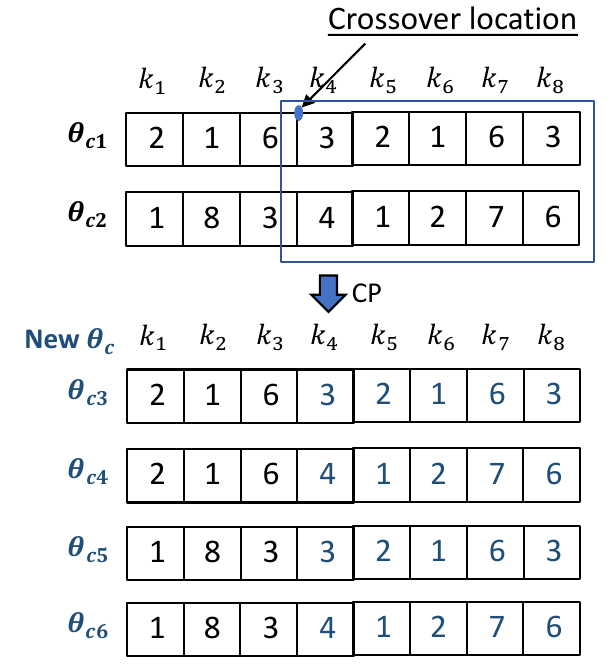} 
\caption{Example of generating new $\bm{\theta_c}$ candidates} 
\label{fig:crossover_q}
\end{figure}

The following theoretical result sheds light on subQ tuning, indicating that $\bm{\theta_c}$ derived from the subQ-level Pareto optimal solutions can serve as a useful warm-start for generating new $\bm{\theta_c}$ configurations.

\begin{proposition} \label{prop: dominated_c}
	As shown in Figure \ref{fig:red_region}, for all subQs, solutions with the same $\bm{\theta_c}$ located in the red region cannot contribute to query-level Pareto optimal solutions.
\end{proposition}

\begin{proof}
In any subQ $s$ of an arbitrary DAG, the dominated solution $f^s$ (red area) is dominated by ${f'}^s$ with any arbitrary $\bm{\theta_c}$.

Given query-level solution $F$ built from $f^s$, $F'$ built from ${f'}^s$, 
supposing $F$ is non-dominated with $F'$ or dominates $F'$, there must be at least one $f^s$ with lower latency or cost than ${f'}^s$, which is impossible and concludes the proof.  

\end{proof}

To uncover unexplored $\bm{\theta_c}$ configurations, drawing inspiration from the crossover operation in evolutionary algorithms, we introduce a heuristic method termed \textit{$\bm{\theta_c}$ crossover}. The core concept involves utilizing the \textit{Cartesian Product} (CP) operation to generate new $\bm{\theta_c}$ candidates from the existing pool. This operation is executed by randomly selecting a crossover location to divide the initial $\bm{\theta_c}$ configurations into two parts. The following example illustrates this process.

\underline{Example}
In Figure \ref{fig:crossover_q}, $\bm{\theta_{c1}}$ and $\bm{\theta_{c2}}$ represent initial $\bm{\theta_{c}}$ candidates obtained from the subQ-level tuning, each consisting of 8 variables (e.g., $k_1, ..., k_8$). The values within the boxes denote the configurations of $\bm{\theta_{c}}$. The blue point denotes a randomly generated crossover location.
Based on this location, $\bm{\theta_{c1}}$ and $\bm{\theta_{c2}}$ are divided into two parts: $k_1, k_2, k_3$ and $k_4, k_5, k_6, k_7, k_8$, delineated by a blue rectangle. A Cartesian Product (CP) is then applied to these two parts. In this example, there are two distinct configurations for $k_1, k_2, k_3$ and two distinct configurations for $k_4, k_5, k_6, k_7, k_8$. Consequently, the CP generates four configurations, represented as $\bm{\theta_{c3}}$ to $\bm{\theta_{c6}}$.
It is noteworthy that \textit{$\bm{\theta_c}$ crossover} generates new $\bm{\theta_c}$ configurations (e.g., $\bm{\theta_{c4}}$, $\bm{\theta_{c5}}$) without discarding the initial $\bm{\theta_c}$ candidates (e.g., $\bm{\theta_{c3}},\bm{\theta_{c6}}$).

\cut{
\subsection{DAG Aggregation}

\minip{A general method}

\begin{figure*}
  \centering
  \includegraphics[width=11cm,height=3cm]{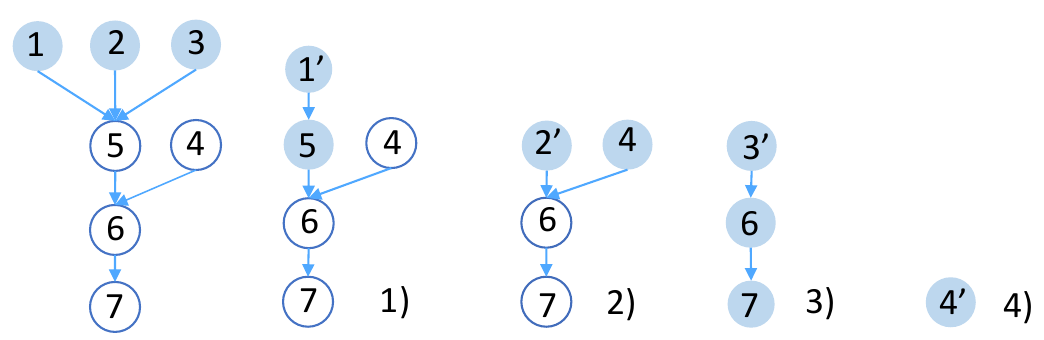}
  \caption{Example of graduallay compressing DAG}
  \label{fig:eg_compressing_tree}
\end{figure*}

\qi{I would propose to remove the following Approach Overview, as our problem is solved based on a simplified DAG. So it would be more clear if we first explain the DAG is simplified to a list structure in our problem due to the sum aggregator in our predictive model, and then present that we could have three methods to solve the optimization problem for this list structure.}

\qi{The Approach Overview shows a framework of how to deal with the scenario that the latency is calculated as the maximum among parallel nodes in the DAG. I would propose to add it to the appendix to show that we have a more general method to deal with different DAG structures. Although it is not applicable in our current problem, but it could be a feasible approach in other situations, e.g. latency is the maximum among parallel nodes rather than the summation}

\minip{Approach Overview.}
Figure \ref{fig:eg_compressing_tree} illustrates an example of the approach to solve the combinatorial MOO problem over a DAG structure. In this instance, the original LQP of the query comprises 7 subQs, with each node representing one subQ containing subQ-level Pareto optimal solutions under a fixed $\bm{\theta_c}$ candidate.

The central concept of DAG aggregation involves compressing parallel subQs (e.g., subQs 1, 2, 3) and serial subQs (e.g., subQs 6, 7). Following compression, the resulting compressed node is treated as a Pseudo-subQ (e.g., Pseudo-subQ 1', 2', 3', and 4') for further computation. Ultimately, the DAG can be compressed into a single node, where the optimal solutions of this node represent the final optimal solutions of the query.

In essence, if each compression operation is guaranteed to be efficient and maintains optimality, it ensures that the optimization over the DAG meets the requirements of both optimality and efficiency.
}

\cut{
\minip{General Divide-and-conquer (GD)}

The core operation in GD is the \textit{merge}, which enumerates over all subQ-level solutions. The following are the theoretical proof. For the simplicity of representation for subQ solutions, here we only use the subQ-level objective values of two nodes without their configurations to show how the function \textit{merge} guarantees to return a full set of Pareto optimal solutions, i.e. $\mathcal{F}_G$ and $\mathcal{F}_H$. In the following context, we use the "Linked List" to represent the simplfied DAG in our problem.

\subsubsection{Claims}

\begin{claim}
    \label{claim:list included}
    Let 
    $$+: (F_1, F_2) \in \R^d \times \R^d \mapsto F_1 + F_2 \in \R^d$$
    be a commutative and associative operation such that, for every graph \\ 
    $G = (V, E)$ that is either a `Linked List' or a list of children in a `Join',
    the evaluation function can be expressed as 
    $$ E_G(\omega) \mapsto \sum_{v \in V} \omega(v) $$
    (note that in the sum we use the addition operation from above).
    Let $A = (V, E)$ be some graph that is either a `Linked List' or a `Join'
    and let $H, G \subseteq V$ be two sets of vertices such that $H \cup G = A$,
    and $H \cap G = \emptyset$. Then 
    $$ Pf(\mathcal{F}_H  \oplus \mathcal{F}_G) \subseteq \mathcal{F}_A$$
    Where by $ \oplus $ we denote Minkowski sum of sets 
    $$ X \oplus Y = \{ x + y ,\, \forall (x, y) \in X \times Y \}$$
\end{claim}
\begin{proof}
    Let $F \in Pf(\mathcal{F}_H \oplus \mathcal{F}_G)$. By definition of the Minkowski sum, we know that 
    there exists a pair of points $F_H \in \mathcal{F}_H$, $F_G \in \mathcal{F}_G$ such that 
    $$F_H + F_G = F.$$ \\
    Finally, choose $\omega_H \in E^{-1}_{H}(F_H)$ and $\omega_{G} \in E^{-1}_{G}(F_G)$ two configurations on $H$ and 
    $G$ respectively that lead to points $F_H, F_G$. \\

    We will prove by contradiction that $p \in \mathcal{F}_A$. \\
    Assume $p \notin \mathcal{F}_A$. Define 
    $$ \omega_{A}: v \mapsto \begin{cases}
        \omega_{H}(v) & \text{if } v \in H \\
        \omega_{G}(v) & \text{if } v \in G
    \end{cases}$$
    , we know that $\omega_A$ is a valid configuration on $A$ since 
    $H \cup G = A$, and $H \cap G = \emptyset$.
    Since $F$ is just the sum $F_H$ and $F_G$, we also deduce that $F = E_G(\omega_{A}) \in \Phi_A$. \\
    Hence $F \in \Phi_A \setminus \mathcal{F}_A$ (so $F$ is an objective that is not Pareto optimal), therefore there must exist a point $F' \in \mathcal{F}_A$ that Pareto dominates $F$. 
    We can decompose the configuration $\omega' \in E^{-1}_{{A}}(F')$ 
    into two configurations $\omega'_{H} = \restr{\omega'}{H}$ and $\omega'_{G} = \restr{\omega'}{G}$ on $H$ and $G$ respectively.
    This also yields two points $F_{H}' = E_{H}(\omega_{H}') \in \mathcal{F}_H$, $F_{G}' = E_{G}(\omega_{G}') \in \mathcal{F}_G$ such that 
    $$F_{H}' + F_{G}' = F'$$  
    We get that $F_{H}' + F_{G}'$ dominates $F_{H} + F_{G}$. \\
    However, $F_{H}' + F_{G}' \in \mathcal{F}_H \oplus \mathcal{F}_G$
    and $F_{H} + F_{G} \in Pf(\mathcal{F}_H \oplus \mathcal{F}_G)$.
    By definition of the $Pf$ function, $F = F_{H} + F_{G}$ cannot be dominated
    by $F' = F_{H}' + F_{G}' \in \mathcal{F}_H \oplus \mathcal{F}_G$, which contradicts 
    the definition of $F'$ and concludes the proof.
\end{proof}

\begin{claim}
    \label{claim:list includes}
    Let 
    $$+: (F_1, F_2) \in \R^d \times \R^d \mapsto F_1 + F_2 \in \R^d$$
    be a commutative and associative operation such that, for every graph \\ 
    $G = (V, E)$ that is either a `Linked List' or a `Join',
    the evaluation function can be expressed as 
    $$ E_G(\omega) \mapsto \sum_{v \in V} \omega(v) $$
    (note that in the sum we use the addition operation from above).
    Let $A = (V, E)$ be some graph that is either a `Linked List' or a `Join'
    and let $H, G \subseteq V$ be two sets of vertices such that $H \cup G = A$,
    and $H \cap G = \emptyset$. Then 
    $$ \mathcal{F}_A \subseteq Pf(\mathcal{F}_H \oplus \mathcal{F}_G) $$
    Where by $ \oplus $ we denote Minkowski sum of sets 
    $$ X \oplus Y = \{ x + y ,\, \forall (x, y) \in X \times Y \}$$
\end{claim}
\begin{proof}
    Let $F \in \mathcal{F}_A$ and let $\omega \in E^{-1}_{A}(F)$ be some configuration that leads to $F$. 
    Let $\omega_{H} = \restr{\omega}{H}$ and $\omega_{G} = \restr{\omega}{G}$
    be the projections of $\omega$ on $H$ and $G$.
    Finally, set $F_{H} = E_{H}(\omega_{H})$ and $F_{G} = E_{H}(\omega_{H})$. 
    By definition, we know that  
    \begin{align*}
        &F = F_{H} + F_{G} \\
        &F_{H} + F_{G} \in \mathcal{F}_H \oplus \mathcal{F}_G
    \end{align*}
    and thus $F \in \mathcal{F}_H \oplus \mathcal{F}_G$. \\
    We now will prove by contradiction that $F \in Pf(\mathcal{F}_H \oplus \mathcal{F}_G)$.\\

    Indeed, assume $F \in (\mathcal{F}_H \oplus \mathcal{F}_G) \setminus Pf(\mathcal{F}_H \oplus \mathcal{F}_G)$. Then there exists a point 
    $F' \in Pf(\mathcal{F}_H \oplus \mathcal{F}_G)$ that dominates $F$. However, by \autoref{claim:list included}, we know 
    that $F' \in \mathcal{F}_A$. We have
    \begin{align*}
        & F \in \mathcal{F}_A \\
        & F' \in \mathcal{F}_A \\
        & F' \text{ dominates } F
    \end{align*}
    This is impossible, and hence the initial assumption is False. This concludes the proof.
\end{proof}

\begin{theorem}
    \label{claim:list dnc correct}
    \autoref{alg:general_div_and_conq} always output the full Pareto front of the input `Linked List'
\end{theorem}
\begin{proof}
    We will prove this by induction on the size of the list $n$.
    First, assume $n = 1$. This case is trivial as each node stores 
    the Pareto front of the underlying stage. \\
    Now assume $n \geq 1$ and we want to compute the Pareto front of a list 
    of size $n + 1$. First, we split the list in two lists of size at most $\ceil{\frac{n + 1}{2}} \leq n$.
    Thus, by the induction hypothesis, both $\mathcal{F}_H$ and $\mathcal{F}_G$ (as defined in the algorithm) will contain the full pareto fronts of the `upper' and `lower' parts of $\mathcal{S}$, respectively.
    Observe that the return value of \autoref{alg:general_div_and_conq} is equivalent in our notation to $ Pf(\mathcal{F}_H + \mathcal{F}_G)$, 
    where $+$ is defined on $\R^2 \times \R^2$ as 
    $$ \begin{pmatrix}
        latency(p_1) \\
        cost(p_1)
    \end{pmatrix} + 
    \begin{pmatrix}
        latency(p_2) \\
        cost(p_2)
    \end{pmatrix} = \begin{pmatrix}
        latency(p_1) + latency(p_2) \\
        cost(p_1) + cost(p_2)
    \end{pmatrix}.$$  
    We now use \autoref{claim:list included} and \autoref{claim:list includes} 
    (noting that $+$ is trivially associative and commutative). 
    From there we conclude that 
    $$Pf(\mathcal{F}_H \oplus \mathcal{F}_G) = \mathcal{F}_A \qquad (*).$$ 
    Observe that the LHS of $(*)$ is the ouput of \autoref{alg:list dnc}, and the RHS 
    is the full Pareto front of $A$. Hence $(*)$ concludes the proof.
\end{proof}
}
\cut{
\minip{WS-based Approximation}

Line 1 initializes the query-level objective values and configurations. Lines 3-9 address the Weighted Sum (WS) method to generate the query-level optimal solution for each weight pair. Specifically, Lines 5-8 apply the WS method to obtain the optimal solution choice for each subQ. Upon iterating through all weights, a Pareto solution set is derived after the necessary filtering.

\begin{algorithm}[t]
    \caption{Compressing\_list\_nodes}
    \label{alg:compress_list}
    \small
    \begin{algorithmic}[1]  
        \REQUIRE {nodes\_list, ws\_pairs}.
        \STATE PO = [ ], conf = \{\}
        \FOR{$[w_l, w_c]$ \textbf{in} ws\_pairs}
            \STATE po\_n = [ ], conf\_n = \{\}
            \FOR{$\bm{n}$ \textbf{in} nodes\_list}
                \STATE ${\bm{n}.po}'$ = normalize($\bm{n}.po$)
                \STATE opt\_ind, opt\_ws = minimize\_ws($[w_l, w_c]$, ${\bm{n}.po}'$)
                \STATE results$.append$($\bm{n}.po$[opt\_ind])
                \STATE conf\_n[$\bm{n}.id$] = $\bm{n}.conf[opt\_ind]$
            \ENDFOR
            \STATE PO$.append$(sum(results)), conf$.append$(conf\_n)
        \ENDFOR
        \RETURN filter(PO, conf)
    \end{algorithmic}
\end{algorithm}

\yanlei{What is the result of Philip's work? The WS algorithm comes out of nowhere. We should articulate exactly how many algorithms we have and their pros and cons.} \qi{-->added}

\input{Qi_compile_time_opt/theoretical_analysis_dag_opt.tex}
}

\minip{Analysis on Boundary-based Approximation}

\begin{figure}[h]
	\centering
		\subfigure[\small{Pareto frontier of TPCH Q2}]{\label{fig:eg_pareto}\includegraphics[height=3.5cm,width=0.23\textwidth]{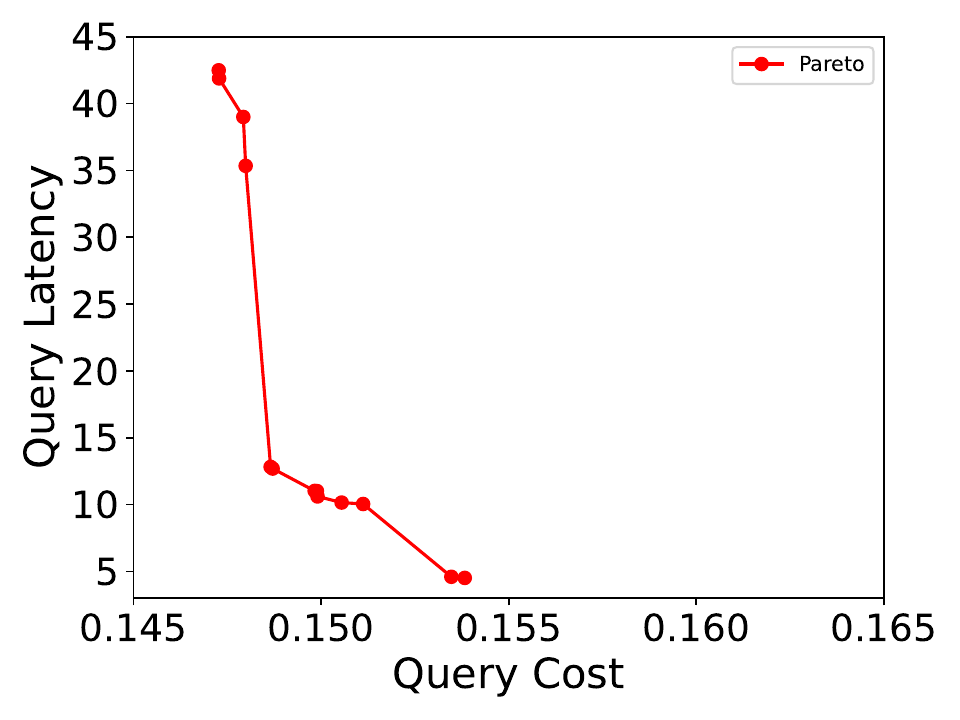}}
		\subfigure[\small{Pareto frontier of TPCH Q2 with different total resources}]	{\label{fig:eg_comp_so}\includegraphics[height=3.5cm,width=0.23\textwidth]{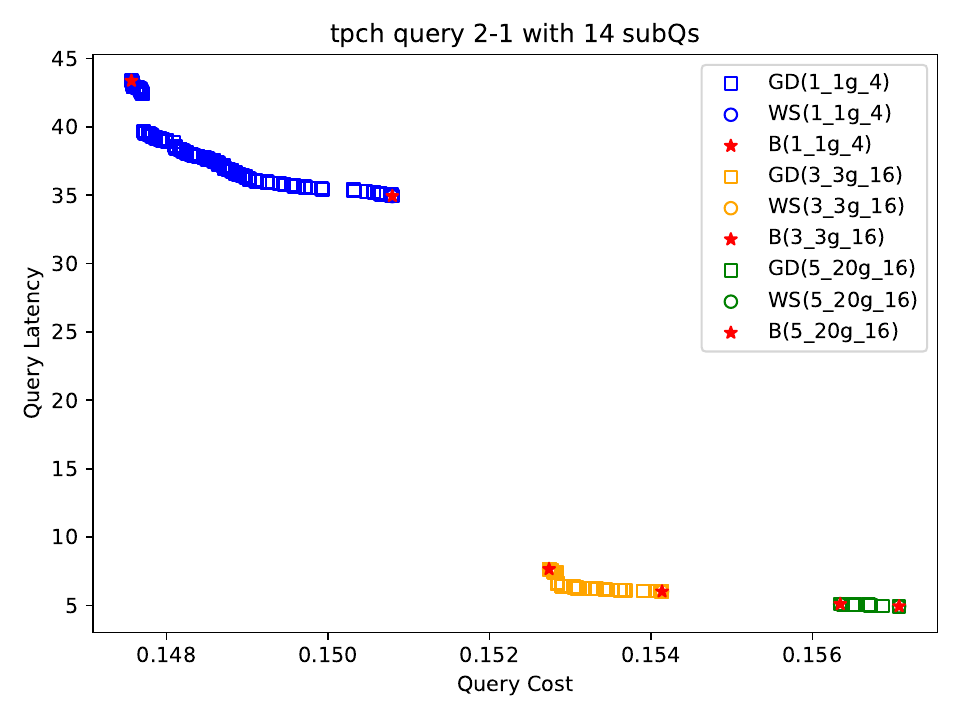}}
\vspace{-0.1in}
\caption{Objective space under differen resources}
\label{fig:diff_resources_hmooc3}
\end{figure}

Since different $\bm{\theta_c}$ values result in diverse total resources and cover various regions of the Pareto frontier, it's noteworthy that all DAG optimization methods produce the same effective set, 
as confirmed by Figure \ref{fig:diff_resources_hmooc3}. 
In scenarios with multiple $\bm{\theta_c}$ candidates, the boundary-based method achieves comparable hypervolume to the others. 

%% file: Qi_compile_time_opt/theoretical_analysis_dag_opt.tex
\minip{Theorectical Analysis}
\todo{the same as the proof in our VLDB2022 paper}

\begin{lemma} \label{lemma: lemma_proof_5.1} 
	For a DAG optimization problem with $k$ objectives that use the \texttt{sum} operator only, Algorithm~\ref{alg:compress_list} guarantees to find a subset of query-level Pareto optimal points under a fixed $\bm{\theta_c}$ candidate.
\end{lemma}

In proving Lemma \ref{lemma: lemma_proof_5.1}, we observe that Algorithm~\ref{alg:compress_list} is essentially a Weighted Sum procedure over Functions (WSF). Indeed we will prove the following two Propositions: 1) each solution returned by WSF is Pareto optimal; 2) the solution returned by the Algorithm~\ref{alg:compress_list} is equivalent to the solution returned by WSF. Then it follows that the solution returned by Algorithm~\ref{alg:compress_list} is Pareto optimal.

To introduce WSF, we first introduce the indicator function $x_{ij}$, $i \in [1, ..., m], j \in [1, ..., p_i]$, to indicate that the $j$-th solution in $i$-th LQP-subtree is selected to contribute to the query-level solution. $\sum_{j=1}^{p_i} x_{ij} = 1$ means that only one solution is selected for each LQP-subtree. Then $x = [x_{1j_1}, ... x_{mj_m}]$ represents the 0/1 selection for all $m$ LQP-subtrees to construct a query-level solution.

So for the $v$-th objective, its query-level value could be represented as the function $H$ applied to $x$: 
\begin{equation}
\begin{aligned}
& F_v = H_v (x) = \sum_{i=1}^{m} \sum_{j=1}^{p_i} x_{ij} * f_{ijv}, \\
& \quad where \sum_{j=1}^{p_i} x_{ij} = 1, i \in [1, ..., m], j \in [1, ..., p_i], v \in [1, ... k]
\end{aligned}
\end{equation}

Now we introduce the Weighted Sum over Functions (WSF) as: 

\begin{align}
& \arg\!\min_{x} (\sum_{v=1}^{k} w_v * H_v(x)) \\
\mbox{s.t.} & \;\; \sum_{v=1}^{v} w_v = 1 
\end{align}

Next, we prove for Lemma \ref{lemma: lemma_proof_5.1}. As stated before, It is done in two steps.

\begin{proposition}
	The solution constructed using $x$ returned by WSF is Pareto optimal.
\end{proposition}

\begin{proof}
	~\\
	\indent Assume that $x^*$ (correspoding to $[F_1^*, ..., F_{k}^*]$ ) is the solution of WSF.
	Suppose that an existing solution $[F_1^{'}, ..., F_{k}^{'}]$ (correspoding to $x^{'}$)  dominates $[F_1^*, ..., F_{k}^*]$. This means that $\sum_{v=1}^{k} w_v * H_v({x^{'}})$ is less than that of $x^*$. \par
	This contradict that  $x^*$ is the solution of WSF. So there is no $[F_1^{'}, ..., F_{k}^{'}]$ dominating $[F_1^*, ..., F_{k}^*]$. Thus, $[F_1^*, ..., F_{k}^*]$ is Pareto optimal.
\end{proof}

\begin{proposition}
	The optimal solution returned by the Algorithm~\ref{alg:compress_list} is equivalent to the solution constructed using $x$ returned by WSF.
\end{proposition}

\begin{proof}
	~\\
	\indent Suppose $x^{'}$ is returned by WSF. The corresponding query-level solution is $[F_1^{'}, ..., F_{k}^{'}]$
	\begin{equation}
	\begin{split}
	x^{'} & = \arg\!\min(\sum_{v=1}^{k} w_v * H_v(x))\\
	& = \arg\!\min(\sum_{v=1}^{k} w_v * (\sum_{i=1}^{m} \sum_{j=1}^{p_i} x_{ij} * f_{ijv}))\\
	& = \arg\!\min(\sum_{i=1}^{m} (\sum_{v=1}^{k} \sum_{j=1}^{p_i} (w_v * f_{ijv}) * x_{ij}))
	\end{split}
	\end{equation}
	
	For the solution $[F_1^{''}, ..., F_{k}^{''}]$ returned by Algorithm~\ref{alg:compress_list}, $x^{''}$ represents the corresponding selection. It is achieved by minimizing the following formula:
	
	\begin{equation}
	\begin{split}
	& \sum_{i=1}^{m} (WS_{ij} | j\in[1, p_i])\\
	& = \sum_{i=1}^{m} (\sum_{v=1}^{k} w_v * f_{ijv} | j\in[1, p_i])\\
	& = \sum_{i=1}^{m} (\sum_{v=1}^{k} \sum_{j=1}^{p_i}
	(w_v * f_{ijv}) * x_{ij})\\
	\end{split}
	\end{equation} 
	where $WS_{ij} = \sum_{v=1}^{k} w_v * f_{ijv}$. 
	
	So, we have:
	\begin{equation}
	\begin{split}
	x^{''} & = \arg\!\min(\sum_{i=1}^{m} (\sum_{v=1}^{k} \sum_{j=1}^{p_i} (w_v * f_{ijv}) * x_{ij}))
	\end{split}
	\end{equation}
	\noindent 
	 Since they are of the same form and achieve min at the same time, so we have $[F_1^{'}, ..., F_{k}^{'}] = [F_1^{''}, ..., F_{k}^{''}]$
	
\end{proof}
 
With these two propositions, we finish the proof of Lemma \ref{lemma: lemma_proof_5.1}.

Algorithm \ref{alg:compress_list} varies $w$ weight vectors to generate multiple query-level solutions. And under each weight vector, it takes $O(m \cdot p_{max})$ to select the optimal solution for each LQP-subtree based on WS, where $p_{max}$ is the maximum number of solutions among $m$ LQP-subtrees. Thus, the overall time complexity of one $\bm{\theta_c}$ candidate is $O(w \cdot (m \cdot p_{max}))$.   

%% file: appendix/app-runtime-opt.tex
\subsection{Additional Materials for Runtime Optimization}
\label{appendix:runtime-optimization}

\subsubsection{More on $\thetabm_p$ and $\thetabm_s$ Aggregation}
Ideally, one could copy $\thetabm_p$ and $\thetabm_s$ from the initial subQ, allowing the runtime optimizer to adjust them by adapting to the real statistics. 

Given the constraint that Spark takes only one copy of $\thetabm_p$ and $\thetabm_s$ at query submission time, we intelligently aggregate the fine-grained $\thetabm_p$ and $\thetabm_s$ from compile-time optimization to initialize the runtime process. In particular, Spark AQE can convert a sort-merge join (SMJ) to a shuffled hash join (SHJ) or a broadcast hash join (BHJ), but not vice versa. Thus, imposing high thresholds ($s_3, s_4$ in Table~\ref{tab:spark-params}) to force SHJ or BHJ based on inaccurate compile-time cardinality can result in suboptimal plans (as shown in Figure~\ref{fig:tpch-9-pqp}). 
On the other hand, setting these thresholds to zero at SQL submission might overlook opportunities for applying BHJs, especially for joins rooted in scan-based subQs with small input sizes.
To mitigate this, we initialize $\thetabm_p$ with the smallest threshold among all join-based subQs, enabling more informed runtime decisions. 
In addition, we cap these thresholds (25 MB for broadcast threshold $s_4$ and 0 MB for shuffle hash threshold $s_5$) at their default values to ensure BHJs are not missed for small scan-based subQs.

\subsubsection{More on Pruning Optimization Requests}
To address this, we established rules to prune unnecessary requests based on the runtime semantics of parametric rules: LQP parametric rules are used to decide join algorithms and QS parametric rules are used to re-balance the data partitions in a post-shuffle QS.
Therefore, we bypass requests for non-join operations and defer requests for \collap\ containing join operators until all input statistics are available, thereby avoiding decisions based on inaccurate cardinality estimations. 
Additionally, we skip all the scan-based QSs and only send the requests when the input size of a QS is larger than the target partition size (configured by $s_1$).
By applying the above rules, we substantially reduce the total number of optimization calls by 86\% and 92\% for TPC-H and TPC-DS respectively.

\subsection{Scalability Issues}

The dimensionality of the parameter space is considered in three aspects in this work.
\begin{enumerate}
\item \underline{Initial feature selection}: We clarify that Spark has over 100 parameters and we selected 19 features as follows: (a) We keep query plan parameters $\thetabm_p$ and partition related parameters $\thetabm_s$, which can be replicated to have one copy for each query stage in our fine-grained tuning approach. (b) Context parameters: The dimensionality of the context parameter space ($\thetabm_c$) can reach up to hundreds~\cite{Locat_2022}. To address the issue, we use the LASSO-path feature selection approach (as in our prior work~\cite{spark-moo-icde21} and in Ottertune~\cite{VanAken:2017:ADM}) and integrate domain knowledge from Spark documentation, resulting in a list of 8 context parameters. More details are in \techreport{Appendix~\ref{appendix:parameter-selection}}.
\item \underline{Feature importance score (FIS) based parameter filtering}: \\ We conducted additional experimental to compute the feature importance scores (FIS) of the 19 parameters in our trained latency and data shuffle models. The FIS of a parameter here is defined to be the increase of the model error if the values of this parameter are randomly shuffled in the training set~\cite{permutation-importance}.results in Figure~\ref{fig:fi-tpch-lat-all-parameters}-\ref{figs:fi-norm} show the FIS of each parameters in a long tail distribution. The parameters at the tail, with a cumulative FIS less than 5\% in WMAPE, do not affect the model performance by much and hence can be filtered out. For instance, the threshold of 5\% cumulative FIS means an increment of at most 0.7\% in WMAPE for TPC-H and 1.2\% for TPC-DS in the subQ model.
\item \underline{Sampling with FIS-based parameter filtering}: Then we introduced three sampling methods that could leverage FIS-based selection: 
(a) random sampling with or without parameter filtering; 
(b) Latin hypercube sampling with or without parameter filtering; 
(c) adaptive grid search with inherent feature filtering, as we described in the new paragraph starting with ``Sampling methods'' in Section~\ref{subsec:subq-tuning}.
\item \underline{Experimental results}: Results in Figure~\ref{fig:hv_tpch_tpcds}-\ref{fig:time_tpch_tpcds} show performance metrics (hypervolume and solving time) of the above three sampling methods with parameter filtering. Among them, we report the adaptive grid search to be an overall winner, achieving the highest hypervolume among all FIS-based methods while keeping the solving time under 2s for most queries tested.
\item \underline{Runtime adaptive scheme for adaptive grid search}: \\ If the predicted latency under the default configuration is over 10s, we use the median sampling rate; otherwise, a low sampling rate. Our analyses show that further increasing the sampling rate does not improve performance but increases solving time, hence negatively impacting query performance.
\end{enumerate}

\subsubsection{More on Adaptive Grid Search}\label{appendix:adaptive-grid-search}

\begin{figure*}[t]
    \centering
    \includegraphics[width=.99\linewidth]{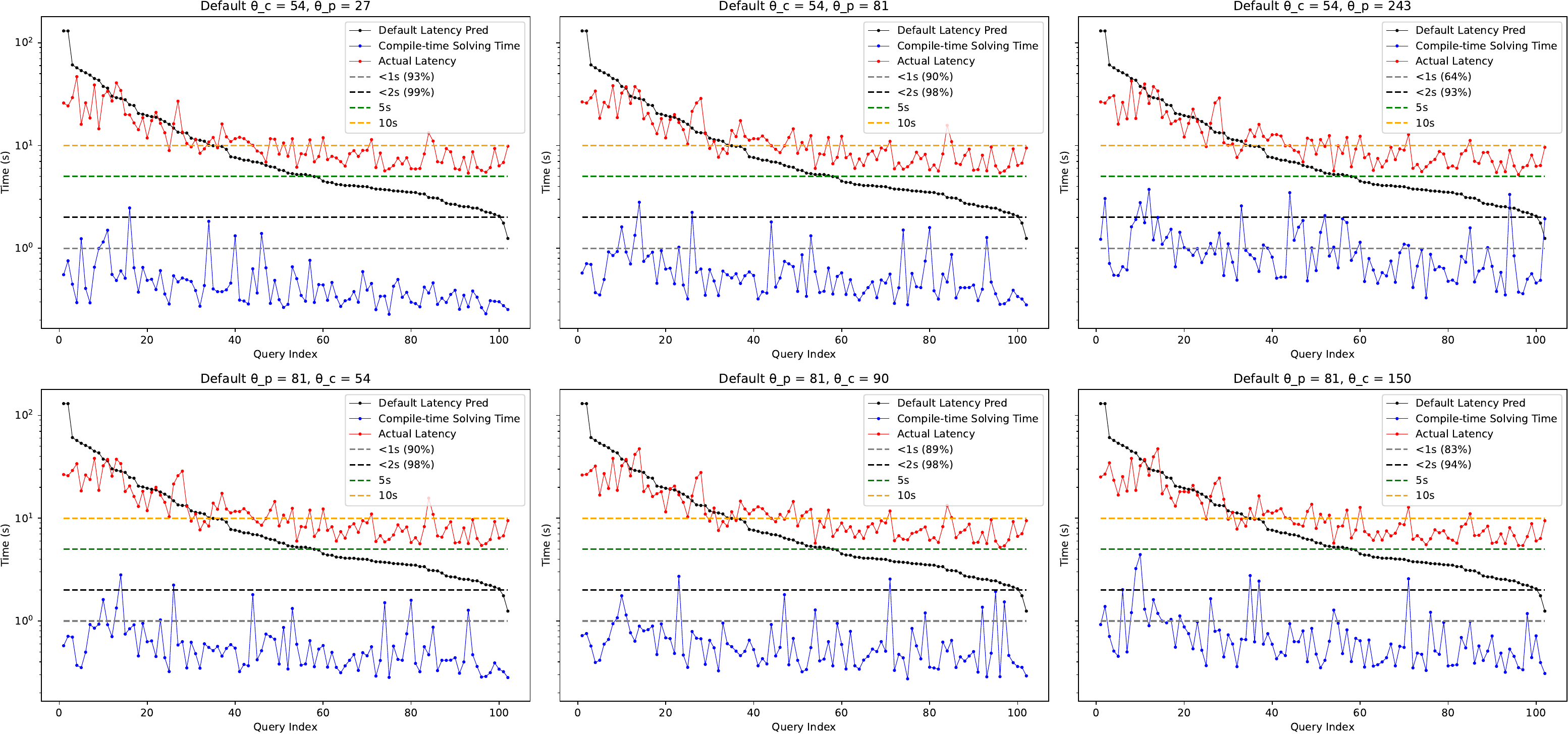}
    \caption{Solving time, predicted latency, and actual latency under different sampling rates for ($\thetabm_c$ and $\thetabm_p$)}
    \label{fig:agrid-sampling-rates}    
\end{figure*}

\begin{figure*}
	\centering
		\subfigure[\small{Actual Time}]	{\label{fig:diff-sampling-rates-actual-time}\includegraphics[width=0.45\textwidth]{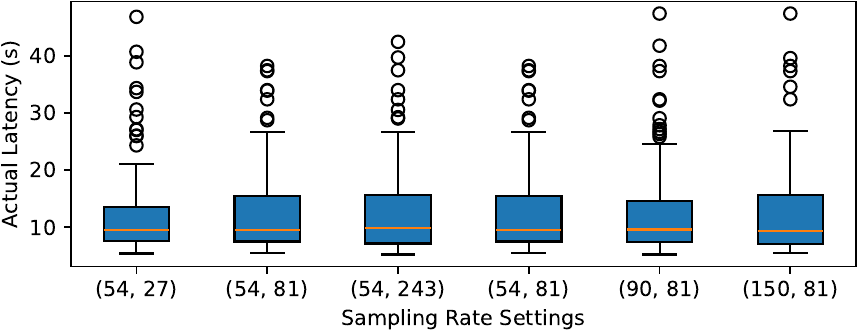}}
		\subfigure[\small{Solving Time}]{\label{fig:diff-sampling-rates-solving-time}\includegraphics[width=0.45\textwidth]{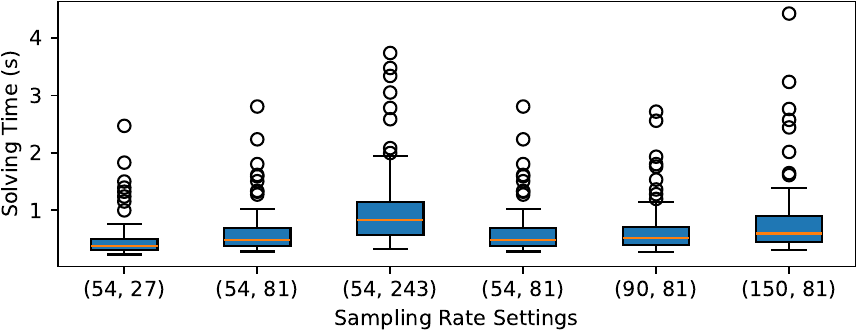}}        
\caption{Solving time and actual runtime under different sampling rates for TPC-DS}
\label{fig:diff_resources}
\end{figure*}

\begin{figure*}[t]
    \centering
    \includegraphics[width=.9\linewidth]{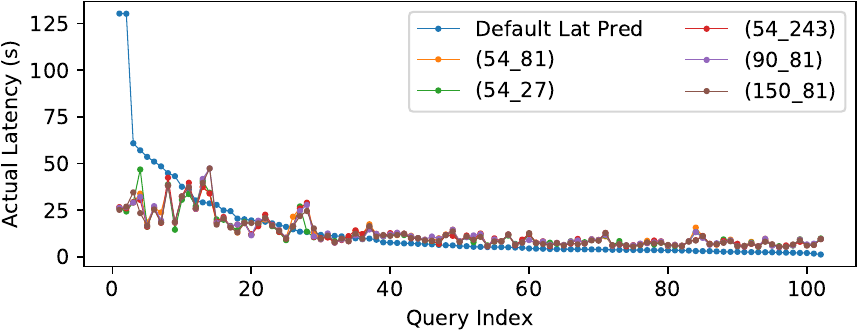}
    \caption{Actual latency under different sampling rates for ($\thetabm_c$ and $\thetabm_p$)}
    \label{fig:agrid-sampling-rates-actual-lat}
\end{figure*}

Figure~\ref{fig:agrid-sampling-rates} presents the latency prediction, solving time, and actual latency under different sampling rates for $\thetabm_c$ and $\thetabm_p$. 
As summarized in Figure~\ref{fig:diff-sampling-rates-actual-time}, when increasing the sampling rate, the actual runtime does not necessarily improve, but using a low sampling rate can lead to more instable performance for longer-running queries, as shown in Figure~\ref{fig:agrid-sampling-rates-actual-lat}.
As summarized in Figure~\ref{fig:diff-sampling-rates-solving-time}, the solving time increases from the low sampling rate to high sampling rate for both $\thetabm_c$ and $\thetabm_p$.

Therefore, we use medium sampling rate (54, 81) for longer running queries (with predicted latency over 10s) and low sampling rate (27, 54) for shorter running queries (with predicted latency under 10s) to achieve the best performance.

%% file: appendix/app-experiments.tex
\section{Additional Experimental Details}

\subsection{More Setup}\label{appendix:expt-more-setup}

\subsubsection{Hardware}
We use two 6-node Spark 3.5.0 clusters with runtime optimization plugins. 
Each node is CentOS-based with 2 16-core Intel Xeon Gold 6130 processors, 768 GB of RAM, and RAID disks, connected with 100 Gbps Ethernet.

\subsubsection{Parametric Optimization Rules}
\label{appendix:optimization-rules}
\begin{figure}[t]
	\centering
	\captionsetup{justification=centering}
	\includegraphics[width=.5\linewidth,height=3cm]{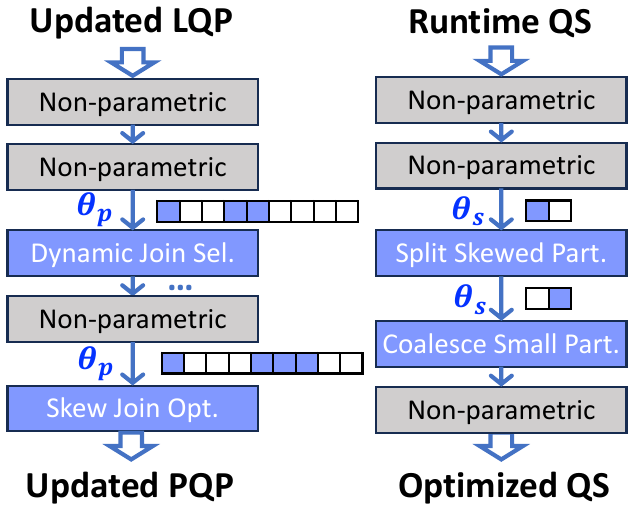}
	\captionof{figure}{\small{Runtime Opt. Rules}}
	\label{fig:opt-rules}
\end{figure}

Figure~\ref{fig:opt-rules} illustrates the transformations of a collapsed LQP and a runtime QS, which both pass through a pipeline of parametric rules (blue) and non-parametric rules (gray). 

\subsubsection{Spark Parameters Details}\label{appendix:spark-params}

We list our 19 selected Spark parameters in Table~\ref{tab:spark-params}, which are categorized into three groups: context parameters ($\thetabm_c$), logical query plan parameters ($\thetabm_p$), and query stage parameters ($\thetabm_s$). 
The default configuration is set to Spark's default values shown in the table.

\begin{table*}[ht]
    \ra{1.1}
    \footnotesize
    \centering
    \caption{\small (Selected) Spark parameters in three categories}
    \label{tab:spark-params-all}
    \begin{tabular}{p{0.1cm}p{8cm}p{1cm}}
    \midrule
    $\thetabm_c$ &{\bf Context Parameters} & {\bf Default}  \\
    \midrule
    $k_1$ & \verb|spark.executor.cores| & 1 \\
    $k_2$ & \verb|spark.executor.memory| & 1G \\
    $k_3$ & \verb|spark.executor.instances| & - \\
    $k_4$ & \verb|spark.default.parallelism| & - \\
    $k_5$ & \verb|spark.reducer.maxSizeInFlight| & 48M \\
    $k_6$ & \verb|spark.shuffle.sort.bypassMergeThreshold| & 200\\
    $k_7$ & \verb|spark.shuffle.compress| & True \\
    $k_8$ & \verb|spark.memory.fraction| & 0.6 \\
    \midrule
    $\thetabm_p$ &{\bf Logical Query Plan Parameters} & {\bf Default}\\
    \midrule
    $s_1$ & \verb|spark.sql.adaptive.advisoryPartitionSizeInBytes| & 64M \\
    $s_2$ & \verb|spark.sql.adaptive.nonEmptyPartitionRatioForBroad| \verb|castJoin| & 0.2 \\
    $s_3$ & \verb|spark.sql.adaptive.maxShuffledHashJoinLocalMapThr| \verb|eshold| & 0b \\
    $s_4$ & \verb|spark.sql.adaptive.autoBroadcastJoinThreshold| & 10M\\
    $s_5$ & \verb|spark.sql.shuffle.partitions| & 200 \\
    $s_6$ & \verb|spark.sql.adaptive.skewJoin.skewedPartitionThresh| \verb|oldInBytes| & 256M \\
    $s_7$ & \verb|spark.sql.adaptive.skewJoin.skewedPartitionFactor| & 5.0 \\
    $s_8$ & \verb|spark.sql.files.maxPartitionBytes| & 128M \\
    $s_9$ & \verb|spark.sql.files.openCostInBytes| & 4M\\
    \midrule
    $\thetabm_s$ &{\bf Query Stage Parameters} & {\bf Default}\\
    \midrule
    $s_{10}$ & \verb|spark.sql.adaptive.rebalancePartitionsSmallParti| \verb|tionFactor| & 0.2 \\
    $s_{11}$ & \verb|spark.sql.adaptive.coalescePartitions.minPartiti| \verb|onSize| & 1M \\
    \midrule
    \end{tabular}
\end{table*}

\rv{\subsection{Tunable Parameter Selection}
\label{appendix:parameter-selection}

\begin{figure}[t]
    \centering
    \begin{tabular}{lr}
        \subfigure[\small{Q10}]
        {\label{fig:lasso-q10}\includegraphics[height=3.3cm,width=.48\linewidth]{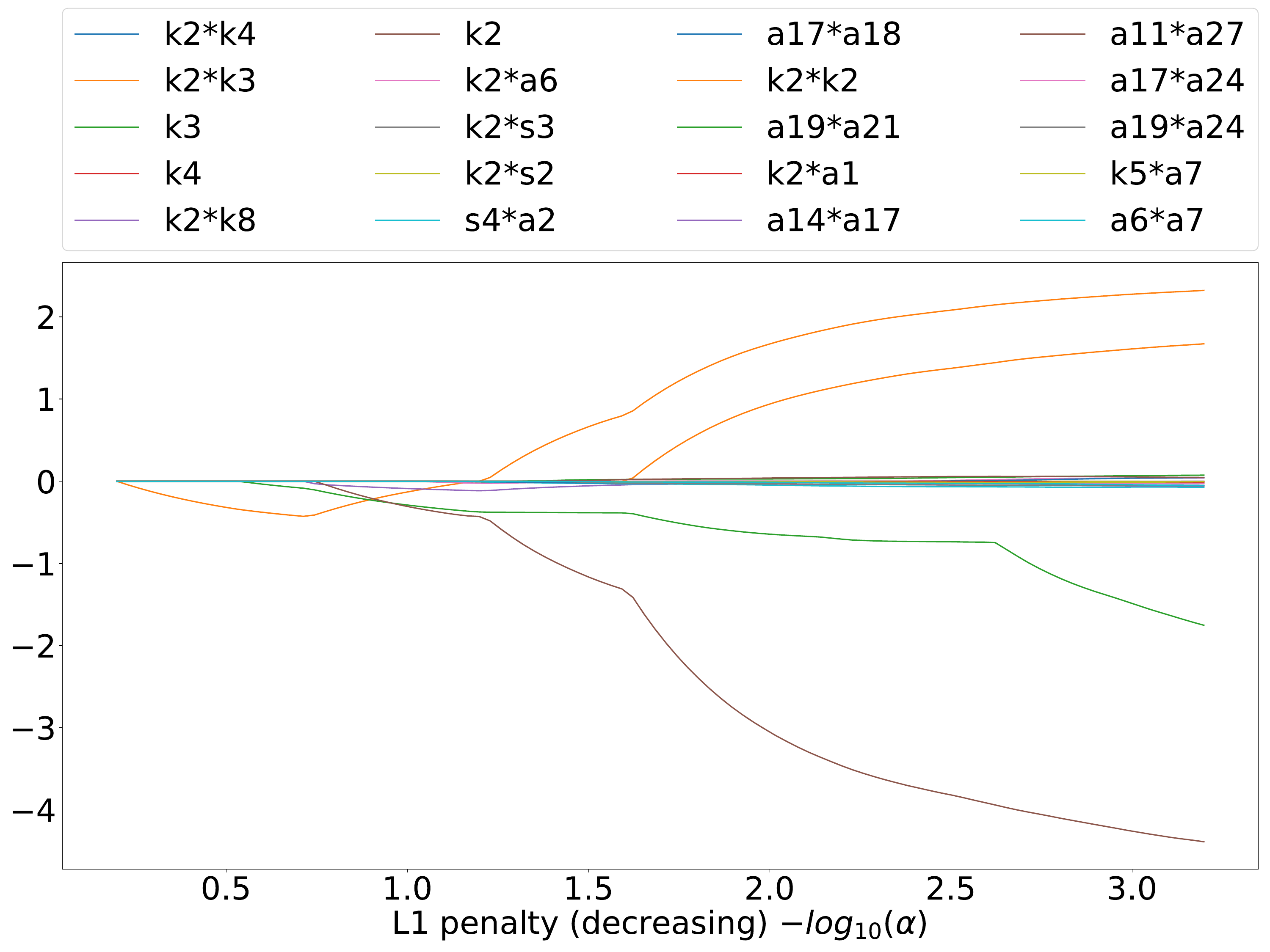}}
        \hfill
        \subfigure[\small{Q13}]
        {\label{fig:lasso-q13}\includegraphics[height=3.3cm,width=.48\linewidth]{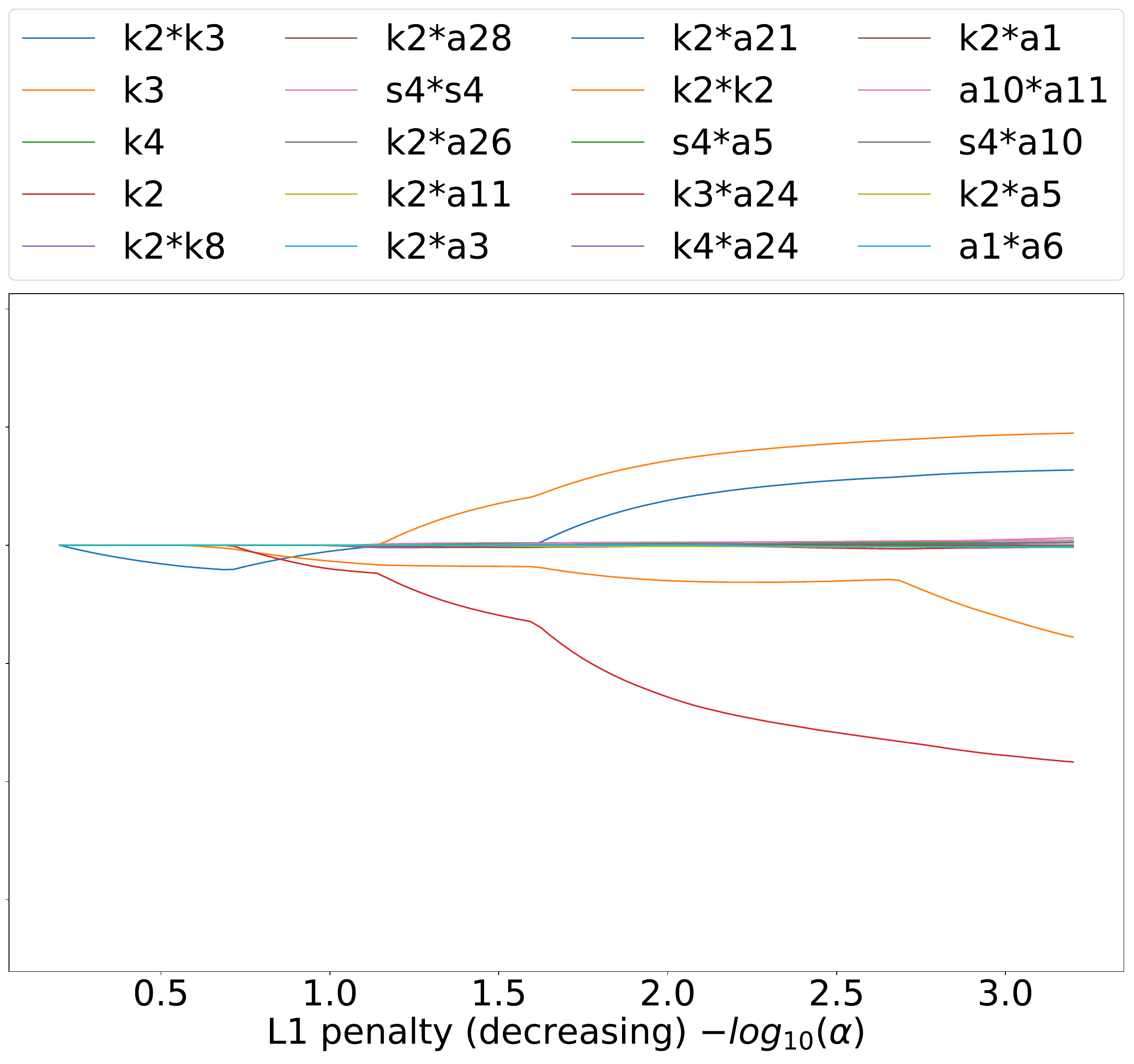}}				
    \end{tabular}
    \caption{\small{LASSO path for different queries, with weights of different parameters against the L1 penalty (varied from $10^{-3}$ to $10^3$}). The importance of each parameter is ordered by the timing of its weight deviating from zero.}
    \label{figs:lasso-path}
\end{figure}

We first select parameters for each query based on the LASSO path approach as our prior work~\cite{spark-moo-icde21} and Ottertune~\cite{VanAken:2017:ADM} did. 
For each query, we prepared ~50 context parameters to form a tunable Spark configuration in high dimension, and run each query over 1000 different configurations. Based on the traces, we trained LASSO regression models for each query and generate the LASSO Path in each model to evaluate the importance of each parameter.
Figure~\ref{figs:lasso-path} shows the LASSO path for different queries. As shown in the figure, important parameters from different queries can vary a lot among each other. Therefore, we further integrate the domain knowledge from the Spark official documentation into the results of the LASSO Path approach to agree on a final list parameters as $\thetabm_c$.  

\subsection{Adapting Important Parameters}\label{appendix:fis-filtering}

We show the feature importance score (FIS), in terms of the permutation importance~\cite{permutation-importance} of  all dimensions of the graph embedding (shown as the non-decision variables), and the selected tunable parameters, shown as the context parameters ($\thetabm_c$) and SQL parameters ($\thetabm_p$ and $\thetabm_s$) in the latency subQ model for TPC-H in Figure~\ref{fig:fi-tpch-lat-all-parameters}. The importance score indicates the performance (WMAPE) drop when the corresponding feature is randomly shuffled. A larger number indicates a more critical feature.

Figure~\ref{figs:fi-tpch-tpcds} shows the importance score of tunable parameters in TPC-H and -DS. We show the importance score of the context parameters ($\thetabm_c$), SQL parameters ($\thetabm_p$ and $\thetabm_s$) in the latency and shuffle size subQ models for TPC-H and -DS. 

To select important parameters from the four diverse models, we first reserve all query structure related parameters (e.g., $s_3$ and $s_4$). Then we normalize the importance score of each parameter in each model by the WMAPE values, and aggregate the normalized scores to show a global importance score of each parameter. 
Figure~\ref{figs:fi-norm} shows the normalized importance scores of all parameters in a long-tail distribution. By further filtering the long tail of parameters whose cumulative FIS is less than a threshold (e.g., 5\%), we get a list of important parameters, including $k_7$, $k_1$, $k_3$, $k_2$ for context parameters, and $s_3, s_4, s_5, s_8, s_9, s_1$ for the SQL parameters.

\begin{figure*}[t]
    \centering
    \includegraphics[width=.9\linewidth]{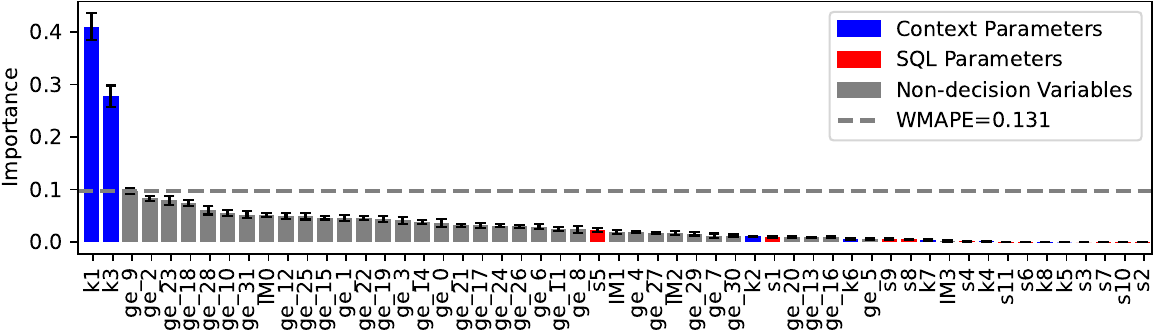}
    \caption{Feature importance scores of all parameters in the latency subQ model for TPC-H}
    \label{fig:fi-tpch-lat-all-parameters}
\end{figure*}

\begin{figure*}[t]
    \centering
    \begin{tabular}{cccc}
        \subfigure[\small{TPC-H, Latency, Zoom out}]
        {\label{fig:fi-ori-tpch-lat}\includegraphics[width=.24\textwidth]{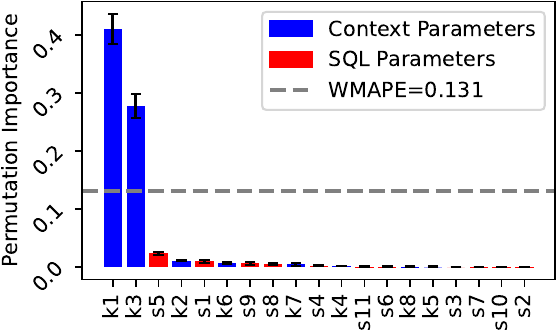}}
        \hfill
        \subfigure[\small{TPC-H, Shuffle(MB), Zoom out}]
        {\label{fig:fi-ori-tpch-io}\includegraphics[width=.24\textwidth]{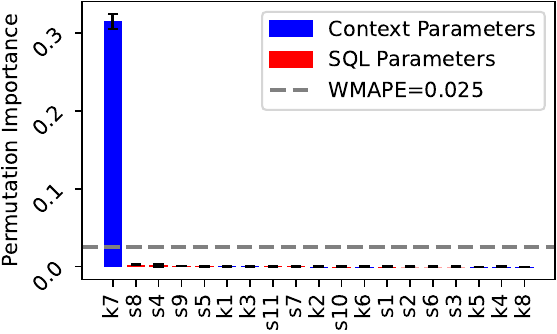}}
        \hfill
        \subfigure[\small{TPC-DS, Latency, Zoom out}]
        {\label{fig:fi-ori-tpcds-lat}\includegraphics[width=.24\textwidth]{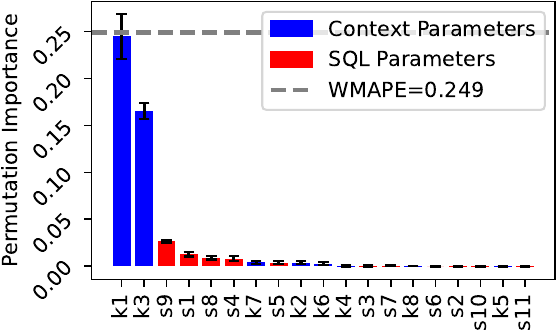}}
        \hfill
        \subfigure[\small{TPC-DS, Shuffle, Zoom out}]
        {\label{fig:fi-ori-tpcds-io}\includegraphics[width=.24\textwidth]{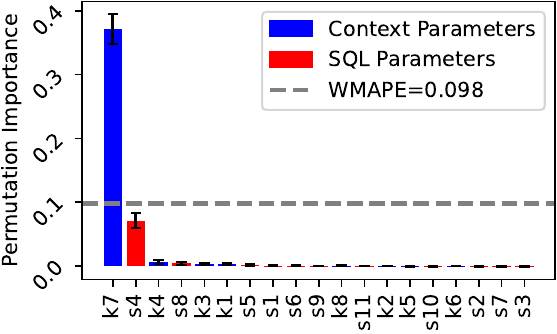}}
        \\
        \subfigure[\small{TPC-H, Latency, Zoom in}]
        {\label{fig:fi-zoomin-tpch-lat}\includegraphics[width=.24\textwidth]{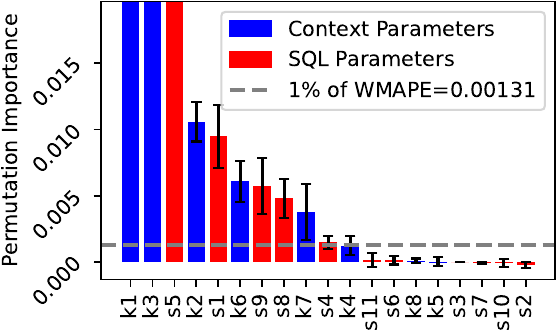}}
        \hfill
        \subfigure[\small{TPC-H, Shuffle(MB), Zoom in}]
        {\label{fig:fi-zoomin-tpch-io}\includegraphics[width=.24\textwidth]{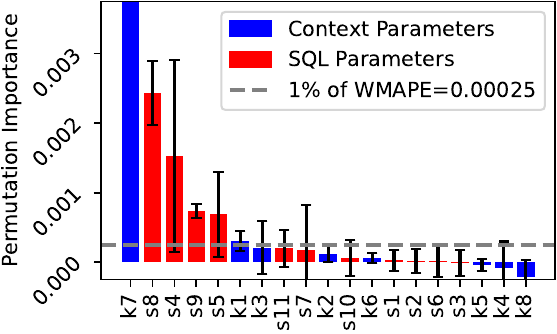}}
        \hfill
        \subfigure[\small{TPC-DS, Latency, Zoom in}]
        {\label{fig:fi-zoomin-tpcds-lat}\includegraphics[width=.24\textwidth]{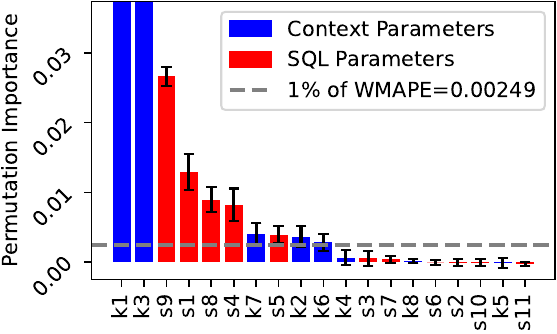}}
        \hfill
        \subfigure[\small{TPC-DS, Shuffle, Zoom in}]
        {\label{fig:fi-zoomin-tpcds-io}\includegraphics[width=.24\textwidth]{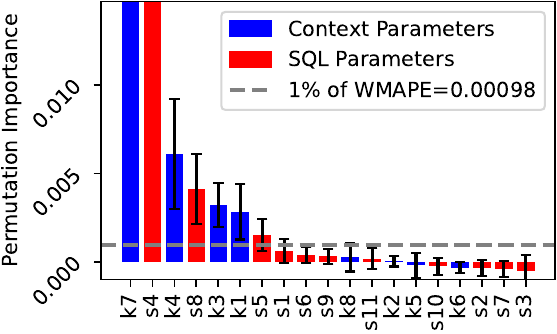}}
    \end{tabular}
    \caption{Feature importance scores of tunable parameters in TPC-H and TPC-DS}
    \label{figs:fi-tpch-tpcds}
\end{figure*}

\begin{figure}[t]
    \centering
    \begin{tabular}{lr}
        \subfigure[\small{Zoom out}]
        {\label{fig:fi-norm-origin}\includegraphics[width=.48\linewidth]{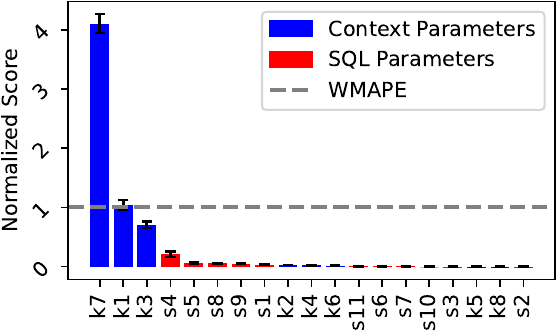}}
        \hfill
        \subfigure[\small{Zoom in}]
        {\label{fig:fi-norm-origin}\includegraphics[width=.48\linewidth]{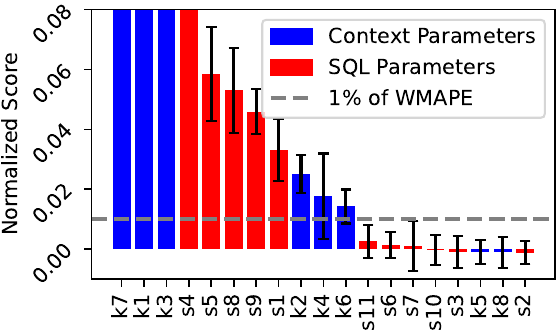}}				
    \end{tabular}
    \caption{\small{Normalized feature importance scores of all parameters across four models}}
    \label{figs:fi-norm}
\end{figure}

}

\rv{
\subsection{Handle query plan related parameters}

As explained in Section~\ref{sec:runtime-opt}, the join threshold parameters \texttt{s4} cannot be set either too small or too big.
Therefore, it is set as the $\max(25MB, \min_{i} s_4^{(i)})$ where $i$ is the indicator of all subQs with at least one join operator.
}




\subsection{Specific Knob Concerns}
\label{appendix:knob-concerns}

\begin{figure}[t]
    \centering
    \begin{tabular}{ll}
        \subfigure[\small{query stage 1 (QS1) and query stage 4 (QS4) do not run in parallel}]
        {\label{fig:eg-stage-trial2}\includegraphics[width=.98\linewidth]{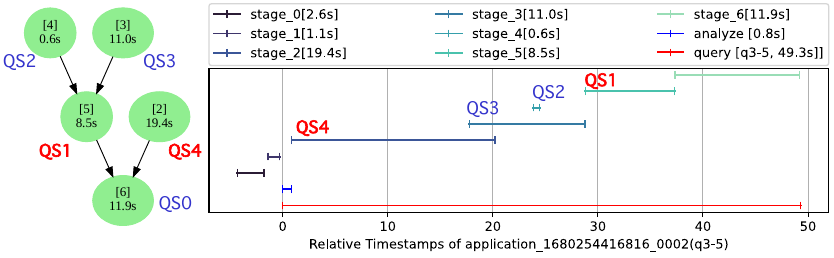}}
        \\
        \subfigure[\small{query stage 1 (QS1) and query stage 4 (QS4) run in parallel}]
        {\label{fig:eg-stage-trial3}\includegraphics[width=.98\linewidth]{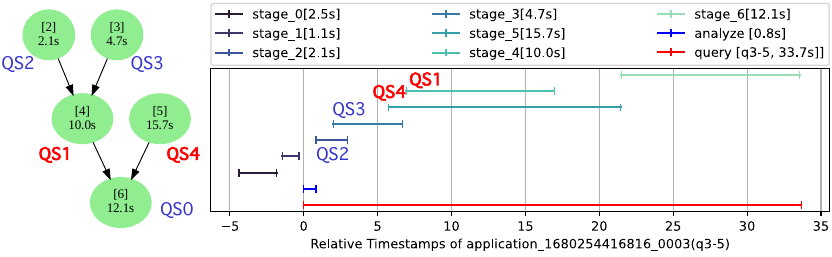}} 					  
    \end{tabular}
    \captionof{figure}{\small{Repeated Runs of TPCH Q3 with the Same Configuration. In Figure~\ref{fig:eg-stage-trial2}, QS4 runs logically ahead of QS2 and QS3 and finishes before running QS1. Therefore, QS1 runs by itself without any resource sharing, and the query takes 49.3s.
In Figure~\ref{fig:eg-stage-trial3}, QS2 and QS3 run logically ahead of QS4, and hence QS1 and QS4 run in parallel by sharing the resources with a 33.7s query latency.}}
    \label{fig:eg-stage-interleaving}
\end{figure}

\underline{spark.sql.adaptive.enable=true}: The parameter was introduced in Spark 1.6 and has been set to true by default since Spark 3.2. We have chosen to enable it for two main reasons.
First, enabling adaptive query execution (AQE) allows us to perform parameter tuning at the stage level. 
Specifically, our runtime optimizer could fine-tune the runtime parameters (at the stage level) while AQE re-optimizes the query plan based on the actual runtime statistics in the middle of query execution.
Second, enabling AQE improves the robustness of query latency. When AQE is disabled, the DAGScheduler asynchronous converts the entire query to a DAG of stages. Consequently, parallel stages can be randomly interleaved during query execution, leading to unpredictable query latencies.
For instance, Figure~\ref{fig:eg-stage-interleaving} demonstrates that disabling AQE can result in different stage interleaving patterns, causing a significant 46\% increase in query latency. 
When AQE is enabled, stages are wrapped in query stages (QSs) that are synchronously created. As a result, the stage interleaving patterns are consistent for running one query, making the query performance more stable and predictable.

\begin{figure*}[t]
    \centering
    \begin{tabular}{lcr}
        \subfigure[\small{Case 1: locality.wait=3s, costs 11.7s}]
        {\label{fig:locality-1}\includegraphics[height=3.3cm,width=.3\textwidth]{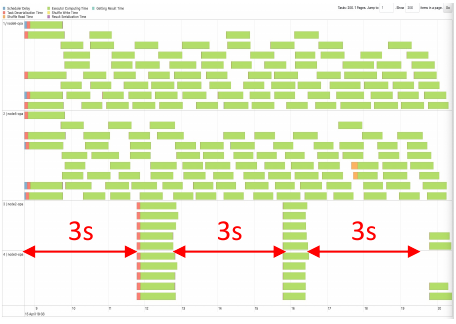}}
        \hfill
        \subfigure[\small{Case 2: locality.wait=3s, cost 18.4s}]
        {\label{fig:locality-2}\includegraphics[height=3.3cm,width=.3\textwidth]{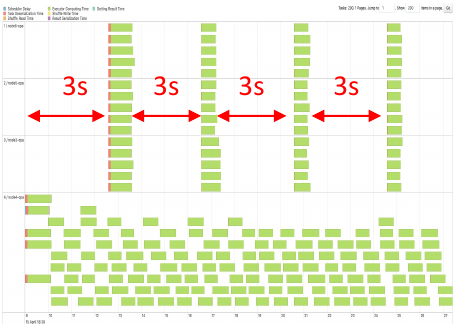}}
        \hfill
        \subfigure[\small{Case 3: locality.wait=0s, cost 7.0s}]
        {\label{fig:locality-3}\includegraphics[height=3.3cm,width=.32\textwidth]{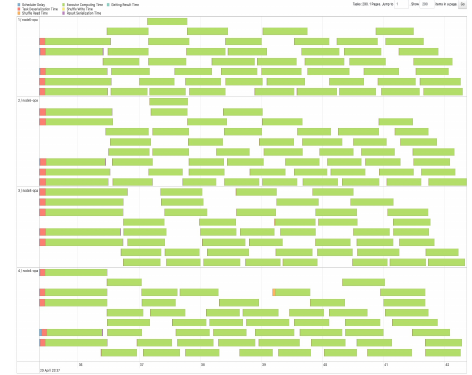}}				
    \end{tabular}
    \caption{The end-to-end stage latency comparison over different settings for spark.locality.wait.}
    \label{figs:locality-waiting}
\end{figure*}

\underline{spark.locality.wait=0s}: The default value of the parameter is 3s, which specifies the wait time for launching a data-local task before giving up and launching it on a less-local node. 
However, the waiting behavior can introduce instability in query performance due to the randomness of locality detection.
As demonstrated in Figure~\ref{fig:locality-1} and Figure~\ref{fig:locality-2}, the latency of a stage can vary significantly (changing from 11.7s to 18.4s) when different locality detection methods are employed. 
To ensure a stable query performance in our workload, we have fixed the \texttt{spark.locality.wait} parameter to 0s, thereby avoiding the waiting time for locality and achieving consistent and better query performance as shown in Figure~\ref{fig:locality-3}.
It is worth noting that in the production environment~\cite{LyuFSSD22}, the impact of locality is mitigated due to the high-speed network cards, which aligns with a near-zero waiting time.

\underline{spark.sql.adaptive.coalescePartitions.parallelismFirst=false}: We respect the recommendation on the Spark official website and set this parameter to false such that the advisory partition size will be respected when coalescing contiguous shuffle partitions.

\underline{spark.sql.adaptive.forceOptimizeSkewedJoin=false}: We follow \\ the default setting for the parameter and avoid optimizing the skewed joins if it requires an extra shuffle. However, if needed, we can set it to true and adjust our tuning process to consider $s_5$-$s_7$ as three additional plan-dependent parameters.

\begin{figure*}[t]
    \centering
    \includegraphics[width=.9\linewidth]{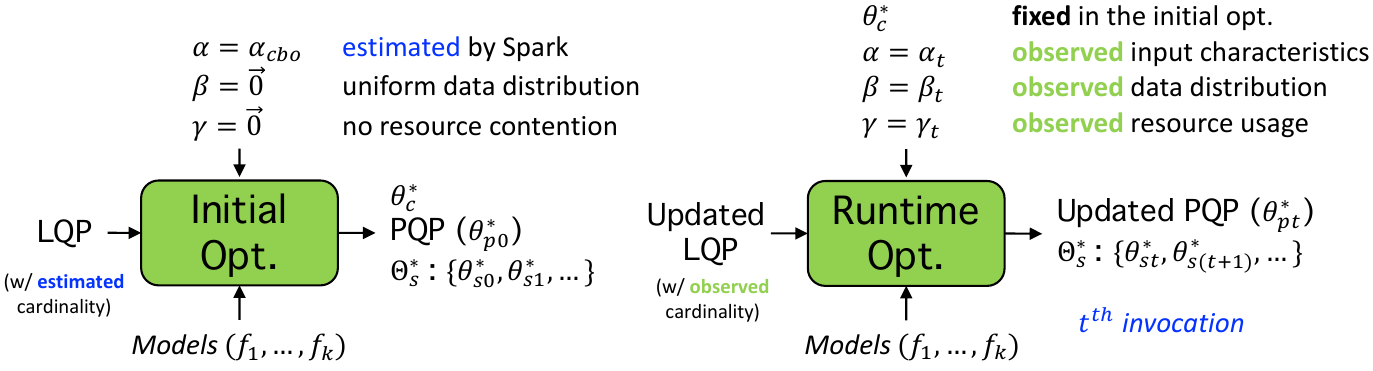}
    \caption{Compile-time Optimization and Runtime Optimization}
    \label{fig:opt-framework}
\end{figure*}

\begin{figure*}[t]
    \centering
    \includegraphics[width=.9\linewidth]{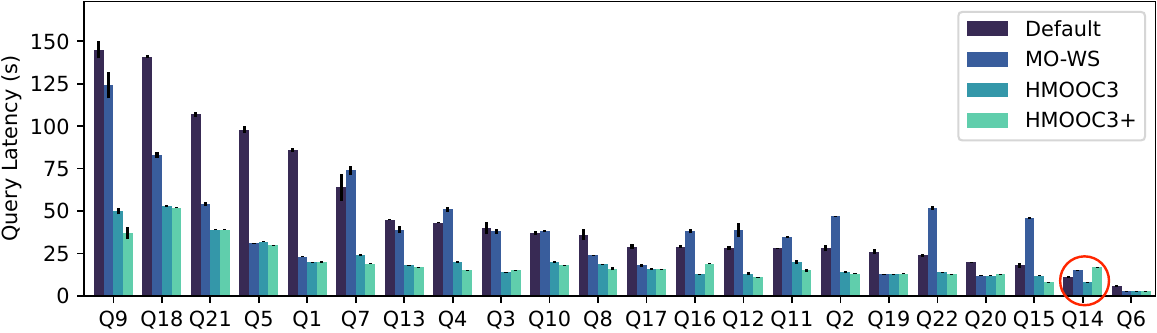}
    \caption{Per-query latency comparison with a strong speed preference in TPCH}
    \label{fig:expt9-e2e-tpch}    
\end{figure*}

\subsection{Model Comparison of Compile-time and Runtime Results}
We then look into the performance differences between 
the runtime QS and its corresponding subQ at compile time.
First, the latency performance in runtime QS is slightly inferior to its corresponding subQ at compile time. This disparity can be attributed to the runtime QS's exposure to more varied and complex query graph structures, which complicates the prediction process. 
Second, the runtime QS consistently surpasses the subQ in \rv{shuffle} prediction. This superior performance is linked to the direct correlation between \rv{shuffle} and input size; the runtime QS benefits from access to actual input sizes, thereby facilitating more precise predictions. In contrast, the subQ must base its predictions on input sizes estimated by the cost-based optimizer (CBO), which introduces more errors.

\begin{figure*}[t]
    \centering
    \begin{tabular}{lcr}
        \subfigure[\small{CDF of QS latency in TPC-H}]
        {\label{fig:lat-cdf-tpch}\includegraphics[height=3.3cm,width=.3\textwidth]{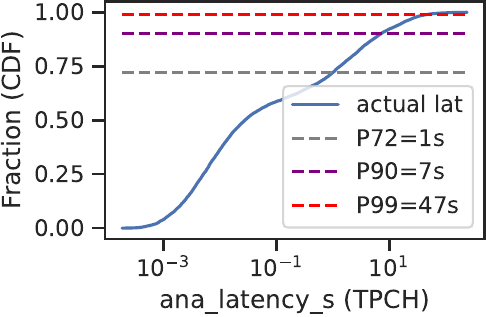}}
        \hfill
        \subfigure[\small{Lat error against skewness ratio in TPC-H}]
        {\label{fig:skew-to-lat-tpch}\includegraphics[height=3.3cm,width=.3\textwidth]{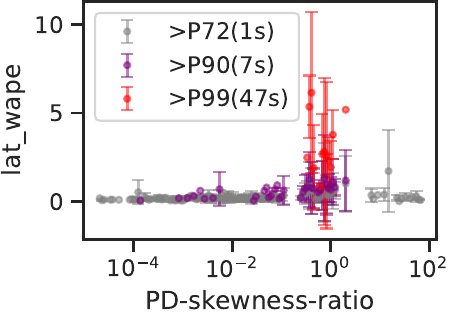}}
        \hfill
        \subfigure[\small{IO error against skewness ratio in TPC-H}]
        {\label{fig:skew-to-io-tpch}\includegraphics[height=3.3cm,width=.3\textwidth]{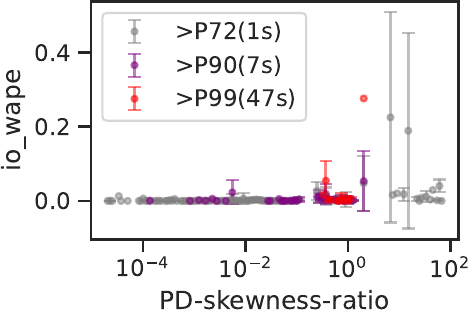}} 
        \\
        \subfigure[\small{CDF of QS latency in TPC-DS}]
        {\label{fig:lat-cdf-tpcds}\includegraphics[height=3.3cm,width=.3\textwidth]{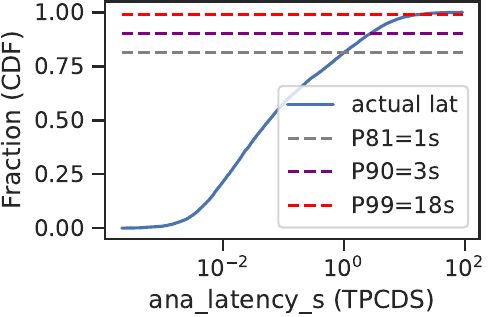}}
        \hfill
        \subfigure[\small{Lat error against skewness ratio in TPC-DS}]
        {\label{fig:skew-to-lat-tpcds}\includegraphics[height=3.3cm,width=.3\textwidth]{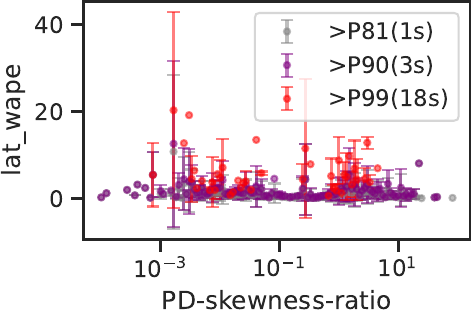}}
        \hfill
        \subfigure[\small{IO error against skewness ratio in TPC-DS}]
        {\label{fig:skew-to-io-tpcds}\includegraphics[height=3.3cm,width=.3\textwidth]{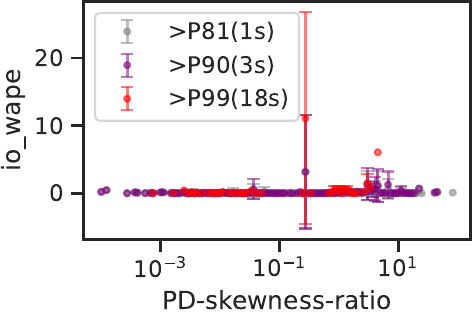}}         
    \end{tabular}
    \caption{More skewness and error analyses in TPC-H and TPC-DS}
    \label{figs:more-skew-analyses}
\end{figure*}

\rv{
\subsection{More Skewness Analyses}
\label{appendix:more-skewness}

Figure~\ref{fig:lat-cdf-tpch} and Figure~\ref{fig:lat-cdf-tpcds} show the CDF of query stage (QS) latencies in TPC-H and TPC-DS, respectively, showing the distribution of the QS latencies and the important percentiles.
The error distributions in different skewness ranges across different latency ranges are shown in Figure~\ref{fig:skew-to-lat-tpch}-\ref{fig:skew-to-io-tpch} and Figure~\ref{fig:skew-to-lat-tpcds}-\ref{fig:skew-to-io-tpcds}, respectively, indicating the skewness does not significantly affect the error distribution. 
Interestingly, we also observed that longer running QSs have higher errors in general, which is consistent with the intuition that longer running queries are more complex.
}

\begin{table*}[t] 
    \ra{0.8}
    \setlength{\belowcaptionskip}{0pt}  
	\setlength{\abovecaptionskip}{0pt} 	   
    \newrobustcmd{\BL}{\color{blue}}
    \newrobustcmd{\B}{\bfseries}
    \centering
    \addtolength{\tabcolsep}{-2.2pt}
    \footnotesize
    \caption{\small{Model performance with Graph+Regressor via transform learning}}
    \begin{tabular}{c|c|cccc|cccc|c}
    \toprule
    & \multirow{2}{*}{Target}       & \multicolumn{4}{c|}{Ana-Latency/{\BL Latency} (s)}    & \multicolumn{4}{c|}{\rv{Shuffle (MB)}}          & Xput \\
    & & \verb|WMAPE| & \verb|P50| & \verb|P90| & \verb|Corr| &\verb|WMAPE| & \verb|P50| & \verb|P90| & \verb|Corr| & K/s        \\ 
    \midrule
    \multicolumn{1}{c|}{\multirow{3}{*}{\makecell{TPC-H \\ (Trained on TPC-H)}}} 
    & subQ & 0.131& 0.029& 0.292& 0.99& 0.025& 0.006& 0.045& 1.00& 70 \\ 
    & QS & 0.149& 0.027& 0.353& 0.98& 0.002& 3e-05& 0.004& 1.00& 86 \\
    & \collap & \BL 0.164& \BL 0.060& \BL 0.337& \BL 0.95& 0.010& 8e-05& 0.002& 1.00& 146 \\ 
    \midrule

    \multicolumn{1}{c|}{\multirow{3}{*}{\rv{\makecell{TPC-H\\ (Directly applying the DS model)}}}}
    & subQ & 0.991 & 0.173 & 2.503 & 0.051 & 1.049 & 0.354 & 2.589 & 0.401 & 17 \\
    & QS & 0.907 & 0.282 & 2.418 & 0.718 & 0.627 & 0.059 & 1.771 & 0.779 & 77 \\
    & \collap & \BL 1.573 & \BL 1.156 & \BL 3.277 & \BL 0.072 & 1.262 & 0.652 & 2.998 & 0.116 & 417 \\
    \midrule

    \multicolumn{1}{c|}{\multirow{3}{*}{\rv{\makecell{TPC-H \\ (Retrain 3 hours based on DS-GTN )}}}}
    & subQ & 0.139 & 0.018 & 0.346 & 0.984 & 0.019 & 0.002 & 0.034 & 0.996 & 45 \\
    & QS & 0.143 & 0.017 & 0.346 & 0.982 & 0.003 & 0.001 & 0.006 & 1.000 & 72 \\
    & \collap & \BL 0.148 & \BL 0.046 & \BL 0.292 & \BL 0.924 & 0.019 & 0.005 & 0.028 & 0.998 & 242 \\




    \bottomrule
    \end{tabular}
    \label{tab:model-perf-xfer}
\end{table*}

\rv{
\subsection{Model Generalization}\label{appendix:more-generalization}
We further explore the generalizability of our trained model to unseen workloads. The first approach applies the latency model trained on TPC-DS directly to TPC-H, or vice versa. However, it leads to high errors due to significant differences in the running environments. The second approach transfers the graph embedding through the GTN model and retrains only the regressor for latency prediction, which is relatively fast. Our results indicate that graph embeddings trained on a workload with a broader range of query operators (e.g., TPC-DS) can be transferred effectively, resulting in a modest 0.008 WMAPE increase for the subQ model and a 0.006 and 0.016 WMAPE decrease for QS and \collap, respectively. 

Table \ref{tab:model-perf-xfer} shows the model performance on TPC-H dataset by (1) a model end-to-end trained over TPC-H, (2) a model end-to-end trained over TPC-DS, and (3) a model retrained over TPC-H based on the graph transformer network (GTN) model trained over TPC-DS. Our results show that applying the DS model directly to TPC-H leads to a significant performance drop, while retraining the model based on the DS-GTN model can achieve a better performance, indicating a good generalization capability of the GTN model.
}

\subsection{More Integration Evaluation}
\label{appendix:more-integration}

The framework of compile-time and runtime optimization is shown in Figure~\ref{fig:opt-framework}.
We show per query latency comparison with a strong speed preference in TPC-H in Figure~\ref{fig:expt9-e2e-tpch}.


\rv{\subsection{More experimental results in compile-time optimization}} \label{appendix:more-expt-sampling}

\rv{
\subsubsection{The sampling rate of sampling methods}:
There are three options of sampling methods, i.e. adaptive grid-search, Latin Hypercube Sampling (LHS) and random sampling.
\cut{
The customized grid-search supports searching from a low sampling rate (e.g. 32 $\bm{\theta_c}$ and 32 $\bm{\theta_p}$ samples) to a higher sampling rate (e.g. 256 $\bm{\theta_c}$ and 512 $\bm{\theta_p}$ samples), where the samples with a higher sampling rate is a superset of the samples with any lower sampling rate. 
The grids are set based on feature domain knowledge. If one feature is important, its minimum and maximum values are chosen as grids, otherwise either minimum or maximum values are chosen as the grid. 
By following an ordered feature list based on feature importance, the  $\bm{\theta_c}$ and $\bm{\theta_p}$ samples are generated with different sampling rate.
LHS and random sampling methods sample parameters randomly, whereas LHS ensures samples well-distributed in each dimension.
}
All sampling choices follow the same sampling rate for a fair comparison of performance. 

Our experiments in compile-time optimization fix the sampling rate for both TPC-H and TPC-DS queries by selecting the one with higher HV under the constraints of solving time (e.g. 1-2s) among different sampling rates. Finally, the sampling rate is set as 54 $\bm{\theta_c}$ and 243 $\bm{\theta_p}$ samples for both  TPC-H queries and TPC-DS queries. 
}

\begin{figure*}[t]
	\centering
	\setlength{\belowcaptionskip}{0pt}  
	\setlength{\abovecaptionskip}{0pt}  
	\begin{tabular}{ccc}
		\subfigure[\small{{Hypervolume}}]
        {\label{fig:hv_w_wo_crossover_adaptive_grid}\includegraphics[width=.24\linewidth]{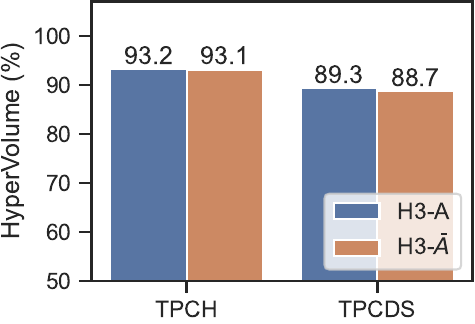}}
		&
		\subfigure[\small{{Hypervolume}}]
        {\label{fig:hv_w_wo_crossover_lhs}\includegraphics[width=.24\linewidth]{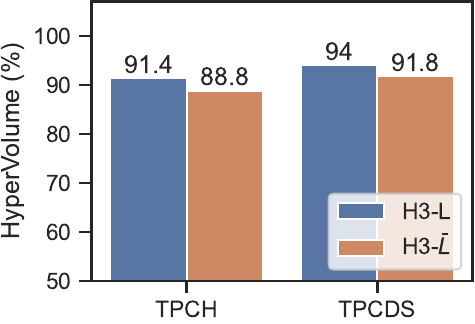}}
		&	
		\subfigure[\small{{Hypervolume}}]
        {\label{fig:hv_w_wo_crossover_random}\includegraphics[width=.24\linewidth]{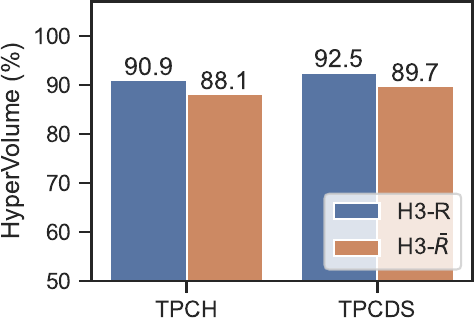}}
		\\
		\subfigure[\small{{Solving time}}]
        {\label{fig:time_w_wo_crossover_adaptive_grid}\includegraphics[width=.24\linewidth]{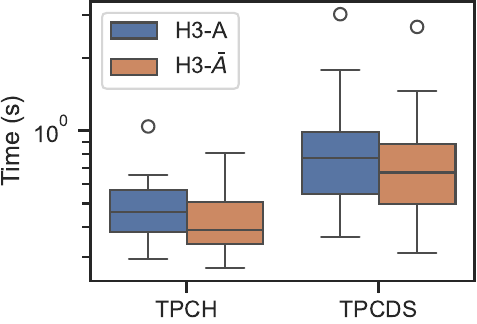}}
		&
		\subfigure[\small{{Solving time}}]
        {\label{fig:time_w_wo_crossover_lhs}\includegraphics[width=.24\linewidth]{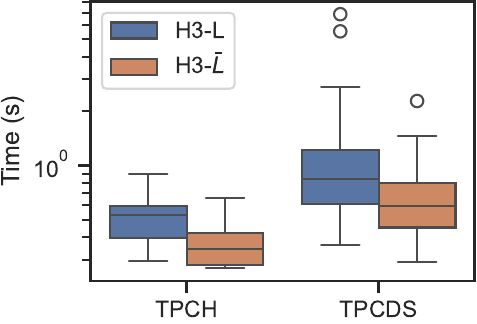}}
		&	
		\subfigure[\small{{Solving time}}]
        {\label{fig:time_w_wo_crossover_random}\includegraphics[width=.24\linewidth]{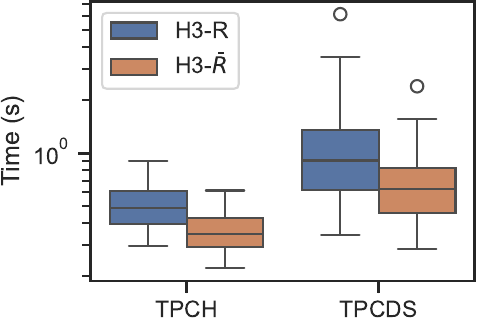}}
	\end{tabular}
	\caption{\small{{Comparison of Hypervolume and solving time with/without new $\bm{\theta_c}$ extension for three sampling methods}}}
	\label{figs:comp_hv_time_w_wo_new_theta_c_all_sampling}
\end{figure*}

\rv{
\subsubsection{Analysis on new $\bm{\theta_c}$ extension}
Figure \ref{figs:comp_hv_time_w_wo_new_theta_c_all_sampling} compares HV and solving time with and without new $\bm{\theta_c}$ extension for all sampling methods, where \texttt{H3-$\bar{A}$}, \texttt{H3-$\bar{L}$} and \texttt{H3-$\bar{R}$} denote results without new $\bm{\theta_c}$ extension from adaptive grid search, LHS and random sampling.

For adaptive grid search, it extends new $\bm{\theta_c}$ by random sampling.
Figures \ref{fig:hv_w_wo_crossover_adaptive_grid} and \ref{fig:time_w_wo_crossover_adaptive_grid} show the HV and solving time of all queries in TPC-H and TPC-DS. 
Without new $\bm{\theta_c}$ extension, HV reduces slightly compared to the HV  with new $\bm{\theta_c}$ extension ($\le$ 0.6\%).
This is because the adaptive grid search uses random sampling to extend new $\bm{\theta_c}$, which works to extend more global Pareto optimal $\bm{\theta_c}$ while with limit capability due to the high-dimensional space.
}

\begin{figure*}[t]
	\centering
	\setlength{\belowcaptionskip}{0pt}  
	\setlength{\abovecaptionskip}{0pt}  
	\begin{tabular}{lccc}
		\subfigure[\small{{TPCH-Q1 with LHS}}]
        {\label{fig:pareto_tpch_q1}\includegraphics[width=.23\linewidth]{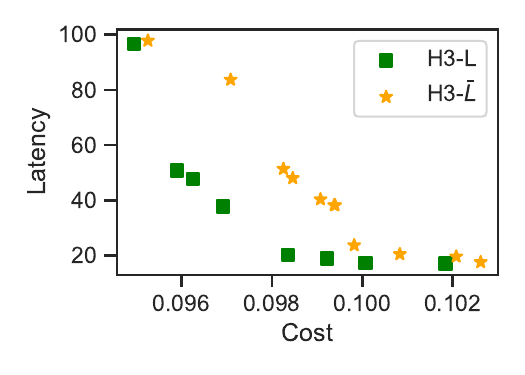}}
        &
        \subfigure[\small{{TPCDS-Q78 with LHS}}]
        {\label{fig:pareto_tpcds_q32}\includegraphics[width=.23\linewidth]{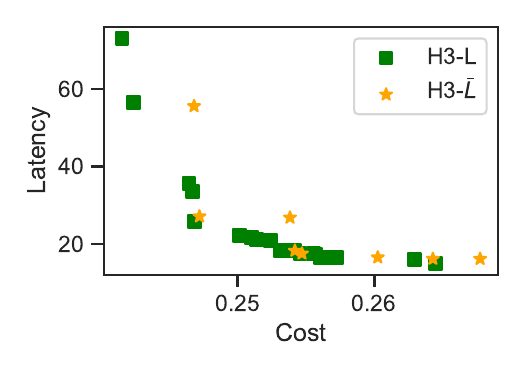}}
        &
        \subfigure[\small{{TPCH-Q1 with random sampling}}]
        {\label{fig:pareto_tpch_q1}\includegraphics[width=.23\linewidth]{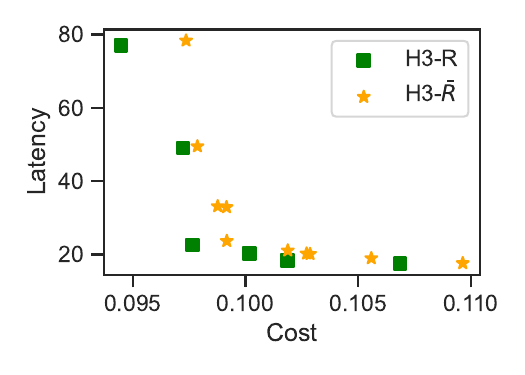}}
        &		
		\subfigure[\small{{TPCDS-Q13 with random sampling}}]
        {\label{fig:pareto_tpcds_q13}\includegraphics[width=.23\linewidth]{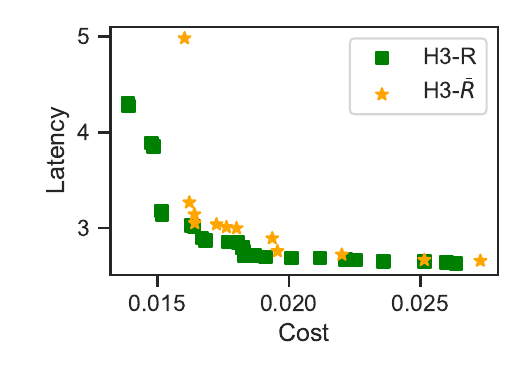}}		
	\end{tabular}
	\caption{\small{{Comparison of Pareto frontiers with/without crossover in LHS and random-sampling}}}
	\label{figs:comp_pareto_w_wo_crossover_lhs_random}
\end{figure*}

\rv{
Figures \ref{fig:hv_w_wo_crossover_lhs} and \ref{fig:time_w_wo_crossover_lhs} compare the performance of LHS with and without the new $\bm{\theta_c}$ extension by crossover. These figures display the HV and solving time for all queries in TPC-H and TPC-DS. Without crossover, HV decreases by up to 2.6\% compared to HV with crossover. This is because the boundary solutions (i.e., solutions with minimum latency or minimum cost) significantly impact HV by setting the boundary of the dominated space of the Pareto front, while crossover helps generate better global optimal solutions in the middle region of the Pareto front, as shown in Figure \ref{figs:comp_pareto_w_wo_crossover_lhs_random}.
The solving time without crossover decreases as it reduces the procedures following the new $\bm{\theta_c}$ extension. Similar observations are seen in Figures \ref{fig:hv_w_wo_crossover_random} and \ref{fig:time_w_wo_crossover_random} for extending new $\bm{\theta_c}$ in random sampling.

} 

\rv{
Figure \ref{figs:comp_pareto_w_wo_crossover_lhs_random} shows the Pareto frontiers of example queries based on the maximum HV differences between results with and without crossover using LHS and random sampling. Green points represent solutions with the crossover operation, while orange points represent solutions without crossover. For both LHS and random sampling as the initial sampling methods, solutions with crossover yield better results in the middle region of the Pareto front.
This improvement is due to the crossover location set in the implementation. In our experiment, the crossover location is set to 3 based on hyperparameter tuning, which splits $\bm{\theta_c}$ into two sets: resource and non-resource parameters. The Cartesian product of these two sets does not generate new values for total resources. Consequently, the crossover explores better solutions in the middle region rather than better boundary solutions with significantly lower latency or cost.
}

\begin{figure*}[t] 
\centering
\includegraphics[height=4.5cm,width=8cm]{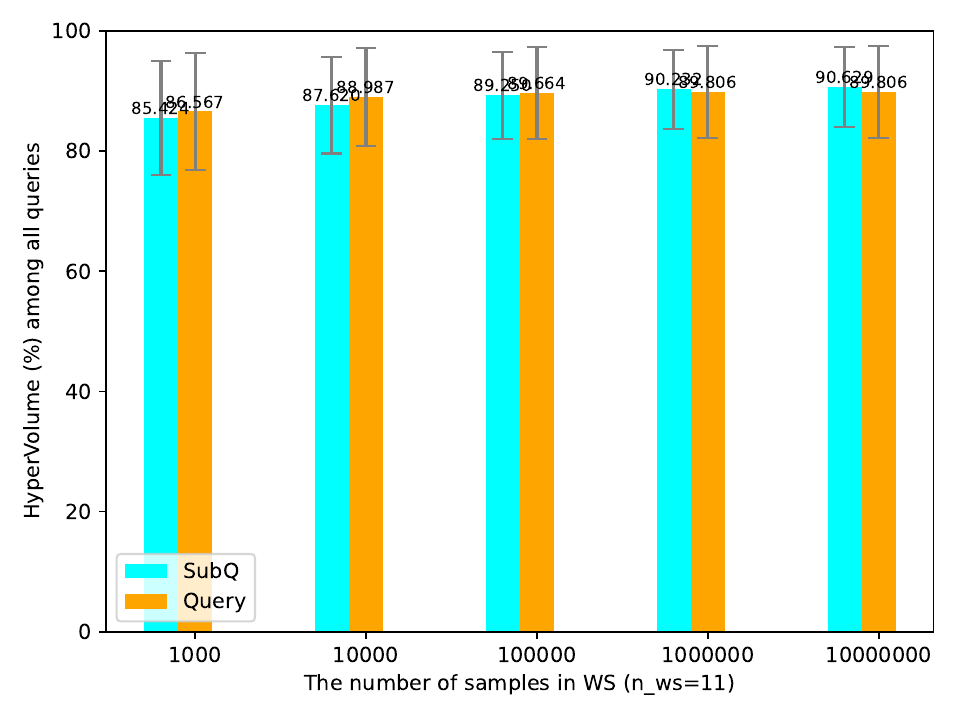} 
\caption{Comparison of query-control and finer-control with smaller searching space} 
\label{fig:smaller_searching}
\end{figure*}

\subsubsection{Analysis on Query-level control}:
It's noteworthy that query-control cannot achieve a higher upper bound than finer-control. To verify this, we implemented a smaller search space (each parameter having only 2 values) for WS to fully explore query-control, where WS performs the best among all baselines for both TPC-H and TPC-DS. Figure \ref{fig:smaller_searching} displays the hypervolumes (HVs) of WS under different numbers of samples, with blue and orange bars representing the HVs of finer-control and query-control, respectively. It is observed that as the number of samples increases, the HV of query-control stops increasing at 1M samples (89.8\%), while the HV of finer-control continues to improve (90.6\%). This demonstrates that finer-control has the potential to achieve better solutions than query-control, illustrating the necessity of finer-control in our problem.



\begin{figure*}[t] 
	\centering
	\includegraphics[height=3.5cm,width=6cm]{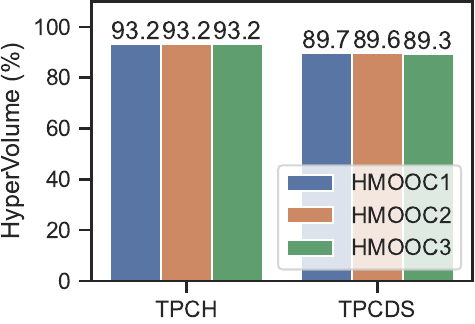} 
	\caption{{Average Hypervolume with different DAG aggregation methods}} 
	\label{fig:new_grids_hv_dag_opt}
\end{figure*}

\rv{
\subsubsection{More results on DAG aggregation methods}: 
Figure \ref{fig:new_grids_hv_dag_opt} compares the HVs of three DAG aggregation methods with adaptive grid-search. The HVs of three methods are close with lower than 0.1\% differences.
}

\begin{figure*}[t]
	\centering
	\setlength{\belowcaptionskip}{0pt}  
	\setlength{\abovecaptionskip}{0pt}  
	\begin{tabular}{lccc}
		\subfigure[\small{\rv{Average Hypervolume of our algorithm and baselines in TPC-H}}]
		{\label{fig:hv_tpch}\includegraphics[height=2.5cm,width=0.23\textwidth]{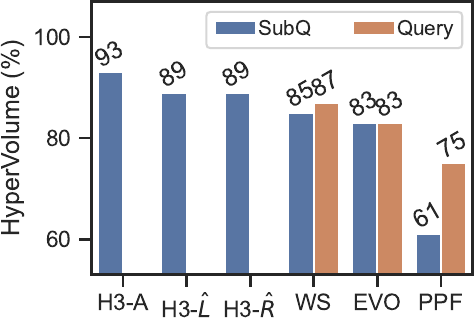}}
		&	
		\subfigure[\small{\rv{Solving time of our algorithm and baselines in TPC-H}}]
		{\label{fig:time_tpch}\includegraphics[height=2.5cm,width=0.23\textwidth]{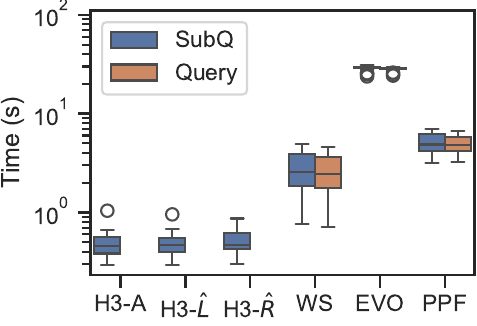}}
		&
		\subfigure[\small{\rv{Average Hypervolume of our algorithm and baselines in TPC-DS}}]
		{\label{fig:hv_tpcds}\includegraphics[height=2.5cm,width=0.23\textwidth]{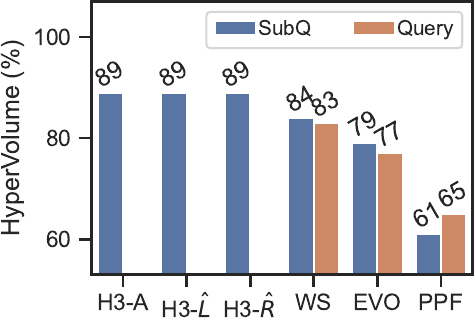}}
		&
		\subfigure[\small{\rv{Solving time of our algorithm and baselines in TPC-DS}}]
		{\label{fig:time_tpcds}\includegraphics[height=2.5cm,width=0.23\textwidth]{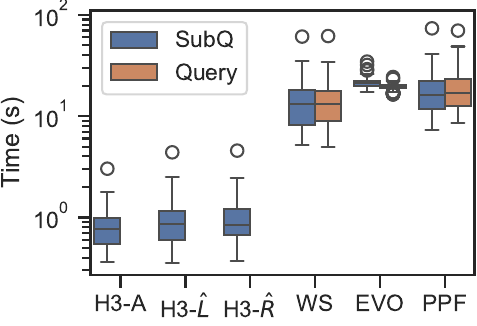}}
	\end{tabular}
	\caption{\small{Analytical performance of our algorithm, compared to the state-of-the-art (SOTA) methods with query-level tuning}}
	\label{fig:expr-in-analyses_with_query_level_tuning}
\end{figure*}

\rv{
\subsubsection{More results on Query-level tuning}
Figure \ref{fig:expr-in-analyses_with_query_level_tuning} shows results of query-level tuning, where TPC-H queries take over an average of at least 2.7s and much lower HV (at most 87\%) than \texttt{HMOOC3} (93\%).
All of TPCDS queries with query-level tuning lose to \texttt{HMOOC3} in HV (at most 83\% v.s. 89\%) and in solving time (the average exceeding 14s v.s. 0.8s). 
}

%% file: appendix/app-related-work.tex
\section{More Related Work}\label{appendix:related-work}

{\bf Relationship with other distributed algorithms for system optimization.}
Li et al.~\cite{LiMRSY19} tunes parameters that determine the computation graph, as well as the executor numbers and degree of parallelism to efficiently minimize the execution time by cutting the search space with a runtime environment independent cost model. However, its parameters are not adaptable during the execution, and its search strategy is hard to be adapted to the MOO setting.
Having said that, Spark AQE is able to take those distributed algorithms as separate query optimization rules, and our framework can then treat the hyperparameters of those algorithms as SQL parameters and optimize those algorithms by tuning their hyperparameters in compile-time/runtime optimization.